\documentclass[10pt]{article}

\usepackage{amsmath}
\usepackage{amsthm}
\usepackage{amssymb}
\usepackage{amscd}
\usepackage{epsf}
\usepackage{rotating}
\usepackage{wrapfig}

\chardef\bslash=`\\ 

\makeatletter
\def\verbatim{\interlinepenalty\@M \@verbatim
  \leftskip\@totalleftmargin\advance\leftskip2pc
  \frenchspacing\@vobeyspaces \@xverbatim}
\makeatother
\theoremstyle{plain}

\newtheorem{defn}{Definition}

\newtheorem{lem}{Lemma}
\newtheorem{prop}{Proposition}

\newtheorem{exercise}{Exercise}

\newtheorem{summary}{Summary}

\theoremstyle{remark}
\newtheorem{rem}{Remark}

\numberwithin{equation}{section}

\def\1I{\relax{\rm 1\kern-.25em \rm l}} 

\newcommand{\unity}{\1I}

\newcommand{\spin}[3]{\omega_{\raisebox{-1.5pt}{\scriptsize $#1$}}{}_{\hat{#2}}{}^{\hat{#3}}}

\newcommand{\talpha}{{\tilde\alpha}}
\newcommand{\tbeta}{{\tilde\beta}}
\newcommand{\tgamma}{{\tilde\gamma}}
\newcommand{\tdelta}{{\tilde\delta}}

\def\Rahmen#1#2#3 {
   \vbox{\hrule height#2
         \hbox{
               \vrule width#2
               \hskip#1
               \vbox{
                     \vskip#1{}
                     \hbox{#3}
                     \vskip#1
                    }%
               \hskip#1
               \vrule width#2
              }
         \hrule height#2
        }}
\def\href#1#2{#2}

\begin{document}

\thispagestyle{empty}
\vspace{2truecm}
\centerline{\bf \Large Basics of M-Theory}
\vspace{2.5truecm}

\newcounter{Institut}
\centerline{\bf Andr\'e
  Miemiec\refstepcounter{Institut}\label{Inst_Andre}{}$^{*_{\theInstitut}}$
  and Igor Schnakenburg\refstepcounter{Institut}\label{Inst_Igor}{}$^{*_{\theInstitut}}$}

\vskip1cm
\parbox{.9\textwidth}{
\parbox{.4\textwidth}{
\centerline{$^{*_{\ref{Inst_Andre}}}$ Institut f\"ur Physik~~}
\centerline{                           Humboldt Universit\"at}
\centerline{                           D-12489 Berlin, Germany}
\centerline{                           Newtonstr. 15}
\centerline{                           miemiec@physik.hu-berlin.de}             
              
}\hfill
\parbox{.4\textwidth}{\centerline{$^{*_{\ref{Inst_Igor}}}$~ Department
of Mathematics}
\centerline{                           King's College London}
\centerline{                           Strand WC2R 2LS, London}
\centerline{                           UK}
\centerline{                           schnake@mth.kcl.ac.uk}
}
}

\vspace{2.0truecm}
\begin{abstract}
\noindent This is a review article of eleven dimensional supergravity in which
we present all necessary calculations, namely the Noether procedure, the
equations of motion (without neglecting the fermions), the Killing spinor
equation, as well as some simple and less simple supersymmetric solutions to
this theory. All calculations are printed in much detail and with explicit
comments as to how they were done. Also contained is a simple approach to
Clifford algebras to prepare the grounds for the harder calculations in spin
space and Fierz identities.   

\end{abstract}
\bigskip \bigskip
\newpage

\tableofcontents

\newpage
\vfill\eject
\section{Preface}
\label{Sec_Preface}

This is a provisional version of lecture notes on M-theory. The etymology  
of the name ``M-Theory'' is explained in \cite{Duff:2004nh} and traced
back to (M)embranes. Nowadays the ``M'' is thought to refer  
to the word M(other) due to the pivotal role it seems to claim in the
unification of string theories. Strictly speaking, M-Theory is as yet
undiscovered. All that is known is its low energy limit, i.e.\ eleven
dimensional supergravity (11d SUGRA).\\ 
       
The construction of eleven dimensional supergravity was performed in 
1978 \cite{Cremmer:1978km}. In this seminal paper the Lagrangian density, the  
equations of motion and the transformation properties of the fields 
with respect to the supercharges were obtained by performing the
Noether procedure. Later the doubled field approach was used to 
rederive these results from a suitably chosen supergroup by 
cohomologial techniques \cite{Auria:1982nx}.  
A new investigation of the formal structure of eleven dimensional 
supergravity was pursued in two papers twenty years 
later \cite{Cremmer:1998ct,Cremmer:1998px}.
By combining two different techniques, i.e.\ the doubled field
approach and the method of nonlinear realisations, it
became possible to understand the equations of motion of 
the bosonic p-form fields as generalised self duality conditions. The field
strengths can be obtained from algebraic considerations. This idea can
be extended to also include gravity thereby identifying a huge
symmetry algebra as was first laid out in
\cite{West:2000ga,West:2001as}. Apparently, this approach can be
extended to more supergravity theories \cite{Schnakenburg:2001ya} 
and possibly be
used for an a posteriori understanding of some simple solutions \cite{Schnakenburg:2003qw}. Supposedly, these new symmetries can be used to further
our insight into M-theory. However, those recent developments shall not be part
of this lecture. \\ 

Much is known about eleven dimensional supergravity, but it seems hard to
find most of the calculations in a {\it single} article using the {\it same}
conventions throughout. Hence, it is our intension to give a readable account
of all typical calculations concerning eleven dimensional supergravity in a
single work. A detailed discussion of the equations of motion (section 3), the 
derivation of the Lagrangian, the supervariations (section 4) and 
the Killing spinor equation (section 5) is presented. A similarly detailed
discussion of the most typical supersymmetric solutions is given as well
(section 5, 6, and 7). Hereby, we put weight on going through the calculations
in full length. Some easy parts of these calculations are given as
homeworks. The reader interested in more general questions on $p$-brane
solutions is referred to \cite{Stelle:1998xg,Duff:1995an}.\\ 

The claimed pivotal role of M-theory is due to the established
relations to the ten dimensional supergravity theories that are low energy
limits of superstring theory. The advantage of the eleven dimensional theory
over its ten dimensional siblings is its uniqueness and its simplicity - it
contains only three different particles: the graviton, the gravitino, and a
gauge potential. 
A particularly simple relation to ten dimensional IIA supergravity
is given by dimensional reduction on a circle. This reduction is also
explicitly treated in this paper (section 8). In general, Kaluza Klein
reductions are an involved subject since one has to decide whether the lower
dimensional theory is consistent. This is always the case for circle
compactifications. The more general case is discussed in \cite{Cvetic:2000dm}
or the very understandable article by Chris Pope \cite{Pope}.\\ 

As for the prerequisites, we merely assume the readers familiarity with   
the tensor calculus of general relativity. Good and legible 
introductions into general relativity are \cite{Carol,Blau}. In addition, 
we assume some general knowledge of the basic concepts underlying 
supersymmetry. A valuable introduction to supersymmetry is \cite{Wipf}.
Good sources of information about supergravity in four dimensions 
are \cite{Nieuwenhuizen,Theiss}. There are more reviews on higher dimensional 
supergravities. A very useful one is \cite{Duff:1999rk}.\\


When discussing supersymmetric solutions to supergravity theories 
one will immediately come across the topics of calibrations
\cite{HL,Gauntlett:1998vk,Townsend:1999hi} and 
$G$-structures \cite{Gauntlett:2002fz,Gauntlett:2003wb}. 
Here these topics are omitted entirely but they will be treated in 
a later article \cite{Miemiec2} by one of us.\\

Since we do not wish to omit fermions (as is typical in SUGRA
calculations) more than necessary, we start with a simple introduction
into Clifford algebras (section 2) - the essential tool for dealing with
fermions and their transformation properties\footnote
%
{
  An excellent review of Clifford algebras in general
  is \cite{Figueroa-O'Farrill}.
}. 
%
This makes our review essentially self contained. Again our 
presentation of Clifford algebras is both fairly explicit 
and intuitive.\\[1cm]

\centerline{We would appreciate any comments, suggestions, etc.}

\vfill\eject
\section{Clifford Algebras}\label{Sec_Cliff_Algebra}
\label{Sec_Clifford}

We will derive Clifford algebras by looking at representations of typical
rotation groups arising in mathematical physics.

\subsection{An explicit example: ${\rm Cliff}(\mathbb{R}^3)$}
It is clear that the usual scalar product of ${\mathbb{R}}^3$  with
metric tensor $\eta_{ab}\,=\,{\rm diag}(1,1,1)$ possesses the
invariance group $SO(3)$. This is the group of rotation and hence
preserves the length of a vector. From a formal point of view one can
define $SO(3)$ as the group of transformations preserving the
symmetric bilinear form $B(\circ,\circ)$, defined by\\  
\begin{eqnarray}\label{BiF_B}
       B(\vec{x},\vec{y}) &=& \eta_{ab}\,x^a\,y^b\,.   
\end{eqnarray}\\
Assuming $A$ to be a rotation matrix, the preservation of the bilinear form
can be written as\\
\begin{equation}
    (A\vec{x},\, A\vec{y}) = (\vec{x},\,\vec{y})\,.
\end{equation}\\
The left hand side can also be rewritten as\\
\begin{equation}
    (A\vec{x},\, A\vec{y}) = (\vec{x},\,A^T\,A\,\vec{y})\,,
\end{equation}\\
and it therefore follows simply that\\
\begin{equation}
   A^T\,A \equiv 1 \quad \rightarrow \quad A^T = A^{-1}\,,
\end{equation}\\
which is the condition that $A$ has to be an orthogonal matrix. Using $\det
(A^T) = \det (A)$, the condition on the determinant of $A$ can be found to be
$\det (A)^2 = 1$ which leaves $\det (A) = \pm 1$. Choosing the positive sign
entails preservation of orientation which is typically the case for
rotations. Allowing both signs, the more general group is called $O(3)$ rather
than $SO(3)$. \\

\noindent
An obvious question is whether there are more general bilinear forms
possessing $SO(3)$ as part of their invariance group. To find such a
generalised form we introduce a formal product $\circ$ and define a new
symmetric bilinear form just by\\ 
\begin{eqnarray}
       V(\vec{x},\vec{y}) &=& \sum\limits_{i\leq
       j}x^i\,y^j\,\left(e_i\circ e_j+e_j\circ e_i\right).   
\end{eqnarray}\\
This obviously contains $B$\\
\begin{eqnarray}\label{sym bilinear form}
       V(\vec{x},\vec{y}) &=& 2\,\underbrace{\sum\limits_{i}x^i\,y^i\,e_i\circ
       e_i}_{(1)}\,
     +\,\underbrace{\sum\limits_{i<j}x^i\,y^j\,\left(e_i\circ
         e_j+e_j\circ e_i\right)}_{(2)}\,,   
\end{eqnarray}\\
the first part can be identified with eq.~(\ref{BiF_B}) by letting
$e_i\circ e_i\,=\,B(e_i,e_i)\cdot \unity$. The unit matrix has been
added to give us more freedom: in particular using the $\circ$ product
of two basis vectors does not result in a scalar. With this product
the basis vectors form a closing algebra (so any product is
well-defined). Any algebra by definition contains a unit element. In
the above product we have hence defined that the $\circ$ product of a
basis vector with itself be proportional to unity and the factor of
proportionality is a scalar, namely the scalar product or the length
of the relevant basis vector.

\noindent
In order to ensure that $SO(3)$ is still part of the invariance group
of  $V(\vec{x},\vec{y})$ the second part must vanish. This is achieved
by\\   
\begin{eqnarray}
         e_i\circ e_j+e_j\circ e_i &=& 0\,.
\end{eqnarray}\\
Both conditions can be summarised in the defining formula of a
Clifford algebra\\
\begin{eqnarray}\label{Clifford} 
         e_i\circ e_j+e_j\circ e_i &=& 2\cdot B(e_i,e_j)\cdot\unity\,.
\end{eqnarray}\\
Let us compare the different products. The usual scalar product of
orthogonal basis vectors is of course\\
\begin{equation}\label{scalar product flat metric}
    e_i \cdot e_j = \eta_{ij}\,,
\end{equation}\\
where $\eta$ contains -1 or +1 on its diagonal depending on the
signature of the underlying $\mathbb{R}^{p,q}$; all other entries are
zero. 

\noindent
In relation (\ref{Clifford}) the basis vectors are considered as
elements of an algebra - the Clifford algebra. The symmetric
combination of the $\circ$ product of two basis vectors vanishes if
they are different. Taking the square of a basis element results in an
algebra element proportional to unit element.\\

\noindent
It is easy to see that the basis elements of the Clifford algebra over
$\mathbb{R}^3$ can be taken to be the Pauli matrices, viz.\\
\begin{align}\label{Pauli}
  e_1 \mapsto \tau_1 &= \left(\begin{array}{cc}
                     0 & 1\\
                     1 & 0\\
                   \end{array}\right) &
  e_2 \mapsto \tau_2 &= \left(\begin{array}{cc}
                     0 & -i\\
                     i & 0\\
                   \end{array}\right) &
  e_3 \mapsto \tau_3 &= \left(\begin{array}{cc}
                     1 & 0\\
                     0 & -1\\
                   \end{array}\right)~. &
\end{align}\\
The full Clifford algebra is generated by all
linear combinations and powers of those three elements. The square of all
Pauli matrices is the unit matrix which therefore naturally occurs
in the Clifford algebra. Hence there are four generators for this Clifford
algebra: $\dim({\rm Cliff}(\mathbb{R}^3))= 4$. In
particular, the product of two basis elements in the Clifford algebra is
proportional to the remaining basis vector, e.g.\\ 
\begin{equation}\label{Clifford relation of basis elements}
   e_1 \circ e_2 = i e_3\, \quad{({\rm and\,\, not\,\, 0\,\, as\,\, in\,\,
       the\,\, scalar\,\, product})}\,. 
\end{equation}\\

\centerline{\bf Is the invariance group of $V(\circ,\circ)$ larger than the
  one of $B(\circ,\circ)$?} \vspace{3ex}

\noindent
As a starting point for answering this questions serves the assumption\\
\begin{equation}\label{metric = bilinear}
   V(x,\, y) = B(\vec{x},\,\vec{y})\cdot \unity\,.
\end{equation}\\
Some comments are in place here. On the left hand side of the last equation,
there are no arrows above the vectors $x,\,y$ since $x$ and $y$ are
considered as elements of the Clifford algebra. Usually (as on the right hand side) the vector is
expanded into the basis vectors by\\ 
\begin{equation}\label{vector expansion basis}
    \vec{x} = x^1\, e_1 + x^2\, e_2 + x^3\,e_3\,,
\end{equation}\\
where basis vectors are multiplied by a scalar product (\ref{scalar product
  flat metric}). On the left hand side of the fore last equation, however, the
expansion is\\ 
\begin{equation}\label{vector expansion Clifford}
    x = x^1\, \tau_1 + x^2\, \tau_2 + x^3\,\tau_3\,.
\end{equation}\\
Hence the ``vector'' $x$ is expanded into the Clifford algebra and therefore
is a two-by-two matrix now. In the bilinear form $V$ the product of
two elements using $\circ$ is obtained by usual matrix
multiplication. Since $V$ is symmetric,
one has to take the symmetrised matrix product as specified in (\ref{sym
  bilinear form}) to reproduce (\ref{metric = bilinear}).\\

\noindent
Since both bilinear forms are supposed to be invariant under rotations,
equation (\ref{metric = bilinear}) must also hold in the form\\
\begin{equation}
   V(A\,x,\, A\,y) = B(A\,\vec{x},\,A\,\vec{y})\cdot \unity =
   B(\vec{x},\,\vec{y})\cdot \unity = V(x,\,y)\,. 
\end{equation}\\
Again, it is easy to understand how the rotation matrix $A\in SO(3)$
acts on the vectors $\vec{x}$ or $\vec{y}$ in this equation: simply in the fundamental representation of $SO(3)$ in
terms of three-by-three matrices. The leftmost side is more subtle,
since here an action on two-by-two matrices $x$ and $y$ is required
which necessarily has to leave the invariance property intact. Since 
it is known that the {\bf adjoint} representation of $SU(2)$ is three
dimensional it occurs natural to try $g\, \tau_i\, g^{-1}= A\, e_i$,
where $g\in SU(2)$ is a two-by-two matrix and can therefore easily
'rotate' the Pauli matrices as does the $SO(3)$ matrix $A$ with the
basis vectors $e_i$ of $\mathbb{R}^3$.

\noindent
More explicitly: it is possible to expand any vector into linear combinations
of basis vectors of $\mathbb{R}^3$ (\ref{vector expansion basis}). Under
$SO(3)$ rotations these basis vectors transform in the obvious way leaving the
bilinear form invariant $B(e_i,\, e_j)= B(A\,e_i,\,A\,e_j)$. Using the other
symmetric bilinear form in which the vector takes the form of a two-by-two
matrix the same invariance can be obtained by using the fact that the adjoint
representation of $SU(2)$ is three dimensional and acts on two-by-two
matrices. If the $SO(3)$ rotation matrix $A$ is given, the corresponding
element $g \in SU(2)$ can be calculated via\\
\begin{equation}\label{adjoint versus fundamental}
    g\, \tau_i\, g^{-1} = A \,e_i\,,
\end{equation}\\
meaning that the action on the respective basis elements coincide. The last
equation looks simple, but has to be understood algebraically: on the left
hand side there is a product of three different matrices while on the right
hand side there is a rotation matrix acting on a vector. The point is that the
$\tau_i$ despite of having a representation as two-by-two matrices from an
algebraic point of view are basis vectors in the same way as the $e_i$. In
this way, the above correspondence is precise.\\

\noindent
By presenting the whole calculation, we shall see that the invariance group of
the symmetric bilinear form $V(x,\,y)$ is twice as big as the one of
$B(\vec{x},\, \vec{y})$. We start by computing the left hand side of the last
equation. A general group element of $SU(2)$ can be expressed by $g= \exp (a^i
\, t_i) $, where the $a^i$ are the relevant coefficients and $t_i =
\frac{i}{2}\tau_i$. The factor of $\frac{i}{2}$ is conventional and will be
commented on in due course. Obviously it does not in principal spoil the
algebraic relation fulfilled by the Pauli matrices $\tau_i$. An obvious change
is that in comparison to equation (\ref{Clifford relation of basis elements})
the imaginary unit disappears: ${\mathfrak{t}}_1\,{\mathfrak{t}}_2=
-{\mathfrak{t}}_3$. The generators $t_i$ of the Lie algebra $\mathfrak{su}(2)$
are then\\    
\begin{eqnarray*}
   {\mathfrak{t}}_1 
       ~=~  \left(\, 
                    \begin{array}{cc}
                      0 & \frac{i}{2} \\ [2ex]
                      \frac {i}{2} & 0
                    \end{array}\,
            \right)\hspace{5ex} 
   {\mathfrak{t}}_2
       ~=~  \left(\,
                   \begin{array}{cc}
                      0 & \frac{1}{2}  \\ [2ex]
                      -\frac{1}{2}  & 0
                   \end{array}\,
            \right)\hspace{5ex} 
   {\mathfrak{t}}_3
       ~=~  \left(\, 
                   \begin{array}{cc}
                      \frac{i}{2} & 0 \\ [2ex] 
                       0 & -\frac{i}{2}
                   \end{array}\,
            \right) \,.
\end{eqnarray*}\\
The three one parameter subgroups generated by these 
${\mathfrak{t}}_1\ldots {\mathfrak{t}}_3$ are\\
\begin{eqnarray}\label{one parameter groups of SU(2)}
   g_1 &=& \left(\,
                   \begin{array}{cc}
                       \cos(\frac{\theta_1}{2}) & i\,\sin(\frac{\theta_1}{2})\\
                    i\,\sin(\frac{\theta_1}{2}) & \cos(\frac{\theta_1}{2})
                   \end{array}\,
           \right)\\[2ex]
   g_2 &=& \left(\,\begin{array}{cc}
                       \cos(\frac{\theta_2}{2}) & \sin(\frac{\theta_2}{2})\\
                      -\sin(\frac{\theta_2}{2}) & \cos(\frac{\theta_2}{2})
                   \end{array}\,
           \right)\\[2ex]
   g_3 &=& \left(\,\begin{array}{cc}
                      \exp(i\frac{\theta_3}{2}) & 0 \\ 
                       0 & \exp(-i\frac{\theta_3}{2})
                   \end{array}\,
           \right)\,.
\end{eqnarray}\\
It is a straightforward task to calculate the adjoint action of these elements
on the basis elements $\mathfrak{t}_i$\\
\begin{eqnarray}
   \pi(g_1)\;{\mathfrak{t}}_1 &=& \hspace{28.75ex}
       ~=~ \phantom{-\,}{\mathfrak{t}}_1 \nonumber \\[2ex] 
   \pi(g_1)\;{\mathfrak{t}}_2
       &=& \left(\begin{array}{rr}
            -\,\frac{i}{2}\,\sin\theta_1 & \phantom{-\,}
               \frac{1}{2}\,\cos\theta_1 \\ 
            -\,\frac{1}{2}\,\cos\theta_1 & \frac{i}{2}\,\sin\theta_1 
           \end{array}\right)    
       ~=~  \cos\theta_1 \cdot {\mathfrak{t}}_2 ~-~
            \sin\theta_1 \cdot {\mathfrak{t}}_3\label{g1 on ti}\\[2ex] 
    \pi(g_1)\;{\mathfrak{t}}_3
       &=& \left(\begin{array}{rr}
            \frac{i}{2}\,\cos\theta_1 & \frac{1}{2}\,\sin\theta_1 \\ 
              -\,\frac{1}{2}\,\sin\theta_1 &  -\,\frac{i}{2}\,\cos\theta_1
           \end{array}\right)\,    
       ~=~ \sin\theta_1 \cdot {\mathfrak{t}}_2 ~+~ 
               \cos\theta_1 \cdot {\mathfrak{t}}_3\nonumber  
\end{eqnarray}\\
Similarly for $g_2$ and $g_3$\\
\begin{eqnarray}
~~~\pi(g_2)\;{\mathfrak{t}}_1 
          &=& \left(\begin{array}{rr}
                 \phantom{-\,}\frac{i}{2}\,\sin\theta_2 & 
                 \frac{i}{2}\,\cos\theta_2  \\
                 \frac{i}{2}\,\cos\theta_2 & 
                 -\,\frac{i}{2}\,\sin\theta_2
              \end{array}\right)\hspace{1ex}
          ~=~ \phantom{-\,} \cos\theta_2\cdot {\mathfrak{t}}_1
            ~+~\sin\theta_2\cdot {\mathfrak{t}}_3\nonumber\\[1ex] 
\pi(g_2)\;{\mathfrak{t}}_2 &=& \hspace{28.75ex}\, 
          ~=~ \phantom{-\,}{\mathfrak{t}}_2\label{g2 on ti}\\[1ex] 
\pi(g_2)\;{\mathfrak{t}}_3 
          &=& \left(\begin{array}{rr}
                 \frac{i}{2}\,\cos\theta_2 & - \frac{i}{2}\,\sin\theta_2\\ 
                 - \frac{i}{2}\,\sin\theta_2 & - \frac{i}{2}\,\cos\theta_2
              \end{array}\right)\hspace{1.5ex}
          ~=~ -\,\sin\theta_2\cdot {\mathfrak{t}}_1
           ~+~\cos\theta_2\cdot {\mathfrak{t}}_3\nonumber
\end{eqnarray}\\
\begin{eqnarray}
~~~\pi(g_3)\;{\mathfrak{t}}_1 
          &=& \left(\,\begin{array}{cc}
                  0 & \frac{i}{2}\,e^{-i\theta_3}\nonumber \\
                  \frac{i}{2}\,e^{i\theta_3} & 0
              \end{array}\right)\hspace{8ex}
          ~=~ \phantom{-\,}
              \cos\theta_3\cdot {\mathfrak{t}}_1 ~+~
              \sin\theta_3\cdot {\mathfrak{t}}_2\nonumber\\[2ex] 
\pi(g_3)\;{\mathfrak{t}}_2 
          &=& \left(\,\begin{array}{rr}
                 0 & \frac{1}{2}\,e^{-i\theta_3}\\
                -\frac{1}{2}\,e^{i\theta_3} & 0 
              \end{array}\,\right)\hspace{5.5ex}\,
          ~=~ -\,\sin\theta_3\cdot {\mathfrak{t}}_1 ~+~ 
                 \cos\theta_3\cdot {\mathfrak{t}}_2\label{g3 on ti} \\[2ex] 
\pi(g_3)\;{\mathfrak{t}}_3 &=& \hspace{29.4ex} 
          ~=~ \phantom{-\,}{\mathfrak{t}}_3\,.\nonumber
\end{eqnarray}\\
The first set of equation (\ref{g1 on ti}) can be rewritten in matrix form in
the following way\\
\begin{eqnarray}
   \pi(g_1)\;\vec{{\mathfrak{t}}} 
          =  \left(\,\begin{array}{ccc}
                  1 & 0 & 0\\
                  0 & \cos\theta_1 & -\sin\theta_1\\
                  0 & \sin\theta_1 & \cos\theta_1
              \end{array}\right)\cdot 
              \left(\,\begin{array}{c}
                  t_1\\ t_2\\ t_3 \end{array}\right) =
                  \left(\,\begin{array}{ccc} 
                   1 & 0 & 0\\
                  0 & \cos\theta_1 & -\sin\theta_1\\
                  0 & \sin\theta_1 & \cos\theta_1
              \end{array}\right)\cdot 
              \left(\,\begin{array}{c}
                  e_1\\ e_2\\ e_3 \end{array}\right) \,,
\end{eqnarray}\\
which is the explicit form of formula (\ref{adjoint versus fundamental}) and
hence provides the relation between the $SU(2)$ and $SO(3)$ representation in
their action on the respective basis elements. The relevant formulae for the 
other basic rotations are\\
\begin{equation}
   \pi(g_2)\;\vec{{\mathfrak{t}}} 
          = \left(\,\begin{array}{ccc}
                  \cos\theta_2 & 0 & \sin\theta_2\\
                  0 & 1 & 0 \\
                  -\sin\theta_2 & 0 & \cos\theta_2
              \end{array}\right)\cdot 
              \left(\,\begin{array}{c}
                  e_1\\ e_2\\ e_3 \end{array}\right)
\end{equation}\\
\begin{equation}
   \pi(g_3)\;\vec{{\mathfrak{t}}} 
          = \left(\,\begin{array}{ccc}
                  \cos\theta_2 & \sin\theta_2 & 0\\
                  -\sin\theta_2 & \cos\theta_2& 0\\
                    0 & 0 & 1
              \end{array}\right)\cdot 
              \left(\,\begin{array}{c}
                  e_1\\ e_2\\ e_3 \end{array}\right)
\end{equation}\\ 
These equations show that the adjoint action of $SU(2)$ on the basis of the
Clifford algebra  is precisely the same as the fundamental action of $SO(3)$
on the vector space basis. By construction it is clear that the $SU(2)$ action
leaves invariant the bilinear form $V(x,\, y)$. 

\noindent However, the invariant groups are of different size since we
have to assume that there exist objects which transform in the
fundamental of $SU(2)$ (while we have used the adjoint to match with
the rotation of vectors). The three one parameter subgroups of $SU(2)$
in (\ref{one parameter groups of SU(2)}) clearly contain half
the angle of the corresponding $SO(3)$ rotation though. An object that
rotates around 360 degrees in the fundamental representation of
$SU(2)$ will appear like an object that was rotated by 720 degrees in
the vector space basis. Or the other way: if we rotate a vector around
360 degrees in the usual vector space basis, it will have come back to
itself. An element transforming in the fundamental of $SU(2)$ in the
Clifford algebra (these elements are called spinors) will only appear
to have been rotated by 180 degrees. Let's consider two spinors rotated
by 90 degrees and 270 respectively. In the corresponding $SO(3)$
rotation they correspond to vectors rotated by 180 and 540 degrees,
respectively. However, a vector rotated around 180 degrees is
naturally indistinguishable from one that was rotated around 540
degrees (difference 360 degrees) but the spinors are obviously
differently oriented. This fact can be referred to by saying that
$SU(2)$ is the double cover of $SO(3)$. The point for this argument is
that a sign change in the $SU(2)$ element is not transported to the
relevant $SO(3)$ element since in the adjoint representation required
to calculate the relevant $SO(3)$ element the two signs cancel each
other. Two different $SU(2)$ elements thus correspond to the same
$SO(3)$ element. \\

\subsection{General dimensions and general signature}
\label{SubSec_Cliff_General}

The explicit example was given over the real vector space
$\mathbb{R}^3$, but the idea of Clifford algebras extends to general
vector spaces $\mathbb{R}^{p,q}$ or $\mathbb{C}^n$. 

\noindent
Instead of using $\tau_i$ which refer to the Pauli matrices, for the
general case the Greek $\Gamma$ is used to refer to the representation of
the Clifford algebra\\
\begin{eqnarray}\label{defining relation Gamma}
   \{\,\Gamma_i,\,\Gamma_j\,\} ~=~
   \Gamma_i\,\Gamma_j + \Gamma_j\,\Gamma_i ~=~
    2\; \eta_{ij}\cdot \unity.
\end{eqnarray}\\
However, the Pauli matrices play a crucial role in constructing the
$\Gamma$-matrices which form representations of Clifford algebras of
higher dimensional vector spaces. Since the application will be in
physics, we will not discuss the case of Clifford algebras over
complex numbers (leave alone quaternions), but concentrate on Clifford
algebras over $\mathbb{R}^{p,q}$ still allowing arbitrary
signature. The obvious question is how to find concrete matrices
$\Gamma_i$ fulfilling the defining relations of Clifford algebras
(\ref{defining relation Gamma}).

\noindent In order to answer this it is helpful to look at the
Clifford algebra over $\mathbb{R}^2$ with $\eta = {\rm
  diag}(1,1)$. The point is that the first two Pauli matrices
$\tau_{1,2}$ obviously furnish a representation since they do for the
three dimensional case ${\rm Cliff}(\mathbb{R}^3)$. The only
difference is that the third Pauli matrix does not represent a basis
element of the vector space anymore (we have only got two in
$\mathbb{R}^2$). It turns out\\    
\begin{equation}
   {\rm Cliff}(\mathbb{R}^2)= {\rm Cliff}(\mathbb{R}^3)\,.
\end{equation}\\
While in the three dimensional case the third Pauli matrix represents
the third basis vector in the Clifford algebra (and hence squares to
one because we assumed the metric to be positive definite), in the two
dimensional case the third Pauli matrix can be used as a projector
(since it squares to one). The important role of this projector will
be discussed later when we introduce the notion of chirality of spinors.

\noindent This last observation is generalisable in the sense that
always the Clifford algebra over an even-dimensional real field is the
same as the one over the real field in one dimension higher\\
\begin{equation}\label{taking product}
    {\rm Cliff}(\mathbb{R}^{2k}) = {\rm Cliff}(\mathbb{R}^{2k+1})\,.
\end{equation}\\
Since this has worked so nicely, it might be tempting to try to get
the Clifford algebra over $\mathbb{R}^4$ by multiplying all basis
elements over $\mathbb{R}^3$, i.e. introduce $e_4 \sim e_1 \circ e_2
\circ e_3$ as an independent basis vector. As demonstrated above, the
$\circ$ product turns into matrix multiplication if the basis elements
are represented by Pauli matrices. The product of all Pauli matrices
is proportional to the unit matrix and hence does not result in a new
independent element of the algebra. This is also obvious from the fact
that the two-by-two matrices have four degrees of freedom only and can
therefore be expanded as\\
\begin{equation}
   \begin{pmatrix} a & b\\c & d\end{pmatrix}= \sum_{i=0}^3 a_i \tau_i\,,
\end{equation}\\
where we have defined $\tau_0 = \unity$. This is a complete basis of
the two-by-two matrices, so there is not space for an independent
basis element. Upshot: the Clifford algebra over $\mathbb{R}^4$ cannot
be represented in two-by-two matrices.\\

\noindent Look at the following expressions involving tensor products
of Pauli matrices (to increase the degrees of freedom)\\
\begin{eqnarray}
   \Gamma_1 &=& Id \otimes \tau_1 
            = \left(\begin{array}{crrr}
                   0&1&0&0\\
                   1&0&0&0\\ 
                   0&0&0&\phantom{-}1\\ 
                   0&0&1&0
              \end{array}\right),\quad 
   \Gamma_2 = Id \otimes \tau_2 
            = \left(\begin{array}{rrrr}
                   0&-i&0&0\\ 
                   i&0&0&0\\ 
                   0&0&0&-i\\
                   0&0&i&0
              \end{array}\right)\\[2ex] 
   \Gamma_3 &=& \tau_1 \otimes T 
            = \left(\begin{array}{rrrr}
                   0&0&-1&0\\
                   0&0&0&1\\ 
                  -1&0&0&0\\ 
                   0&1&0&0
              \end{array}\right),\quad 
   \Gamma_4 = \tau_2 \otimes T 
            = \left(\begin{array}{rrrr}
                   0&0&i&0\\ 
                   0&0&0&-i\\  
                  -i&0&0&0\\ 
                   0&i&0&0
              \end{array}\right)\,,
\end{eqnarray}\\
where $T = i \tau_1\tau_2$.
It is easy to check that\\ 
\begin{equation}
   \Gamma_1^2 = Id \otimes \tau_1 \cdot Id \otimes \tau_1 = Id^2
   \otimes \tau_1^2 = Id \otimes Id = Id_{4\times 4}\,,
\end{equation}\\
and similarly all other $\Gamma$'s square to the four dimensional unit
matrix (you can check by using the matrix representation and multiply
four-by-four matrices explicitly).

\noindent Let us check anticommutators\\
\begin{equation}
    \{\Gamma_1,\,\Gamma_2\} = 2\,Id \otimes (\tau_1\tau_2+\tau_2\tau_1)
    = 0\qquad {\rm
      relying\,\,on\,\,the\,\,two\,\,dim\,\,Clifford\,\,algebra}\,. 
\end{equation}\\
Also very simple are\\
\begin{eqnarray}
    \{\Gamma_1,\,\Gamma_3\} & = & 2 \tau_1 \otimes (\tau_1T+T\tau_1) = 2
    \tau_1 \otimes (\tau_1T-\tau_1T) = 0\quad {\rm
      merely\,\,using\,\,2d\,\,properties} \\
    \{\Gamma_1,\,\Gamma_4\} & = & 2 \tau_2 \otimes (\tau_1T+T\tau_1) =
    0\hspace{23.5ex} {\rm merely\,\,using\,\,2d\,\,properties}\,.  
\end{eqnarray}\\
Similarly -try as an exercise- all other anticommutators of different
$\Gamma_i$, where $i=1,\ldots, 4$, vanish. In fact, the matrices given
above form a representation of ${\rm Cliff}(\mathbb{R}^4)$, and all
that was required are the anticommutation relations of Pauli matrices.

\noindent A result previously derived can be checked again. Defining
$\Gamma_5$ to be proportional to the product of all four dimensional
$\Gamma_i$, we find\\
\begin{equation}\label{Gamma 5}
    \Gamma_5 \sim \Gamma_1\cdot \Gamma_2\cdot \Gamma_3\cdot\Gamma_4
    = \tau_1\tau_2\otimes \tau_1 \tau_2 TT = \tau_1\tau_2\otimes
    \tau_1 \tau_2 =\begin{pmatrix}-1&0&0&0\\
     0&1&0&0\\ 0&0&1&0\\ 0&0&0&-1\end{pmatrix} \,,
\end{equation}\\
which squares to unity without being proportional to the unit element
itself. $\Gamma_5$ is linearly independent and hence is used to
construct ${\rm Cliff}(\mathbb{R}^5)$ which coincides with ${\rm
  Cliff}(\mathbb{R}^4)$.

\noindent Please note that we have discussed the case of positive
definite metric. If indefinite metrics are considered some of the
basis elements have to be dressed with factors of $i$ when represented
by matrices in the Clifford algebra. For example, if we want to
consider the Clifford algebra  ${\rm Cliff}(\mathbb{R}^{1,3})$ with
signature $(-,+,+,+)$ then all we need to do is to use
$\tilde\Gamma_1=i\Gamma_1$. This does not change the anticommutators
but it does change the square $\tilde\Gamma_1^2 = -1$. In order to
preserve $\Gamma_5^2 =1$ we need to introduce another imaginary unit
$\Gamma_5 = i\tilde\Gamma_1\cdot \Gamma_2\cdot \Gamma_3\cdot\Gamma_4$,
hence\\    
\[
   \Gamma_5^2 = -\tilde\Gamma_1\cdot \Gamma_2\cdot \Gamma_3\cdot
   \Gamma_4\cdot\tilde\Gamma_1 \cdot\Gamma_2\cdot \Gamma_3\cdot \Gamma_4 =
   \tilde\Gamma_1^2 \Gamma_2\cdot \Gamma_3\cdot\Gamma_4 \cdot\Gamma_2\cdot
   \Gamma_3\cdot\Gamma_4 = - \Gamma_2^2 \cdot \Gamma_3\cdot\Gamma_4
   \Gamma_3\cdot\Gamma_4 = \Gamma_3^2\cdot\Gamma_4^2\,, 
\]\\
which indeed is the unit matrix.\\

\noindent
The idea of taking tensor products of Pauli matrices is
generalisable. In fact, it is again easy to check that for a general
Clifford algebra over $\mathbb{R}^{2k}$ the basis elements can be
defined via\\ 
\begin{equation}\label{basis elements of CA}
   e_j \rightarrow Id \otimes \ldots \otimes Id \otimes
   \tau_{\alpha(j)}\otimes \underbrace{T\otimes \ldots \otimes
     T}_{[\frac{j-1}{2}]-{\rm times}},  
\end{equation}\\
where\\ 
\[
   T= i\cdot\tau_1\tau_2\qquad \alpha(j) = 1, {\rm if}\; j\; {\rm
     is\; odd, \,or}\quad \alpha(j) = 2\; {\rm otherwise}\,.
\]\\
In this way, the properties of the Pauli matrices get inherited to any
dimensionality. Let us finally give a representation of the eleven
basis elements of ${\rm Cliff}(\mathbb{R}^{11})$ just to ensure that
the above formula looks more complicated than it actually is. The
first ten basis elements are simply\\
\begin{eqnarray}
   e_1 & \sim   Id \otimes Id \otimes Id \otimes Id \otimes \tau_1\,,
   \qquad & e_2  \sim  Id \otimes Id \otimes Id \otimes Id \otimes
   \tau_2\,, 
   \nonumber\\
   e_3 & \sim  Id \otimes Id \otimes Id \otimes \tau_1 \otimes T\,,
   \qquad & e_4  \sim  Id \otimes Id \otimes Id \otimes \tau_2 \otimes
   T\,,
   \nonumber\\
   e_5 & \sim Id \otimes Id \otimes \tau_1 \otimes T \otimes T\,,
   \qquad & e_6  \sim  Id \otimes Id \otimes \tau_2 \otimes T \otimes
   T\,,
   \label{Eucl_11d_Cl_Al}\\
   e_7 & \sim Id \otimes \tau_1 \otimes T \otimes T \otimes T\,,
   \qquad  & e_8  \sim  Id \otimes \tau_2 \otimes T \otimes T \otimes
   T\,,
   \nonumber\\
   e_9 & \sim  \tau_1 \otimes T \otimes T \otimes T \otimes T\,,
   \qquad & e_{10} \sim \tau_2 \otimes T \otimes T \otimes T \otimes
   T\,.
   \nonumber
\end{eqnarray}\\
The eleventh basis element is taken to be the product of all other
basis elements and hence is\\
\begin{equation}\label{basis element e_11}
   e_{11} \sim i \tau_1\tau_2 \otimes \tau_1\tau_2 \otimes\tau_1\tau_2
   \otimes\tau_1\tau_2 \otimes\tau_1\tau_2 \,.
\end{equation}\\
It is straight forward to check that all these elements square to
unity (in $e_{11}$ we have introduced the imaginary unit to achieve
this) while the anticommutator of different basis elements
vanishes. Taking tensor products of five two-by-two matrices,we note
that the Clifford algebra over $\mathbb{R}^{11}$ is represented in
32-by-32 matrices. In the next section, the Clifford algebra over
$\mathbb{R}^{1,10}$ is constructed in a slightly different
way.\\

\noindent To slowly find our way towards calculation involving
Clifford algebras it is natural to derive the algebra of products of
two $\Gamma$ matrices which is defined by\\
\[
   \Gamma_{ab} = \Gamma_{[a}\,\Gamma_{b]}\,.
\]\\
Please check in detail the following calculation\\
\begin{eqnarray*}
   \Gamma_{ab}\Gamma_{cd} = \Gamma_{abcd} - \eta_{ac} \Gamma_{bd} +
   \eta_{ad}\Gamma_{bc} - \eta_{bd} \Gamma_{ac} +
   \eta{_bc}\Gamma_{ad}\,,
\end{eqnarray*}\\
where $\Gamma_{abcd}$ is the antisymmetrised product of four
$\Gamma$'s. The antisymmetrisation ensures that all indices are
different. Since the square of $\Gamma$ matrices is defined by the
Clifford relation to be proportional to unity, only the shortest
(simplest) representative of a particular Clifford algebra element is
used, for example: instead of writing $\Gamma_1\cdot \Gamma_1\cdot
\Gamma_2$ we use $\pm\Gamma_2$ since $\Gamma_1\cdot \Gamma_1 =\pm
\unity$ (depending on the case).   

\noindent The last equation expressed in words: either all indices of
the product of two $\Gamma_{ab}$ are different (in which case we get a
four indexed $\Gamma$ matrix, i.e. the first term on the right hand
side), or two indices coincide and can be taken to be proportional to
the metric (and we have written all possible indices that can
coincide). Relabelling indices the last equation takes the form\\
\begin{eqnarray*}
   \Gamma_{cd}\Gamma_{ab} = \Gamma_{cdab} - \eta_{ca} \Gamma_{db} +
   \eta_{cb}\Gamma_{da} - \eta_{ca} \Gamma_{ca} +
   \eta_{da}\Gamma_{cb}\,.
\end{eqnarray*}\\
Therefore\\
\begin{eqnarray*}
   \Gamma_{ab}\Gamma_{cd} -\Gamma_{cd}\Gamma_{ab} = \,
   [\Gamma_{ab},\,\Gamma_{cd}] & = & \Gamma_{abcd} - \eta_{ac} \Gamma_{bd} +
   \eta_{ad}\Gamma_{bc} - \eta_{bd} \Gamma_{ac} +
   \eta_{bc}\Gamma_{ad} \\ & - & (\Gamma_{cdab} - \eta_{ca} \Gamma_{db} +
   \eta_{cb}\Gamma_{da} - \eta_{ca} \Gamma_{ca} +
   \eta_{da}\Gamma_{cb}) \,.
\end{eqnarray*}\\
The last expression can be simplified by considering the antisymmetry
in the product of Clifford elements and also using the symmetry in the
metric $\eta_{ac}=\eta_{ca}$ to give\\
\begin{eqnarray*}
   [\Gamma_{ab},\,\Gamma_{cd}] & = & - 2 \eta_{ac} \Gamma_{bd} + 2
   \eta_{ad}\Gamma_{bc} - 2 \eta_{bd} \Gamma_{ac} + 2
   \eta_{bc}\Gamma_{ad}  \,.
\end{eqnarray*}\\
Defining $\Gamma_{ab} = 2J_{ab}$ ($\Gamma$ matrices transform with
twice the angle) this can be re-expressed as\\
\begin{eqnarray}\label{algebra Lorentz rotation}
  [J_{ab},\,J_{cd}] & = & -\,\eta_{ac} J_{bd}\,+\,\eta_{ad}J_{bc}\, 
                      -\,\eta_{bd} J_{ac}\,+\,\eta_{bc}J_{ad}  \,,
\end{eqnarray}\\
which turns out to be the algebra of the rotation generators $J_{ab}$.\\

\begin{summary}
We have made a full circle and have thus shown that everything is
consistent. We have seen that the scalar product can be embedded into
an algebra where the basis vectors fulfil slightly generalised
relations (the product of two basis elements is non-trivial) but
maintain the length of a vector (the square of basis vectors is one -
or rather $\unity$ since in the new algebra). By definition is the
product of two different basis elements {\it in the Clifford algebra}
antisymmetric. We have also just seen that this product of two basis
elements except for a factor of two fulfils the algebraic relations
typical of rotations. Hence $\frac{1}{2}\Gamma_{ab}$ can be considered
as generators of rotations.\\
\noindent Objects transforming under Clifford algebra elements in the
fundamental representation, i.e. matrix multiplication, are
called spinors. We have argued how to get explicit representations of
all Clifford elements in all dimensions and all signatures.
\end{summary}

\subsection{The Groups PIN and SPIN.}
We have derived all relations between the algebras of rotations and
spin groups but we will now show that everything can even be
understood on the level of the corresponding groups. Hence we define:
Be $x\in \mathbb{R}^n$ a vector, and $C_n$ the Clifford algebra with
positive definite metric $\eta= diag(1,\ldots,1)$. Accordingly,
$x\cdot x = ||x||^2$ can be used to find an inverse element\\ 
\[
   x^{-1} = \frac{x}{||x||^2}\,.
\]\\
The crucial point is that the basis vectors are naturally elements of
a vector space but at the same time also elements of the Clifford
algebra. In the Clifford algebra it is also possible to consider all
products of basis elements with respect to the Clifford multiplication 
$\circ$, for example, if $\Gamma_i$
represents the basis element $e_i$ in the Clifford algebra, then also
$\Gamma_{[ij]}$ or $\Gamma_{[ijk]}$ are in the Clifford algebra. In
particular the Clifford algebra is generated by all the antisymmetric 
products of gamma matrices. The
highest possible number of indices is the number of basis elements,
i.e. the dimensionality of the vector space $\mathbb{R}^n$ under
consideration. Adding all possible combinations of any number of
indices gives the result\\ 
\begin{equation}
    \sum_{p=0}^n \binom{n}{p} = (1+1)^n = 2^n\,.
\end{equation}\\ 
In even dimensions these $2^n$ elements are linearly independent.  As was
demonstrated above, the odd dimensional case has the same
dimensionality as the next lower even dimensional case since the
product of all basis elements just forms the representative for the
additional direction (for example $\Gamma_3 \sim \Gamma_{12}$). We
note that in principle this number of degrees of freedom could be
nicely represented by $2^{|\frac{n}{2}|}$-by-$2^{|\frac{n}{2}|}$
matrices. These would naturally act on vectors (better spinors) with
dimension $2^{|\frac{n}{2}|}$. This is indeed true for most of the
cases.\\

\noindent
We are ready to define 
\begin{defn}\label{Def Pin} $PIN(n)\subset C_n$ is the group which is
  multiplicatively generated by all vectors $x\in S^{n-1}$ (sphere).
\end{defn}
\noindent
Before moving on to the definition of the $SPIN(n)$ group, we want to
motivate its construction by considering the involution on the vector
space $\mathbb{R}^n$ which simply changes the sign of all vectors $V
\ni v \rightarrow -v\in V$. In the Clifford algebra this means that
all elements generated by an even number of basis elements are inert
under this mapping, but not the vectors themselves. Nor are the
elements inert that are generated by an odd number of basis
elements. The $PIN(n)$-group is spanned by all multiplicatively
generated elements independently of whether the number of basis
elements is even or odd. The $SPIN(n)$ group is defined to be the
subgroup of $PIN(n)$ which is inert under the sign swap of vectors
\begin{defn}\label{Def Spin} $SPIN(n)\subset C_n = PIN(n)\cap
  C_0$, where $C_0$ is inert under sign swaps of single basis
  elements. 
\end{defn}
\noindent
An anti-involution $\gamma$ can be defined by inverting the order of
the basis elements:\\
\begin{equation}
   \gamma (x_1\circ\ldots \circ x_m) 
    ~=~ \gamma(x_m)\circ \ldots  \circ \gamma(x_1)
    \quad \& \quad \gamma(x_j)=x_j,\quad\forall\, x_j\in {\mathbb{R}}^n~.
\end{equation}\\
$\gamma$ can be understood as the inversion in $PIN(n)$, e.g.\\ 
\begin{eqnarray}
    \gamma( e_i \circ e_j \circ e_k \circ e_l) &=& e_l \circ e_k \circ e_j \circ e_i\,,
\end{eqnarray}\\
since now the square simply reads\\
\begin{equation}
    (e_i \circ e_j \circ e_k \circ e_l) \circ (e_l \circ e_k \circ e_j
    \circ e_i) \equiv \unity\,,
\end{equation}\\
which is obviously $\unity$. $\gamma$ can be used to define reflexions
and is used to define the adjoint representation of $PIN(n)$ 
(cf. (\ref{adjoint versus fundamental})):\\
\begin{lem} If $y\in \mathbb{R}^n\subset C_n$ and $x\in
  PIN(n)\subset C_n$, then $x\cdot y \cdot \gamma(x)$ is again in
  $\mathbb{R}^n\subset C_n$.
\end{lem}
\noindent
To see this we can restrict ourselves to the case of one basis element
of the $PIN$-group $x= e_1$; all other cases can be derived
analogously. A simple calculation with a general vector $y= \sum_1^n
y_ie_i$ yields\\
\begin{eqnarray*}
   x \cdot y \cdot \gamma(x) & = & e_1(\sum_1^n y_ie_i)e_1= e_1y_1e_1e_1 -
   \sum_2^n y_ie_i \\
   & = & y_1e_1 - \sum_2^n y_ie_i,
\end{eqnarray*}\\
where we have merely used $e_1^2 = 1$ as defined from the quadratic
form. This, however, is obviously a reflexion of the vector $y$.
It is easy to see that taking a composite element of $PIN$ will
only result in successive reflexions. 

\noindent
Correspondingly, $SPIN$ will necessarily generate an even number of
reflexions. It is worthwhile to state an old fact which will
bring out the full significance of Clifford algebras in physics:
{\sc Every rotation can be understood as an even number of reflexions.}
We have come a long way from abstract mathematics down to rotations
and reflexions in space or space-time.

\subsection{Spinors, Majorana and Weyl conditions}
The final paragraph on abstract Clifford algebras is dedicated to the
objects that actually transform under Clifford elements. Since the
complex Pauli matrices (and their tensor products) give rise to
representations of Clifford algebras it is natural to define spinors
as complex valued vectors.\\
\begin{defn}\label{complex spinors}
   The vector space of complex $n$-spinors is\\
\[
   \Delta_n: = \mathbb{C}^{2^k} = \mathbb{C}^2 \otimes \ldots \otimes
   \mathbb{C}^2 \qquad {\rm for}\quad n = 2k, 2k+1.
\]\\
Elements of $\Delta_n$ are called complex spinors or Dirac spinors. 
\end{defn}
\noindent
Some properties of spinors can be derived from simple considerations
on Clifford products of basis elements. Let us for example consider
the case of an even dimensional vector space $\mathbb{R}^{2k}$. In
this case the product of all basis elements\\ 
\begin{equation}\label{Gammad+1}
   \Gamma_{2k+1} = \pm a\,\Gamma_1 \cdot \ldots \cdot \Gamma_d
\end{equation}\\
(the name $\Gamma_{2k+1}$ is just convention and originates from 4
dimensions where the product of all $\Gamma$-matrices was usually
called $\Gamma_5$) results in an element which is not proportional to
$\unity$, and generally is linearly independent of all other basis
elements and their products. This is a consequence of the above
equation (\ref{basis elements of CA}) \footnote{Compare the specific
  example (\ref{basis element e_11}).}. 
Taking the square of any product of basis elements in the Clifford
algebra will result in an element proportional to one. 
Depending on the particular case, we introduce an
imaginary unit into the product of all basis elements to 
ensure that it squares to $+\,\unity$ (hence $a=\{1,i\})$. Since it
squares to $\unity$ while not representing a basis vector, this
element can be taken as an involution. Being a product of an even
number of basis element, it will also commute with all other products
containing an even number of basis vectors, as for example the
elements of $SPIN(2k)$. This involution can therefore be used to split
the vector space (i.e. the space of Dirac spinors) into two parts:
those with positive, respectively negative eigenvalue under
$\Gamma_{2k+1}$. Explicitly: take a spinor $\lambda$ and measure its 
eigenvalue under $\Gamma^{2k+1}$. Since
Lorentz generators commute with $\Gamma^{2k+1}$, this eigenvalue will
not change under Lorentz rotations and hence serves as an invariant
which is called handedness or chirality. It therefore makes sense to
define\\ 
\begin{defn}\label{Def Weyl} The elements of the sub-spaces with
  positive or negative eigenvalue under $\Gamma_{2k+1}$ are called
  (positive or negative) Weyl-spinors.  
\end{defn}
\noindent
Weyl-spinors are elements of irreducible modules, i.e. sub vector
spaces that do not mix under the action of the spin group. 
By construction it
is clear that Weyl-spinors exist in all even dimensions and never in
odd dimensions.

\noindent
Before describing another involution which can be used to find
irreducible sub-modules, we need to state a fundamental theorem of
Clifford algebras without proving it. Its content is simple. It
merely states the fact that two different representations of a
Clifford algebra over the same base space are isomorphic and hence
related by a similarity transformation. The proof is not difficult and
the isomorphism can be constructed explicitly, however, in order to
keep the pace we refer the interested reader to \cite{Friedrich}. We
are now ready for the following considerations.\\

\noindent
It can be seen from (\ref{basis elements of CA}) that despite
considering real vector spaces $\mathbb{R}^d$, the representation of
the Clifford algebra may involve imaginary units and thus acts on
complex-valued spinors (as laid out in Definition \ref{complex
  spinors}). It is easy to see, however, that if a set of
$\Gamma$-matrices fulfils the Clifford relations (\ref{defining
  relation Gamma}) then so does the set $\Gamma^*$, i.e. the complex
conjugate. In other words both are representation over the same
quadratic form. Due to the fundamental Lemma just mentioned, they must
be isomorphic. Therefore a matrix $B$ exists such that\\
\begin{equation}\label{Bconj}
   \Gamma^*_\mu = B \,\Gamma_\mu\, B^{-1}, \qquad {\rm and\; thus\;
     also} \qquad \Gamma^*_{\mu\nu} = B \,\Gamma_{\mu\nu}\, B^{-1}.
\end{equation}\\
Since we have introduced the notion of complex conjugation on the
Clifford algebra, one can now pose the question whether it might be
possible to find purely real-valued representations of its
elements. One way is to just go and try. But there is a more elegant
way leading to {\bf Majorana} spinors. These are defined through the
observation that also $\Gamma^T$ (transposed matrices) furnish a
representation if the original $\Gamma$'s do. Again the same argument
as above applies, and there exists a matrix $C$ such that\\
\begin{eqnarray}
   \Gamma^T_\mu = -C \,\Gamma_\mu\, C^{-1}, \qquad {\rm and\; thus\;
     also} \qquad \Gamma^T_{\mu\nu} = -C \,\Gamma_{\mu\nu}\, C^{-1}
   \label{Gamma_Igor}
\end{eqnarray}\\
where we have introduced a minus sign which physically allows us to
identify the matrix $C$ with the charge conjugation matrix. Next,
using the fact that a spinor transforms under a $SPIN$-transformation
as (additional factor one half in comparison to above formula
(\ref{algebra Lorentz rotation}) because of antisymmetry in the
indices)\\ 
\[
   \delta \lambda = \frac{1}{4}\,\omega^{mn}\,\Gamma_{mn}\,\lambda
\]\\
we calculate\\
\begin{small}
\begin{equation}
   \delta (B^{-1}\lambda^*) = B^{-1}\delta \lambda^*= B^{-1}
   \left(\frac{1}{4}\,\omega^{mn}(\Gamma_{mn}\lambda)^*\right)
   = \frac{1}{4}\,\omega^{mn}B^{-1}B\Gamma_{mn}B^{-1}\lambda^*=
   \frac{1}{4}\,\omega^{mn}\Gamma_{mn}(B^{-1}\lambda^*). 
\end{equation}
\end{small}\\
This equation means that also $B^{-1}\lambda^*$ transforms under
$SPIN(d)$ in the same way that $\lambda$ does. Hence, it might be the
case the the complex conjugate of a spinor can just be calculated by
$\lambda^*= B\lambda$. This can then be understood as some kind of
reality condition of the spinor.

Using the fundamental lemma once more  there must be a relation
between the matrices $B$ and $C$, since both representations must also
be isomorphic. The precise relation between these matrices depends
crucially one the signature of the vector space $\mathbb{R}^{p,q}$
under consideration. 

Having argued that the matrix $B$ might relate the spinor $\lambda$
to its complex conjugate spinor $\lambda^* = B\,\lambda$, we take the
complex conjugate of this equation and plug it into itself to find\\ 
\[
   \lambda ~=~ B^*B\,\lambda ~=~ \epsilon\,\lambda\,,
\]\\
since $B^*B$ commutes with the representation when taking the complex
conjugate twice (\ref{Bconj}) as can be seen from the following 
computation:\\
\[
  \Gamma_m ~=~ (\Gamma_m^*)^* ~=~ B^*B\,\Gamma_m\, B^{-1}B^{-1*}.
\]\\
By Schur's Lemma $B^*B$ will thus have to be a multiple of the unit
matrix  $B^*B=\epsilon \;Id$. One can work out the precise value for
$\epsilon$ in various space-time dimensions and signatures. Below we
will give a table for the specific case of one time dimension and
$d-1$ space dimensions. Majorana spinors then exist if $\epsilon =1$
(since then $B$ becomes an involution) which can be shown to be the
case if $d= 2,4 \;{\rm mod}\, 8$.  \\

\noindent Finally, one could ask in which dimensions -assuming
'physical' signature $\mathbb{R}^{1,d-1}$- the Majorana condition is
compatible with the condition for Weyl spinors. This merely depends on
whether or not we had to introduce the imaginary unit into
$\Gamma^{2k+1}$ (\ref{Gammad+1}) to ensure its squaring to $\unity$
\footnote
{In the physical signature example underneath (\ref{Gamma
    5}), we had to introduce the imaginary unit. You can check by
  hand that you would also have to introduce it when defining 
  $\Gamma^9$ in $\mathbb{R}^{1,7}$.  You can also check explicitly
  that you don't need to introduce the imaginary unit for $\Gamma^3$
  in $\mathbb{R}^{1,1}$ or $\Gamma^7$ in $\mathbb{R}^{1,5}$ or
  $\Gamma^{11}$ in $\mathbb{R}^{1,9}$.
}. 
%
It turns out that for $d= 2
\;{\rm mod}\, 4$ no imaginary unit is required. Hence is the Majorana
condition compatible with the Weyl condition in
$d=2,\,10,\,18,\,26,\,\ldots$. In the table below, we have indicated
the reducibility of the spinors in physical signature up to eleven
dimensions. 

The first case, $d=2$, happens to be the dimension of the world-sheet
of a string, the second case, $d=10$, is the dimensionality of super
string theories, and the case $d=26$ is the critical dimension of
bosonic string theory.\\[3ex] 
\begin{small}
\begin{tabbing}
\hspace{6ex}\= \qquad d \quad\=\qquad 2 \quad \=\qquad 3\quad \= \qquad 4\quad
  \=\qquad 5\quad \= \qquad 6\quad \=\qquad 7\quad \=\qquad 8 \quad
  \=\qquad 9 \quad \=\qquad 10\quad \= \qquad 11 \\ \\
\>  \>\quad M-W \>\qquad M \>\quad M/W \>\qquad - \>\qquad W
  \>\qquad - \>\qquad W \>\qquad M \>\quad M-W \>\qquad M.
\end{tabbing}
\vskip4ex
\centerline{\it M stands for Majorana, W for Weyl, M/W stands for
  either Majorana or Weyl,} 
\centerline{\it M-W for Majorana and Weyl, correspondingly
  a dash means neither M nor W.}
\end{small}
\vspace{1cm}

\begin{summary}
This finishes the overview into Clifford algebras. From now on we
will present explicit calculations. After having understood the basic
ideas behind Clifford algebras, the reader might be tempted to assume
that calculations are easy to do. This will turn out to be a false
assumption since the representation of ${\rm
  Cliff}(\mathbb{R}^{1,10})$ is in terms of 32-by-32 matrices. The
authors therefore thought to present some explicit calculations in
order to see how they are done in general. 

\noindent
Unfortunately, many calculations are depending very much
on conventions. The first choice effecting the representation of the
Clifford algebras is of course whether mostly plus or mostly minus
quadratic forms are used. Secondly, we have given an explicit
representation of basis elements in (\ref{basis elements of CA}) where
others are possible but of course isomorphic (due to the fundamental
theorem on Clifford algebras). Multiplying purely real $\Gamma_i$
matrices by the imaginary unit results in purely imaginary basis
elements. Since eleven dimensions indeed do admit a Majorana condition
imposed on the spinors, some papers talk of purely imaginary, others
of purely real representations of Clifford algebras. 
\end{summary}

\subsection{Eleven dimensional  Clifford Algebra}

In subsection \ref{SubSec_Cliff_General} we have already constructed the 
eleven dimensional Euclidean Clifford algebra (\ref{Eucl_11d_Cl_Al}).
In eleven dimensional SUGRA we also need a realisation of an 
eleven dimensional Clifford algebra of a pseudo Euclidean 
space with Minkowskian signature. The fundamental anticommutation 
relation of the corresponding  Clifford algebra reads\\
\begin{eqnarray}\label{CliffordAlgebra}
     \{\,\Gamma^a,\Gamma^b\,\} &=& 2\,\eta^{ab}\,\unity_{32},\hspace{2cm}
     \eta^{ab}\,=\,{\rm diag}(-1,1,\ldots,1)
\end{eqnarray}\\
A representation by real matrices $\Gamma^a$ can be obtained by 
taking appropriate tensor products of the Pauli matrices (\ref{Pauli})
and $\epsilon=i\tau_2$:\\
\begin{align}
   \Gamma^0 &= -\epsilon\otimes\unity\otimes\unity\otimes\unity\otimes\tau_3 &
   \Gamma^1 &= \epsilon\otimes\epsilon\otimes\epsilon\otimes\epsilon\otimes\tau_1\nonumber\\  
\Gamma^2 &= \epsilon\otimes \unity\otimes\tau_1\otimes\epsilon\otimes\tau_1 &
\Gamma^3 &= \epsilon\otimes \unity\otimes\tau_3\otimes\epsilon\otimes\tau_1\nonumber\\
\Gamma^4 &= \epsilon\otimes \tau_1\otimes\epsilon\otimes\unity\otimes\tau_1 &
\Gamma^5 &= \epsilon\otimes \tau_3\otimes\epsilon\otimes\unity\otimes\tau_1\label{PEGA11d}\\
\Gamma^6 &= \epsilon\otimes \epsilon\otimes\unity\otimes\tau_1\otimes\tau_1 &
\Gamma^7 &= \epsilon\otimes \epsilon\otimes\unity\otimes\tau_3\otimes\tau_1\nonumber\\
\Gamma^8 &= \epsilon\otimes \unity\otimes\unity\otimes\unity\otimes\epsilon &
   \Gamma^9 &= \tau_1\otimes\unity_{16}\nonumber\\
   \Gamma^{10} &= \tau_3\otimes\unity_{16}\nonumber
\end{align}\\[-3ex]

\noindent
It turns out that the algebra generated by the eleven gamma matrices
given above and all antisymmetrised products of them is isomorphic to 
the space of real $32\times 32$ matrices, i.e.\\
\begin{eqnarray}\label{isom}
     {\rm Cliff}({\mathbb{R}}^{1,10}) &\equiv&
     Mat(32,\mathbb{R})\oplus Mat(32,\mathbb{R})~.
\end{eqnarray}\\
In supergravity we are concerned with spinorial representations
only. As discussed in  a previous sections this restricts the
consideration to the even dimensional Clifford algebra or picking 
up one of the two $Mat(32,\mathbb{R})$. In this sense we sometimes 
claim ${\rm Cliff}({\mathbb{R}}^{1,10})\,=\,Mat(32,\mathbb{R})$.\\

\noindent
In the following we work in this representation of the
$\Gamma$-matrices. For most computations apart from those in 
section \ref{Sec_Exotic} the details of the representation are 
insignificant. But of utmost importance in this note are 
the following relations\\
\begin{eqnarray}\label{AdjointTranspose}
   (\Gamma^a)^\dagger \,=\, \Gamma^0\,\Gamma^a\,\Gamma^0  
   \hspace{1cm}\Leftrightarrow\hspace{1cm}
   (\Gamma^a)^T \,=\, \Gamma^0\,\Gamma^a\,\Gamma^0
\end{eqnarray}\\[-4ex]
\begin{prop} All antisymmetric products of $\Gamma$-matrices not equal 
            to $\unity$ are traceless.
\end{prop} 
\begin{proof}
For $\Gamma^{a_1..a_{2i}}$ one has\\
\begin{eqnarray*}
     Tr(\,\Gamma^{a_1..a_{2i}}\,) 
     &=& Tr(\,\Gamma^{a_1}\cdot\ldots\cdot\Gamma^{a_{2i}}\,)\\
     &=& Tr(\,\Gamma^{a_2}\cdot\ldots\cdot\Gamma^{a_{2i}}\,\Gamma^{a_1}\,)\\
     &=& (-1)^{2i-1}\,Tr(\,\Gamma^{a_1..a_{2i}}\,)\hspace{5ex}
\end{eqnarray*}\\
For $\Gamma^{a_1..a_{2i+1}}$ with $2i+1\,<\,11$ there exists an 
$a_j \notin (a_1,\ldots,a_{2i+1})$\\ 
\begin{eqnarray*}
     Tr(\,\Gamma^{a_1..a_{2i+1}}\,) 
     &=& Tr(\,(\Gamma^{a_j}\Gamma_{a_j})\,\Gamma^{a_1}\cdot\ldots\cdot
              \Gamma^{a_{2i+1}}\,)\\
     &=& Tr(\,\Gamma_{a_j}\,\Gamma^{a_1}\cdot\ldots\cdot\Gamma^{a_{2i+1}}\,
              \Gamma^{a_j}\,)\\
     &=& (-1)^{2i+1}\,Tr(\,\Gamma^{a_1..a_{2i+1}}\,)
\end{eqnarray*}
\end{proof}

\noindent
\begin{prop}  We note the following useful identity without proof:\\
\begin{eqnarray}\label{Prop_14}
    \Gamma^{a_j\ldots a_1}\Gamma_{b_1\ldots b_k} 
    &=& \sum\limits_{l=0}^{{\rm min}(j,k)}\,l!\,
        \left(\begin{array}{c} j\\l\end{array}\right)\,
        \left(\begin{array}{c} k\\l\end{array}\right)\,
        \delta^{[a_1}_{[b_1}\cdots\delta^{a_l}_{b_l}\,
        \Gamma^{a_j\ldots a_{l+1}]}{}_{b_{l+1}\ldots b_k]}~.
\end{eqnarray}\\
\end{prop}

\noindent
\begin{prop} The following identity holds\\
            \begin{eqnarray}\label{FourierExpansion}
                \frac{1}{32}\,Tr
                \left(\,
                         \Gamma^{a_n\ldots a_1}\Gamma_{b_1\ldots b_n}\,
                \right) 
                &=&  \delta^{a_1\ldots a_n}_{b_1\ldots b_n} 
            \end{eqnarray}\\[-3ex]
\end{prop}
\begin{proof}
The proof is a direct consequence of computing the trace of the 
identity (\ref{Prop_14}) combined with the tracelessness of 
antisymmetrised $\Gamma$-matrices.\\[2ex]
\end{proof}

\noindent
\begin{defn} With $\psi_\mu\,=\,\psi_{\mu}^\alpha$ a Majorana spinor
1-form is  given by

\begin{eqnarray*}
    \psi &=& \psi_{\mu}\,dx^{\mu}~.
\end{eqnarray*}
\end{defn}

\noindent
\begin{defn} Dirac/Majorana conjugate
\begin{eqnarray}
     \bar{\psi}^D &=& \psi^\dagger\,A\\
     \bar{\psi}^M &=& \psi^T\,C\label{MajoranaConj}
\end{eqnarray}
\end{defn}
\noindent
Here $A$ and $C$ are intertwiners of representations of the
$\Gamma$-algebra\footnote
%
{
   This is consistent with (\ref{Gamma_Igor}) due to 
   $(\Gamma^0)^{-1}\,=\,-\Gamma^0$.
}\\
%
\begin{eqnarray*}
    (\Gamma^a)^\dagger \,=\, A\,\Gamma^a\,A  
   \hspace{1cm}\Leftrightarrow\hspace{1cm}
   (\Gamma^a)^T \,=\, C\,\Gamma^a\,C 
\end{eqnarray*}\\
Comparison with eq.~(\ref{AdjointTranspose}) leads to $A\,=\,\Gamma^0$ and
$C\,=\,\Gamma^0$.\\

\noindent
\begin{rem}
Exchanging two spinors produces a sign (due to the Fermi statistics of 
fermions).
\end{rem}

\noindent
\begin{prop}
   \begin{eqnarray}\label{Prop21}
       \bar\psi\,\Gamma^{a_1\,\ldots\,a_i}\wedge\psi \,=\, 0~,
       \hspace{1cm} i\,=\,0,3,4,7,8,11
   \end{eqnarray}
\end{prop}

\vspace{1ex}

\begin{proof} Write eq.~(\ref{Prop21}) in coordinates, i.e.\\
\begin{eqnarray*}
  \bar\psi\,\Gamma^{a_1\,\ldots\,a_i}\wedge\psi 
  &=& \left(\,
               \bar\psi_{[\mu}\,\Gamma^{a_1\,\ldots\,a_i}\,\psi_{\nu]}\,
      \right)\,dx^\mu\,\wedge\,dx^\nu
\end{eqnarray*}\\ 
Now performing a transposition in the spin-space, the ``scalar''
$~\bar\psi\,\Gamma^{a_1\,\ldots\,a_i}\wedge\psi~$ does not transform 
and we obtain using eq.~(\ref{AdjointTranspose}) and the anticommuting 
property of spinors:\\  
\begin{eqnarray}
      \bar\psi_{[\mu}\,\Gamma^{a_1\,\ldots\,a_i}\,\psi_{\nu]} 
      &\buildrel !\over =& 
          \left(\,
               \bar\psi_{[\mu}\,\Gamma^{a_1\,\ldots\,a_i}\,\psi_{\nu]}\,
          \right)^T\nonumber\\
      &=& -\,\psi_{[\nu}^T\,(\Gamma^{a_i})^T\,\ldots\,(\Gamma^{a_1})^T\,
                       (\Gamma^{0})^T\psi_{\mu]}\nonumber\\
      &=&
      (-1)^i\,\bar\psi_{[\nu}\,\Gamma^{a_i\,\ldots\,a_1}\,\psi_{\mu]}
      \nonumber\\
      &=& -\,(-1)^{\frac{i(i+1)}{2}}\,
           \bar\psi_{[\mu}\,\Gamma^{a_1\,\ldots\,a_i}\,\psi_{\nu]}
      \label{SwitchPsiPsi}
\end{eqnarray}\\
The value of the sign on the right hand side of the last equation 
is depicted in the following table:\\
\begin{center}
\begin{tabular}{|c|c|c|}
\hline
    & &\\[-1.5ex]
$i$ & $-\,(-1)^{\frac{i(i+1)}{2}}$ &
      $\bar\psi\,\Gamma^{a_1\,\ldots\,a_i}\wedge\psi$\\[0.5ex]
\hline
    &      &     \\[-1.5ex]
0   &  -1  &   0 \\
1   &  +1  &     \\
2   &  +1  &     \\
3   &  -1  &   0 \\
4   &  -1  &   0 \\
5   &  +1  &     \\
6   &  +1  &     \\
7   &  -1  &   0 \\
8   &  -1  &   0 \\
9   &  +1  &     \\
10  &  +1  &     \\
11  &  -1  &   0 \\[0.5ex]
\hline
\end{tabular}
\end{center}
\vspace{2ex}
Due to the type of signs the terms pick up during the manipulation
some of them can be seen to vanish identically. In fact, up to
duality, just the terms with one, two or five Gamma matrices 
persist.\\[2ex] 
\end{proof}

\noindent
\begin{prop}  A consequence is the elementary Fierz identity:\\[1ex]
\begin{equation}\label{Prop23}
    \psi)\,\wedge\,(\bar\psi 
    \,=\, \frac{1}{32}\,
          \left(\,
                \Gamma^{c_1}\,(\bar\psi\,\Gamma_{c_1}\,
                 \wedge\,\psi) \,+\,
                \frac{\Gamma^{c_2c_1}}{2!}\,
                (\bar\psi\,\Gamma_{c_1c_2}\,\wedge\,\psi) \,+\,
                \frac{\Gamma^{c_5\ldots c_1}}{5!}\,
                (\bar\psi\,\Gamma_{c_1\ldots c_5}\,\wedge\,\psi)\,
          \right)~.
\end{equation}\\
\end{prop}

\begin{proof}
According to (\ref{isom}) a generic real $32\times 32$-matrix $\Gamma$ 
can be expanded in a basis of the Clifford algebra as\\
\begin{eqnarray}\label{FierzTrafo0}
     \Gamma &=& \sum\limits_{i=0}^5 C_{|i|}\cdot \Gamma^{|i|}
\end{eqnarray}\\  
with $\Gamma^{|i|}\,=\,\Gamma^{a_1\ldots a_i}$.
Using the identity (\ref{FourierExpansion}) one obtains\\ 
\begin{eqnarray*}
  Tr(\Gamma\,\Gamma_{|k|}) &=& \sum\limits_{i=0}^5 C_{|i|}\cdot 
                               Tr(\Gamma^{|i|}\,\Gamma_{|k|})\\
                           &=& \sum\limits_{i=0}^5 C_{|i|}\;32\,
                               (-1)^{\frac{i(i-1)}{2}}
                               \delta^{|i|}{}_{|k|}\\
                           &=& k!\cdot C_{|k|}\;32\,
                               (-1)^{\frac{k(k-1)}{2}}
\end{eqnarray*}\\
or\\ 
\begin{eqnarray*}
    C_{|k|} &=& \frac{(-1)^{\frac{k(k-1)}{2}}}{32\cdot k!}\,
                Tr(\Gamma\,\Gamma_{|k|})
\end{eqnarray*}\\
This leads to\\ 
\begin{eqnarray}\label{FierzTrafo1}
     \Gamma &=& \frac{1}{32}\,
                \sum\limits_{i=0}^5\frac{(-1)^{\frac{i(i-1)}{2}}}{i!}\,
                Tr(\Gamma\,\Gamma_{|i|})\cdot \Gamma^{|i|}
\end{eqnarray}\\ 
Choosing 
$(\Gamma)^{\alpha\beta}\,=\,\psi^{\alpha}_{[\mu})\,(\bar\psi^{\beta}_{\nu]}$,
i.e. the left hand side of eq.~(\ref{Prop23}) one obtains\\ 
\begin{eqnarray*}
 \psi^{\alpha}_{[\mu})\,(\bar\psi^{\beta}_{\nu]} 
 &=& \frac{1}{32}\,
                \sum\limits_{i=0}^5\frac{(-1)^{\frac{i(i-1)}{2}}}{i!}\,
                \underbrace{
                             Tr(\psi^{\gamma}_{[\mu})\,
                               (\bar\psi^{\delta}_{\nu]}\,
                                (\Gamma_{|i|})_{\delta\gamma})
                           }_{(\bar\psi_{[\mu}\,
                \Gamma_{|i|}\psi_{\nu]})}\cdot \Gamma^{|i|}\\
 &=& \frac{1}{32}\,
                \sum\limits_{i=0}^5\frac{(-1)^{\frac{i(i-1)}{2}}}{i!}\,
                (\bar\psi_{[\mu}\,
                \Gamma_{|i|}\psi_{\nu]})\cdot \Gamma^{|i|}\\
\end{eqnarray*}\\
Due to eq.~(\ref{Prop21}) only three terms can contribute to the sum. 
Absorbing the signs in the reordering of the indices of the Gamma-matrices
one finally obtains\\ 
\begin{equation}\label{Basic_FI}
 \psi^{\alpha}_{[\mu})\,(\bar\psi^{\beta}_{\nu]} 
  \,=\, \frac{1}{32}\,
      \left(\,
               (\bar\psi_{[\mu}\,
               \Gamma_{c_1}\psi_{\nu]})\cdot \Gamma^{c_1}
               \,+\, (\bar\psi_{[\mu}\,
               \Gamma_{c_1c_2}\psi_{\nu]})\cdot 
               \frac{\Gamma^{c_2c_1}}{2!}
               \,+\, (\bar\psi_{[\mu}\,
               \Gamma_{c_1\ldots c_5}\psi_{\nu]})\cdot 
               \frac{\Gamma^{c_5\ldots c_1}}{5!}
      \right)
\end{equation}\\
which is the expression (\ref{Prop23}) in coordinates.\\[2ex] 
\end{proof}

\begin{defn}{Open/Closed Terms:} Terms of the type 
$\bar\psi_{[\mu}\,\Gamma^{(j)}\,\psi_{\nu ]}$, i.e.\ 
with all spinor indices contracted, are called closed 
terms. Opposite to that we call terms of the type 
$\Gamma^{(i)}\psi_{[\mu}\,\bar\psi_{\nu ]}\Gamma^{(j)}$ open.
\end{defn}

\noindent
It is the nature of Fierz identities to express open terms in terms of 
closed terms and vice versa. 

\noindent
\begin{prop} The duality relation of the 11d Clifford algebra
             reads\footnote
             { \label{EpsConv}
                From (\ref{PEGA11d}) it follows
                $\Gamma^{0\ldots 10}\,=\,-\unity_{32}$
                and correspondingly 
                we define  $\varepsilon^{0\ldots 10}\,=\,-1$.
                
             }\\
             %
            \begin{eqnarray}\label{DualityRel}
                \Gamma^{a_1\ldots a_p} 
                    &=& \frac{(-1)^{\frac{(p+1)(p-2)}{2}}}{(11-p)!}\;
                        \varepsilon^{a_1\ldots a_p
                                     a_{p+1}\ldots a_{11}}\;
                                  \Gamma_{a_{p+1}\ldots a_{11}}~.
            \end{eqnarray}\\[-3ex]
\end{prop}
\begin{proof} We start again from (\ref{FierzTrafo0}) and chose
                $\Gamma\,=\,\Gamma^{a_1\ldots a_p}$
\begin{eqnarray}
     \Gamma^{a_1\ldots a_p} &=& \sum\limits_{i=0}^5 C^{|i|}\cdot
     \Gamma_{|i|}~.
\end{eqnarray}\\ 
Now we multiply by $\Gamma^{|k|}$ with the length $|k|\,=\,11-p$. We obtain 
\begin{eqnarray}\label{DR1}
     \Gamma^{a_1\ldots a_p}\Gamma^{a_{p+1}\ldots a_{11}} 
     &=& \sum\limits_{i=0}^5 C^{|i|}\cdot
     \Gamma_{|i|}\Gamma^{|k|}~.
\end{eqnarray}\\
If all indices are different, then due to (\ref{PEGA11d}) we have 
$\Gamma^{a_1\ldots a_pa_{p+1}\ldots a_{11}}\,=\,-\,\unity_{32}$ and 
after taking the trace the left hand side of (\ref{DR1}) reads (cf. 
 footnote on page \pageref{EpsConv}):\\
\begin{eqnarray}
     Tr\,(\,\Gamma^{a_1\ldots a_p}\Gamma^{a_{p+1}\ldots a_{11}}\,) 
     &=& 32\,\varepsilon^{a_1\ldots a_{11}}~,
\end{eqnarray}\\
while the right hand side reads 
\begin{eqnarray}
     Tr(\,\sum\limits_{i=0}^5 C^{|i|}\cdot
     \Gamma_{|i|}\Gamma^{|k|}\,)
     &=& 32\,C^{b_{11}\ldots b_{p+1}}\,\delta_{b_{p+1}\ldots
     b_{11}}^{a_{p+1}\ldots a_{11}}~=~ 32\,(11-p)!\,C^{a_{11}\ldots a_{p+1}}~.
\end{eqnarray}\\
Comparing both results with each other one obtains\\
\begin{eqnarray}
   C^{a_{11}\ldots a_{p+1}} 
   &=& \frac{1}{(11-p)!}\,\varepsilon^{a_1\ldots a_{11}}~,
\end{eqnarray}\\
and finally
\begin{eqnarray}
  \Gamma^{a_1\ldots a_p} &=& \frac{1}{(11-p)!}\,\varepsilon^{a_1\ldots
  a_{11}}\,\Gamma_{a_{11}\ldots a_{p+1}}
\end{eqnarray}
Ordering the indices produces a sign
$(-1)^{\frac{(11-p-1)(11-p)}{2}}$, which using modulo 4 arithmetic can
be written as 
\begin{eqnarray}
    (-1)^{\frac{(11-p-1)(11-p)}{2}} &=& (-1)^\frac{(p+1)(p-2)}{2}~.
\end{eqnarray} 

\end{proof}

\subsubsection*{Homework:}

\begin{exercise}\label{Exercise1} 
                 Prove in analogy to (\ref{SwitchPsiPsi}) the identity. 
                 Here $\epsilon$ and $\psi$ commute and 
                 $\Gamma^{(j)}\,=\,\Gamma^{a_1\ldots a_j}$:                
    \begin{eqnarray}
           \bar{\psi}_{
\mu}\,\Gamma^{(j)}\,\epsilon 
           &=& -\,(-1)^{\frac{j(j+1)}{2}}\,\bar\epsilon\,
               \Gamma^{(j)}\,\psi_\mu~.
     \label{SwitchEpsPsi}
    \end{eqnarray}
\end{exercise}

\newpage

\subsection{Cremmer-Julia-Scherk Fierz Identity}

In the application of the eleven dimensional Clifford algebra to 
supergravity the efficient handling of Fierz identities becomes 
essential. The topic is technically involved. A good reference 
for a deeper study  is \cite{Fierzidentities}. Another source 
of information is \cite{Bergshoeff:1981um}. In this section 
we want to prove the most important Fierz identity appearing 
in computations within eleven dimensional supergravity and providing  
a taste of the technicalities showing up in this context. 
It is dubbed the CJS Fierz identity and given below:\\ 
\begin{eqnarray}\label{CJSFierIdentity}
    \frac{1}{8}\,\Gamma^{\mu\nu\alpha\beta\gamma\delta}\,
    \psi_\nu\bar\psi_\alpha\,\Gamma_{\beta\gamma} ~-~
    \frac{1}{8}\,\Gamma_{\beta\gamma}\,\psi_\nu\bar\psi_\alpha\,
    \Gamma^{\mu\nu\alpha\beta\gamma\delta}\nonumber\\
    ~-~
    \frac{1}{4}\,\Gamma^{\mu\nu\alpha\beta\delta}
    \psi_\nu\bar\psi_\alpha\Gamma_\beta ~+~
    \frac{1}{4}\,\Gamma_\beta
    \psi_\nu\bar\psi_\alpha\Gamma^{\mu\nu\alpha\beta\delta}\nonumber\\
    ~-~
    2\,g^{\beta[\alpha}\Gamma^{\delta\mu\nu]}\psi_\nu
         \bar\psi_\alpha\Gamma_\beta
    ~-~ 
    2\,\Gamma_\beta\psi_\nu\bar\psi_\alpha 
         g^{\beta[\alpha}\Gamma^{\delta\mu\nu]}\nonumber\\
    ~+~
    2\,g^{\beta[\alpha}\Gamma^{\delta\mu\nu]}\,
       \bar\psi_\alpha\Gamma_\beta\psi_\nu ~=~ 0 
\end{eqnarray}\\[-3ex]
\begin{proof}
Inserting the basic Fierz identity eq.~(\ref{Basic_FI}) 
into eq.~(\ref{CJSFierIdentity}) and ordering 
after closed terms one obtains:\\
\begin{eqnarray*}
    \hbox{Eq.~(\ref{CJSFierIdentity})} 
    &=& \frac{1}{32}\,
        \left\{\vbox{\vspace{2.5ex}}\right.\,
               1^{\rm st}-{\rm term}\, 
        \left.\vbox{\vspace{2.5ex}}\right\}\,
        \psi_{[\nu}\,\Gamma_{|c_1|}\,\psi_{\alpha]}
    ~+~ \frac{1}{32}\,
        \left\{\vbox{\vspace{2.5ex}}\right.\,
             2^{\rm nd}-{\rm term}\,
        \left.\vbox{\vspace{2.5ex}}\right\}\,
        \frac{1}{2!}\,\psi_{[\nu}\,\Gamma_{|c_1c_2|}\,\psi_{\alpha]}\\
    &+& \frac{1}{32}\,
        \left\{\vbox{\vspace{2.5ex}}\right.\,
             3^{\rm th}-{\rm term}\,
        \left.\vbox{\vspace{2.5ex}}\right\}\,
        \frac{1}{5!}\,\psi_{[\nu}\,\Gamma_{|c_1\ldots c_5|}\,
        \psi_{\alpha]}
\end{eqnarray*}\\
Each of the three terms must vanish separately. There is 
a small difference in the treatment of the first and the 
last two terms. This is due to the presence of  closed term
$2\,g^{\beta[\alpha}\Gamma^{\delta\mu\nu]}\,
\bar\psi_\alpha\Gamma_\beta\psi_\nu$ in eq.~(\ref{CJSFierIdentity}), 
which contributes only to the first but not to the last two terms. 
The part which is common to all three terms and which  
must be rewritten by Fierzing consists of the first six terms in 
 eq.~(\ref{CJSFierIdentity}) and we are going to rewrite it 
now.
A unified notation allowing to handle the common part (c.p.) of all 
three terms at once is ( $\Gamma^{(j)}$ denotes a totally antisymmetric
matrix with $j$ indices )\\  
\begin{eqnarray}
  \left\{\vbox{\vspace{2.5ex}}\right.\,
       j^{\rm th}-{\rm term,~c.p.}
  \left.\vbox{\vspace{2.5ex}}\right\}
        &=&    \underbrace{
               \frac{1}{8}\,\Gamma^{\mu\nu\alpha\beta\gamma\delta}\,
               \Gamma^{(j)}\,\Gamma_{\beta\gamma}
               \,-\,
               \frac{1}{8}\,\Gamma_{\beta\gamma}\,\Gamma^{(j)}\,
               \Gamma^{\mu\nu\alpha\beta\gamma\delta}
               }_{(1)}\nonumber\\
               &&\underbrace{
               \,-\,
               \frac{1}{4}\,\Gamma^{\mu\nu\alpha\beta\delta}\,
               \Gamma^{(j)}\,\Gamma_{\beta}
               \,+\,
               \frac{1}{4}\,\Gamma_{\beta}\,\Gamma^{(j)}\,
               \Gamma^{\mu\nu\alpha\beta\delta}
               }_{(2)}\label{Acht0}\\
               &&\underbrace{
               \,-\,
               2\,g^{\beta[\alpha}\,\Gamma^{\delta\mu\nu]}\,
               \Gamma^{(j)}\,\Gamma_{\beta}
               \,-\,
               2\,\Gamma_{\beta}\,\Gamma^{(j)}\,g^{\beta[\alpha}\,
               \Gamma^{\delta\mu\nu]}
               }_{(3)}~.\hspace{20ex}\nonumber
\end{eqnarray}\\
To simplify the terms on the right hand side we need two identities 
of Clifford matrices, 
which are easy to prove. The first by direct computation and the
second by iterating the first one twice:\\
\begin{eqnarray}
      \Gamma_\beta\,\Gamma_{a_1\ldots a_j}\,\Gamma^\beta\,~ &=&
      \hspace{6ex}(-1)^j\,(D-2\,j)\;\Gamma_{a_1\ldots a_j}\label{SUM1}\\
      \Gamma_{\beta\gamma}\,\Gamma_{a_1\ldots a_j}\,\Gamma^{\beta\gamma} &=&
      -\,\left[\,(D-2\,j)^2\,-\,D\,\right]\,\Gamma_{a_1\ldots a_j}~.
        \label{SUM2}
\end{eqnarray}\\[-3ex]
\begin{proof} of (\ref{SUM2})\\
\begin{eqnarray*}
 \Gamma_{\beta\gamma}\,\Gamma_{a_1\ldots a_j}\,\Gamma^{\beta\gamma} 
 &=& -\,\left(\,\Gamma_{\beta}\Gamma_{\gamma}-\eta_{\beta\gamma}\,\right)\,
        \Gamma_{a_1\ldots a_j}\,
     \left(\,\Gamma^{\gamma}\Gamma^{\beta}-\eta^{\gamma\beta}\,\right)\\
 &=& -\,(D-2\,j)^2\,\Gamma_{a_1\ldots a_j}\,+\,D\,\Gamma_{a_1\ldots
 a_j}\,+\,D\,\Gamma_{a_1\ldots a_j}\,-\,D\,\Gamma_{a_1\ldots a_j}\\
 &=& -\,\left[\,(D-2\,j)^2\,-\,D\,\right]\,\Gamma_{a_1\ldots a_j}
\end{eqnarray*}\\[-3ex]
\end{proof}
\noindent
In eleven dimensional SUGRA we must specify  $D\,=\,11$, obviously.

\subsubsection*{Ad (1) :} Instead of (1) we consider\\ 
\begin{eqnarray}
     \frac{1}{8}\,\Gamma^{\mu\nu\alpha\delta}\Gamma_{\beta\gamma}\,
     \Gamma^{(j)}\,\Gamma^{\beta\gamma}
     \,-\,
     \frac{1}{8}\,\Gamma^{\beta\gamma}\,\Gamma^{(j)}\,\Gamma_{\beta\gamma}
     \Gamma^{\mu\nu\alpha\delta}\label{Acht1}
\end{eqnarray}\\
which using eq.~(\ref{SUM2}) reads\\
\begin{eqnarray}
     (\ref{Acht1})
     &=& -\,\frac{1}{8}\,\left[\,(D-2\,j)^2-D\,\right]\,\left[\,\Gamma^{\mu\nu\alpha\delta},\,\Gamma^{(j)}\,\right]~.\hspace{18ex}\label{Acht2}
\end{eqnarray}\\
Alternatively, using (\ref{Prop_14}), it can be rewritten as\\ 
\begin{eqnarray}
     (\ref{Acht1})
     &=& \frac{1}{8}\,
         \left\{\,
                   \Gamma^{\mu\nu\alpha\delta}{}_{\beta\gamma}
                   \,+\,8\,
                   \delta^{[\delta}_{[\beta}\Gamma^{\mu\nu\alpha
     ]}{}_{\gamma ]}
                   \,+\,2\cdot 6\cdot
                   \delta^{[\delta}_{[\beta}\delta^\alpha_{\gamma
     ]}\Gamma^{\mu\nu ]}\,
         \right\}\,
         \Gamma^{(j)}\,\Gamma^{\beta\gamma}\nonumber\\
      &-&\frac{1}{8}\,\Gamma^{\beta\gamma}\,\Gamma^{(j)}\,
         \left\{\,
                   \Gamma_{\beta\gamma}{}^{\mu\nu\alpha\delta}
                   \,+\,8\,
                   \delta^{[\mu}_{[\gamma}\Gamma_{\beta ]}{}^{\nu\alpha\delta
     ]}
                   \,+\,2\cdot 6\cdot
                   \delta^{[\mu}_{[\gamma}\delta^\nu_{\beta
     ]}\Gamma^{\alpha\delta ]}\,
         \right\}\,\label{Acht3}
\end{eqnarray}\\
Comparing (\ref{Acht2}) with (\ref{Acht3}) one obtains\\
\begin{eqnarray*}
     \frac{1}{8}\,\Gamma^{\mu\nu\alpha\delta}{}_{\beta\gamma}\,
               \Gamma^{(j)}\,\Gamma^{\beta\gamma}
               \,-\,
               \frac{1}{8}\,\Gamma^{\beta\gamma}\,\Gamma^{(j)}\,
               \Gamma^{\mu\nu\alpha\delta}{}_{\beta\gamma}
     &=&
               -\,\frac{1}{8}\,\left[\,(D-2\,j)^2-D\,\right]\,\left[\,\Gamma^{\mu\nu\alpha\delta},\,\Gamma^{(j)}\,\right]\\
      &&-\,
\frac{1}{8}\,
         \left\{\,
                   8\,
                   \delta^{[\delta}_{[\beta}\Gamma^{\mu\nu\alpha
     ]}{}_{\gamma ]}
                   \,+\,2\cdot 6\cdot
                   \delta^{[\delta}_{[\beta}\delta^\alpha_{\gamma
     ]}\Gamma^{\mu\nu ]}\,
         \right\}\,
         \Gamma^{(j)}\,\Gamma^{\beta\gamma}\nonumber\\
      &&+\,\frac{1}{8}\,\Gamma^{\beta\gamma}\,\Gamma^{(j)}\,
         \left\{\,
                   8\,
                   \delta^{[\mu}_{[\gamma}\Gamma_{\beta ]}{}^{\nu\alpha\delta
     ]}
                   \,+\,2\cdot 6\cdot
                   \delta^{[\mu}_{[\gamma}\delta^\nu_{\beta
     ]}\Gamma^{\alpha\delta ]}\,
         \right\}\,
\end{eqnarray*}\\
The left hand side is exactly what we denoted by (1) in (\ref{Acht0}).
The two terms with the factors ``$2\cdot 6$'' cancel each other and one
obtains\\
\begin{eqnarray}
     (1)
     &=&
               -\,\frac{1}{8}\,\left[\,(D-2\,j)^2-D\,\right]\,\left[\,\Gamma^{\mu\nu\alpha\delta},\,\Gamma^{(j)}\,\right]\nonumber\\
      &&-\,\underbrace{
                        \delta^{[\delta}_{[\beta}
                        \Gamma^{\mu\nu\alpha ]}{}_{\gamma ]}\,
                        \Gamma^{(j)}\,\Gamma^{\beta\gamma}
                      }_{
                        \Gamma^{[\mu\nu\alpha |}{}_{\gamma}\,
                        \Gamma^{(j)}\,\Gamma^{|\delta ]\gamma}
                      }
      \,+\,\underbrace{
                        \Gamma^{\beta\gamma}\,\Gamma^{(j)}\,
                        \delta^{[\mu}_{[\gamma}\Gamma_{\beta ]}
                        {}^{\nu\alpha\delta ]}
                      }_{
                        \Gamma^{\gamma [\mu |}\,\Gamma^{(j)}\,
                        \Gamma_{\gamma}{}^{|\nu\alpha\delta ]}
                      }\hspace{2ex}\label{Acht4}
\end{eqnarray}\\
To get rid of the contractions over $\gamma$ in the last two terms 
we obtain by (\ref{SUM1})\\ 
\begin{eqnarray*}
    -\,\Gamma^{[\mu\nu\alpha |}{}_{\gamma}\,
    \Gamma^{(j)}\,\Gamma^{|\delta ]\gamma}
    &=& \left(\,
                    \Gamma^{[\mu\nu\alpha |}\,\Gamma_\gamma\,-\,
                    3\,\delta^{[\alpha}_\gamma \Gamma^{\mu\nu}\,
           \right)\,\Gamma^{(j)}\,
           \left(\,
                   \Gamma^\gamma\Gamma^{|\delta ]} 
                  \,-\,\eta^{\gamma |\delta ]}
           \right)\\
    &=& (-1)^j\,(D-2\,j)\,\Gamma^{[\mu\nu\alpha |}\,
        \Gamma^{(j)}\,\Gamma^{|\delta ]}
        \,-\,\Gamma^{\mu\nu\alpha\delta}\,\Gamma^{(j)}
        \,-\,3\,\delta^{[\alpha}_\gamma
              \Gamma^{\mu\nu}\,\Gamma^{(j)}\,
              \Gamma^\gamma\Gamma^{|\delta ]} 
        \,+\,0
\end{eqnarray*}\\ 
and analogously\\ 
\begin{eqnarray*}
    \Gamma^{\gamma [\mu |}\,\Gamma^{(j)}\,
    \Gamma_{\gamma}{}^{|\nu\alpha\delta ]}
    &=& -\,(-1)^j\,(D-2\,j)\,\Gamma^{[\mu |}\,\Gamma^{(j)}\,
    \Gamma^{|\nu\alpha\delta ]}
    \,+\,3\,\Gamma^{[\mu |}\,\Gamma^\gamma\,\Gamma^{(j)}\,
    \delta_{\gamma}^{|\nu}\Gamma^{\alpha\delta ]}
    \,+\,\Gamma^{(j)}\,\Gamma^{\mu\nu\alpha\delta}~.
\end{eqnarray*}\\
Plugging the last two results back into (\ref{Acht4}) the two terms with
the factor ``3'' cancel each other and one obtains\\
\begin{eqnarray}
     (1)
     &=&
               -\,\frac{1}{8}\,\left[\,(D-2\,j)^2-D+8\,\right]\,
                \left[\,
                         \Gamma^{\mu\nu\alpha\delta},\,\Gamma^{(j)}\,
                \right]\nonumber\\
      && +\,(-1)^j\,(D-2\,j)\,
          \left\{\,
                      \Gamma^{[\mu\nu\alpha |}\,
                      \Gamma^{(j)}\,\Gamma^{|\delta ]}\,-\,
                      \Gamma^{[\mu |}\,
                      \Gamma^{(j)}\,\Gamma^{|\nu\alpha\delta ]}
          \right\}\label{Acht5}
\end{eqnarray}\\
It is useful to observe, that the expression in curly brackets has the
same tensor structure as (3) in (\ref{Acht0}). 

\subsubsection*{Ad (2) :}
In order to simplify the terms in (2) we proceed similarly to the 
previous calculations:\\
\begin{eqnarray}
 (2) &=&  \,-\,\frac{1}{4}\,
               \Gamma^{\mu\nu\alpha}{}_\beta{}^{\delta}\,
               \Gamma^{(j)}\,\Gamma^{\beta}
               \,+\,
               \frac{1}{4}\,
               \Gamma^{\beta}\,\Gamma^{(j)}\,
               \Gamma^{\mu\nu\alpha}{}_\beta{}^{\delta}\nonumber\\
     &=&       \frac{1}{4}\,
               \left\{\,
                          \Gamma^{\mu\nu\alpha\delta}\,\Gamma_{\beta}\,
                        -4\,\delta^{[\delta}_\beta
                           \Gamma^{\mu\nu\alpha ]}\,
               \right\}\,
               \Gamma^{(j)}\,\Gamma^{\beta}
               \,-\,
               \frac{1}{4}\,
               \Gamma^{\beta}\,\Gamma^{(j)}\,
               \left\{\,
                          \Gamma_{\beta}\,\Gamma^{\mu\nu\alpha\delta}\,
                        -4\,\delta^{[\mu}_\beta
                           \Gamma^{\nu\alpha\delta ]}\,
               \right\}\nonumber\\
      &=& \frac{(-1)^j\,(D-2\,j)}{4}\,
          \left[\,\Gamma^{\mu\nu\alpha\delta},\,\Gamma^{(j)}\,\right] 
          \,-\,\Gamma^{[\mu\nu\alpha |}\,\Gamma^{(j)}\,
               \Gamma^{|\delta ]}
          \,+\,\Gamma^{[\mu |}\,\Gamma^{(j)}\,
               \Gamma^{|\nu\alpha\delta ]}\label{Vier1}
\end{eqnarray}

\subsubsection*{Final checks:}
With the help of (\ref{Acht5}) and (\ref{Vier1}) equation (\ref{Acht0})
now reads\\
\begin{eqnarray}
  \left\{\vbox{\vspace{2.5ex}}\right.\,
       j^{\rm th}-{\rm term,~c.p.}
  \left.\vbox{\vspace{2.5ex}}\right\}
        &=& 
               -\,\frac{1}{8}\,\left[\,(D-2\,j)^2-D+8-2\,(-1)^j\,(D-2\,j)\,\right]\,
                \left[\,
                         \Gamma^{\mu\nu\alpha\delta},\,\Gamma^{(j)}\,
                \right]\nonumber\\
      && +\,\left[\,(-1)^j\,(D-2\,j)\,-\,1\,+\,2\,\right]\,
          \left\{\,
                      \Gamma^{[\mu\nu\alpha |}\,
                      \Gamma^{(j)}\,\Gamma^{|\delta ]}\,-\,
                      \Gamma^{[\mu |}\,
                      \Gamma^{(j)}\,\Gamma^{|\nu\alpha\delta ]}
          \right\}\hspace{8ex}\label{j_term_cp}
\end{eqnarray}\\
This must now be evaluated for all the three values $j$ might take on.\\ 

\noindent
\underline{\underline{$j\,=\,1$\,:}}\hspace{2ex} $\Rightarrow~$
$~\Gamma^{(j)}\,=\,\Gamma_{c_1}$ ~and (\ref{j_term_cp}) reads\\
\begin{eqnarray*}
  \left\{\vbox{\vspace{2.5ex}}\right.\,
       1^{\rm st}-{\rm term,~c.p.}
  \left.\vbox{\vspace{2.5ex}}\right\}
        &=& 
               -\,12\,
                \left[\,
                         \Gamma^{\mu\nu\alpha\delta},\,\Gamma_{c_1}\,
                \right]
          \,-\,8\,
          \left\{\,
                      \Gamma^{[\mu\nu\alpha}\,
                      \Gamma_{c_1}\,\Gamma^{\delta ]}\,-\,
                      \Gamma^{[\mu}\,
                      \Gamma_{c_1}\,\Gamma^{\nu\alpha\delta ]}
          \right\}
\end{eqnarray*}\\
With\\ 
\begin{eqnarray*}
     \left[\,
           \Gamma^{\mu\nu\alpha\delta},\,\Gamma_{c_1}\,
     \right]
     &=& -\,8\,\delta_{c_1}^{[\mu}\,\Gamma^{\nu\alpha\delta ]}
\end{eqnarray*}\\
\begin{eqnarray*}
  \left\{\vbox{\vspace{2.5ex}}\right.\,
       1^{\rm st}-{\rm term,~c.p.}
  \left.\vbox{\vspace{2.5ex}}\right\}
        &=& 
               96\,\delta_{c_1}^{[\mu}\,\Gamma^{\nu\alpha\delta ]}
          \,-\,8\,
          \left\{\,
                      \Gamma^{[\mu\nu\alpha}\,
                      \Gamma_{c_1}\,\Gamma^{\delta ]}\,-\,
                      \Gamma^{[\mu}\,
                      \Gamma_{c_1}\,\Gamma^{\nu\alpha\delta ]}
          \right\}\\
        &=& 
               96\,\delta_{c_1}^{[\mu}\,\Gamma^{\nu\alpha\delta ]}
          \,-\,8\,
          \left\{\vbox{\vspace{2.5ex}}\right.\,
                      \Gamma^{[\mu\nu\alpha}\,
                      \left(\,
                                -\,\Gamma^{\delta ]}\,\Gamma_{c_1}
                                \,+\,2\,\delta^{\delta ]}_{c_1}
                      \right)
                      \,-\,
                      \left(\,
                                -\,\Gamma_{c_1}\Gamma^{[\mu }\,
                                +\,2\,\delta_{c_1}^{[\mu}\,
                      \right)\,\Gamma^{\nu\alpha\delta ]}
          \left.\vbox{\vspace{2.5ex}}\right\}\\
      &=& 
          \left(\,96+16+16-64\,\right)\,\delta_{c_1}^{[\mu}\,\Gamma^{\nu\alpha\delta ]}
      ~=~ 64\,\delta_{c_1}^{[\mu}\,\Gamma^{\nu\alpha\delta ]}
\end{eqnarray*}\\
This $64$ exactly matches with opposite sign the only closed term presented
in (\ref{CJSFierIdentity}).\\

\noindent
\underline{\underline{$j\,=\,2$\,:}}\hspace{2ex} $\Rightarrow~$
$~\Gamma^{(j)}\,=\,\Gamma_{c_1c_2}$ ~and (\ref{j_term_cp}) reads\\
\begin{eqnarray*}
  \left\{\vbox{\vspace{2.5ex}}\right.\,
       2^{\rm nd}-{\rm term,~c.p.}
  \left.\vbox{\vspace{2.5ex}}\right\}
        &=& -\,4\,
                \left[\,
                         \Gamma^{\mu\nu\alpha\delta},\,\Gamma_{c_1c_2}\,
                \right]
          \,+\,8\,
          \left\{\vbox{\vspace{2.5ex}}\right.\,
                  \underbrace{
                      \Gamma^{[\mu\nu\alpha |}\,
                      \Gamma_{c_1c_2}\,\Gamma^{|\delta ]}\,-\,
                      \Gamma^{[\mu |}\,
                      \Gamma_{c_1c_2}\,\Gamma^{|\nu\alpha\delta ]}
                  }_{(\ast)}\,
          \left.\vbox{\vspace{2.5ex}}\right\}
\end{eqnarray*}\\
With\\ 
\begin{eqnarray*}
     \left[\,
           \Gamma^{\mu\nu\alpha\delta},\,\Gamma_{c_1c_2}\,
     \right]
     &=& 16\,\delta^{\mu}_{[c_1}\,\Gamma_{c_2]}{}^{\nu\alpha\delta ]}
\end{eqnarray*}\\
and\\
\begin{eqnarray*}
   (\ast)
   &=&
   -\,g^{\beta [\alpha}\,\Gamma^{\delta\mu\nu ]}\,\Gamma_{c_1c_2}\,\Gamma_\beta
   \,-\,
   \Gamma_\beta\,\Gamma_{c_1c_2}\,g^{\beta
   [\alpha}\,\Gamma^{\delta\mu\nu ]}\\
   &=&
   -\,g^{\beta [\alpha}\,\Gamma^{\delta\mu\nu ]}\,
    \left\{\,
              \Gamma_{c_1c_2\beta}\,
         +\,2\,\eta_{\beta[c_2}\Gamma_{c_1 ]}\,
    \right\}
   \,-\,
   \left\{\,
            \Gamma_{\beta c_1c_2}\,+\,2\,\eta_{\beta[c_1}\Gamma_{c_2 ]}\,
   \right\}\,g^{\beta
   [\alpha}\,\Gamma^{\delta\mu\nu ]}\\
   &=&
    -\,g^{\beta [\alpha}\,
    \left\{\,
             \Gamma^{\delta\mu\nu ]}{}_{c_1c_2\beta}\,
            +\,9\,\delta_{[c_1}^{\nu}
                  \Gamma^{\delta\mu ]}{}_{c_2\beta ]}\,
            +\,18\,\delta_{[c_1}^{\nu}\delta^\mu_{c_2}
                  \Gamma^{\delta ]}{}_{\beta ]}\,
            +\,6\,\delta_{[c_1}^{\nu}\delta^\mu_{c_2}
                  \delta^{\delta ]}_{\beta ]}\,
    \right\}\\
   &&
   -\,g^{\beta
   [\alpha}\,
   \left\{\,
            \Gamma_{\beta c_1c_2}{}^{\delta\mu\nu ]}\,
            +\,9\,\delta_{[c_2}^{\delta}
                  \Gamma_{\beta c_1 ]}{}^{\mu\nu ]}\,
            +\,18\,\delta_{[c_2}^{\delta}\delta^\mu_{c_1}
                  \Gamma_{\beta ]}{}^{\nu ]}
            +\,6\,\delta_{[c_2}^{\delta}\delta^\mu_{c_1}
                  \delta^{\nu ]}_{\beta ]}\,
   \right\}\\
   &&
   -2\,g^{\beta[\alpha}\,\Gamma^{\delta\mu\nu ]}\,\eta_{\beta
   [c_2}\,\Gamma_{c_1]}\,
   -2\,\eta_{\beta [c_1}\,\Gamma_{c_2 ]}\,
    g^{\beta [\alpha}\,\Gamma^{\delta\mu\nu ]}\\
   &=& -\,18\,g^{\beta [\alpha}\delta_{[c_1}^\nu\,\Gamma^{\delta\mu
   ]}{}_{c_2\beta ]}
   -2\,g^{\beta[\alpha}\,\Gamma^{\delta\mu\nu ]}\,\eta_{\beta
   [c_2}\,\Gamma_{c_1]}\,
   -2\,\eta_{\beta [c_1}\,\Gamma_{c_2 ]}\,
    g^{\beta [\alpha}\,\Gamma^{\delta\mu\nu ]}\\
   &=& -\,18\,g^{\beta [\alpha}\delta_{[c_1}^\nu\,\Gamma^{\delta\mu
   ]}{}_{c_2\beta ]}
   -2\,g^{\beta[\alpha}\,\eta_{\beta [c_2}\,
   \left\{\,
             \Gamma^{\delta\mu\nu ]}{}_{c_1]}\,
             +\,3\,\delta^\nu_{c_1]}\Gamma^{\delta\mu ]}\,
   \right\}\\
   &&\hspace{21ex}
   -2\;\eta_{\beta [c_1}\,g^{\beta [\alpha}\;
    \left\{\,
             \Gamma_{c_2 ]}{}^{\delta\mu\nu ]}\,
             +\,3\,\delta^\delta_{c_2]}\Gamma^{\mu\nu ]}\,
    \right\}\\
    &=& -\,18\,g^{\beta [\alpha}\delta_{[c_1}^\nu\,\Gamma^{\delta\mu
   ]}{}_{c_2\beta ]}
    -4\,\delta^{[\alpha}_{[c_2}\,
             \Gamma^{\delta\mu\nu ]}{}_{c_1]}\\
     &=& 12\,\delta_{[c_1}^{[\nu}\,\Gamma^{\delta\mu\alpha]}{}_{c_2]}
    -4\,\delta^{[\alpha}_{[c_2}\,
             \Gamma^{\delta\mu\nu ]}{}_{c_1]}\\
     &=& 8\,\delta_{[c_1}^{[\mu}\,\Gamma_{c_2]}{}^{\nu\alpha\delta]}
\end{eqnarray*}\\
we obtain\\
\begin{eqnarray*}
  \left\{\vbox{\vspace{2.5ex}}\right.\,
       2^{\rm nd}-{\rm term,~c.p.}
  \left.\vbox{\vspace{2.5ex}}\right\}
        &=& \left(\,-\,64\,+\,64\,\right)\,
              \delta^{\mu}_{[c_1}\,\Gamma_{c_2]}{}^{\nu\alpha\delta ]}
        ~~{\buildrel !\over =}~~ 0.
\end{eqnarray*}

\noindent
\underline{\underline{$j\,=\,5$\,:}}\hspace{2ex} $\Rightarrow~$
$~\Gamma^{(j)}\,=\,\Gamma^{c_1\ldots c_5}$\hspace{3ex}
This case is left as an exercise.\\[2ex]
\end{proof}

\subsubsection*{Homework:}

\begin{exercise} 
    Complete the above proof by doing the case $j\,=\,5$.
\end{exercise}

\vfill\eject
\section{Eleven dimensional Supergravity \`a la Cremmer-Julia-Scherk}
\label{Sec_MTheory}

In a seminal paper Cremmer, Julia and Scherk derived the Lagrangian
of the unique 11d supergravity nowadays mostly referred to as the low 
energy limit of M-theory \cite{Cremmer:1978km}.
It is an eleven dimensional theory of gravity involving a set of 
massless fields in eleven dimensions which carry a 
representation of supersymmetry. Since supersymmetry assigns to each
bosonic degree of freedom a corresponding fermionic one and vice versa
a consequence
of the last property is that the number of physical degrees of freedom  
for fermionic and bosonic fields must match. 
Physical degrees of freedom for massless particles can be counted by
choosing light cone gauge, i.e. by classifying the massless fields 
according to the little group $SO(9)$. A very nice physical 
discussion of how this counting works in detail can be found 
in \cite{Theiss}.
Massless fields must transform as
irreducible representations of the little group.

\noindent
Being a theory of gravity, the metric $g_{\mu\nu}$ should appear 
in the set of fields. The same must be true for the  gravitino 
$\psi_\mu^\alpha$, which is the superpartner of the metric. The
corresponding degrees of freedom of metric and gravitino have to form
irreducible representations of tensor products of $SO(9)$. The metric
is a symmetric tensor with two indices and thus contained in\\
\begin{itemize}
\item $[{\bf 9}_V\otimes {\bf 9}_V]_{\rm symm} = [{\bf 1}
\oplus {\bf 36}\oplus {\bf 44}]_{\rm symm} = {\bf 1} \oplus {\bf 44}$.\\
\end{itemize} 
Here the ${\bf 1}$ corresponds to the trace and the ${\bf 44}$ to the 
symmetric traceless tensor. The metric is identified with the ${\bf
44}$, while the ${\bf 1}$, usually called dilaton, does not occur as a 
dynamical field in eleven dimensional supergravity. Similarly the
gravitino is contained in the tensor product of the fundamental vector 
representation and the spinor representation of $SO(9)$, which itself 
is not irreducible but decomposes into two irreducible components:\\
\begin{itemize}
\item ${\bf 16}_{S}\otimes {\bf 9}_V = {\bf 16}_S\oplus {\bf 128}$\\
\end{itemize}
Obviously, the gravtino belongs to the ${\bf 128}$.

\noindent
Supersymmetry requires matching of bosonic and fermionic degrees of
freedom on-shell. The big mismatch between  ${\bf 44}$ and ${\bf 128}$
seems to require the introduction of another bosonic field. It turns
out that the product of three vector representation contains a {\bf 84}
dimensional representation in its totally antisymmetric part\\
\begin{itemize}
\item $[{\bf 9}_V\otimes {\bf 9}_V\otimes {\bf 9}_V]_{\rm antisym} =
  [{\bf 36}\otimes {\bf 9}_V]_{\rm antisym} = [{\bf 84} \oplus {\bf
    9}\oplus {\bf 231}]_{\rm antisym} = {\bf 84}$ \\
\end{itemize}
Using this three form potential, the boson/fermion count now reads\\ 
\begin{eqnarray*}
    {\bf 44}\oplus {\bf 84} = {\bf 128}
\end{eqnarray*}\\
From the point of view of counting the degrees of freedom there is a
chance to find a supersymmetric theory containing the metric, a
gravitino and a three form antisymmetric tensor field $C_{\mu\nu\rho}$
subject to suitable constraints in order to get rid of superfluous
degrees of freedom as discussed before. The construction of the theory 
from scratch will be postponed until section \ref{Sec_Noether}. 
At this point, we make do with giving the final result for the
Lagrangian whose details depends on the conventions chosen. The vast
majority of research papers uses conventions which differ from the ones
used in  the original paper \cite{Cremmer:1978km}.
Using the following redefinitions \\
\begin{itemize}
\item $\kappa\,=\,1$\\[-1.5ex]
\item $F_{\mu\nu\rho\sigma}\,=\,\frac{1}{2}\,G_{\mu\nu\rho\sigma}$\\[-1.5ex]
\item $\Gamma^a\mapsto i\Gamma^a$ \\
\end{itemize}
one obtains from the original Lagrangian (see appendix \ref{App_CJS_Act}) 
the Lagrangian in our conventions\footnote
{
  $\varepsilon^{0123456789\,10}\,=\,-1$
}\\
\begin{eqnarray}\label{MTheory}
   {\cal L} &=& \frac{1}{4}\,eR 
                +\frac{1}{2}\,e\bar{\psi}_\mu\Gamma^{\mu\nu\rho}
                 D_\nu\left(\frac{\omega\,+\,\hat{\omega}}{2}\right)\psi_\rho
                -\frac{1}{4\cdot 48}\,eG_{\mu\nu\rho\sigma}G^{\mu\nu\rho\sigma}
                 \nonumber\\
             && -\frac{1}{4\cdot 48}\,e 
                 \left(
                   \bar{\psi}_\mu\Gamma^{\mu\nu\alpha\beta\gamma\delta}
                   \psi_\nu
                   +12\,\bar{\psi}^\alpha\Gamma^{\gamma\delta}\psi^\beta
                 \right)\, 
                 \left(\,
                          \frac{G_{\alpha\beta\gamma\delta}\,+\,
                          \hat{G}_{\alpha\beta\gamma\delta}}{2}\,
                 \right)\nonumber\\
             && +\frac{1}{4\cdot 144^2}\,\epsilon^{\alpha_1\ldots\alpha_4
                 \beta_1\ldots\beta_4\mu\nu\rho}G_{\alpha_1\ldots\alpha_4}
                 G_{\beta_1\ldots\beta_4}C_{\mu\nu\rho}~.
\end{eqnarray}\\
the action of which is supposed to be invariant under the following 
supersymmetric transformation rules:\\ 
\begin{eqnarray}
  \delta_{Q}e_\mu{}^a &=&  \bar\epsilon\,\Gamma^a\psi_\mu\nonumber\\
  \delta_{Q}C_{\mu\nu\rho} &=& 3\,\bar\epsilon\,
                               \Gamma_{[\mu\nu}\psi_{\rho]}\nonumber\\
  \delta_{Q}\psi_\mu{} &=& D_\mu(\hat{\omega})\epsilon
                           -\frac{1}{2\cdot 144}
                           \left(
                                 \Gamma^{\alpha\beta\gamma\delta}{}_{\mu}
                                  -8\,\Gamma^{\beta\gamma\delta}
                                     \delta_\mu^\alpha
                           \right)\,\epsilon\,\hat{G}_{\alpha\beta\gamma\delta}
                           ~=~ \hat{D}_\mu\epsilon\label{SUSYTRAFO}
\end{eqnarray}\\
The signature of the metric is $\eta_{ab}\,=\,(-1,1,\ldots,1)$ and the 
$\Gamma$-matrices are in a real representation of the 
Clifford algebra\\
\begin{eqnarray*}
         \{\,\Gamma^a,\,\Gamma^b\,\} &=& 2\,\eta^{ab}\,\unity_{32}~.
\end{eqnarray*}\\ 
Apart from the elementary fields (vielbein $e^a{}_\mu$, gravitino
$\psi_\mu$ and antisymmetric tensor field
$A_{\mu\nu\rho}$) the abbreviations in the above Lagrangian have the 
following meaning:\\
\begin{eqnarray*}
   D_\nu(\omega)\psi_\mu &=& \partial_\nu\psi_\mu\,-\,\frac{1}{4}\,
                             \omega_{\nu ab}\,\Gamma^{ab}\,\psi_\mu\\[2ex]
   G_{\mu\nu\rho\sigma} &=& 4\,\partial_{[\mu}\,C_{\nu\rho\sigma]}\\[2ex]
   \hat{G}_{\mu\nu\rho\sigma} &=& G_{\mu\nu\rho\sigma} ~+~ 6\,
                                  \bar{\psi}_{[\mu}\Gamma_{\nu\rho}
                                  \psi_{\sigma]}\\[2ex]
   K_{\mu ab} &=& \frac{1}{4}\,
                  \left[\,
                            \bar{\psi}_{\alpha}\Gamma_{\mu ab}{}^{\alpha\beta}
                            \psi_{\beta} ~-~ 2\,
                            \left(\,
                                     \bar{\psi}_{\mu}\Gamma_{b}\psi_{a}\,
                                  -\,\bar{\psi}_{\mu}\Gamma_{a}\psi_{b}\,
                                  +\,\bar{\psi}_{b}\Gamma_{\mu}\psi_{a}\,
                            \right)\,
                  \right]\quad {\rm (\,contorsion\,)}\\[2ex]
   \omega_{\mu ab} &=& \underbrace{
                                     \omega_{\mu ab}^{(0)}
                                  }_{
                                     \hbox{\scriptsize{Christ}}
                                  } ~+~ K_{\mu ab}\\
   \hat{\omega}_{\mu ab} &=& \omega_{\mu ab} ~-~ \frac{1}{4}\,
                             \bar{\psi}_{\alpha}\,
                             \Gamma_{\mu ab}{}^{\alpha\beta}\,
                             \psi_{\beta}
\end{eqnarray*}

\vfill\eject
\subsection{Equation of Motion}
\label{SubSec_EOM}

\subsubsection{$\psi_{\mu}$}

\begin{eqnarray*}
   {\cal L}_{\psi} 
             &=& \frac{1}{2}\,\bar{\psi}_\mu\Gamma^{\mu\nu\rho}
                 D_\nu\left(\frac{\omega\,+\,\hat{\omega}}{2}\right)\psi_\rho
             ~-~ \frac{1}{4\cdot 48}\, 
                 \left(
                   \bar{\psi}_\mu\Gamma^{\mu\nu\alpha\beta\gamma\delta}
                   \psi_\nu
                   +12\,\bar{\psi}^\alpha\Gamma^{\gamma\delta}\psi^\beta
                 \right)\, 
                 \left(\,
                           \hat{G}_{\alpha\beta\gamma\delta}\,-\,3\,
                           \bar{\psi}_{[\alpha}
                           \Gamma_{\beta\gamma}
                           \psi_{\delta]}
                 \right)
\end{eqnarray*}

\begin{eqnarray*}
    0 &=& \frac{\partial {\mathcal{L}}}{\partial \bar\psi_\xi^n} ~-~
          \frac{\partial}{\partial x^\omega}\,
          \left(\,
                   \frac{\partial {\mathcal{L}}}
                   {\partial\left(\,\partial_\omega \bar\psi^n_\xi\,\right)}\,
          \right)
      ~\buildrel !\over =~
      \frac{\partial {\mathcal{L}}}{\partial \bar\psi_\xi^n}
\end{eqnarray*}

\begin{eqnarray}
    0 &=& \frac{1}{2}\,\Gamma^{\xi\nu\rho}\,
          \left[\,
                  D_\nu(\hat{\omega}) \,-\,\frac{1}{32}\,\bar\psi_\alpha
                  \Gamma_{\nu ab}{}^{\alpha\beta}\,\psi_\beta\,\Gamma^{ab}
          \right]\,\psi_\rho
      ~+~\frac{1}{64}\,\Gamma_{\nu ab}{}^{\xi\beta}\,\psi_\beta\,
         \bar\psi_\mu\,\Gamma^{\mu\nu\rho}\,\Gamma^{ab}\,\psi_\rho
         \nonumber\\
      &&
      ~-~\frac{1}{4\cdot 48}\,
         \left(\,
                 \Gamma^{\xi\nu}{}_{\alpha\beta\gamma\delta}\,
                 \psi_\nu ~+~ 12\,\delta^{\xi}_{\alpha}\,\Gamma_{\gamma\delta}
                 \psi_{\beta}
         \right)\,
         \left(\,
                 \hat{G}^{\alpha\beta\gamma\delta} \,-\,
                 3\,\bar\psi^{[\alpha}\Gamma^{\beta\gamma}\psi^{\delta]}
         \right)\nonumber\\
      &&
      ~-~\frac{3}{4\cdot 48}\,
         \delta^\xi_{[\alpha}
         \Gamma^{\phantom{\xi}}_{\beta\gamma}
         \psi^{\phantom{\xi}}_{\delta]}\,
         \left(\,
                 \bar\psi_\mu\,\Gamma^{\mu\nu\alpha\beta\gamma\delta}\,
                 \psi_\nu ~+~ 12\,\bar\psi^\alpha\,\Gamma^{\gamma\delta}
                 \psi^{\beta}
         \right)\nonumber\\[4ex]
    &=& \frac{1}{2}\,
          \left[\,
                  \Gamma^{\xi\nu\rho}\,
                  D_\nu(\hat{\omega}) \,-\,\frac{1}{96}\,
                  \left(
                         \Gamma^{\xi\rho}{}_{\alpha\beta\gamma\delta}
                         ~+~12\,\delta^{\xi}_{\alpha}\,
                          \Gamma_{\gamma\delta}\delta^\rho_\beta
                  \right)\,\hat{G}^{\alpha\beta\gamma\delta}
          \right]\,\psi_\rho\nonumber\\
      &&
      ~-~\frac{1}{64}\,\Gamma^{\xi\nu\rho}\,\Gamma^{ab}\,\psi_\rho\,
         \bar\psi_\alpha
         \Gamma_{\nu ab}{}^{\alpha\beta}\,\psi_\beta
      ~+~\frac{1}{64}\,\Gamma_{\nu ab}{}^{\xi\beta}\,\psi_\beta\,
         \bar\psi_\mu\,\Gamma^{\mu\nu\rho}\,\Gamma^{ab}\,\psi_\rho
         \nonumber\\
      &&
      ~+~\frac{1}{64}\,
         \left(\,
                 \Gamma^{\xi\nu}{}_{\alpha\beta\gamma\delta}\,
                 \psi_\nu\, 
         \bar\psi^{[\alpha}\Gamma^{\beta\gamma}\psi^{\delta]}
      ~-~
         \delta^\xi_{[\alpha}
         \Gamma^{\phantom{\xi}}_{\beta\gamma}
         \psi^{\phantom{\xi}}_{\delta]}\;
                 \bar\psi_\mu\,\Gamma^{\mu\nu\alpha\beta\gamma\delta}\,
                 \psi_\nu\, 
         \right)
       \label{EOM_PSI}
\end{eqnarray}

\begin{prop}\label{NiceRelation}
\begin{eqnarray*}
    \Gamma^{\mu\nu\rho}\,
    \left(
           \Gamma^{\alpha\beta\gamma\delta}{}_{\nu}
           -8\,\Gamma^{\beta\gamma\delta}\,\delta_\nu^\alpha
    \right)\,\psi_{\rho}\,\hat{G}_{\alpha\beta\gamma\delta} 
    &=& 3\, 
    \left(
           \Gamma^{\mu\rho}{}_{\alpha\beta\gamma\delta}
           ~+~12\,\delta^{\mu}_{\alpha}\,\Gamma_{\gamma\delta}\delta^\rho_\beta
    \right)\,\psi_\rho\,\hat{G}^{\alpha\beta\gamma\delta}
\end{eqnarray*}
\end{prop}

\noindent
\begin{proof}
\noindent
We quote two useful identities here. The first is simply  a  
variant of (\ref{Prop_14}).\\
\begin{eqnarray}
    \Gamma^a\Gamma^{a_1\ldots a_j} &=& \Gamma^{a a_1\ldots a_j} 
    ~+~ j\,\eta^{a[a_1}\Gamma^{a_2\ldots a_j]}\label{Prop13}\\[2ex]
    \Gamma^d\Gamma_{a_1\ldots a_j}\Gamma_d 
    &=& (-1)^j\,(11-2j)\,\Gamma_{a_1\ldots a_j}\label{Prop16}
\end{eqnarray}\\
Using eq.~(\ref{Prop_14}) and eq.~(\ref{Prop13}) repeatedly one obtains\\
\begin{eqnarray*}
    \Gamma^{\mu\nu\rho}\,\Gamma_{\alpha\beta\gamma\delta\nu}
    &=& - \Gamma^{\nu\mu\rho}\,\Gamma_{\alpha\beta\gamma\delta\nu}\\
    &=& -\,
        \left(\vbox{\vspace{2ex}}\right.\,
                \Gamma^\nu\,\Gamma^{\mu\rho}\,-\,
                2\,\eta^{\nu [\mu}\,\Gamma^{\rho ]}
        \left.\vbox{\vspace{2ex}}\right)\,
        \left(\vbox{\vspace{2ex}}\right.\,
                \Gamma_{\alpha\beta\gamma\delta}\Gamma_\nu\,-\,
                4\,\Gamma_{[\alpha\beta\gamma}\eta_{\delta ]\nu}\,
        \left.\vbox{\vspace{2ex}}\right)\\
    &=& -\,\Gamma^\nu\,\Gamma^{\mu\rho}\,
           \Gamma_{\alpha\beta\gamma\delta}\Gamma_\nu
    ~+~ 4\,\Gamma^\nu\,\Gamma^{\mu\rho}\,
               \Gamma_{[\alpha\beta\gamma}\eta_{\delta ]\nu}\\
    &&  +\,2\,\eta^{\nu [\mu}\,\Gamma^{\rho ]}\,
              \Gamma_{\alpha\beta\gamma\delta}\Gamma_\nu
    ~-~ 8\,\eta^{\nu [\mu}\,\Gamma^{\rho ]}\,
              \Gamma_{[\alpha\beta\gamma}\eta_{\delta ]\nu}\\
    &=& -\,\Gamma^\nu\,
        \left\{\vbox{\vspace{2.5ex}}\right.\,
           \Gamma^{\mu\rho}{}_{\alpha\beta\gamma\delta}
           \,+\,8\,\delta^{[\rho}_{[\alpha}\,
                   \Gamma^{\mu ]}{}_{\beta\gamma\delta ]}
           \,+\,12\,\delta^{[\rho}_{[\alpha}\delta^{\mu ]}_{\beta}
                   \Gamma_{\gamma\delta ]}\,
        \left.\vbox{\vspace{2.5ex}}\right\}\,\Gamma_\nu\,\\
        && +\,4\,
              \left\{\vbox{\vspace{2.5ex}}\right.\,
                  \Gamma^{\nu\mu\rho}\,+\,2\,\eta^{\nu [\mu}\,\Gamma^{\rho ]}\,
              \left.\vbox{\vspace{2.5ex}}\right\}\,
               \Gamma_{[\alpha\beta\gamma}\eta_{\delta ]\nu}
        ~+~2\,\eta^{\nu [\mu}\,\Gamma^{\rho ]}\,
              \Gamma_{\alpha\beta\gamma\delta}\Gamma_\nu\\
        &&-\,8\,\eta^{\nu [\mu}\,\Gamma^{\rho ]}\,
              \Gamma_{[\alpha\beta\gamma}\eta_{\delta ]\nu}\\
     &=& +\,\Gamma^{\mu\rho}{}_{\alpha\beta\gamma\delta}
           \,-\,24\,\delta^{[\rho}_{[\alpha}\,
                   \Gamma^{\mu ]}{}_{\beta\gamma\delta ]}
           \,-\,84\,\delta^{[\rho}_{[\alpha}\delta^{\mu ]}_{\beta}
                   \Gamma_{\gamma\delta ]}\,\\
        &&+\,4\,\Gamma^{\nu\mu\rho}\,
               \Gamma_{[\alpha\beta\gamma}\eta_{\delta ]\nu}
        ~+~2\,\eta^{\nu [\mu}\,\Gamma^{\rho ]}\,
              \Gamma_{\alpha\beta\gamma\delta}\Gamma_\nu
\end{eqnarray*}\\
Now contracting with $\hat{G}^{\alpha\beta\gamma\delta}$ one obtains\\
\begin{eqnarray*}
    \Gamma^{\mu\nu\rho}\,\Gamma_{\alpha\beta\gamma\delta\nu}\,
    \hat{G}^{\alpha\beta\gamma\delta}
    &=& \left\{\vbox{\vspace{2.5ex}}\right.
           \Gamma^{\mu\rho}{}_{\alpha\beta\gamma\delta}
           \,-\,24\,\delta^{[\rho}_{\alpha}\,
                   \Gamma^{\mu ]}{}_{\beta\gamma\delta}
           \,-\,84\,\delta^{[\rho}_{\alpha}\delta^{\mu ]}_{\beta}
                   \Gamma_{\gamma\delta}\,\\
        &&-\,4\,\Gamma^{\nu\mu\rho}\,
               \Gamma_{\beta\gamma\delta}\eta_{\alpha\nu}
        ~+~2\,\eta^{\nu [\mu}\,\Gamma^{\rho ]}\,
              \Gamma_{\alpha\beta\gamma\delta}\Gamma_\nu
       \left.\vbox{\vspace{2.5ex}}\right\}\,
       \hat{G}^{\alpha\beta\gamma\delta}
\end{eqnarray*}\\
The explicit antisymmetrisation in $(\alpha,\beta,\gamma,\delta)$ of the 
second, third and fourth term could be neglected, since we contract
via an complete antisymmetric tensor $\hat{G}^{\alpha\beta\gamma\delta}$. 
We consider now\\ 
\begin{eqnarray}\label{ZwischenResultat}
   \Gamma^{\mu\nu\rho}\left(
           \Gamma_{\alpha\beta\gamma\delta\nu}
           -8\,\Gamma_{\beta\gamma\delta}\,\eta_{\nu\alpha}
    \right)\,\hat{G}^{\alpha\beta\gamma\delta}
    &=& \left\{\vbox{\vspace{2.5ex}}\right.
           \Gamma^{\mu\rho}{}_{\alpha\beta\gamma\delta}
           \,-\,24\,\delta^{[\rho}_{\alpha}\,
                   \Gamma^{\mu ]}{}_{\beta\gamma\delta}
           \,-\,84\,\delta^{[\rho}_{\alpha}\delta^{\mu ]}_{\beta}
                   \Gamma_{\gamma\delta}\,\\
        &&-\,4\,\Gamma^{\mu\nu\rho}\,
               \Gamma_{\beta\gamma\delta}\eta_{\alpha\nu}
        ~+~2\,\eta^{\nu [\mu}\,\Gamma^{\rho ]}\,
              \Gamma_{\alpha\beta\gamma\delta}\Gamma_\nu
       \left.\vbox{\vspace{2.5ex}}\right\}\,
       \hat{G}^{\alpha\beta\gamma\delta}\nonumber
\end{eqnarray}\\
Expanding the fourth term of eq.~(\ref{ZwischenResultat}) via 
eq.~(\ref{Prop_14}) one obtains:\\
\begin{eqnarray*}
   \Gamma^{\mu\nu\rho}\Gamma_{\beta\gamma\delta}\,\eta_{\nu\alpha}
   &=& \Gamma^{\mu}{}_{\alpha}{}^{\rho}{}_{\beta\gamma\delta}
       ~+~ 9\,\eta_{\nu\alpha}\,\delta^{[\rho}_{[\beta}
           \Gamma^{\mu\nu]}{}_{\gamma\delta]}
       ~+~ 18\,\eta_{\nu\alpha}\,\delta^{[\rho}_{[\beta}\delta^\nu_\gamma
           \Gamma^{\mu]}{}_{\delta]}
       ~+~ 6 \eta_{\nu\alpha}\,\delta^{[\mu}_{[\beta}\delta^\nu_\gamma
                               \delta^{\rho ]}_{\delta ]}
\end{eqnarray*}\\
The last term of eq.~(\ref{ZwischenResultat}) can be expanded as\\
\begin{eqnarray*}
     2\,\eta^{\nu [\mu}\,\Gamma^{\rho ]}\,
       \Gamma_{\alpha\beta\gamma\delta}\Gamma_\nu\,
     \hat{G}^{\alpha\beta\gamma\delta}
     &=&
     2\,\eta^{\nu [\mu}\,\Gamma^{\rho ]}\,
     \left\{\vbox{\vspace{2.5ex}}\right.
         \Gamma_{\alpha\beta\gamma\delta\nu}\,+\,
         4\,\Gamma_{[\alpha\beta\gamma}\eta_{\delta ] \nu}
     \left.\vbox{\vspace{2.5ex}}\right\}\,
     \hat{G}^{\alpha\beta\gamma\delta}\\
     &=&
     2\,\eta^{\nu [\mu}\,\delta^{\rho]}_{\kappa}\,\Gamma^{\kappa }\,
     \left\{\vbox{\vspace{2.5ex}}\right.
         \Gamma_{\alpha\beta\gamma\delta\nu}\,+\,
         4\,\Gamma_{\alpha\beta\gamma}\eta_{\delta\nu}
     \left.\vbox{\vspace{2.5ex}}\right\}\,
     \hat{G}^{\alpha\beta\gamma\delta}\\
     &=&
     2\,\eta^{\nu [\mu}\,\delta^{\rho]}_{\kappa}\,
     \left\{\vbox{\vspace{2.5ex}}\right.
         \Gamma^{\kappa}{}_{\alpha\beta\gamma\delta\nu}\,+\,
         5\delta^\kappa_{[\alpha}\,\Gamma_{\beta\gamma\delta\nu]}\,+\,
         4\,
         \left(\,
                   \Gamma^{\kappa}{}_{\alpha\beta\gamma}\,+\,
                   3\delta^{\kappa}_{[\alpha}\Gamma_{\beta\gamma ]}\,
         \right)\eta_{\delta\nu}
     \left.\vbox{\vspace{2.5ex}}\right\}\,
     \hat{G}^{\alpha\beta\gamma\delta}\\
     &=&
     2\,\eta^{\nu [\mu}\,\delta^{\rho]}_{\kappa}\,
     \left\{\vbox{\vspace{2.5ex}}\right.
         \Gamma^{\kappa}{}_{\alpha\beta\gamma\delta\nu}\,+\,
         4\,
         \left(\,
                   \Gamma^{\kappa}{}_{\alpha\beta\gamma}\,+\,
                   3\delta^{\kappa}_{\alpha}\Gamma_{\beta\gamma}\,
         \right)\eta_{\delta\nu}
         \,+\,
         \left(\vbox{\vspace{2ex}}\right.\,
                 \delta^\kappa_{\alpha}\,\Gamma_{\beta\gamma\delta\nu}\,+\,
                 \underbrace{
                    \delta^\kappa_{\nu}\,\Gamma_{\alpha\beta\gamma\delta}
                 }_{0}\\
         &&\hspace{11ex}             
                 +\,
                 \delta^\kappa_{\delta}\,\Gamma_{\nu\alpha\beta\gamma}\,+\,
                 \delta^\kappa_{\gamma}\,\Gamma_{\delta\nu\alpha\beta}\,+\,
                 \delta^\kappa_{\beta}\,\Gamma_{\gamma\delta\nu\alpha}\,
         \left.\vbox{\vspace{2ex}}\right)
     \left.\vbox{\vspace{2.5ex}}\right\}\,
     \hat{G}^{\alpha\beta\gamma\delta}\\
     &=& \left\{\vbox{\vspace{2.5ex}}\right.\,
             -\,2\,\Gamma^{\mu\rho}{}_{\alpha\beta\gamma\delta}
             \,+\,
             8\,\delta_{\delta}^{[\mu}\,\delta^{\rho]}_{\kappa}\,
             \left(\,
                   \Gamma^{\kappa}{}_{\alpha\beta\gamma}\,+\,
                   3\delta^{\kappa}_{\alpha}\Gamma_{\beta\gamma}\,
             \right)
             \,+\,8\,\eta^{\nu [\mu}\,\delta^{\rho]}_{\kappa}\,
             \delta^\kappa_{[\alpha}\,\Gamma_{\beta\gamma\delta] \nu}\,
         \left.\vbox{\vspace{2.5ex}}\right\}\,
         \hat{G}^{\alpha\beta\gamma\delta}\\
     &=& \left\{\vbox{\vspace{2.5ex}}\right.\,
             -\,2\,\Gamma^{\mu\rho}{}_{\alpha\beta\gamma\delta}
             \,+\,
             8\,\delta_{\delta}^{[\mu}\,
                   \Gamma^{\rho ]}{}_{\alpha\beta\gamma}\,+\,
             24\,\delta_{\delta}^{[\mu}\,\delta^{\rho]}_{\alpha}\,
               \Gamma_{\beta\gamma}
             \,-\,8\,\delta_{\delta}^{[\mu}\,
                  \Gamma^{\rho ]}{}_{\alpha\beta\gamma}\,
         \left.\vbox{\vspace{2.5ex}}\right\}\,
         \hat{G}^{\alpha\beta\gamma\delta}\\
      &=& \left\{\vbox{\vspace{2.5ex}}\right.\,
             -\,2\,\Gamma^{\mu\rho}{}_{\alpha\beta\gamma\delta}
             \,+\,
             24\,\delta_{\delta}^{[\mu}\,\delta^{\rho]}_{\alpha}\,
               \Gamma_{\beta\gamma}
         \left.\vbox{\vspace{2.5ex}}\right\}\,
         \hat{G}^{\alpha\beta\gamma\delta}
\end{eqnarray*}\\
Putting things together eq.~(\ref{ZwischenResultat}) reads\\
\begin{eqnarray}\label{Step1}
   \Gamma^{\mu\nu\rho}\left(
           \Gamma_{\alpha\beta\gamma\delta\nu}
           -8\,\Gamma_{\beta\gamma\delta}\,\eta_{\nu\alpha}
    \right)\,\hat{G}^{\alpha\beta\gamma\delta}
    &=& \left\{\vbox{\vspace{2.5ex}}\right.
           \Gamma^{\mu\rho}{}_{\alpha\beta\gamma\delta}
           \,-\,24\,\delta^{[\rho}_{\alpha}\,
                   \Gamma^{\mu ]}{}_{\beta\gamma\delta}
           \,-\,84\,\delta^{[\rho}_{\alpha}\delta^{\mu ]}_{\beta}
                   \Gamma_{\gamma\delta}\,\nonumber\\
        &&-\,4\,
          \left(\vbox{\vspace{2.5ex}}\right.\,
             \Gamma^{\mu}{}_{\alpha}{}^{\rho}{}_{\beta\gamma\delta}
             ~+~ 9\,\eta_{\nu\alpha}\,\delta^{[\rho}_{[\beta}
             \Gamma^{\mu\nu]}{}_{\gamma\delta]}
             ~+~ 18\,\eta_{\nu\alpha}\,\delta^{[\rho}_{[\beta}
                 \delta^\nu_\gamma\Gamma^{\mu]}{}_{\delta]}
             ~+~ \underbrace{
                               6 \eta_{\nu\alpha}\,
                               \delta^{[\mu}_{[\beta}\delta^\nu_\gamma
                               \delta^{\rho ]}_{\delta ]}
                            }_{0}
          \left.\vbox{\vspace{2.5ex}}\right)\nonumber\\
        &&+\,\left(\,
                  -\,2\,\Gamma^{\mu\rho}{}_{\alpha\beta\gamma\delta}
                  \,+\,
                  24\,\delta_{\delta}^{[\mu}\,\delta^{\rho]}_{\alpha}\,
                      \Gamma_{\beta\gamma}
              \right)\,
       \left.\vbox{\vspace{2.5ex}}\right\}\,
       \hat{G}^{\alpha\beta\gamma\delta}\nonumber\nonumber\\[2ex]
&=& \left\{\vbox{\vspace{2.5ex}}\right.
           3\,\Gamma^{\mu\rho}{}_{\alpha\beta\gamma\delta}
           \,-\,84\,\delta^{[\rho}_{\alpha}\delta^{\mu ]}_{\beta}
                   \Gamma_{\gamma\delta}
           \,-\, 72\,\eta_{\nu\alpha}\,\delta^{[\rho}_{\beta}
                 \delta^\nu_\gamma\Gamma^{\mu]}{}_{\delta}
           \,+\,24\,\delta_{\delta}^{[\mu}\,\delta^{\rho]}_{\alpha}\,
                      \Gamma_{\beta\gamma}\nonumber\\
        &&\,-\,24\,\delta^{[\rho}_{\alpha}\,
                   \Gamma^{\mu ]}{}_{\beta\gamma\delta}
           \,-\,36\,\eta_{\nu\alpha}\,\delta^{[\rho}_{\beta}
             \Gamma^{\mu\nu]}{}_{\gamma\delta}
       \left.\vbox{\vspace{2.5ex}}\right\}\,
       \hat{G}^{\alpha\beta\gamma\delta}
\end{eqnarray}\\
Now a case by case study of the remaining terms follows\\
\begin{eqnarray*}
   -\,36\,\eta_{\nu\alpha}\,\delta^{[\rho}_{\beta}
    \Gamma^{\mu\nu]}{}_{\gamma\delta}\,\hat{G}^{\alpha\beta\gamma\delta}
   &=&-\,12\,\eta_{\nu\alpha}\,
       \left[\vbox{\vspace{2.5ex}}\right.\,
              \delta^{\rho}_{\beta}
              \Gamma^{\mu\nu}{}_{\gamma\delta}
          ~+~ \delta^{\mu}_{\beta}
              \Gamma^{\nu\rho}{}_{\gamma\delta}
          ~+~ \delta^{\nu}_{\beta}
              \Gamma^{\rho\mu}{}_{\gamma\delta}
       \left.\vbox{\vspace{2.5ex}}\right]\,
       \hat{G}^{\alpha\beta\gamma\delta}\\
   &=&-\,12\,
       \left[\vbox{\vspace{2.5ex}}\right.\,
              \delta^{\rho}_{\beta}
              \Gamma^{\mu}{}_{\alpha\gamma\delta}
          ~-~ \delta^{\mu}_{\beta}
              \Gamma^{\rho}{}_{\alpha\gamma\delta}
          ~+~ \underbrace{
                           \eta_{\alpha\beta}
                           \Gamma^{\rho\mu}{}_{\gamma\delta}
                         }_{0}
       \left.\vbox{\vspace{2.5ex}}\right]\,
       \hat{G}^{\alpha\beta\gamma\delta}\\
    &=& 12\,\Gamma^{\mu}{}_{\alpha\gamma\delta}
        \hat{G}^{\rho\alpha\gamma\delta}
          ~-~ 
        12\,\Gamma^{\rho}{}_{\alpha\gamma\delta}
        \hat{G}^{\mu\alpha\gamma\delta}\\
    &=& 24\,\Gamma^{[\mu}{}_{\alpha\gamma\delta}
        \hat{G}^{\rho] \alpha\gamma\delta}
    ~=~ 24\,\Gamma^{[\mu}{}_{\alpha\gamma\delta}
        \delta^{\rho]}_\beta \hat{G}^{\beta\alpha\gamma\delta}
    ~=~ 24\,\delta^{[\rho}_\alpha\Gamma^{\mu ]}{}_{\beta\gamma\delta}
        \hat{G}^{\alpha\beta\gamma\delta}
\end{eqnarray*}\\
\begin{eqnarray*}
  -\,72\,\eta_{\nu\alpha}\,\delta^{[\rho}_{\beta}\delta^\nu_\gamma
           \Gamma^{\mu]}{}_{\delta}\,\hat{G}^{\alpha\beta\gamma\delta}
   &=& -\,72\,\eta_{\nu\alpha}\,\frac{1}{3}\,
        \left[\vbox{\vspace{2.5ex}}\right.\,
             \delta^{\rho}_{\beta}\delta^\nu_\gamma
             \Gamma^{\mu}{}_{\delta}
         ~+~ \delta^{\nu}_{\beta}\delta^\mu_\gamma
             \Gamma^{\rho}{}_{\delta}
         ~+~ \delta^{\mu}_{\beta}\delta^\rho_\gamma
             \Gamma^{\nu}{}_{\delta}
        \left.\vbox{\vspace{2.5ex}}\right]\, 
        \hat{G}^{\alpha\beta\gamma\delta}\\
   &=& -\,72\,\eta_{\nu\alpha}\,\frac{1}{3}\,
        \left[\vbox{\vspace{2.5ex}}\right.\,
             \underbrace{\delta^{\rho}_{\beta}\delta^\nu_\gamma
             \Gamma^{\mu}{}_{\delta}}_{0}
         ~+~ \underbrace{\delta^{\nu}_{\beta}\delta^\mu_\gamma
             \Gamma^{\rho}{}_{\delta}}_{0}
         ~+~ \delta^{\mu}_{\beta}\delta^\rho_\gamma
             \Gamma^{\nu}{}_{\delta}
        \left.\vbox{\vspace{2.5ex}}\right]\, 
        \hat{G}^{\alpha\beta\gamma\delta}\\
   &=& -\,72\,\frac{1}{3}\,
             \delta^\rho_\gamma\delta^{\mu}_{\beta}
             \Gamma_{\alpha\delta}\, 
             \hat{G}^{\alpha\beta\gamma\delta}\\
   &=& 24\;\delta^\rho_\alpha\delta^{\mu}_{\beta}\,
             \Gamma_{\gamma\delta}\, 
             \hat{G}^{\alpha\beta\gamma\delta}
\end{eqnarray*}\\
Plugging this expressions in eq.~(\ref{Step1}) one obtains\\ 
\begin{eqnarray}
    \Gamma^{\mu\nu\rho}\left(
           \Gamma_{\alpha\beta\gamma\delta\nu}
           -8\,\Gamma_{\beta\gamma\delta}\,\eta_{\nu\alpha}
    \right)\,\hat{G}^{\alpha\beta\gamma\delta}
    &=& 3\cdot\left(\vbox{\vspace{2ex}}\right.\,
           \Gamma^{\mu\rho}{}_{\alpha\beta\gamma\delta}
           \,+\,12\,\delta^{\mu}_{\alpha}\,
                    \Gamma_{\gamma\delta}\,
                    \delta^{\rho}_{\beta}\,
        \left.\vbox{\vspace{2ex}}\right)\, 
        \hat{G}^{\alpha\beta\gamma\delta}\label{GammaFormel}
\end{eqnarray}\\
A contraction of this equation with $\psi_\rho$ gives the desired result.\\[2ex]
\end{proof}

\begin{prop}{\label{Psi_Fierz}}
The last four terms in eq.~(\ref{EOM_PSI}) vanish identically.
\end{prop}

\begin{proof}
The basic steps of the proof are outlined 
in \cite{Bergshoeff:1981um} and boil down to 
checking that the Fierz identity (\ref{CJSFierIdentity}) holds.\\[2ex]
\end{proof}

\noindent
If one inserts the results of Proposition 1 and 2 in 
eq.~(\ref{EOM_PSI}) and defines\\ 
\begin{eqnarray*}
   \hat{D}_\nu\psi_{\rho} &=& D_\nu(\hat{\omega})\psi_{\rho}
                           -\frac{1}{2\cdot 144}
                           \left(
                                 \Gamma^{\alpha\beta\gamma\delta}{}_{\nu}
                                  -8\,\Gamma^{\beta\gamma\delta}
                                     \delta_\nu^\alpha
                           \right)\,\psi_{\rho}\,
                           \hat{G}_{\alpha\beta\gamma\delta}
\end{eqnarray*}\\
the equation of motion finally reads\\ 
\begin{eqnarray}\label{PSI_EOM}
    \Gamma^{\mu\nu\rho}\hat{D}_\nu\psi_{\rho} &=& 0
\end{eqnarray}\\

\begin{rem}
\noindent
In practical applications it is very cumbersome to work with the
fermionic fields. So one mostly restricts considerations to the 
set of bosonic fields and sets the gravitino explicitly to zero.  
To make the vanishing of the gravitino 
consistent with the presence of supersymmetry one has to impose 
the constraint that the supervariations acting on such a bosonic 
background cannot restore a non vanishing 
gravitino, i.e. the supervariation of the gravitino evaluated in the 
bosonic background must vanish 
identically:\\
\begin{eqnarray}\label{KSP_Eq}
    \delta_{Q}\psi_\mu{} &=& D_\mu({\omega})\epsilon
                           -\frac{1}{2\cdot 144}
                           \left(
                                 \Gamma^{\alpha\beta\gamma\delta}{}_{\mu}
                                  -8\,\Gamma^{\beta\gamma\delta}
                                     \delta_\mu^\alpha
                           \right)\,\epsilon\,{G}_{\alpha\beta\gamma\delta}
                           ~\buildrel !\over =~ 0~.
\end{eqnarray}\\  
This is known as the Killing spinor equation of eleven dimensional
supergravity. Following the common practice we just work out the
bosonic equations of motion of the metric $g_{\mu\nu}$ and three 
form potential $C_{\mu\nu\rho}$, now.  
\end{rem}

\vfill\eject
\subsubsection{$g_{\mu\nu}$}

\begin{eqnarray}\label{EinsHilbert}
   S &=& \int\,d^Dx\,\underbrace{\sqrt{g}\,R}_{(I)}~-~\frac{1}{48}\,\int\,d^Dx\,\underbrace{\sqrt{g}\,G_{\mu\nu\rho\sigma}\,G^{\mu\nu\rho\sigma}}_{(II)}
\end{eqnarray}\\
We will now derive the Euler-Lagrange equations following from the 
Einstein-Hilbert action (I), coupled to a four form field strength (II). 
First we study the variation of the Einstein-Hilbert action (I). 
Since the Ricci scalar is 
$R\,=\,g^{\mu\nu}R_{\mu\nu}$, it is simpler to consider variations 
of $g^{\mu\nu}$, i.e. the inverse metric, instead of $g_{\mu\nu}$.\\
\begin{eqnarray}\label{VarFirstStep}
   \delta S_{(I)} &=& \int\,d^Dx\,
                \left[\vbox{\vspace{3ex}}\right.\,
                          \delta\sqrt{g}\,g^{\mu\nu}\,R_{\mu\nu} ~+~
                          \sqrt{g}\,\delta g^{\mu\nu}\,R_{\mu\nu} ~+~
                          \sqrt{g}\,g^{\mu\nu}\,\delta R_{\mu\nu}\,
                \left.\vbox{\vspace{3ex}}\right]
\end{eqnarray}\\  
We must evaluate the variation of the determinant in the the first
term. We start from the identity\\ 
\begin{eqnarray*}
       g &\equiv& -\det\,g ~=~ -\,\varepsilon^{i_1\ldots i_n}\,
                                g_{1i_1}{\cdot\ldots\cdot}g_{ni_n}
\end{eqnarray*}\\
and  compute\\
\begin{eqnarray*}
   \frac{\partial g}{\partial x^l} &=& -\,\sum\limits_{j=1}^n\,
                                        \varepsilon^{i_1\ldots i_n}\,
                                        g_{1i_1}{\cdot\ldots\cdot}\frac{
                                        \partial g_{ji_j}}{\partial x^l}
                                        {\cdot\ldots\cdot}g_{ni_n}~.\hspace{11ex}
\end{eqnarray*}\\
The derivative can be rewritten via\\ 
\begin{eqnarray*}
   \frac{\partial g_{ji_j}}{\partial x^l} &=& \delta_{i_j}^k\,
                                        \frac{\partial g_{jk}}{\partial x^l}
                                    ~=~ g_{i_jm}g^{mk}\,
                                        \frac{\partial g_{jk}}{\partial x^l}~.
\end{eqnarray*}\\
Inserting this back into the previous equation we obtain\\
\begin{eqnarray}
   \frac{\partial g}{\partial x^l} &=& -\,\sum\limits_{j=1}^n\,
                                        \varepsilon^{i_1\ldots i_n}\,
                                        g_{1i_1}{\cdot\ldots\cdot}
                                        \left(\vbox{\vspace{3ex}}\right.\,
                                             \underbrace{
                                                g_{i_jm}g^{mk}\,
                                                \frac{\partial g_{jk}}
                                                     {\partial x^l}
                                             }_{m\,\buildrel !\over =\,j}\,
                                        \left.\vbox{\vspace{3ex}}\right)
                                        {\cdot\ldots\cdot}g_{ni_n}\nonumber\\
   &=& g^{jk}\,\frac{\partial g_{jk}}{\partial x^l}\cdot g\label{Var1}
\end{eqnarray}\\
Algebraically $\delta$ behaves like a derivation. Therefore we use  
eq.~(\ref{Var1}) and just replace the partial derivative by $\delta$
to obtain\\ 
\begin{eqnarray}\label{Var_detg}
   \delta\sqrt{g} &=& \frac{1}{2}\,\frac{1}{\sqrt{g}}\,\delta{g}
                  ~=~ \frac{1}{2}\,\frac{1}{\sqrt{g}}\,g\,g^{jk}\,
                      \delta{g_{jk}}
                  ~=~ \frac{1}{2}\,\sqrt{g}\,g^{jk}\,\delta{g_{jk}}
                  ~=~ -\,\frac{1}{2}\,\sqrt{g}\,g_{jk}\,\delta{g^{jk}}
\end{eqnarray}\\
and eq.~(\ref{VarFirstStep}) simplifies to\\
\begin{eqnarray}\label{VarSecondStep}
   \delta S_{(I)} &=& \int\,d^Dx\,\sqrt{g}\,
                \left[\vbox{\vspace{3ex}}\right.\,
                         -\,\frac{1}{2}\,g_{\mu\nu}\,R ~+~ R_{\mu\nu}\,
                \left.\vbox{\vspace{3ex}}\right]\,\delta g^{\mu\nu}
                ~+~
                \int\,d^Dx\,\sqrt{g}\,g^{\mu\nu}\,\delta R_{\mu\nu}
\end{eqnarray}\\
The first two terms in this expression already form the Einstein 
tensor. Therefore we must prove only that the third term does not
contribute. To this purpose we have to evaluate the variation of the 
Ricci tensor, who is given by\\ 
\begin{eqnarray*}
   R_{\mu\nu} &=& R^\rho{}_{\mu\rho\nu} 
                    ~=~ \frac{\partial\Gamma^{\lambda}{}_{(\mu\nu)}}
                             {\partial x^\lambda}
                    ~-~ \frac{\partial\Gamma^{\lambda}{}_{(\mu\lambda)}}
                             {\partial x^\nu}
                    ~+~ \Gamma^\lambda{}_{(\lambda\rho)}\,
                        \Gamma^\rho{}_{(\nu\mu)}
                    ~-~ \Gamma^\lambda{}_{(\nu\rho)}\,
                        \Gamma^\rho{}_{(\lambda\mu)}~.
\end{eqnarray*}\\
There is an easy way to obtain the variation with respect to the metric.
For this we compute the variation of $R_{\mu\nu}$ in terms of the 
variations $\delta\Gamma^\mu{}_{\nu\lambda}$ induced by the
variations of the metric. So $\delta R_{\mu\nu}$ becomes\\
\begin{eqnarray*}
 \delta R_{\mu\nu} &=& \frac{\partial\delta\Gamma^{\lambda}{}_{(\mu\nu)}}
                            {\partial x^\lambda}
                   ~-~ \frac{\partial\delta\Gamma^{\lambda}{}_{(\mu\lambda)}}
                            {\partial x^\nu}
                   ~+~ \delta\Gamma^\lambda{}_{(\lambda\rho)}\,
                       \Gamma^\rho{}_{(\nu\mu)}
                   ~+~ \Gamma^\lambda{}_{(\lambda\rho)}\,
                       \delta\Gamma^\rho{}_{(\nu\mu)}\\[2ex]
                   &&\hspace{23ex}
                   ~-~ \delta\Gamma^\lambda{}_{(\nu\rho)}\,
                       \Gamma^\rho{}_{(\lambda\mu)}
                   ~-~ \Gamma^\lambda{}_{(\nu\rho)}\,
                       \delta\Gamma^\rho{}_{(\lambda\mu)}
\end{eqnarray*}\\
Now the crucial observation is that $\delta\Gamma^\mu{}_{(\nu\lambda)}$ 
is a tensor. The Christoffel symbols themselves are not tensors, because of 
the inhomogeneous (second derivative) term appearing in the transformation 
rule under coordinate transformations. But this term is independent of the 
metric. Thus the metric variation of the Christoffel symbols indeed 
transforms as a tensor, and it turns out that $\delta R_{\mu\nu}$ can be 
written rather compactly in terms of covariant derivatives of 
$\delta\Gamma^\mu{}_{(\nu\lambda)}$, namely as\\
\begin{eqnarray}\label{VarOfRicci}
   \delta R_{\mu\nu} &=& \nabla_{\lambda}\delta\Gamma^\lambda{}_{(\mu\nu)}
                    ~-~ \nabla_{\nu}\delta\Gamma^\lambda{}_{(\lambda\mu)}
\end{eqnarray}\\ 
To establish (\ref{VarOfRicci}), one simply has to use the definition of 
the covariant derivative.
What we really need is $g^{\mu\nu}\delta R_{\mu\nu}$. But since 
the metric is covariant constant one obtains\\
\begin{eqnarray*}
     g^{\mu\nu}\delta R_{\mu\nu}  &=& 
     \nabla_{\lambda}\left(\,
                             g^{\mu\nu}\delta\Gamma^\lambda{}_{(\mu\nu)}\,
                     \right)
     ~-~ \nabla_{\nu}\left(\,
                             g^{\mu\nu}\delta\Gamma^\lambda{}_{(\lambda\mu)}\,
                     \right)
\end{eqnarray*}\\
Since all indices are contracted we may rename them in each single term. 
Renaming $\nu$ to $\lambda$ and vice versa one obtains\\
\begin{eqnarray*}
     g^{\mu\nu}\delta R_{\mu\nu}  &=& 
     \nabla_{\lambda}\left(\,
                             g^{\mu\nu}\delta\Gamma^\lambda{}_{(\mu\nu)}
                             ~-~
                             g^{\mu\lambda}\delta\Gamma^\nu{}_{(\nu\mu)}\,
                     \right),
\end{eqnarray*}\\
i.e. a total divergence. The third term in (\ref{VarSecondStep}) does
not contribute and we obtain\\
\begin{eqnarray}\label{VarFinalResult}
   \delta S_{(I)} &=& \int\,d^Dx\,\sqrt{g}\,
                \left[\vbox{\vspace{3ex}}\right.\,
                       R_{\mu\nu} ~-~\frac{1}{2}\,g_{\mu\nu}\,R\,
                \left.\vbox{\vspace{3ex}}\right]\,\delta g^{\mu\nu}
\end{eqnarray}\\
Using eq.~(\ref{Var_detg}) we easily obtain the variation of the second part (II):\\
\begin{eqnarray}
 \delta S_{(II)} &=&
 -\,\frac{1}{48}\,\int\,d^Dx\,\left[\vbox{\vspace{3ex}}\right.\,
     \delta\sqrt{g}\,G_{\mu\nu\rho\sigma}\,G^{\mu\nu\rho\sigma}\,+\,
    \sqrt{g}\,\delta\left(G_{\mu\nu\rho\sigma}\,G^{\mu\nu\rho\sigma}\right)\,
    \left.\vbox{\vspace{3ex}}\right]\nonumber\\
 &=&
-\,\frac{1}{48}\,\int\,d^Dx\,\left[\vbox{\vspace{3ex}}\right.\,
     -\frac{1}{2}\,\sqrt{g}\,g_{\alpha\beta}\,G_{\mu\nu\rho\sigma}\,G^{\mu\nu\rho\sigma}\,+\,4\,
    \sqrt{g}\,G_{\alpha\nu\rho\sigma}\,G_{\beta}{}^{\nu\rho\sigma}\,
    \left.\vbox{\vspace{3ex}}\right]\,\delta\,g^{\alpha\beta}
\end{eqnarray}\\
The equation of motion now reads\\
\begin{eqnarray} 
R_{\alpha\beta} ~-~\frac{1}{2}\,g_{\alpha\beta}\,R &=& \frac{1}{48}\,\left(\vbox{\vspace{3ex}}\right.\,
     -\frac{1}{2}\,g_{\alpha\beta}\,G_{\mu\nu\rho\sigma}\,G^{\mu\nu\rho\sigma}\,+\,4\,G_{\alpha\nu\rho\sigma}\,G_{\beta}{}^{\nu\rho\sigma}\,
    \left.\vbox{\vspace{3ex}}\right)\,
\end{eqnarray}\\
Contracting the whole equation with $g^{\beta\alpha}$ we obtain\\
\begin{eqnarray} 
    \frac{1}{2}\,\,R &=&  \frac{1}{48}\,\left(\vbox{\vspace{3ex}}\right.\,
     \frac{1}{6}\,G_{\mu\nu\rho\sigma}\,G^{\mu\nu\rho\sigma}\,
    \left.\vbox{\vspace{3ex}}\right)\,
\end{eqnarray}\\
or\\
\begin{eqnarray}\label{Einstein} 
R_{\alpha\beta} &=&
  \frac{1}{12}\,\left(\vbox{\vspace{3ex}}\right.\,
     G_{\alpha\nu\rho\sigma}\,G_{\beta}{}^{\nu\rho\sigma}\,
     -\frac{1}{12}\,g_{\alpha\beta}\,G_{\mu\nu\rho\sigma}\,G^{\mu\nu\rho\sigma}\,
    \left.\vbox{\vspace{3ex}}\right)\,
\end{eqnarray}

\vfill\eject
\subsubsection{$C_{\mu\nu\rho}$}

\underline{Attention:} This calculation is done in a flat metric background 
                       $\eta_{ij}$. But the modifications to a nontrivial 
                       metric are simple.

\begin{eqnarray*}
  {\mathcal{L}}_{C} 
                &=& -\,\frac{1}{48}\;G_{\mu\nu\rho\sigma}\,G^{\mu\nu\rho\sigma}
                    ~+~ \frac{1}{144^2}\,\varepsilon^{\alpha_1\ldots\alpha_4
                                                      \beta_1\ldots\beta_4
                                                      \gamma_1\gamma_2\gamma_3}
                     \,G_{\alpha_1\ldots\alpha_4}\,G_{\beta_1\ldots\beta_4}\,
                       C_{\gamma_1\gamma_2\gamma_3}
\end{eqnarray*}

\begin{eqnarray*}
     0~=~ \frac{\partial {\mathcal L}}{\partial \phi} 
      ~-~ \partial_\xi\,\left(
                               \frac{\partial {\mathcal L}}
                                    {\partial(\,\partial_\xi\phi\,)}
                        \right)
\end{eqnarray*}

\noindent
If one chooses $\phi\,=\,C_{ijk}$ and uses the identities 
\begin{eqnarray*}
   \frac{\partial\,G_{\mu\nu\rho\sigma}}{\partial(\,\partial_\xi C_{ijk}\,)}
    &=& \delta_{\mu\nu\rho\sigma}^{\xi ijk}\\[1ex]
    \delta_{\beta_1\ldots\beta_4}^{\alpha_1\ldots\alpha_4}\,
    G^{\beta_1\ldots\beta_4} &=& 4!\cdot G^{\alpha_1\ldots\alpha_4}
\end{eqnarray*}
frequently one can compute by brute force:

\begin{eqnarray}
   0 &=& \frac{1}{144^2}\,\varepsilon^{\alpha_1\ldots\alpha_4
                                                      \beta_1\ldots\beta_4
                                                      \gamma_1\gamma_2\gamma_3}
                     \,G_{\alpha_1\ldots\alpha_4}\,G_{\beta_1\ldots\beta_4}\,
                     \delta_{\gamma_1\gamma_2\gamma_3}^{ijk}\nonumber\\
     &-&\frac{\partial}{\partial x^\xi}
        \left[\vbox{\vspace{3ex}}\right.\,
               -\,\frac{1}{48}\,
                 \left\{\vbox{\vspace{2.5ex}}\right.\,
                         \delta_{\mu\nu\rho\sigma}^{\xi ijk}\,
                         G^{\mu\nu\rho\sigma} ~+~ g^{\mu\tau_1} 
                         g^{\nu\tau_2} g^{\rho\tau_3} g^{\sigma\tau_4}
                         G_{\mu\nu\rho\sigma}\,
                         \delta_{\tau_1\ldots\tau_4}^{\xi ijk}\,
                 \left.\vbox{\vspace{2.5ex}}\right\}\nonumber\\
     &&\hspace{6ex}~+~ \frac{2}{144^2}\;\varepsilon^{\alpha_1\ldots\alpha_4
                                                 \beta_1\ldots\beta_4
                                                 \gamma_1\gamma_2\gamma_3}
                     \,\delta^{\xi ijk}_{\alpha_1\ldots\alpha_4}\,
                       G_{\beta_1\ldots\beta_4}\,
                       C_{\gamma_1\gamma_2\gamma_3}\,
          \left.\vbox{\vspace{3ex}}\right]\nonumber\\[2ex]
   0 &=& \partial_\xi G^{\xi ijk} ~+~ 
         \frac{3!}{144^2}\,\varepsilon^{\alpha_1\ldots\alpha_4
                                        \beta_1\ldots\beta_4 ijk}
                     \,G_{\alpha_1\ldots\alpha_4}\,G_{\beta_1\ldots\beta_4}\,
                     \nonumber\\
         &-& \frac{2}{144^2}\;\varepsilon^{\alpha_1\ldots\alpha_4
                                                 \beta_1\ldots\beta_4
                                                 \gamma_1\gamma_2\gamma_3}
                     \,\delta^{\xi ijk}_{\alpha_1\ldots\alpha_4}\,
                       \underbrace{G_{\beta_1\ldots\beta_4}\,
                                   \partial_\xi C_{\gamma_1\gamma_2\gamma_3}
                                  }_{
                                        {\rm since}\; dG\,=\,0
                                  }\nonumber\\[2ex]
   0 &=& \partial_\xi G^{\xi ijk} ~+~ 
         \frac{3!}{144^2}\,\varepsilon^{\alpha_1\ldots\alpha_4
                                        \beta_1\ldots\beta_4 ijk}
                     \,G_{\alpha_1\ldots\alpha_4}\,G_{\beta_1\ldots\beta_4}\,
                     \nonumber\\
         &+& \frac{2\cdot 4!}{144^2}\;\varepsilon^{\beta_1\ldots\beta_4\xi
                                                 \gamma_1\gamma_2\gamma_3 ijk}
                     \,G_{\beta_1\ldots\beta_4}\,
                   \underbrace{\partial_\xi
         C_{\gamma_1\gamma_2\gamma_3}}_{\partial_{[\xi}
         C_{\gamma_1\gamma_2\gamma_3]}}\nonumber\\[2ex]
   0 &=& \partial_\xi G^{\xi ijk} ~+~ 
         \left( 
                \frac{3!}{144^2}~+~\frac{2\cdot 3!}{144^2}
         \right)\,\varepsilon^{\alpha_1\ldots\alpha_4\beta_1\ldots\beta_4 ijk}
         \,G_{\alpha_1\ldots\alpha_4}\,G_{\beta_1\ldots\beta_4}
         \label{EqofMotAmunurho}
\end{eqnarray}

\subsubsection*{Language of Diff. forms}
\noindent
To find the corresponding expression in the language of differential forms, 
one has to rewrite each of the terms as a 3-form (number of free indices !).\\

\noindent
The first term must be of the form $\ast d \ast G$:

\begin{eqnarray*}
      G &=& \frac{1}{4!}\;G_{\mu_1\mu_2\mu_3\mu_4}\,dx^{\mu_1}\wedge\ldots
            \wedge dx^{\mu_4}\\
   (\ast G)_{\mu_5\ldots\mu_{11}} &=& \frac{1}{4!}\,
             \varepsilon_{\mu_5\ldots\mu_{11}\mu_1\ldots\mu_4}\,
             G^{\mu_1\ldots\mu_4}\\
   d(\ast G)_{\nu_1\ldots\nu_8} &=& 8\cdot\partial_{[\nu_1}\,\frac{1}{4!}
                                    \varepsilon_{\nu_2\ldots\nu_8]\alpha_1
                                \ldots\alpha_4}\,G^{\alpha_1\ldots\alpha_4}\\
 \ast d(\ast G)_{\nu_1\ldots\nu_3} &=& \frac{1}{8!}\,
                                       \varepsilon_{\nu_1\ldots\nu_{11}}\, 
                                        8\,\partial^{[\nu_4}\,\frac{1}{4!}
                                    \varepsilon^{\nu_5\ldots\nu_{11}]\alpha_1
                                \ldots\alpha_4}\,G_{\alpha_1\ldots\alpha_4}\\
 \ast d(\ast G)_{\nu_1\ldots\nu_3} &=& \frac{1}{7!\cdot 4!}\,
                                       \varepsilon_{\nu_1\ldots\nu_{11}}\, 
                               \varepsilon^{\alpha_1\ldots\alpha_4}
 \underbrace{{}^{[\nu_5\ldots\nu_{11}}\,\partial^{\nu_4]}}_{\nu_4\in\{\alpha_1              \ldots\alpha_4\}}\,G_{\alpha_1\ldots\alpha_4}\\
 \ast d(\ast G)_{\nu_1\ldots\nu_3} &=& \frac{1}{7!\cdot 3!}\,
                                      \underbrace{
                                         \varepsilon_{\nu_1\ldots\nu_{11}}\, 
                                         \varepsilon^{\alpha_1\ldots\alpha_3
                                                      \nu_4\ldots\nu_{11}}
                                       }_{-8!\,\delta_{\nu_1\ldots\nu_3}^{\alpha_1\ldots\alpha_3}}\,\frac{1}{8}\,\partial^{\nu_4}\,G_{\alpha_1\ldots\alpha_3\nu_4}\\
    &=& \partial^\xi G_{\xi\nu_1\nu_2\nu_3} 
\end{eqnarray*}

\noindent
The second term is related to $\ast (G\wedge G)$:
\begin{eqnarray*}
    G \wedge G &=& \frac{1}{(4!)^2}\;G_{\alpha_1\ldots\alpha_4}\,
                   G_{\beta_1\ldots\beta_4}\,dx^{\alpha_1}\wedge\ldots\wedge
                   dx^{\alpha_4}\wedge 
                   dx^{\beta_1}\wedge\ldots\wedge dx^{\beta_4}\\
               &=& \frac{1}{8!}\,
                   \left(\,
                            \frac{8!}{(4!)^2}\;
                            G_{\alpha_1\ldots\alpha_4}\,
                            G_{\beta_1\ldots\beta_4}\,
                   \right)\,dx^{\alpha_1}\wedge\ldots\wedge dx^{\alpha_4}
                   \wedge dx^{\beta_1}\wedge\ldots\wedge dx^{\beta_4}\\
  \left[\,\ast (G\wedge G)\,\right]_{\nu_1\nu_2\nu_3} &=& \frac{1}{8!}\,
                       \varepsilon_{\nu_1\ldots\nu_3\alpha_1\ldots
                                    \alpha_4\beta_1\ldots\beta_4}\;
                        \frac{8!}{(4!)^2}\;
                        G^{\alpha_1\ldots\alpha_4}\,
                        G^{\beta_1\ldots\beta_4}\\
  \left[\,\ast (G\wedge G)\,\right]_{\nu_1\nu_2\nu_3} &=& \frac{1}{(4!)^2}\,
                       \varepsilon_{\nu_1\ldots\nu_3\alpha_1\ldots
                                    \alpha_4\beta_1\ldots\beta_4}\;
                        G^{\alpha_1\ldots\alpha_4}\,
                        G^{\beta_1\ldots\beta_4}\\
\end{eqnarray*}

\noindent
Using this both coordinate expressions we can rewrite 
eq.~(\ref{EqofMotAmunurho}) as follows:
\begin{eqnarray}
    0 &=& \ast d(\ast G) ~+~ (4!)^2\,\left(\,
                                            \frac{3!}{144^2}~+~
                                            \frac{2\cdot 3!}{144^2}\,
                                     \right)\,\ast(G\wedge G)\nonumber\\
    0 &=& d(\ast G) ~+~ \frac{1}{2}\; G\wedge G\label{4Form}
\end{eqnarray}


\subsubsection*{Homework:}

\begin{exercise} Compute the Ricci tensor of the following diagonal
                 metric in D dimensions:
                 \begin{eqnarray*}
                      g_{\mu\nu} &=& N\cdot \delta_{\mu\nu}
                 \end{eqnarray*}

\end{exercise}

\vfill\eject
\section{Constructing the Lagrangian}
\label{Sec_Noether}

In Chapter \ref{Sec_MTheory} the Lagrangian of eleven dimensional 
supergravity was introduced. Now we are going to discuss its 
structure in more detail by reviewing some of the basic steps of 
its construction along the lines of the original 
paper \cite{Cremmer:1978km}. We performing the construction of the 
complete bosonic action, which requires the construction up to second
order of the 
fermionic terms in the action. The starting 
point is an action where we simply added together 
an Einstein-Hilbert describing pure gravity with a Rarita-Schwinger 
action describing a Majorana gravitino, two of the three fields
forming the representation of $N\,=\,1$ supersymmetry in $D\,=\,11$:\\  
\begin{eqnarray}\label{M_linear}
   S &=& \int dx^D\,\sqrt{g}\,\left[\,\frac{1}{4}\,R 
                +\frac{1}{2}\,\bar{\psi}_\mu\Gamma^{\mu\nu\rho}
                 D_\nu\left(\omega\right)\psi_\rho\,\right]~.
\end{eqnarray}\\
An ansatz for the linearised  transformation law of supersymmetry is:\\  
\begin{eqnarray}
  \delta_{Q}\psi_\mu{}~ &=& D_\mu(\omega)\epsilon\nonumber\\
  \delta_{Q}e_\mu{}^a &=&  \bar\epsilon\,\Gamma^a\psi_\mu
  \quad\Rightarrow\quad 
  \delta_{Q}g_{\mu\nu} ~=~ \delta_{Q}\left(\,\eta_{ab}\,e^a_{\mu}\,e^b_{\nu}\,\right)~=~ 2\,\bar\epsilon\,\Gamma_\mu\psi_\nu~.
  \label{Q_linear}
\end{eqnarray}\\
Performing the variation of the action according to 
a symmetry $\delta_Q$ and using the result eq.~(\ref{VarFinalResult}) 
one obtains\\
\begin{eqnarray*}
  \delta_Q S &=& \int dx^D\,\sqrt{g}\,
                 \left[\vbox{\vspace{3ex}}\right.\,\frac{1}{4}\,
                \left(\vbox{\vspace{3ex}}\right.\,
                       R_{\mu\nu} ~-~\frac{1}{2}\,g_{\mu\nu}\,R\,
                \left.\vbox{\vspace{3ex}}\right)\,\delta_Q g^{\mu\nu}\\
             && \hspace{1.7cm}
               +\underbrace{\frac{1}{2}\,\delta_Q\bar{\psi}_\mu\Gamma^{\mu\nu\rho}
                 D_\nu\left(\omega\right)\psi_\rho}_{\rm (2nd)}\,
                -\underbrace{\frac{1}{2}\,\bar{\psi}_\mu\Gamma^{\mu\nu\rho}
                 D_\nu\left(\omega\right)\delta_Q\psi_\rho}_{\rm (3th)}\,\\
             &&\hspace{1.7cm} 
                +\underbrace{\frac{3}{2}\,\bar{\psi}_{[\mu}\delta_Q
                 g^{\mu\alpha}\Gamma_{\alpha}{}^{\nu\rho}
                 D_\nu\left(\omega\right)\psi_{\rho ]}}_{\rm (4th)}\, 
                 \left.\vbox{\vspace{3ex}}\right]
\end{eqnarray*}\\
The fourth term is of third order in the fermionic fields and must be
cancelled by other terms containing three fermions yet to be added. 
Therefore, on
this level, i.e. ${\mathcal{O}}(\bar\epsilon\psi G^0)$, it can be neglected. 
Now we proceed to simplify the second term in (\ref{M_linear}), i.e.\\ 
\begin{eqnarray}
 (2{\rm nd})
 &=&\frac{1}{2}\,\overline{D_\mu\epsilon}\,\Gamma^{\mu\nu\rho}
                 D_\nu\left(\omega\right)\psi_\rho~.\hspace{2ex}
\end{eqnarray}\\
By partial integration it reads\\ 
\begin{eqnarray}
 (2{\rm nd})
 &=&-\,\frac{1}{2}\,\bar\epsilon\,\Gamma^{\mu\nu\rho}
                 D_\mu\,D_\nu\left(\omega\right)\psi_\rho~.
\end{eqnarray}\\
Due to\\ 
\begin{eqnarray}
    D_{[\mu}D_{\nu]}\epsilon &=& \frac{1}{8}\,R_{\mu\nu}{}^{\alpha\beta}\,\Gamma_{\alpha\beta}\,\epsilon\hspace{0ex}
\end{eqnarray}\\
we obtain\\ 
\begin{eqnarray}
 (2{\rm nd})
 &=&-\,\frac{1}{16}\,R_{\mu\nu}{}^{\alpha\beta}\,\bar\epsilon\,\Gamma^{\mu\nu\rho}
             \,\Gamma_{\alpha\beta} \psi_\rho~.\hspace{44ex}
\end{eqnarray}\\
The right hand side can be expanded in the Clifford algebra and gives:\\
\begin{eqnarray*}
  R_{\mu\nu}{}^{\alpha\beta}\,\bar{\epsilon}\,
  \Gamma^{\mu\nu\rho}\Gamma_{\alpha\beta}\,\psi_\rho
  &=&
  R_{\mu\nu}{}^{\alpha\beta}\,\bar{\epsilon}
  \left\{\vbox{\vspace{2.5ex}}\right.\,\Gamma^{\mu\nu\rho}{}_{\alpha\beta}+6\,\delta^{[\rho}_{[\alpha}\Gamma^{\mu\nu]}{}_{\beta]}+6\,\delta^{[\rho}_{[\alpha}\delta^{\nu}_{\beta]}\Gamma^{\mu]}\,\left.\vbox{\vspace{2.5ex}}\right\}\,\psi_\rho\\[3ex]
  &=& R_{\mu\nu}{}^{\alpha\beta}\,\bar{\epsilon}
  \left\{\vbox{\vspace{2.5ex}}\right.\,\Gamma^{\mu\nu\rho}{}_{\alpha\beta}+6\,\delta^{[\rho}_{[\alpha}\Gamma^{\mu\nu]}{}_{\beta]}\,\left.\vbox{\vspace{2.5ex}}\right\}\,\psi_\rho\,+\\
 && R_{\mu\nu}{}^{\alpha\beta}\,\bar{\epsilon}
  \left\{\vbox{\vspace{2.5ex}}\right.\,\delta^{\rho}_{\alpha}\delta^{\nu}_{\beta}\Gamma^{\mu}-\delta^{\rho}_{\alpha}\delta^{\mu}_{\beta}\Gamma^{\nu}+\delta^{\mu}_{\alpha}\delta^{\rho}_{\beta}\Gamma^{\nu}\\
  &&\hspace{9ex}-\delta^{\mu}_{\alpha}\delta^{\nu}_{\beta}\Gamma^{\rho}+\delta^{\nu}_{\alpha}\delta^{\mu}_{\beta}\Gamma^{\rho}-\delta^{\nu}_{\alpha}\delta^{\rho}_{\beta}\Gamma^{\mu}\,\left.\vbox{\vspace{2.5ex}}\right\}\,\psi_\rho\\[3ex]
&=& R_{\mu\nu}{}^{\alpha\beta}\,\bar{\epsilon}
  \left\{\vbox{\vspace{2.5ex}}\right.\,\Gamma^{\mu\nu\rho}{}_{\alpha\beta}+6\,\delta^{[\rho}_{[\alpha}\Gamma^{\mu\nu]}{}_{\beta]}\,\left.\vbox{\vspace{2.5ex}}\right\}\,\psi_\rho\,+\,\bar{\epsilon}
  \left\{\vbox{\vspace{2.5ex}}\right.\,
      4\,R_{\mu}{}^{\rho}\,\Gamma^{\mu}\,-2\,R\,\Gamma^{\rho}\,
  \left.\vbox{\vspace{2.5ex}}\right\}\,\psi_\rho\\[3ex]
&=&R_{\mu\nu}{}^{\alpha\beta}\,\bar{\epsilon}
  \left\{\vbox{\vspace{2.5ex}}\right.\,\Gamma^{\mu\nu\rho}{}_{\alpha\beta}+6\,\delta^{[\rho}_{[\alpha}\Gamma^{\mu\nu]}{}_{\beta]}\,\left.\vbox{\vspace{2.5ex}}\right\}\,\psi_\rho\,+\,4\,
  \left\{\vbox{\vspace{2.5ex}}\right.\,
      R^{\mu\rho}\,-\,\frac{1}{2}\,R\,g^{\mu\rho}\,
  \left.\vbox{\vspace{2.5ex}}\right\}\,\bar{\epsilon}\,\Gamma_\mu\,\psi_\rho~.
\end{eqnarray*}\\
The second term finally reads\\ 
\begin{eqnarray}
 (2{\rm nd})
 &=& -\,\frac{1}{16}\,R_{\mu\nu}{}^{\alpha\beta}\,\bar{\epsilon}
  \left\{\vbox{\vspace{2.5ex}}\right.\,\Gamma^{\mu\nu\rho}{}_{\alpha\beta}+6\,\delta^{[\rho}_{[\alpha}\Gamma^{\mu\nu]}{}_{\beta]}\,\left.\vbox{\vspace{2.5ex}}\right\}\,\psi_\rho\,-\,\frac{1}{4}\,
  \left\{\vbox{\vspace{2.5ex}}\right.\,
      R^{\mu\rho}\,-\,\frac{1}{2}\,R\,g^{\mu\rho}\,
  \left.\vbox{\vspace{2.5ex}}\right\}\,\bar{\epsilon}\,\Gamma_\mu\,\psi_\rho~.
\end{eqnarray}\\
The same steps must be applied to the third term and we obtain:\\
\begin{eqnarray}
 (3{\rm th})
  &=& \frac{1}{16}\,R_{\nu\rho}{}^{\alpha\beta}\,\bar{\psi}_\mu
  \left\{\vbox{\vspace{2.5ex}}\right.\,\Gamma^{\mu\nu\rho}{}_{\alpha\beta}+6\,\delta^{[\rho}_{[\alpha}\Gamma^{\mu\nu]}{}_{\beta]}\,\left.\vbox{\vspace{2.5ex}}\right\}\,\epsilon\,+\,\frac{1}{4}\,
  \left\{\vbox{\vspace{2.5ex}}\right.\,
      R^{\rho\mu}\,-\,\frac{1}{2}\,R\,g^{\mu\rho}\,
  \left.\vbox{\vspace{2.5ex}}\right\}\,\bar{\psi}_\mu\,\Gamma_\rho\,\epsilon~.
\end{eqnarray}\\
To compare the second with the third term in the latter one one has to 
switch the position of $\epsilon$ and $\psi$ via the identity 
(\ref{SwitchEpsPsi}). Then the term with $\Gamma^{(3)}$ picks up a 
sign. Subtracting the third from the second term we end up with:\\
\begin{eqnarray*}
(2{\rm nd})\,-\,(3{\rm th})
&=& \phantom{-}\frac{1}{16}\,(\,-1\,-\,1\,)\,R_{\mu\nu\alpha\beta}\,\bar{\epsilon}\,
                             \Gamma^{\mu\nu\rho\alpha\beta}\,\psi_\rho\,\\
&&-\,\frac{6}{16}\,(\,\phantom{-}1\,-\,1\,)\,R_{\mu\nu}{}^{\alpha\beta}\,\bar{\epsilon}\,
  \delta^{[\rho}_{[\alpha}\Gamma^{\mu\nu]}{}_{\beta]}\,\psi_\rho\,\\
&&\;+\,\frac{1}{4}\;(\,-1\,-\,1\,)\,
  \left\{\vbox{\vspace{2.5ex}}\right.\,
      R^{\mu\rho}\,-\,\frac{1}{2}\,R\,g^{\mu\rho}\,
  \left.\vbox{\vspace{2.5ex}}\right\}\,\bar{\epsilon}\,\Gamma_\mu\,\psi_\rho~.
\end{eqnarray*}\\
Obviously, on the right hand side the second line vanishes and the 
first line does too, due to a symmetry of the curvature tensor ($R_{\mu[\nu\alpha\beta]}\,=\,0$):\\
\begin{eqnarray}
   R_{\mu\nu\alpha\beta}\,\Gamma^{\mu\nu\rho\alpha\beta}
   &=& \frac{1}{3}\,
       \left(\vbox{\vspace{2.5ex}}\right.\,
          \underbrace{
           R_{\mu\nu\alpha\beta}\,+
           R_{\mu\beta\nu\alpha}\,+\,
           R_{\mu\alpha\beta\nu}\,
           }_{0}
       \left.\vbox{\vspace{2.5ex}}\right)\,\Gamma^{\mu\nu\rho\alpha\beta}~.
\end{eqnarray}\\ 
Plugging these results back into the variation of the action it reads\\
\begin{eqnarray*}
  \delta_Q S &=& \int dx^D\,\sqrt{g}\,\left[\vbox{\vspace{3ex}}\right.\,
                \frac{1}{2}\,
                \left(\vbox{\vspace{3ex}}\right.\,
                       R^{\mu\nu} ~-~\frac{1}{2}\,g^{\mu\nu}\,R\,
                \left.\vbox{\vspace{3ex}}\right)\,
                \bar\epsilon\,\Gamma_{(\mu}\psi_{\nu)}
                \,-\,\frac{1}{2}\,
                \left(\vbox{\vspace{3ex}}\right.\,
                       R^{\mu\nu} ~-~\frac{1}{2}\,g^{\mu\nu}\,R\,
                \left.\vbox{\vspace{3ex}}\right)\,
                \bar\epsilon\,\Gamma_{(\mu}\,\psi_{\nu)}\,
                \left.\vbox{\vspace{3ex}}\right]~.
\end{eqnarray*}\\
Here we added the two symmetrisation symbols on $\mu$ and $\nu$, 
because the contraction with the symmetric Einstein tensor picks out  
the symmetric combination.
To this order the variation of the action vanishes identically. Up to
here and to the chosen order ${\mathcal{O}}(\bar\epsilon\,\psi\,G^0)$ 
the action (\ref{M_linear}) and the supersymmetry 
variation (\ref{Q_linear}) are
consistent. Now we add in the kinetic term for the three form
potential, i.e.\\
\begin{eqnarray}
   S &=& \int dx^D\,\sqrt{g}\,\left[\,\frac{1}{4}\,R 
                +\frac{1}{2}\,\bar{\psi}_\mu\Gamma^{\mu\nu\rho}
                 D_\nu\left(\omega\right)\psi_\rho\,
                -\frac{1}{4\cdot 48}\,G_{\mu\nu\rho\sigma}
                 G^{\mu\nu\rho\sigma}\,\right]~.
\end{eqnarray}\\
On the same level of the fermionic fields considered before the
variation of the additional term with respect to $g^{\mu\nu}$ 
produces now a new contribution of 
$-1/24\,\bar\epsilon\,\Gamma^\mu\psi^\nu\,(G^2)_{\mu\nu}$, which must
be balanced by the following modification of the action\\
\begin{eqnarray}\label{Noether2}
   S &=& \int dx^D\,\sqrt{g}\,\left[\,\frac{1}{4}\,R 
                +\frac{1}{2}\,\bar{\psi}_\mu\Gamma^{\mu\nu\rho}
                 D_\nu\left(\omega\right)\psi_\rho+\bar\psi_\mu(XG)^{\mu\rho}\,\psi_\rho\,
                -\frac{1}{4\cdot 48}\,G_{\mu\nu\rho\sigma}
                 G^{\mu\nu\rho\sigma}\,\right]
\end{eqnarray}\\
and subsequently by the following  modified supersymmetry transformations\\
\begin{eqnarray}
  \delta_{Q}g_{\mu\nu} &=& 2\,\bar\epsilon\,\Gamma_\mu\psi_\nu\nonumber\\
  \delta_{Q}\psi_\mu{}~ &=& D_\mu(\omega)\epsilon + (ZG)_\mu\epsilon
  ~=~ \hat{D}_\mu(\omega)\epsilon~.\label{LinSUSY}
\end{eqnarray}\\
The actual determination of $(XG)^{\mu\nu}$ is a bit messy 
but straightforward and is in fact a generalisation of the 
calculations done to prove Proposition \ref{NiceRelation} 
in subsection \ref{EOM_PSI}. The logic is first to write down 
all possible terms consistent with the tensor structure
$(XG)^{\mu\nu}$. The building blocks are the fields and 
Gamma matrices.
There are just five possibilities to write a tensor 
$\bar\psi_\mu(XG)^{\mu\rho}\,\psi_\rho$ in the Clifford
algebra. Depending on the symmetry in $\mu$ and $\nu$ one obtains:\\    
\begin{eqnarray*}
  (XG)^{\mu\nu} &=& \underbrace{a\,\Gamma^{(\mu}{}_{\alpha_1\ldots \alpha_3}\,
                    G^{\nu )\alpha_1\ldots\alpha_3}
                  +b\,g^{\mu\nu}\,\Gamma_{\alpha_1\ldots\alpha_4}G^{\alpha_1\ldots\alpha_4}}_{\rm symm.}\\
   &+&
                  \underbrace{c\,\Gamma^{\mu\nu}{}_{\alpha_1\ldots \alpha_4}\,
                    G^{\alpha_1\ldots\alpha_4}
                  +d\,\Gamma_{\alpha_1\alpha_2}\,G^{\mu\nu\alpha_1\alpha_2}
                  +e\,\Gamma^{[\mu}{}_{\alpha_1\alpha_2\alpha_3}\,G^{\nu ]\alpha_1\alpha_2\alpha_3}}_{\rm antisymm.}~. 
\end{eqnarray*}\\ 
When we derived the equation of motion of the gravitino
(\ref{PSI_EOM}), we observed its close relationship to the operator 
governing the supersymmetry variation of the gravitino. This
observation can be utilised to determine $(ZG)_\mu$ directly. Just
compute the equation of motion of the gravitino and read of the
definition of  $(ZG)_\mu$, again. Now $(ZG)_\mu$ also becomes a tensor 
in the five real unknowns $a$, $b$, $c$, $d$, $e$. Using the so determined 
tensors one has to perform a supervariation of the action in the
appropriate level of fermions and obtains a set of equations to fix 
the five unknowns. The computation one has to do is straightforward 
but tedious. 
To cut the story short it finally turns out that only two 
antisymmetric terms occur with $c=-(8\cdot 4!)^{-1}$  and 
$d=-12\cdot(8\cdot 4!)^{-1}$.
According to Proposition \ref{NiceRelation} on page
\pageref{NiceRelation} the modification of the supersymmetry variation 
is then given by\\ 
\begin{eqnarray}
     (ZG)_\mu &=& -\frac{1}{2\cdot 144}
                           \left(
                                 \Gamma^{\alpha\beta\gamma\delta}{}_{\mu}
                                  -8\,\Gamma^{\beta\gamma\delta}
                                     \delta_\mu^\alpha
                           \right)\,G_{\alpha\beta\gamma\delta}~.
\end{eqnarray}\\
\underline{\bf Plan:}
Now we want to recompute $\delta_Q S$ up to order
${\mathcal{O}}(\bar\epsilon\psi)G^2$ in order to 
check if the action (\ref{Noether2}) constructed so 
far is consistent. We will discover the need to add in 
an additional term, the so called Chern-Simons-Term.\\

\noindent
The terms in the variation of the action (\ref{Noether2}) of order 
${\mathcal{O}}(\bar\epsilon\psi)G^2$ are generated from the two pieces\\ 
\begin{eqnarray}\label{Noether3}
   \delta_QS &=& \int dx^D\,\left[\vbox{\vspace{3ex}}\right.\,
                 \underbrace{\delta_Q\left(\,-\,\frac{e}{4\cdot 48}\,
                                 \bar\psi_\mu(XG)^{\mu\rho}\,\psi_\rho\,\right)
                            }_{(II)} ~+~
                 \underbrace{\delta_Q\left(\,-\,\frac{e}{4\cdot 48}\,
                              G_{\mu\nu\rho\sigma}
                              G^{\mu\nu\rho\sigma}\,\right)
                            }_{(I)}\,
                 \left.\vbox{\vspace{3ex}}\right]~.
\end{eqnarray}\\
Using (\ref{Var_detg}) and (\ref{LinSUSY}) one obtains\\
\begin{eqnarray}
  (I) &=& \delta\,        
        \left(\,
                -\,\frac{e}{4\cdot 48}\,
                   G_{\alpha_1\ldots\alpha_4}\,
                   g^{\alpha_1\beta_1}\,\cdots
                   g^{\alpha_4\beta_4}\,
                   G_{\beta_1\ldots\beta_4}\,
        \right)\nonumber\\[2ex]
   &{\buildrel {\bar\epsilon\psi G^2} \over = }&
         -\,\frac{1}{4\cdot 48}\;
          G^2\cdot
        \left(\,
               -\frac{1}{2}\,e\,g_{jk}\,2\,\bar\epsilon\,\Gamma^{(j}\,
               \psi^{k)}\,
        \right)
      -\,\frac{4\cdot e}{4\cdot 48}\,
          (G^2)_{\alpha_1\beta_1}\,
          2\,\bar\epsilon\,\Gamma^{(\alpha_1}\,\psi^{\beta_1)}
      \nonumber\\[2ex]
  &=&
      -\,\frac{e}{24}\,
         \left(\,
          (G^2)_{\alpha_1\beta_1}\,          
      -\,\frac{1}{8}\;
          G^2\cdot g_{\alpha_1\beta_1}\,\right)\,\bar\epsilon\,\Gamma^{(\alpha_1}\,\psi^{\beta_1)}~.\label{CSPartI}
\end{eqnarray}\\
The evaluation of the second piece is more complicated:\\
\begin{eqnarray}
 (II) &=& \delta\,
        \left(\,
                 -\,\frac{e}{4\cdot 48}\,
                 \left(\,
                         \bar\psi_\mu\,\Gamma^{\mu\nu\alpha\beta\gamma\delta}
                         \psi_\nu
                         \,+\,
                         12\,
                         \bar\psi^\alpha\,\Gamma^{\gamma\delta}
                         \psi^\beta
                 \right)\,G_{\alpha\beta\gamma\delta}\,
        \right)\hspace{14ex}\nonumber\\
   &{\buildrel {\bar\epsilon\psi G^2} \over = }&
   -\,\frac{e}{4\cdot 48}\,
     \delta\,
                 \left(\,
                         \bar\psi_\mu\,\Gamma^{\mu\nu\alpha\beta\gamma\delta}
                         \psi_\nu
                         \,+\,
                         12\,
                         \bar\psi^\alpha\,\Gamma^{\gamma\delta}
                         \psi^\beta
                 \right)\,G_{\alpha\beta\gamma\delta}\label{CSPartII}
\end{eqnarray}\\
The first problem is to determine the variation of $\bar\psi$.\\[4ex]
\underline{$1^{\rm st}$ Auxiliary calculation:}\\
\begin{eqnarray*}
   \delta\bar\psi_\mu 
   &=& \delta\left(\,\psi_\mu^T\,\Gamma^0\,\right)
   ~=~ \left(\,\delta\psi_\mu\,\right)^T\,\Gamma^0\\
   &=& \left(\,D_\mu\epsilon\,-\,\frac{1}{2\cdot
   144}\,\left[\,\Gamma^{\alpha\beta\gamma\delta}{}_{\mu}\,-\,8\,\Gamma^{\beta\gamma\delta}\,\delta^\alpha_\mu\,\right]\,G_{\alpha\beta\gamma\delta}\,\epsilon\,\right)^T\,\Gamma^0\\
   &=& \left(\,D_\mu\epsilon\,\right)^T\,\Gamma^0\,-\,\frac{1}{2\cdot 144}\,\left(\,\overline{\Gamma^{\alpha\beta\gamma\delta}{}_{\mu}\,\epsilon}\,-\,8\,\overline{\Gamma^{\beta\gamma\delta}\,\epsilon}\,\delta^\alpha_\mu\,\right)\,G_{\alpha\beta\gamma\delta}
\end{eqnarray*}\\
\parbox{\textwidth}
{
 \parbox{.1\textwidth}
 {
   \phantom{AAA}
 }\hfill
 \parbox{0.85\textwidth}
 {
\underline{$2^{\rm nd}$ Auxiliary calculation:}\\
\begin{eqnarray*}
   \overline{\Gamma^{(j)}\,\epsilon} 
   &=& \left(\,\Gamma^{(j)}\,\epsilon\,\right)^T\,\Gamma^0\\
   &=& \epsilon^T\,(\Gamma^{a_j})^T\ldots (\Gamma^{a_1})^T\,\Gamma^0\\
   &=&
   (-1)^j\,\epsilon^T\,(\Gamma^0)^2\,(\Gamma^{a_j})^T\,(\Gamma^0)^2\ldots\,(\Gamma^0)^2\,(\Gamma^{a_1})^T\,\Gamma^0\\
   &=&
   (-1)^j\,\bar\epsilon\,\Gamma^{a_j}\ldots\Gamma^{a_1}\\
   &=&
   (-1)^j\,(-1)^{\frac{(j-1)j}{2}}\,\bar\epsilon\,\Gamma^{a_1}\ldots\Gamma^{a_j}\\
   &=&
   (-1)^{\frac{(j+1)j}{2}}\,\bar\epsilon\,\Gamma^{(j)}\\
\end{eqnarray*}
}
}\\
\begin{eqnarray*}\hspace{-8ex}\Rightarrow\quad
   \delta\bar\psi_\mu 
   &=& \left(\,D_\mu\epsilon\,\right)^T\,\Gamma^0\,+\,\frac{1}{2\cdot 144}\,\bar\epsilon\,\left(\,\Gamma^{\alpha\beta\gamma\delta}{}_{\mu}\,+\,8\,\Gamma^{\beta\gamma\delta}\,\delta^\alpha_\mu\,\right)\,G_{\alpha\beta\gamma\delta}
\end{eqnarray*}\\
Inserting the variation of $\psi$ and $\bar\psi$ and keeping in mind 
the sign from anticommuting the supervariation with the spinor  
one obtains:\\
\begin{eqnarray}
 (II) &=& 
   -\,\frac{e}{4\cdot 48}\,
     \left\{\,
                \delta\bar\psi_\mu
                 \left[\,
                         \Gamma^{\mu\nu\alpha\beta\gamma\delta}
                         \psi_\nu
                         \,+\,
                         12\,
                         g^{\mu [\alpha}\,\Gamma^{\gamma\delta}
                         \psi^{\beta ]}\,
                 \right]
               ~-~ \left[\,
                         \bar{\psi}_\mu\Gamma^{\mu\nu\alpha\beta\gamma\delta}
                         \,+\,
                         12\,
                         \bar{\psi}^{[\alpha}\,\Gamma^{\gamma\delta}
                         g^{\beta ]\nu}\,
                 \right]\,\delta\psi_\nu\,
    \right\}\,G_{\alpha\beta\gamma\delta}\nonumber\\
   &{\buildrel {\bar\epsilon\psi G^2} \over = }& 
   -\,\frac{e}{32\cdot (12)^3}\,
     \left\{\vbox{\vspace{2.5ex}}\right.\,
                 \bar\epsilon\,
                 \left(\,
                           \Gamma^{\talpha\tbeta\tgamma\tdelta}{}_{\mu}\,
                          +\,8\,\Gamma^{\tbeta\tgamma\tdelta}\,
                           \delta^\talpha_\mu\,
                 \right)\,
                 \left[\,
                         \Gamma^{\mu\nu\alpha\beta\gamma\delta}
                         \,+\,
                         12\,
                         g^{\mu [\alpha}\,\Gamma^{\gamma\delta}
                         g^{\beta ]\nu}\,
                 \right]\,\psi_\nu\nonumber\\
           &&\hspace{10ex}~-~ \bar{\psi}_\nu\,\left[\,
                         \Gamma^{\mu\nu\alpha\beta\gamma\delta}
                         \,-\,
                         12\,
                         g^{\nu [\alpha}\,\Gamma^{\gamma\delta}
                         g^{\beta ]\mu}\,
                 \right]\,
                 \left(\,
                           \Gamma^{\talpha\tbeta\tgamma\tdelta}{}_{\mu}\,
                          -\,8\,\Gamma^{\tbeta\tgamma\tdelta}\,
                           \delta^\talpha_\mu\,
                 \right)\,\epsilon\,
    \left.\vbox{\vspace{2.5ex}}\right\}\,G_{\talpha\tbeta\tgamma\tdelta}\,
                                         G_{\alpha\beta\gamma\delta}\nonumber
\end{eqnarray}\\ 
In section \ref{SubSec_EOM} we proved formula
(\ref{GammaFormel}). Performing a simple transposition and reordering
the Gamma matrices one obtains the identity 
(exercise \ref{exercise_transpo})\\ 
\begin{eqnarray}
    \left(
           \Gamma_{\alpha\beta\gamma\delta\nu}
           +8\,\Gamma_{\beta\gamma\delta}\,\eta_{\nu\alpha}
    \right)\,\Gamma^{\mu\nu\rho}\,\hat{G}^{\alpha\beta\gamma\delta}
    &=& 3\cdot\left(\vbox{\vspace{2ex}}\right.\,
           \Gamma^{\mu\rho}{}_{\alpha\beta\gamma\delta}
           \,+\,12\,\delta^{\mu}_{\alpha}\,
                    \Gamma_{\gamma\delta}\,
                    \delta^{\rho}_{\beta}\,
        \left.\vbox{\vspace{2ex}}\right)\, 
        \hat{G}^{\alpha\beta\gamma\delta}~.\label{GammaFormelB}
\end{eqnarray}\\
Inserting (\ref{GammaFormel}) and (\ref{GammaFormelB}) into II 
one obtains:\\
\begin{eqnarray}
 (II) &=&    -\,\frac{e}{3\cdot 32\cdot (12)^3}\,
     \left\{\vbox{\vspace{2.5ex}}\right.\,
                 \bar\epsilon\,
                 \left(\,
                           \Gamma^{\talpha\tbeta\tgamma\tdelta}{}_{\mu}\,
                          +\,8\,\Gamma^{\tbeta\tgamma\tdelta}\,
                           \delta^\talpha_\mu\,
                 \right)\,
                 \Gamma^{\mu\rho\nu}\,
                 \left(\,
                           \Gamma^{\talpha\tbeta\tgamma\tdelta}{}_{\rho}\,
                          -\,8\,\Gamma^{\tbeta\tgamma\tdelta}\,
                           \delta^\talpha_\rho\,
                 \right)\,
                 \psi_\nu\nonumber\\
           &&\hspace{12ex}~+~ \bar{\psi}_\nu\,
                 \left(\,
                           \Gamma^{\talpha\tbeta\tgamma\tdelta}{}_{\rho}\,
                          +\,8\,\Gamma^{\tbeta\tgamma\tdelta}\,
                           \delta^\talpha_\rho\,
                 \right)\,
                 \Gamma^{\nu\rho\mu}
                 \left(\,
                           \Gamma^{\talpha\tbeta\tgamma\tdelta}{}_{\mu}\,
                          -\,8\,\Gamma^{\tbeta\tgamma\tdelta}\,
                           \delta^\talpha_\mu\,
                 \right)\,\epsilon\,
    \left.\vbox{\vspace{2.5ex}}\right\}\,G_{\talpha\tbeta\tgamma\tdelta}\,
                                         G_{\alpha\beta\gamma\delta}\nonumber
\end{eqnarray}\\
The advantage of this representation of (II) is that the symmetry is 
made explicit.  
In exercise \ref{Exercise1} we developed formula (\ref{SwitchEpsPsi}), 
which tells us now, that out of the above expressions just the Gamma
matrices with 1,2 and 5 antisymmetrised indices and the corresponding 
Hodge duals (6,9,10) survive. Keeping just the first line, the corresponding terms are
doubled:\\
\begin{eqnarray}
 (II) &=&    -\,\underbrace{\frac{2\cdot e}{3\cdot 32\cdot
     (12)^3}}_{e\,\frac{1}{4\cdot (12)^4}}\,
     \left\{\vbox{\vspace{2.5ex}}\right.\,
                 \bar\epsilon\,
                 \left(\,
                           \Gamma^{\talpha\tbeta\tgamma\tdelta}{}_{\mu}\,
                          +\,8\,\Gamma^{\tbeta\tgamma\tdelta}\,
                           \delta^\talpha_\mu\,
                 \right)\,
                 \Gamma^{\mu\rho\nu}\,
                 \left(\,
                           \Gamma^{\talpha\tbeta\tgamma\tdelta}{}_{\rho}\,
                          -\,8\,\Gamma^{\tbeta\tgamma\tdelta}\,
                           \delta^\talpha_\rho\,
                 \right)\,
                 \psi_\nu\,
    \left.\vbox{\vspace{2.5ex}}\right\}_{1,2,5}\,
    G_{\talpha\tbeta\tgamma\tdelta}\,
    G_{\alpha\beta\gamma\delta}\nonumber
\end{eqnarray}\\
Using eq.~(\ref{GammaFormel}) ones more we obtain again\\
\begin{eqnarray}
 (II) &=&    -\,\frac{3\,e}{4\cdot (12)^4}\,
     \left\{\vbox{\vspace{2.5ex}}\right.\,
                    \bar\epsilon\,
                 \left(\,
                           \Gamma^{\talpha\tbeta\tgamma\tdelta}{}_{\mu}\,
                          +\,8\,\Gamma^{\tbeta\tgamma\tdelta}\,
                           \delta^\talpha_\mu\,
                 \right)\,
                 \left[\,
                         \Gamma^{\mu\nu\alpha\beta\gamma\delta}
                         \,+\,
                         12\,
                         g^{\mu [\alpha}\,\Gamma^{\gamma\delta}
                         g^{\beta ]\nu}\,
                 \right]\,\psi_\nu\,
    \left.\vbox{\vspace{2.5ex}}\right\}_{1,2,5}\,
    G_{\talpha\tbeta\tgamma\tdelta}\,
    G_{\alpha\beta\gamma\delta}\nonumber\\
  &=& -\,\frac{3\,e}{4\cdot (12)^4}\,
     \left\{\vbox{\vspace{2.5ex}}\right.\,
       \bar\epsilon\,\Gamma^{\talpha\tbeta\tgamma\tdelta}{}_{\mu}\,
       \Gamma^{\mu\nu\alpha\beta\gamma\delta}\,\psi_\nu\,     
       +\,12\,\bar\epsilon\,\Gamma^{\talpha\tbeta\tgamma\tdelta}{}_{\mu}\,
         g^{\mu [\alpha}\,\Gamma^{\gamma\delta}\,
         g^{\beta ]\nu}\,\psi_\nu\,\nonumber\\
  &&  \hspace{12ex}+\,8\,\bar\epsilon\,\Gamma^{\tbeta\tgamma\tdelta}\,
         \delta^\talpha_\mu\,
         \Gamma^{\mu\nu\alpha\beta\gamma\delta}\,\psi_\nu\,
      +\,96\,\bar\epsilon\,\Gamma^{\tbeta\tgamma\tdelta}\,
         \delta^\talpha_\mu\,g^{\mu [\alpha}\,\Gamma^{\gamma\delta}\,
         g^{\beta ]\nu}\,\psi_\nu\,
     \left.\vbox{\vspace{2.5ex}}\right\}_{1,2,5}\,
     G_{\talpha\tbeta\tgamma\tdelta}\,
     G_{\alpha\beta\gamma\delta}\nonumber     
\end{eqnarray}\\
To further simplify this expression one needs the following identity,\\
\begin{eqnarray}
   \Gamma^{\talpha\tbeta\tgamma\tdelta}{}_{\mu}
   \Gamma^{\mu\nu\alpha\beta\gamma\delta}
   &=& (D-9)\,\Gamma^{\talpha\tbeta\tgamma\tdelta}\,
           \Gamma^{\nu\alpha\beta\gamma\delta}\,
      +\,20\,\Gamma^{[\talpha\tbeta\tgamma}\,g^{\tdelta
   ][\nu}\Gamma^{\alpha\beta\gamma\delta ]}~,\label{Formel900}
\end{eqnarray}\\
the proof of which is left as an exercise. One obtains\\ 
\begin{eqnarray}
 (II) 
  &=& -\,\frac{3\,e}{4\cdot (12)^4}\,
     \left\{\vbox{\vspace{2.5ex}}\right.\,
       (D-9)\,
       \bar\epsilon\,\Gamma^{\talpha\tbeta\tgamma\tdelta}\,
       \Gamma^{\nu\alpha\beta\gamma\delta}\,\psi_\nu\,
       +\,20\,\bar\epsilon\,\Gamma^{[\talpha\tbeta\tgamma}g^{\tdelta ][\nu}\,
       \Gamma^{\alpha\beta\gamma\delta ]}\,\psi_\nu\nonumber\\[2ex]
  && \hspace{12ex}
      +\,12\,\bar\epsilon\,\Gamma^{\talpha\tbeta\tgamma\tdelta}{}_{\mu}\,
         g^{\mu [\alpha}\,\Gamma^{\gamma\delta}\,
         g^{\beta ]\nu}\,\psi_\nu\,
      +\,8\,\bar\epsilon\,\Gamma^{\tbeta\tgamma\tdelta}\,
         \delta^\talpha_\mu\,
         \Gamma^{\mu\nu\alpha\beta\gamma\delta}\,\psi_\nu\,\nonumber\\[2ex]
 && \hspace{12ex}  
      +\,96\,\bar\epsilon\,\Gamma^{\tbeta\tgamma\tdelta}\,
         \delta^\talpha_\mu\,g^{\mu [\alpha}\,\Gamma^{\gamma\delta}\,
         g^{\beta ]\nu}\,\psi_\nu\,
     \left.\vbox{\vspace{2.5ex}}\right\}_{1,2,5}\,
     G_{\talpha\tbeta\tgamma\tdelta}\,
     G_{\alpha\beta\gamma\delta}\hspace{5cm}\nonumber     
\end{eqnarray}\\
Another useful identity is\\ 
\begin{eqnarray}
8\,\bar\epsilon\,\Gamma^{\tbeta\tgamma\tdelta}\,
         \delta^\talpha_\mu\,
         \Gamma^{\mu\nu\alpha\beta\gamma\delta}\,\psi_\nu\,
  &=& 8\,\bar\epsilon\,\Gamma^{\tbeta\tgamma\tdelta\talpha}\,
         \Gamma^{\nu\alpha\beta\gamma\delta}\,\psi_\nu\,
     +\,40\,\bar\epsilon\,\Gamma^{[\talpha\tbeta\tgamma}\,g^{\tdelta ][\nu}\,
         \Gamma^{\alpha\beta\gamma\delta ]}\,\psi_\nu\,
\end{eqnarray}\\
and inserting this one obtains
\begin{eqnarray}
 (II) 
  &=& -\,\frac{3\,e}{4\cdot (12)^4}\,
     \left\{\vbox{\vspace{2.5ex}}\right.\,
       (D-9-8)\,
       \bar\epsilon\,\Gamma^{\talpha\tbeta\tgamma\tdelta}\,
       \Gamma^{\nu\alpha\beta\gamma\delta}\,\psi_\nu\,
       +\,60\,\bar\epsilon\,\Gamma^{[\talpha\tbeta\tgamma}g^{\tdelta ][\nu}\,
       \Gamma^{\alpha\beta\gamma\delta ]}\,\psi_\nu\nonumber\\[2ex]
  && \hspace{12ex}
      +\,12\,\bar\epsilon\,\Gamma^{\talpha\tbeta\tgamma\tdelta}{}_{\mu}\,
         g^{\mu [\alpha}\,\Gamma^{\gamma\delta}\,
         g^{\beta ]\nu}\,\psi_\nu\,  
      +\,96\,\bar\epsilon\,\Gamma^{\tbeta\tgamma\tdelta}\,
         \delta^\talpha_\mu\,g^{\mu [\alpha}\,\Gamma^{\gamma\delta}\,
         g^{\beta ]\nu}\,\psi_\nu\,
     \left.\vbox{\vspace{2.5ex}}\right\}_{1,2,5}\,
     G_{\talpha\tbeta\tgamma\tdelta}\,
     G_{\alpha\beta\gamma\delta}\nonumber     
\end{eqnarray}\\
The remaining four terms must be expanded in a Clifford basis by brute
force. The result is 
\begin{eqnarray}
 (II) 
  &=& -\,\frac{3\,e}{4\cdot (12)^4}\,
     \left\{\vbox{\vspace{2.5ex}}\right.\,\nonumber\\
  &&\hspace{14ex}
       (D-9-8)\,
       \bar\epsilon\,
       \left[\vbox{\vspace{2.5ex}}\right.
                 \Gamma_{\talpha\tbeta\tgamma\tdelta}
                  {}^{\nu\alpha\beta\gamma\delta}\,
               +\,20\,\delta_{[\tdelta}^{[\nu}\Gamma_{\talpha\tbeta\tgamma]}\,
                 {}^{\alpha\beta\gamma\delta ]}\,
               +\,120\,
                 \delta_{[\tdelta}^{[\nu}\delta_\tgamma^\alpha
                 \Gamma_{\talpha\tbeta ]}{}^{\beta\gamma\delta ]}\nonumber\\
             &&\hspace{29ex}
               +\,240\,\delta_{[\tdelta}^{[\nu}\delta_\tgamma^\alpha
                       \delta_\tbeta^\beta
                 \Gamma_{\talpha ]}{}^{\gamma\delta ]}\,
               +\,120\,\delta_{[\tdelta}^{[\nu}\delta_\tgamma^\alpha
                       \delta_\tbeta^\beta
                 \delta_{\talpha ]}^{\gamma}\Gamma^{\delta ]}
       \left.\vbox{\vspace{2.5ex}}\right]\,\psi_\nu\,\nonumber\\[2ex]
  &&\hspace{12ex}
       +\,60\,\bar\epsilon\,\delta_{\tdelta}^{[\nu}
       \left[\vbox{\vspace{2.5ex}}\right.\,
         \Gamma_{\talpha\tbeta\tgamma}{}^{\alpha\beta\gamma\delta ]}\,
         +\,12\,\delta^{\alpha}_{\tgamma}
           \Gamma_{\talpha\tbeta}{}^{\beta\gamma\delta ]}\,
         +\,36\,\delta^{\alpha}_{\tgamma}\delta^{\beta}_{\tbeta}
           \Gamma_{\talpha}{}^{\gamma\delta ]}\,
         +\,24\,\delta^{\alpha}_{\tgamma}\delta^{\beta}_{\tbeta}\delta^{\gamma}_{\talpha}
           \Gamma^{\delta ]}\,
       \left.\vbox{\vspace{2.5ex}}\right]\,\psi_\nu\nonumber\\[2ex]
  && \hspace{12ex}
      +\,12\,\bar\epsilon\,g^{\mu\alpha}\,
      \left[\vbox{\vspace{2.5ex}}\right.\,
         \Gamma_{\talpha\tbeta\tgamma\tdelta\mu}{}^{\gamma\delta}\,
        +\,10\,\delta^{\gamma}_{[\mu}\Gamma_{\talpha\tbeta\tgamma\tdelta ]}\,
         {}^{\delta}\,
        +\,20\,\delta^{\gamma}_{[\mu}\delta^{\delta}_{\delta}\Gamma_{\talpha\tbeta\tgamma ]}\,
      \left.\vbox{\vspace{2.5ex}}\right]\,g^{\beta\nu}\,\psi_\nu\,
       \nonumber\\[2ex]
  &&\hspace{12ex}  
      +\,96\,\bar\epsilon\,\delta^{\alpha}_{\talpha}
       \left[\vbox{\vspace{2.5ex}}\right.
         \Gamma_{\tbeta\tgamma\tdelta}{}^{\gamma\delta}\,
        +\,6\,\delta_{\tdelta}^\gamma\Gamma_{\tbeta\tgamma}{}^{\delta}\,
        +\,6\,\delta_\tdelta^\gamma\delta_\tgamma^\delta \Gamma_{\tbeta}\,
       \left.\vbox{\vspace{2.5ex}}\right]\,\psi^\beta\nonumber\\[2ex]
  &&\hspace{11ex}
     \left.\vbox{\vspace{2.5ex}}\right\}_{1,2,5}\,
     G_{\talpha\tbeta\tgamma\tdelta}\,
     G^{\alpha\beta\gamma\delta}\label{IIBrute}   
\end{eqnarray}\\

\pagebreak
Collecting the terms in (\ref{IIBrute}) with just one Gamma matrix one finds\\ 
\begin{eqnarray*} 
  \bar\epsilon\Gamma^{(1)}\psi &=&-\frac{3}{4\cdot (12)^4}\,
   \left\{\vbox{\vspace{2.5ex}}\right.\,
            -\,6\cdot 120\,\bar\epsilon\,\delta^{[\alpha}_{\tdelta}
             \delta^{\beta}_{\tgamma}\delta^{\gamma}_{\tbeta}
             \delta^{\delta}_{\talpha}\Gamma^{\nu ]}\psi_{\nu}\,
            +\,12\cdot 120\,\bar\epsilon\,\delta^{[\alpha}_{\tdelta}
             \delta^{\beta}_{\tgamma}\delta^{\gamma}_{\tbeta}
             \delta^{\delta}_{\talpha}\Gamma^{\nu ]}\psi_{\nu}\,\\
     &&\hspace{16ex}       
           +\,96\cdot 6\,\bar\epsilon\,\delta_{\talpha}^{\alpha}
             \delta_{\tdelta}^{\gamma}\delta_{\tgamma}^{\delta}
             \Gamma_{\tbeta}\psi^{\beta}\,
   \left.\vbox{\vspace{2.5ex}}\right\}\,
    G^{\talpha\tbeta\tgamma\tdelta}\,
    G_{\alpha\beta\gamma\delta}\\
  &=&-\frac{9}{2\cdot (12)^3}\,
   \left\{\vbox{\vspace{2.5ex}}\right.\,
             10\,\bar\epsilon\,\delta^{[\alpha}_{\tdelta}
             \delta^{\beta}_{\tgamma}\delta^{\gamma}_{\tbeta}
             \delta^{\delta}_{\talpha}\Gamma^{\nu ]}\psi_{\nu}\,
           +\,8\,\bar\epsilon\,\delta_{\talpha}^{\alpha}
             \delta_{\tdelta}^{\gamma}\delta_{\tgamma}^{\delta}
             \Gamma_{\tbeta}\psi^{\beta}\,
   \left.\vbox{\vspace{2.5ex}}\right\}    
    G^{\talpha\tbeta\tgamma\tdelta}\,
    G_{\alpha\beta\gamma\delta}\\
  &=&-\frac{9}{2\cdot (12)^3}\,
   \left\{\vbox{\vspace{2.5ex}}\right.\,
             -\,8\,(G^2)_{\alpha\beta}\,\bar\epsilon\,\Gamma_{\alpha}\psi^{\beta}\,
             +\,10\,\bar\epsilon\,\delta^{[\alpha}_{\tdelta}
             \delta^{\beta}_{\tgamma}\delta^{\gamma}_{\tbeta}
             \delta^{\delta}_{\talpha}\Gamma^{\nu ]}\psi_{\nu}\,    
             G^{\talpha\tbeta\tgamma\tdelta}\,
             G_{\alpha\beta\gamma\delta}
   \left.\vbox{\vspace{2.5ex}}\right\}\\
&=&-\frac{9}{2\cdot (12)^3}\,
   \left\{\vbox{\vspace{2.5ex}}\right.\,
             -\,16\,(G^2)_{\alpha\beta}\,\bar\epsilon\,\Gamma_{\alpha}\psi^{\beta}\,
             +\,2\,(G^2)\,g_{\alpha\beta}\,\bar\epsilon\,\Gamma_{\alpha}\psi^{\beta}\,
   \left.\vbox{\vspace{2.5ex}}\right\}\\
&=&\frac{1}{24}\,
   \left\{\vbox{\vspace{2.5ex}}\right.\,
             (G^2)_{\alpha\beta}\,\,
             -\,\frac{1}{8}\,(G^2)\,g_{\alpha\beta}\,
   \left.\vbox{\vspace{2.5ex}}\right\}\,
   \bar\epsilon\,\Gamma_{\alpha}\psi^{\beta}~.
\end{eqnarray*}\\
\noindent
Collecting the terms in (\ref{IIBrute}) with just two (or nine) Gamma matrices 
and using (\ref{DualityRel}) one finds\\ 
\begin{eqnarray*}
\bar\epsilon\Gamma^{(2)}\psi 
   &=& -\,\frac{3}{4\cdot (12)^3}\;(-6)\;\bar\epsilon\,
        \Gamma^{\talpha\tbeta\tgamma\tdelta\nu\alpha\beta\gamma\delta}\,
         \psi_\nu\,
         G_{\talpha\tbeta\tgamma\tdelta}\,G_{\alpha\beta\gamma\delta}\\
   &=& -\,\frac{3\cdot 6}{4\cdot (12)^3}\;\bar\epsilon\,
        \frac{1}{2!}\,\varepsilon^{\talpha\tbeta\tgamma\tdelta\nu\alpha\beta\gamma\delta\mu\rho}\,\Gamma_{\mu\rho}\,\psi_\nu\,
         G_{\talpha\tbeta\tgamma\tdelta}\,G_{\alpha\beta\gamma\delta}\\
   &=& -\,\frac{9}{4\cdot (12)^3}\;\varepsilon^{\alpha_1\ldots
   \alpha_4\beta_1\ldots \beta_4\mu\nu\rho}\,\bar\epsilon\,\Gamma_{[
   \mu\nu}\,\psi_{\rho ]}\,
         G_{\alpha_1\ldots\alpha_4}\,G_{\beta_1\ldots\beta_4}~.\hspace{19ex}
\end{eqnarray*}\\
Collecting the terms in (\ref{IIBrute}) with just five Gamma matrices, while
skipping the two $G_{[4]}$ factors one finds\\ 
\begin{eqnarray*}
\bar\epsilon\Gamma^{(5)}\psi &=& -\,\frac{3}{4\cdot (12)^3}\,
   \bar\epsilon\,\left\{\vbox{\vspace{2.5ex}}\right.\,
         -\,6\cdot 120\,
          \delta^{[\nu}_{[ \tdelta}\,\delta^\alpha_\tgamma\,
          \Gamma_{\talpha\tbeta ]}{}^{\beta\gamma\delta
   ]}\,\psi_{\nu}\,
         +\,6\cdot 120\,
          \delta^{[\nu}_{[ \tdelta}\,\delta^\alpha_\tgamma\,
          \Gamma_{\talpha\tbeta ]}{}^{\beta\gamma\delta
   ]}\,\psi_{\nu}\,\\
   && +\,120\,g_{\mu\alpha}\,\delta^{[\mu}_{[\gamma}\,
        \Gamma^{\talpha\tbeta\tgamma\tdelta ]}{}_{\delta
   ]}\,\delta^\nu_\beta\,\psi_\nu\,
   +\,96\,\delta^{\talpha}_{\alpha}\,
         \Gamma^{\tbeta\tgamma\tdelta}{}_{\gamma\delta}\,
         \delta^\nu_\beta\,\psi_\nu\,
        \left.\vbox{\vspace{2.5ex}}\right\}\\
  &=& -\,\frac{3}{4\cdot (12)^2}\,
   \bar\epsilon\,\left\{\vbox{\vspace{2.5ex}}\right.\,
      10\,g_{\mu\alpha}\,\delta^{[\mu}_{[\gamma}\,
        \Gamma^{\talpha\tbeta\tgamma\tdelta ]}{}_{\delta
   ]}\,
   +\,8\,\delta^{\talpha}_{\alpha}\,
      \Gamma^{\tbeta\tgamma\tdelta}{}_{\gamma\delta}\,
   \left.\vbox{\vspace{2.5ex}}\right\}\,\psi_\beta\\
  &=& 0~.\hspace{10.75cm}
\end{eqnarray*}\\

\pagebreak
Inserting all partial results for (I) and (II) back into
(\ref{Noether3}) one obtains:\\
\begin{eqnarray}
   \delta_QS &=& \int dx^D\,\left[\vbox{\vspace{3ex}}\right.\,
                 \frac{e}{24}\,
                                \left(\,
                                        (G^2)_{\alpha_1\beta_1}\,          
                                         -\,\frac{1}{8}\;
                                         G^2\cdot g_{\alpha_1\beta_1}\,
                                \right)\,\bar\epsilon\,
                                \Gamma^{(\alpha_1}\,\psi^{\beta_1)}\nonumber\\[2ex]
                        && \hspace{9ex}-\,\frac{9}{4\cdot (12)^4}\,
                \varepsilon^{\alpha_1\ldots\alpha_4
                             \beta_1\ldots\beta_4\mu\nu\rho}\,
                G_{\alpha_1\ldots\alpha_4}\,
                G_{\beta_1\ldots\beta_4}\,\bar\epsilon\,\Gamma_{[\mu\nu}\,\psi_{\rho]}\nonumber\\[2ex]
                        && \hspace{9ex} -\,\frac{e}{24}\,
                                \left(\,
                                        (G^2)_{\alpha_1\beta_1}\,          
                                         -\,\frac{1}{8}\;
                                         G^2\cdot g_{\alpha_1\beta_1}\,
                                \right)\,\bar\epsilon\,
                                \Gamma^{(\alpha_1}\,\psi^{\beta_1)}\,
                 \left.\vbox{\vspace{3ex}}\right]\nonumber\\[2ex]
 &=&\int dx^D\,\left[\vbox{\vspace{3ex}}\right.\,
   -\,\frac{9}{4\cdot (12)^4}\,
                \varepsilon^{\alpha_1\ldots\alpha_4
                             \beta_1\ldots\beta_4\mu\nu\rho}\,
                G_{\alpha_1\ldots\alpha_4}\,
                G_{\beta_1\ldots\beta_4}\,\bar\epsilon\,\Gamma_{[\mu\nu}\,\psi_{\rho]}\left.\vbox{\vspace{3ex}}\right]\label{CSTA}
\end{eqnarray}\\
The variation of the action does not vanish. This must be cured by  
adding to the action (\ref{Noether3}) a suitable term, whose variation
produces the same result but with opposite sign.
One makes the following ansatz for a compensating term, called
Chern-Simons term,\\ 
\begin{eqnarray}
   S_{CS} &=& \int
   dx^{D}\;\frac{1}{4\cdot (12)^4}\,\varepsilon^{\alpha_1\ldots\alpha_4\beta_1\ldots\beta_4\mu\nu\rho}\,
           G_{\alpha_1\ldots\alpha_4}\,G_{\beta_1\ldots\beta_4}\,C_{\mu\nu\rho}
\end{eqnarray}\\
and for the super transformation of the 3-form potential\\ 
\begin{eqnarray}\label{SUSY_3F}
   \delta_QC_{\mu\nu\rho} 
   &=& a\,\bar\epsilon\,\Gamma_{[\mu\nu}\,\psi_{\rho ]}~.
\end{eqnarray}\\
Performing the variation of the new term with respect to supersymmetry
one obtains:\\ 
\begin{eqnarray}
   \delta_QS_{CS} &=& \int
   dx^{D}\;\left\{\vbox{\vspace{2.5ex}}\right.\,
                \frac{2}{4\cdot (12)^4}\,
                \varepsilon^{\alpha_1\ldots\alpha_4
                             \beta_1\ldots\beta_4\mu\nu\rho}\,
                \delta_Q\,G_{\alpha_1\ldots\alpha_4}\,
                G_{\beta_1\ldots\beta_4}\,C_{\mu\nu\rho}\,\nonumber\\
             &&\hspace{9ex} +\,\frac{1}{4\cdot (12)^4}\,
                \varepsilon^{\alpha_1\ldots\alpha_4
                             \beta_1\ldots\beta_4\mu\nu\rho}\,
                G_{\alpha_1\ldots\alpha_4}\,
                G_{\beta_1\ldots\beta_4}\,\delta_QC_{\mu\nu\rho}\,
          \left.\vbox{\vspace{2.5ex}}\right\}\nonumber
\end{eqnarray}\\
and by inserting (\ref{SUSY_3F}) it reads\\
\begin{eqnarray*}
   \delta_QS_{CS}  &=& \int
   dx^{D}\;\left\{\vbox{\vspace{2.5ex}}\right.\,
                \frac{2}{4\cdot (12)^4}\,
                \varepsilon^{\alpha_1\ldots\alpha_4
                             \beta_1\ldots\beta_4\mu\nu\rho}\,
                4\,\partial_{[\alpha_1}\,a\,\bar\epsilon\,
                   \Gamma_{\alpha_2\alpha_3}
                   \psi_{\alpha_4 ]}\,
                G_{\beta_1\ldots\beta_4}\,C_{\mu\nu\rho}\,\nonumber\\
             &&\hspace{9ex} +\,\frac{1}{4\cdot (12)^4}\,
                \varepsilon^{\alpha_1\ldots\alpha_4
                             \beta_1\ldots\beta_4\mu\nu\rho}\,
                G_{\alpha_1\ldots\alpha_4}\,
                G_{\beta_1\ldots\beta_4}\,a\,\bar\epsilon\,\Gamma_{[\mu\nu}\,\psi_{\rho]}\,
          \left.\vbox{\vspace{2.5ex}}\right\}~.
\end{eqnarray*}\\
A partial integration can be done and due to\\
\begin{eqnarray*}
   \varepsilon^{\alpha_1\ldots\alpha_4
                             \beta_1\ldots\beta_4\mu\nu\rho}\,
                4\,a\,\partial_{[\alpha_1}\bar\epsilon
                   \Gamma_{\alpha_2\alpha_3}
                   \psi_{\alpha_4 ]}\,
                G_{\beta_1\ldots\beta_4}\,C_{\mu\nu\rho}
   &=& 4\,\partial_{\alpha_1}
          \left(\vbox{\vspace{2.5ex}}\right.
                \varepsilon^{\alpha_1\ldots\alpha_4
                             \beta_1\ldots\beta_4\mu\nu\rho}\,
                      a\,\bar\epsilon
                   \Gamma_{\alpha_2\alpha_3}
                   \psi_{\alpha_4}\,
                G_{\beta_1\ldots\beta_4}\,C_{\mu\nu\rho}\,
         \left.\vbox{\vspace{2.5ex}}\right)\\
   &-& 4\,
                \varepsilon^{\alpha_1\ldots\alpha_4
                             \beta_1\ldots\beta_4\mu\nu\rho}\,
                   a\,\bar\epsilon
                   \Gamma_{\alpha_2\alpha_3}
                   \psi_{\alpha_4 ]}\,
                G_{\beta_1\ldots\beta_4}\,\partial_{[\alpha_1}C_{\mu\nu\rho ]}
        \\
   &\sim&\varepsilon^{\alpha_1\ldots\alpha_4
                             \beta_1\ldots\beta_4\mu\nu\rho}\,
                G_{\alpha_1\ldots\alpha_4}
                G_{\beta_1\ldots\beta_4}\,a\,\bar\epsilon\Gamma_{[\mu\nu}\psi_{\rho]}
\end{eqnarray*}\\
the variation of the Chern-Simons-term reads\\ 
\begin{eqnarray}
   \delta_Q{\mathcal{L}}_{CS}
             &=& \frac{3}{4\cdot (12)^4}\,
                \varepsilon^{\alpha_1\ldots\alpha_4
                             \beta_1\ldots\beta_4\mu\nu\rho}\,
                G_{\alpha_1\ldots\alpha_4}\,
                G_{\beta_1\ldots\beta_4}\,a\,\bar\epsilon\,\Gamma_{[\mu\nu}\,\psi_{\rho]}~.
\end{eqnarray}\\
This must be compared with the outcome of the variation of the action
constructed so far, i.e. with (\ref{CSTA}). Relating the coefficients to each
other fixes a to be 
\begin{eqnarray}
      a &=& 3~.
\end{eqnarray}
This is consistent with the choice made for the supersymmetric 
transformation law of the three form (\ref{SUSYTRAFO}) of section 
\ref{Sec_MTheory} and basically proves it.\\

\noindent
The action now reads\\ 
\begin{eqnarray}\label{Noether4}
   {\cal L} &=& \frac{1}{4}\,eR 
                +\frac{1}{2}\,e\bar{\psi}_\mu\Gamma^{\mu\nu\rho}
                 D_\nu\left(\omega\right)\psi_\rho
                -\frac{1}{4\cdot 48}\,eG_{\mu\nu\rho\sigma}G^{\mu\nu\rho\sigma}
                 \nonumber\\
             && -\frac{1}{4\cdot 48}\,e 
                 \left(
                   \bar{\psi}_\mu\Gamma^{\mu\nu\alpha\beta\gamma\delta}
                   \psi_\nu
                   +12\,\bar{\psi}^\alpha\Gamma^{\gamma\delta}\psi^\beta
                 \right)\,G_{\alpha\beta\gamma\delta}\nonumber\\
             && +\frac{1}{4\cdot 144^2}\,\epsilon^{\alpha_1\ldots\alpha_4
                 \beta_1\ldots\beta_4\mu\nu\rho}G_{\alpha_1\ldots\alpha_4}
                 G_{\beta_1\ldots\beta_4}C_{\mu\nu\rho}~.
\end{eqnarray}\\
Here we are missing out all four fermion terms, which can be included 
strangely enough by a simple substitution (c.f. (\ref{MTheory}) ).\\

\noindent
Finally one has to check once again 
that the so constructed Lagrangian density with all four fermion terms 
is invariant under the full action of supersymmetry. The explicit 
calculation is done in \cite{Fierzidentities} and boils down to 
checking certain Fierz identities sometimes even more complicated 
than (\ref{CJSFierIdentity}).

\subsubsection*{Homework:}

\begin{exercise}\label{exercise_transpo}
                 Prove (\ref{GammaFormelB}) by transposing 
                 (\ref{GammaFormel}). 
\end{exercise}

\begin{exercise}
      Prove (\ref{Formel900}). 
\end{exercise}

\vfill\eject
\section{Elementary M-Brane Solutions}

This chapter is devoted to the presentation of simple solutions to 11d
supergravity. We show explicitly that the Einstein equations are
satisfied and present some techniques typical of this kind of 
calculation.

\subsection{M5-Brane}
\label{SubSec_M5_Brane}

The M5-brane solution \cite{Gueven:1992hh} has the form 
\begin{eqnarray}\label{SugraM5}
      ds^2 &=& N^{-1/3}\left(-dt^2+dx_1^2+\ldots +dx_5^2 \right)
            ~+~N^{2/3}\left(dx_6^2+\ldots +dx_{10}^2\right)\nonumber\\
     G_{\alpha_1\ldots\alpha_4} &=& c\,N^{2/3}\,
     \epsilon_{\alpha_1\ldots\alpha_5}\,
     \partial^{\alpha_5}\,N(x_6,\ldots,x_{10}),\hspace{2ex} c~=~\pm1,
     \hspace{5ex} \Delta N~=~0
\end{eqnarray}
and it is convenient to consider the fundamental M5-brane solution as 
corresponding  to the choice $N~=~1+\frac{a}{r^3}$.\\

\noindent
For transparency the metric is written in matrix form:
\begin{eqnarray}
    g_{\mu\nu} &=& \left(
                         \begin{array}{cccccc}
                             -\,N^{-1/3} & & & & & \\
                                      & \ddots &&&&\\
                                      & & N^{-1/3} &&&\\
                             & & & N^{2/3} & &\\
                             & & & & \ddots & \\
                             & & & & & N^{2/3}(x_6\ldots x_{10})
                         \end{array}
                   \right)
\end{eqnarray}

\subsubsection{Christoffel Symbols}

Now we are going to compute the Christoffel symbols of the M5-Brane,
defined by
\begin{eqnarray}\label{ChristoffelSymbolLower}
    \Gamma_{\rho (\nu\lambda)} &=&
    \frac{1}{2}\,\left(\,g_{\rho\nu,\lambda}+g_{\rho\lambda,\nu} - g_{\nu\lambda,\rho}\,\right)~.
\end{eqnarray}

\noindent
Splitting the index $\mu\in\{0..10\}$ into $a\in\{0..5\}$ and $i\in
\{6..10\}$ the Christoffel symbols read:
\begin{eqnarray}
    \Gamma_{a(bc)} &=& 0\nonumber\\
    \Gamma_{i(bc)} &=& -\frac{1}{2}\,g_{bc,i} ~=~ -\frac{1}{2}\,
                        \frac{\partial N^{-1/3}}{\partial x^i}\,\eta_{bc}
                   ~=~ \frac{1}{6}\,\frac{\partial_i N}{N^{4/3}}\,
                       \eta_{bc} \nonumber\\
    \Gamma_{b(ic)} &=& \frac{1}{2}\,g_{bc,i} ~=~ \frac{1}{2}\,
                        \frac{\partial N^{-1/3}}{\partial x^i}\,\eta_{bc}
                   ~=~ -\frac{1}{6}\,\frac{\partial_i N}{N^{4/3}}\,
                        \eta_{bc}\nonumber\\
    \Gamma_{a(ij)} &=& \frac{1}{2}\left(g_{ai,j}+g_{aj,i}\right) ~=~
    0\\
    \Gamma_{i(aj)} &=& \frac{1}{2}\left(g_{ia,j}-g_{aj,i}\right) ~=~
    0\nonumber\\
    \Gamma_{i(jk)} &=&
    \frac{1}{2}\left(g_{ij,k}+g_{ik,j}-g_{jk,i}\right)
                   ~=~ \frac{1}{2}\,
                       \left(\,
                          \frac{\partial N^{2/3}}{\partial x^k}\,\delta_{ij} +
                          \frac{\partial N^{2/3}}{\partial x^j}\,\delta_{ik} -
                          \frac{\partial N^{2/3}}{\partial x^i}\,\delta_{jk}\,
                       \right)\nonumber\\ 
                   &=& \frac{1}{3}\,\frac{1}{N^{1/3}}\,
                       \left(\,
                                  \partial_k N\,\delta_{ij} +
                                  \partial_j N\,\delta_{ik} -
                                  \partial_i N\,\delta_{jk}\,
                       \right)\nonumber 
\end{eqnarray}

\noindent
It follows that
\begin{eqnarray}\label{ChristoffelSymbolUpper}
   \Gamma^{\mu}{}_{(\nu\lambda)} &=&
   g^{\mu\rho}\,\Gamma_{\rho(\nu\lambda)}
   ~=~ g^{\mu a}\Gamma_{a(\nu\lambda)} + g^{\mu i}\Gamma_{i(\nu\lambda)}
\end{eqnarray}
becomes
\begin{eqnarray}
    \Gamma^a{}_{(bc)} &=& g^{ad}\underbrace{\Gamma_{d(bc)}}_{0}+\underbrace{g^{ai}}_{0}\Gamma_{i(bc)} ~=~ 0\nonumber\\
    \Gamma^i{}_{(bc)} &=& g^{ij}\Gamma_{j(bc)} ~=~ 
     g^{ij}\frac{1}{6}\,\frac{\partial_j
    N}{N^{4/3}}\,\eta_{bc} ~=~ \frac{1}{6}\frac{\partial^i N}{N^{4/3}}
    \eta_{bc}\nonumber\\
    \Gamma^a{}_{(ib)} &=& g^{ac}\Gamma_{c(ib)} ~=~
    N^{1/3}\eta^{ac}\left(-\frac{1}{6}\,\frac{\partial_i
    N}{N^{4/3}}\,\eta_{bc}\,\right) ~=~ -\frac{1}{6}
    \frac{\partial_i N}{N} \delta^a_{b}\label{Chr_M5}\\
    \Gamma^i{}_{(jb)} &=& g^{ik}\Gamma_{k(jb)} ~=~ 0\nonumber\\
    \Gamma^a{}_{(ij)} &=& g^{ab}\Gamma_{b(ij)} ~=~ 0\nonumber\\
    \Gamma^i{}_{(jk)} &=& g^{il}\Gamma_{l(jk)} ~=~ N^{-2/3}\delta^{il}
                          \frac{1}{3}\,\frac{1}{N^{1/3}}\,
                       \left(\,
                                  \partial_k N\,\delta_{lj} +
                                  \partial_j N\,\delta_{lk} -
                                  \partial_l N\,\delta_{jk}\,
                       \right) \nonumber\\
                      &=& \frac{1}{3}\,\frac{1}{N}\,
                       \left(\,
                                  \partial_k N\,\delta^i_{j} +
                                  \partial_j N\,\delta^i_{k} -
                                  \partial^i N\,g_{jk}\,
                       \right)~.\nonumber 
\end{eqnarray}
Of special importance is the partial contraction of the Christoffel
symbols, which we add for the convenience of the reader here:
\begin{eqnarray}
   \Gamma^\mu{}_{(j\mu)} &=& \Gamma^a{}_{(ja)} + \Gamma^i{}_{(ji)}
                       ~=~ -\frac{\partial_j N}{N} +
                       \frac{5}{3}\,\frac{\partial_j N}{N}
                       ~=~ \frac{2}{3}\,\frac{\partial_j N}{N}~.\label{TraceChr}
\end{eqnarray}

\subsubsection{Ricci Tensor}

The Riemann curvature tensor is defined by   
\begin{eqnarray}
     R^\alpha{}_{\beta\gamma\delta} 
       &=& \frac{\partial \Gamma^\alpha{}_{(\beta\delta)}}{\partial x^\gamma} 
        -  \frac{\partial \Gamma^\alpha{}_{(\beta\gamma)}}{\partial x^\delta} 
        +  \Gamma^\alpha{}_{(\eta\gamma)}\Gamma^\eta{}_{(\beta\delta)}
        -  \Gamma^\alpha{}_{(\eta\delta)}\Gamma^\eta{}_{(\beta\gamma)}~.
\end{eqnarray}
A contraction of the first with the third index produces the Ricci tensor:  
\begin{eqnarray}\label{RicciTensor}
    R_{\beta\delta} 
       &=& R^\alpha{}_{\beta\alpha\delta}\nonumber\\
       &=& \frac{\partial \Gamma^\alpha{}_{(\beta\delta)}}{\partial x^\alpha} 
        -  \frac{\partial \Gamma^\alpha{}_{(\beta\alpha)}}{\partial x^\delta} 
        +  \Gamma^\alpha{}_{(\eta\alpha)}\Gamma^\eta{}_{(\beta\delta)}
        -  \Gamma^\alpha{}_{(\eta\delta)}\Gamma^\eta{}_{(\beta\alpha)}~.       
\end{eqnarray}

\noindent
According to the split of the indices we compute the components 
of the Ricci tensor by beginning with a case by case study, now.

\subsubsection*{(ab)-components:}

With eq.~(\ref{TraceChr})
\begin{eqnarray}
    R_{bd} 
       &=& \frac{\partial \Gamma^i{}_{(bd)}}{\partial x^i} 
        -  0 
        +  \Gamma^\alpha{}_{(i\alpha)}\Gamma^i{}_{(bd)}
        -  \Gamma^\alpha{}_{(\eta d)}\Gamma^\eta{}_{(b\alpha)}\nonumber\\     
       &=&  \partial_i
            \left(
                  \frac{1}{6}\frac{\partial^i N}{N^{4/3}}\eta_{bd}
            \right)
        +   \frac{2}{3}\,\frac{\partial_i
       N}{N}\,\frac{1}{6}\frac{\partial^i N}{N^{4/3}}\eta_{bd}-
       \Gamma^\alpha{}_{(\eta d)}\Gamma^\eta{}_{(b\alpha)}\nonumber\\
       &=&  \underbrace{
           \frac{1}{6}\frac{\partial_i\partial^i N}{N^{4/3}}\eta_{bd}}_{(\ast)}
        -   \frac{2}{9}\frac{\partial_i N\partial^i N}{N^{7/3}}\eta_{bd}
        +   \frac{2}{3}\,\frac{\partial_i
       N}{N}\,\frac{1}{6}\frac{\partial^i N}{N^{4/3}}\eta_{bd}-
       \Gamma^\alpha{}_{(\eta d)}\Gamma^\eta{}_{(b\alpha)}
\end{eqnarray}

\begin{eqnarray}
(\ast) &=& \frac{1}{6}\frac{\partial_i\partial^i N}{N^{4/3}}\eta_{bd}
       ~=~ \frac{1}{6}\frac{\partial_i\left(\,N^{-2/3}\delta^{ij}\,\partial_j
       N\,\right)}{N^{4/3}}\eta_{bd}
       ~=~ \frac{1}{6}\frac{N^{-2/3}\Delta N - 2/3 N^{-5/3}\delta^{ij} 
            \partial_i N \partial_j N}{N^{4/3}}\eta_{bd}\nonumber\\
       &=& \frac{1}{6}\frac{0 - 2/3 N^{-1}\partial_i N 
              \partial^i N}{N^{4/3}}\eta_{bd} 
       ~=~ -\frac{1}{9}\frac{\partial_i N \partial^i N}{N^{7/3}}\eta_{bd} 
\end{eqnarray}

\begin{eqnarray}
  \Gamma^\alpha{}_{(\eta d)}\Gamma^\eta{}_{(b\alpha)}
  &=& \underbrace{\Gamma^c{}_{(ad)}\Gamma^a{}_{(bc)}}_{0}
     +\Gamma^i{}_{(ad)}\Gamma^a{}_{(bi)}
     +\Gamma^c{}_{(id)}\Gamma^i{}_{(bc)}
     +\underbrace{\Gamma^j{}_{(id)}\Gamma^i{}_{(bj)}}_{0}\nonumber\\
  &=& \left(\frac{1}{6}\frac{\partial^i N}{N^{4/3}}\eta_{ad}\right)
      \left(-\frac{1}{6}\frac{\partial_i N}{N}\delta^a_{b}\right)
     +\left(-\frac{1}{6}\frac{\partial_i N}{N}\delta^c_{d}\right)
      \left(\frac{1}{6}\frac{\partial^i N}{N^{4/3}}\eta_{bc}\right)
\end{eqnarray}

\begin{eqnarray}
R_{bd}
&=& -   \frac{1}{9}\frac{\partial_i N\partial^i N}{N^{7/3}}\eta_{bd}    
    -   \frac{2}{9}\frac{\partial_i N\partial^i N}{N^{7/3}}\eta_{bd}
        +   \frac{2}{3}\,\frac{\partial_i
       N}{N}\,\frac{1}{6}\frac{\partial^i N}{N^{4/3}}\eta_{bd}\nonumber\\
   && -
       \left(\frac{1}{6}\frac{\partial^i N}{N^{4/3}}\eta_{ad}\right)
      \left(-\frac{1}{6}\frac{\partial_i N}{N}\delta^a_{b}\right)
     -\left(-\frac{1}{6}\frac{\partial_i N}{N}\delta^c_{d}\right)
      \left(\frac{1}{6}\frac{\partial^i N}{N^{4/3}}\eta_{bc}\right)\nonumber\\
&=& \frac{\partial_i N\partial^i N}{N^{7/3}}\eta_{bd}\,
    \left(\,
            -\frac{1}{9}-\frac{2}{9}+\frac{1}{9} + \frac{1}{18}\,
    \right)\nonumber\\
&=& -\frac{1}{6}\,\frac{\partial_i N\partial^i N}{N^{7/3}}\eta_{bd}
\end{eqnarray}

\subsubsection*{(ij)-components}

\begin{eqnarray}
    R_{ij} &=& R^a{}_{iaj}\,+\,R^k{}_{ikj}
\end{eqnarray}

\begin{eqnarray}
   R^a{}_{iaj} 
    &=& \underbrace{\frac{\partial \Gamma^a{}_{(ij)}}{\partial x^a}}_{0} 
        -  \frac{\partial \Gamma^a{}_{(ia)}}{\partial x^j} 
        +  \Gamma^a{}_{(\eta a)}\Gamma^\eta{}_{(ij)}
        -  \Gamma^a{}_{(\eta j)}\Gamma^\eta{}_{(ia)}\nonumber\\
    &=& 0 -  \partial_j\left(\,-\,\frac{\partial_i N}{N}\,\right) 
        +  \Gamma^a{}_{(k a)}\Gamma^k{}_{(ij)}
        -  \Gamma^a{}_{(b j)}\Gamma^b{}_{(ia)}\nonumber\\
    &=& -\partial_j\left(\,-\,\frac{\partial_i N}{N}\,\right) 
        -  \frac{\partial_k N}{N}\,
           \frac{1}{3}\frac{1}{N}
           \left(\,
                   \partial_j N\delta^k_i +
                   \partial_i N\delta^k_j -
                   \partial^k N\,g_{ij}\,
           \right)
        -  \left(\,-\frac{1}{6}\frac{\partial_j N}{N}\delta^a_b\,\right)
           \left(\,-\frac{1}{6}\frac{\partial_i N}{N}\delta^b_a\,\right)
        \nonumber\\
    &=& \frac{\partial_j\partial_i N}{N}
       +\frac{\partial_i N\partial_j N}{N^2}\,\left(\,-1-\frac{1}{3}-\frac{1}{3}-\frac{1}{6}\,\right)
       +\frac{1}{3}\frac{\partial_k N\partial^k N}{N^2} g_{ij}
\end{eqnarray}

\begin{eqnarray}
   R^k{}_{ikj} 
    &=& \underbrace{\frac{\partial \Gamma^k{}_{(ij)}}{\partial x^k}}_{(1)} 
        -  \underbrace{\frac{\partial \Gamma^k{}_{(ik)}}{\partial x^j} }_{(2)}
        +  \underbrace{\Gamma^k{}_{(\eta k)}\Gamma^\eta{}_{(ij)}}_{(3)}
        -  \underbrace{\Gamma^k{}_{(\eta j)}\Gamma^\eta{}_{(ik)}}_{(4)}\nonumber\\
\end{eqnarray}

\begin{eqnarray*}
  (1) &=& \partial_k
          \left(\,
                  \frac{1}{3}\frac{1}{N}\,
                  \left[\,
                     \partial_j N\delta^k_i +
                     \partial_i N\delta^k_j -
                     \partial^k N\,g_{ij}\,
                  \right]
          \right)\nonumber\\
      &=& -\frac{1}{3}\frac{\partial_k N}{N^2}\,
                  \left[\,
                     \partial_j N\delta^k_i +
                     \partial_i N\delta^k_j -
                     \partial^k N\,g_{ij}\,
                  \right]
           +\frac{1}{3}\frac{1}{N}\,
                  \left[\,
                     \partial_k\partial_j N\delta^k_i +
                     \partial_k\partial_i N\delta^k_j -
                     \partial_k\left(\partial^k N\,g_{ij}\right)\,
                  \right]\nonumber\\
     &=& -\frac{1}{3}\frac{1}{N^2}\,
                  \left[\,
                     \partial_i N \partial_j N +
                     \partial_i N \partial_j N -
                     \partial_k N \partial^k N\,g_{ij}\,
                  \right]
           +\frac{1}{3}\frac{1}{N}\,
                  \left[\vbox{\vspace{2.5ex}}\right.\,
                     \partial_i\partial_j N +
                     \partial_j\partial_i N -
                     \underbrace{\partial_k\left(\partial^k N\,g_{ij}\right)}_{0}\,
                  \left.\vbox{\vspace{2.5ex}}\right]\nonumber\\
     &=& -\frac{2}{3}\frac{\partial_i N \partial_j N}{N^2}\, 
         +\frac{1}{3}\frac{\partial_k N \partial^k N}{N^2}\,g_{ij}\,
           +\frac{2}{3}\frac{\partial_i\partial_j N}{N}\hspace{10cm}
\end{eqnarray*}

\begin{eqnarray*}
  (2) &=& \partial_j\left(\,\frac{5}{3}\frac{\partial_i N}{N}\right)
      ~=~ \frac{5}{3}\frac{\partial_j\partial_i N}{N} -
      \frac{5}{3}\frac{\partial_i N \partial_j N}{N^2}\hspace{10cm}
\end{eqnarray*}

\begin{eqnarray*}
  (3) &=& \frac{5}{3}\frac{\partial_l N}{N}\,\frac{1}{3}\frac{1}{N}\,
          \left(\,
                   \partial_j N\delta^l_i +
                   \partial_i N\delta^l_j -
                   \partial^l N\,g_{ij}\,
          \right)\nonumber\\
      &=& \frac{10}{9}\frac{\partial_i N\partial_j N}{N^2} -
          \frac{5}{9}\frac{\partial_l N\partial^l N}{N^2}\,g_{ij}\hspace{10cm}
\end{eqnarray*}

\begin{eqnarray*}
  (4) &=& \frac{1}{3}\frac{1}{N}\,
          \left(\,
                   \partial_j N\delta^k_l +
                   \partial_l N\delta^k_j -
                   \partial^k N\,g_{lj}\,
          \right)\cdot \frac{1}{3}\frac{1}{N}\,
          \left(\,
                   \partial_k N\delta^l_i +
                   \partial_i N\delta^l_k -
                   \partial^l N\,g_{ik}\,
          \right)\nonumber\\
      &=& \frac{1}{9}\frac{1}{N^2}\,
          \left(\,
                 \partial_jN\partial_iN+5\partial_jN\partial_iN
                -\partial_jN\partial_iN\right.\nonumber\\
               && \hspace{.9cm}+\partial_iN\partial_jN+\partial_jN\partial_iN
                -\partial_lN\partial^lN \cdot g_{ij}\nonumber\\
               &&\hspace{.9cm}\left. 
                 -\partial^kN\partial_kN\cdot g_{ij}-\partial_jN\partial_iN 
               + \partial_iN\partial_jN\,
          \right)\nonumber\\
      &=& \frac{1}{9}\frac{1}{N^2}\,
          \left(\,7\partial_iN\partial_jN
          -2\partial_kN\partial^kN\cdot g_{ij}\,\right)\hspace{10cm}
\end{eqnarray*}

\begin{eqnarray*}
   R^k{}_{ikj} 
    &=& -\frac{2}{3}\frac{\partial_i N \partial_j N}{N^2}\, 
         +\frac{1}{3}\frac{\partial_k N \partial^k N}{N^2}\,g_{ij}\,
           +\frac{2}{3}\frac{\partial_i\partial_j N}{N}
       -\frac{5}{3}\frac{\partial_j\partial_i N}{N} +
      \frac{5}{3}\frac{\partial_i N \partial_j N}{N^2}\nonumber\\
    &&+\frac{10}{9}\frac{\partial_i N\partial_j N}{N^2} -
          \frac{5}{9}\frac{\partial_l N\partial^l N}{N^2}\,g_{ij}
    -\frac{1}{9}\frac{1}{N^2}\,
          \left(\,7\partial_iN\partial_jN
          -2\partial_kN\partial^kN\cdot g_{ij}\,\right)\nonumber\\
    &=& \frac{\partial_i N \partial_j N}{N^2}\,
        \left(\,
                 -\frac{2}{3}+\frac{5}{3}+\frac{10}{9}-\frac{7}{9}\,
        \right)
        + \frac{\partial_k N \partial^k N}{N^2}\,g_{ij}\,
        \left(\,
                 \frac{1}{3}-\frac{5}{9}+\frac{2}{9}\,
        \right)
        +\frac{\partial_i\partial_j N}{N}\,
        \left(\,
                 \frac{2}{3}-\frac{5}{3}\,
        \right)\nonumber\\
  &=&   \frac{12}{9}\,\frac{\partial_i N \partial_j N}{N^2}\,
        -\frac{\partial_i\partial_j N}{N}\,
\end{eqnarray*}

\begin{eqnarray}
   R_{ij} &=& -\frac{1}{2}\,\frac{\partial_i N\partial_j N}{N^2}\,
       +\frac{1}{3}\frac{\partial_k N\partial^k N}{N^2} g_{ij}
\end{eqnarray}

\noindent
The result of the computation can be summarised into the matrix:
\begin{eqnarray}
    R_{\mu\nu} &=& \left(\begin{array}{cc}
        -\frac{1}{6}\,\frac{\partial_i N\partial^i N}{N^{7/3}}\eta_{ab}  & 0 \\
                     0 & -\frac{1}{2}\,\frac{\partial_i N\partial_j N}{N^2}\,
       +\frac{1}{3}\frac{\partial_k N\partial^k N}{N^2} g_{ij}
                   \end{array}\right)
\end{eqnarray}

\subsubsection{Symmetric Field Strength Tensor}

We make the following  ansatz for the 4-form field strength\footnote
%
{
  $\varepsilon_{\alpha_1\ldots\alpha_4}$  is just the  signature
  of the indices. No factor of $\sqrt{|g|}$ !  
}
%
\begin{eqnarray}
   G_{\alpha_1\ldots\alpha_4} 
   &=& \frac{3}{5}\cdot c\cdot\varepsilon_{\alpha_1\ldots\alpha_4\beta}\,
                       \partial^\beta\,\sqrt{|g_{(5)}|} ~\sim~
   \ast_{(5)}d\ast_{(5)} {\rm vol}_{(5)}\nonumber\\
   &=& c\cdot N^{2/3}\,\varepsilon_{\alpha_1\ldots\alpha_4\beta}\,
                       \partial^\beta N~,\qquad\qquad c\,=\,\pm 1
\end{eqnarray}

\begin{eqnarray}
   G_{\alpha_1\ldots\alpha_4}G^{\alpha_1\ldots\alpha_4} 
   &=& c^2\cdot N^{4/3}\,\varepsilon_{\alpha_1\ldots\alpha_4\beta}\,
                         \varepsilon^{\alpha_1\ldots\alpha_4\gamma}\,
                         \partial^\beta N\,\partial_\gamma N\nonumber\\
   &=& 4!\cdot c^2\cdot N^{4/3}\,\left(\det\,g_{(5)}\right)^{-1}\,
                         \partial^\beta N\,\partial_\beta N\nonumber\\
   &=& 4!\cdot c^2\cdot N^{-6/3}\,
                         \partial^\beta N\,\partial_\beta N
\end{eqnarray}

\begin{eqnarray}
   G_{i\alpha_1\ldots\alpha_3}G_j{}^{\alpha_1\ldots\alpha_3} 
   &=& c^2\cdot N^{4/3}\,\varepsilon_{i\alpha_1\ldots\alpha_3\beta}\,
                         g_{jk}\varepsilon^{k\alpha_1\ldots\alpha_3\gamma}\,
                         \partial^\beta N\,\partial_\gamma N\nonumber\\
   &=& 3!\cdot c^2\cdot N^{4/3}\,\left(\det\,g_{(5)}\right)^{-1}\,
                         g_{jk}\,\delta^{k\gamma}_{i\beta}\,\partial^\beta N\,\partial_\gamma N\nonumber\\
   &=& 3!\cdot c^2\cdot N^{-6/3}\,g_{jk}\,\left(\,\delta^k_i\delta^\gamma_\beta-\delta^k_\beta\delta^\gamma_i\,\right)
                         \partial^\beta N\,\partial_\gamma N\nonumber\\
   &=& 3!\cdot c^2\cdot N^{-6/3}\,\left(\,g_{ij}\partial^\beta N\,\partial_\beta N-\partial_j N\,\partial_i N\,\right)
\end{eqnarray}

\begin{eqnarray*}
   \frac{1}{12}\,\left(\,(G^2)_{\mu\nu}\,-\,\frac{1}{12}\,g_{\mu\nu}\,G^{2}\,\right)
   &=& \left(
             \begin{array}{cc}
                -\,\frac{1}{6}\,\frac{\partial_k
                    N \partial^kN}{N^{6/3}}\,g_{ab} &  \\
                   &
                -\,\frac{1}{2}\,\frac{\partial_i N\partial_j N}{N^2}\,
                +\,\frac{1}{3}\,\frac{\partial_k N\partial^kN}{N^{2}}\,
                   g_{ij}
             \end{array}
       \right)
\end{eqnarray*}

\subsubsection{Killing Spinor Equation}

We are now going to investigate the Killing spinor equation (\ref{KSP_Eq})
in the background of an M5-Brane (\ref{SugraM5}) and construct the
preserved supersymmetries explicitly, i.e. we study solutions of:\\
\begin{eqnarray}\label{KSP_M5}
   \hat{D}_M\epsilon  &=& D_M({\omega})\epsilon
                           -\frac{1}{2\cdot 144}
                           \left(
                                 \Gamma^{C_1C_2C_3C_4}{}_{M}
                                  -8\,\Gamma^{C_2C_3C_4}
                                     \delta_M^{C_1}
                           \right)\,\epsilon\,{G}_{C_1C_2C_3C_4}
                           ~=~ 0~.
\end{eqnarray}\\  
with\\
\begin{eqnarray} 
      D_M(\omega)\epsilon &=& \partial_M\epsilon\,-\,\frac{1}{4}\,
                                \omega_{M \hat{A}\hat{B}}\,\Gamma^{\hat{A}\hat{B}} 
\end{eqnarray}\\ 
In order to keep control over the different sorts of indices when
splitting into longitudinal and transversal directions to the brane we 
use for eleven dimensional indices capital letters with or without a hat.
Hatted indices refer to the tangent frame, while unhatted ones to the
spacetime frame.
To write out the Killing spinor equation we have to determine 
the spin connection first. Starting from the definition of the action 
of the covariant derivative on a vector in tangent frame coordinates\\  
\begin{eqnarray}
   \nabla\,X &=& \left(\,
                           \nabla_M\,X^{\hat{A}}\,
                 \right)\,
                 dx^M\otimes e_{\hat{A}}
\end{eqnarray}\\ 
one obtains\\ 
\begin{eqnarray*}
   \nabla\,X 
   &=& \left(\,
                \partial_{\raisebox{-.35ex}{\scriptsize $M$}}\,X^{\hat{A}}\,
               +\,\spin{M}{B}{A}\,X^{\hat{B}}\,
       \right)\,
       dx^M\otimes e_{\hat{A}}\\
   &=& \left(\,
               \partial_M\,\left(\,e^{\hat{A}}_Q\,X^Q\,\right)\,
               +\,\spin{M}{B}{A}\,
                  e^{\hat{B}}_P\,X^P\,
       \right)\,
       dx^M\otimes \left(\,e_{\hat{A}}^N\,\partial_N\,\right)\\
   &=& \left(\,
               \partial_M\,\left(\,e^{\hat{A}}_Q\,X^Q\,\right)\,
               +\,\spin{M}{B}{A}\,
                e^{\hat{B}}_P\,X^P\,
       \right)\,
       dx^M\otimes \left(\,e_{\hat{A}}^N\,\partial_N\,\right)\\
   &=& e_{\hat{A}}^N\,
       \left(\,
               e^{\hat{A}}_Q\,\partial_M\,X^Q\,
              +\,\partial_M\,e^{\hat{A}}_Q\,X^Q\,
               +\,\spin{M}{B}{A}\,
                e^{\hat{B}}_P\,X^P\,
       \right)\,
       dx^M\otimes\partial_N\\
   &=& \left(\,
               \partial_M\,X^N\,
              +\,\left(\,
                       e_{\hat{A}}^N\,\partial_M\,e^{\hat{A}}_P\,
                      +\,e_{\hat{A}}^N\,e^{\hat{B}}_P\,
                       \spin{M}{B}{A}\,
                 \right)\,X^P
       \right)\,
       dx^M\otimes\partial_N\\
   &=& \left(\,
               \partial_M\,X^N\,
               +\,\Gamma^{N}{}_{MP}\,X^P
       \right)\,
       dx^M\otimes\partial_N~.
\end{eqnarray*}\\
Comparing the different lines it is straightforward to read off the 
transformation rule, which expresses the spin connection in terms 
of the Christoffel connection\footnote
{
  The same result can also be obtained from the identity
  $\nabla_\mu\,e_\nu^a\,=\,\partial_\mu\,e_\nu^a\,-\,\Gamma^{\rho}_{(\mu\nu)}e_\rho^a\,+\,\omega_{\mu b}{}^{a}\,e_\nu^b\,=\,0$.
}:\\
\begin{eqnarray}\label{Spinzsh}
  \omega_{\raisebox{-.5truept}{\scriptsize $M$}\hat{B}}{}^{\hat{A}} 
  &=& e^{\hat{A}}_{N}\,e^{P}_{\hat{B}}\,\Gamma^{N}{}_{MP}
      \,-\,e^{P}_{\hat{B}}\,\partial_{M}\,e^{\hat{A}}_{P}~.
\end{eqnarray}\\[2ex]
The vielbein of the M5-brane is:\\
\begin{eqnarray}\label{Viel_M5}
    e^{\hat{A}}_{M} &=& \left(
                         \begin{array}{cccccc}
                             N^{-1/6} & & & & & \\
                                      & \ddots &&&&\\
                                      & & N^{-1/6} &&&\\
                             & & & N^{1/3} & &\\
                             & & & & \ddots & \\
                             & & & & & N^{1/3}
                         \end{array}
                   \right)~.
\end{eqnarray}\\
The name spin connection is somewhat misleading. It is basically just 
the connection representing the Christoffel connection, when acting on
tangent frame indices. The special name is due to its pivotal role it 
plays in the definition of covariant derivatives on spinorial fields, 
which cannot be defined in terms of the usual Christoffel 
symbols \cite{Cartan}.\\  
Now we consider the split $M=(a,i)$ and $\hat{M}=(\hat{a},\hat{i})$
and compute all components of the spin connection individually.\\ 

\noindent
\underline{First we set $M=a$:}\\
\begin{eqnarray*}
   \omega_{\raisebox{-1.5pt}{\scriptsize $a$}}{}_{\hat{B}}{}^{\hat{A}} 
  &=& e^{\hat{A}}_{N}\,e^{P}_{\hat{B}}\,\Gamma^{N}{}_{aP}~-~ 0\nonumber\\
  &=& e^{\hat{A}}_{b}\,e^{P}_{\hat{B}}\,\Gamma^{b}{}_{aP}
  ~+~ e^{\hat{A}}_{i}\,e^{P}_{\hat{B}}\,\Gamma^{i}{}_{aP}\nonumber\\
  &=& e^{\hat{A}}_{b}\,
      \left(\,
                 e^{c}_{\hat{B}}\,\Gamma^{b}{}_{ac}\,
              +\,e^{j}_{\hat{B}}\,\Gamma^{b}{}_{aj}\,
      \right)
  ~+~ e^{\hat{A}}_{i}\,
      \left(\,
                  e^{c}_{\hat{B}}\,\Gamma^{i}{}_{ac}\,
               +\,e^{j}_{\hat{B}}\,\Gamma^{i}{}_{aj}\,
      \right)
\end{eqnarray*}\\
Inserting here the Christoffel symbols (\ref{Chr_M5}) computed before 
one obtains\\
\begin{eqnarray}
  \omega_{\raisebox{-1.5pt}{\scriptsize $a$}}{}_{\hat{B}}{}^{\hat{A}} 
  &=& -\,\frac{1}{6}\,e^{\hat{A}}_{a}\,e^i_{\hat{B}}\,
         \frac{\partial_i N}{N}
      ~+~\frac{1}{6}\,\eta_{ab}\,e^{\hat{A}}_{i}\,e^b_{\hat{B}}\,
         \frac{\partial^i N}{N^{4/3}}
\end{eqnarray}\\
and by splitting the tangent frame indices one finally obtains 
the following components of the spin connection:\\ 
\begin{eqnarray}
   \omega_{\raisebox{-1.75pt}{\scriptsize $a$}}{}_{\hat{b}}{}^{\hat{a}} &=& 0 \nonumber\\
   \omega_{\raisebox{-2pt}{\scriptsize $a$}}{}_{\hat{j}}{}^{\hat{i}} &=& 0 \nonumber\\
   \omega_{\raisebox{-1.75pt}{\scriptsize $a$}}{}_{\hat{j}}{}^{\hat{a}} 
    &=& -\,\frac{1}{6}\,N^{(\,-\frac{1}{6}-\frac{1}{3}\,)}\,
         \delta^{\hat{a}}_a\,\delta^{i}_{\hat{j}}\,
         \frac{\partial_i N}{N}
    ~=~ -\,\frac{1}{6}\,\frac{\partial_{\hat{j}} N}{N}\,e^{\hat{a}}_a
         \nonumber\\
   \omega_{\raisebox{-1.75pt}{\scriptsize $a$}}{}_{\hat{b}}{}^{\hat{i}} 
    &=& \frac{1}{6}\,e_{a\hat{b}}\,\frac{\partial^{\hat{i}} N}{N}
         \label{M5_SC_1}
\end{eqnarray}\\

\noindent
\underline{Now we set $M=i$:}\\
\begin{eqnarray*}
  \omega_{\raisebox{-1.5pt}{\scriptsize $i$}}{}_{\hat{B}}{}^{\hat{A}} 
  &=& e^{\hat{A}}_{N}\,e^{P}_{\hat{B}}\,\Gamma^{N}{}_{iP}
      \,-\,e^{P}_{\hat{B}}\,\partial_{i}\,e^{\hat{A}}_{P}\\
  &=& e^{\hat{A}}_{a}\,e^{P}_{\hat{B}}\,\Gamma^{a}{}_{iP}\,
   +\,e^{\hat{A}}_{j}\,e^{P}_{\hat{B}}\,\Gamma^{j}{}_{iP}
      \,-\,e^{P}_{\hat{B}}\,\partial_{i}\,e^{\hat{A}}_{P}\\
  &=& e^{\hat{A}}_{a}\,
      \left(\,
               e^{b}_{\hat{B}}\,\Gamma^{a}{}_{ib}\,
            +\,e^{k}_{\hat{B}}\,\Gamma^{a}{}_{ik}\,
      \right)
   +\,e^{\hat{A}}_{j}\,
      \left(\,
               e^{b}_{\hat{B}}\,\Gamma^{j}{}_{ib}\,
            +\,e^{k}_{\hat{B}}\,\Gamma^{j}{}_{ik}\,
      \right)
      \,-\,e^{P}_{\hat{B}}\,\partial_{i}\,e^{\hat{A}}_{P}
\end{eqnarray*}\\
Comparing with the Christoffel symbols (\ref{Chr_M5}) we obtain\\
\begin{eqnarray*}
  \omega_{\raisebox{-1.5pt}{\scriptsize $i$}}{}_{\hat{B}}{}^{\hat{A}}
&=& e^{\hat{A}}_{a}\,e^{b}_{\hat{B}}\,
      \left(\,
               -\frac{1}{6}\,\frac{\partial_i N}{N}\,\delta^{a}_{b}\,
      \right)
   +\,e^{\hat{A}}_{j}\,e^{k}_{\hat{B}}\,
                 \frac{1}{3}\,\frac{1}{N}\,
                       \left(\,
                                  \partial_i N\,\delta^j_{k} +
                                  \partial_k N\,\delta^j_{i} -
                                  \partial^j N\,g_{ik}\,
                       \right)
      \,-\,e^{P}_{\hat{B}}\,\partial_{i}\,e^{\hat{A}}_{P}
\end{eqnarray*}\\
and after splitting the tangent frame indices according to 
$\hat{M}=(\hat{a},\hat{i})$ 
\begin{eqnarray}
  \omega_{\raisebox{-1.5pt}{\scriptsize $i$}}{}_{\hat{b}}{}^{\hat{a}}
    &=& -\,\frac{1}{6}\,\frac{\partial_i N}{N}\,\delta^{\hat{a}}_{\hat{b}}\,
        -\,e^{c}_{\hat{b}}\,\partial_{i}\,e^{\hat{a}}_{c} ~~{\buildrel
    !\over =}~~ 0\nonumber\\
  \omega_{\raisebox{-1.5pt}{\scriptsize $i$}}{}_{\hat{l}}{}^{\hat{k}}
    &=& e^{\hat{k}}_{j}\,e^{k}_{\hat{l}}\,
                 \frac{1}{3}\,\frac{1}{N}\,
                       \left(\,
                                  \partial_i N\,\delta^j_{k} +
                                  \partial_k N\,\delta^j_{i} -
                                  \partial^j N\,g_{ik}\,
                       \right)
      \,-\,e^{n}_{\hat{l}}\,\partial_{i}\,e^{\hat{k}}_{n}\nonumber\\
   &=& \frac{1}{3}\,\frac{1}{N}\,
                       \left(\,
                                  \partial_{\hat{l}} N\,e^{\hat{k}}_{i} -
                                  \partial^{\hat{k}} N\,e_{i\hat{l}}\,
                       \right)\nonumber\\
  \omega_{\raisebox{-1.5pt}{\scriptsize $i$}}{}_{\hat{l}}{}^{\hat{a}}
    &=& 0\nonumber\\
  \omega_{\raisebox{-1pt}{\scriptsize $i$}}{}_{\hat{a}}{}^{\hat{l}} 
    &=& 0\label{M5_SC_2}
\end{eqnarray}\\
\noindent
We now write the Killing spinor equation (\ref{KSP_M5}) according to 
the split of the index $M=(a,i)$ into two separate equations.\\ 

\noindent
\underline{M\,=\,a:} Keeping just the operator with $M\,=\,a$ and skipping $\epsilon$ the 
Killing spinor equation reads\\
\begin{eqnarray}
   \hat{D}_a &=& \partial_a\,
                 -\,\frac{1}{4}\,\omega_{a\hat{A}\hat{B}}\,
                  \Gamma^{\hat{A}\hat{B}}
                 -\,\frac{1}{288}
                  \left(\vbox{\vspace{2.5ex}}\right.\,
                          \Gamma^{i_1i_2i_3i_4}{}_{a}
                                  -8\,\Gamma^{i_2i_3i_4}
                                     \underbrace{\delta_a^{i_1}}_{0}\,
                  \left.\vbox{\vspace{2.5ex}}\right)\,{G}_{i_1i_2i_3i_4}
\end{eqnarray} 
According to (\ref{SugraM5}) the four form field strength $G^{(4)}$ 
just lives in the five dimensional transverse space and we specialised
the indices accordingly. Then the last term in the Killing spinor equation 
does not appear. Using the list of the spin connection
components computed above and the ansatz of the four form field
strength we finally obtain:
\begin{eqnarray*}
   \hat{D}_a &=& \partial_a\,
                 -\,\frac{1}{24}\,\frac{\partial_{\hat{j}} N}{N}\,
                  \Gamma_{a}{}^{\hat{j}}
                 +\,\frac{1}{24}\,\frac{\partial_{\hat{j}} N}{N}\,
                  \Gamma^{\hat{j}}{}_{a}
                 -\,\frac{1}{288}\,
                          \Gamma^{i_1i_2i_3i_4}{}_{a}\,
                  c\,N^{2/3}\,
                  \varepsilon_{i_1i_2i_3i_4i_5}\,\partial^{i_5}\,N
\end{eqnarray*} 
We must still evaluate the term containing the four form field strength. 
To this purpose we consider 
\begin{eqnarray*}
   \Gamma^{i_1\ldots i_4}{}_a\,\varepsilon_{i_1i_2i_3i_4i_5}
   &=&\Gamma_a\,\Gamma^{i_1\ldots i_4}\,\varepsilon_{i_1i_2i_3i_4i_5}\\
   &=&\Gamma_a\,e^{i_1}_{\hat{j}_1}\,e^{i_2}_{\hat{j}_2}\,
                e^{i_3}_{\hat{j}_3}\,e^{i_4}_{\hat{j}_4}\,
                \Gamma^{\hat{j}_1\ldots\hat{j}_4}\,
                \varepsilon_{i_1i_2i_3i_4i_5}\\
   &=&\Gamma_a\,e^{i_1}_{\hat{j}_1}\,e^{i_2}_{\hat{j}_2}\,
                e^{i_3}_{\hat{j}_3}\,e^{i_4}_{\hat{j}_4}\,
                \varepsilon_{i_1i_2i_3i_4i_5}\,
                \frac{-1}{(11-4)!}\,\varepsilon^{\hat{j}_1\ldots\hat{j}_4\hat{B}_1\ldots\hat{B}_7}\,\Gamma_{\hat{B}_1\ldots\hat{B}_7}\\
   &=&\Gamma_a\,e^{i_1}_{\hat{j}_1}\,e^{i_2}_{\hat{j}_2}\,
                e^{i_3}_{\hat{j}_3}\,e^{i_4}_{\hat{j}_4}\,
                \varepsilon_{i_1i_2i_3i_4i_5}\,
                \frac{-7}{(11-4)!}\,\varepsilon^{\hat{j}_1\ldots\hat{j}_4\hat{k}\hat{b}_1\ldots\hat{b}_6}\,\Gamma_{\hat{k}\hat{b}_1\ldots\hat{b}_6}\\
\end{eqnarray*} 
In the last line we simply observed that the seven general indices
($B_i$) can be considered as made out of six belonging to the 
longitudinal directions of the brane ($b_i$)  and one transversal 
to the brane ($k$). In addition for the index structure at hand 
\begin{eqnarray*} 
\varepsilon^{\hat{j}_1\ldots\hat{j}_4\hat{k}\hat{b}_1\ldots\hat{b}_6}
 &=& \varepsilon^{\hat{j}_1\ldots\hat{j}_4\hat{k}}\cdot
     \varepsilon^{\hat{b}_1\ldots\hat{b}_6}
\end{eqnarray*} 
and we obtain 
\begin{eqnarray*}
   \Gamma^{i_1\ldots i_4}{}_a\,\varepsilon_{i_1i_2i_3i_4i_5}
   &=& \frac{-7}{(11-4)!}\,\Gamma_a\,e^{i_1}_{\hat{j}_1}\,e^{i_2}_{\hat{j}_2}\,
                e^{i_3}_{\hat{j}_3}\,e^{i_4}_{\hat{j}_4}\,
                \varepsilon_{i_1i_2i_3i_4i_5}\,
                \varepsilon^{\hat{j}_1\ldots\hat{j}_4\hat{k}}\cdot \varepsilon^{\hat{b}_1\ldots\hat{b}_6}\,\Gamma_{\hat{k}\hat{b}_1\ldots\hat{b}_6}\\
   &=& \frac{7\cdot 6!}{(11-4)!}\,\Gamma_a\,e^{i_1}_{\hat{j}_1}\,e^{i_2}_{\hat{j}_2}\,
                e^{i_3}_{\hat{j}_3}\,e^{i_4}_{\hat{j}_4}\,
                \varepsilon_{i_1i_2i_3i_4i_5}\,
                \varepsilon^{\hat{j}_1\ldots\hat{j}_4\hat{k}}
                \,e^{k}_{\hat{k}}\,\Gamma_{k012345}\\
  &=& \frac{7\cdot 6!}{(11-4)!}\,\Gamma_a\,\det(e^{-1}_{(5)})\;
                \varepsilon_{i_1\ldots i_5}\,
                \varepsilon^{i_1\ldots i_4 k}
                \,\Gamma_{k012345}\\
  &=& \frac{7\cdot 6!}{(11-4)!}\,\Gamma_a\,\det(e^{-1}_{(5)})\;
                \delta_{i_1\ldots i_4i_5}^{i_1\ldots i_4 k}
                \,\Gamma_{k012345}\\
  &=& \frac{7\cdot 6!\cdot 4!}{(11-4)!}\,\Gamma_a\,N^{-5/3}\,
                \,\Gamma_{i_5012345}\\
  &=& 24\,\Gamma_{ai_5}\,N^{-5/3}\,
                \,\Gamma_{012345}\\
\end{eqnarray*} 
The appearance of the factor $N^{-5/3}$ and of the gamma matrix 
$\Gamma_{ai_5}$ makes it possible to simplify the Killing spinor 
equation considerably. Plugging this back into the Killing spinor 
equation one obtains:\\
\begin{eqnarray*}
   \hat{D}_a &=& \partial_a\,
                 -\,\frac{1}{12}\,\frac{\partial_{\hat{j}} N}{N}\,
                  \Gamma_{a}{}^{\hat{j}}
                  \left(\vbox{\vspace{2.5ex}}\right.\,
                       1\,+\,c\,\Gamma_{012345}\,
                  \left.\vbox{\vspace{2.5ex}}\right)\,
\end{eqnarray*}\\
Since the solution (\ref{SugraM5}) does not depend of the 
coordinates along the brane it is natural to guess that this 
might hold for the Killing spinor as well. In this case the 
partial derivative action on the Killing spinor $\epsilon$ 
vanishes and the above equation becomes purely algebraic, i.e.\\
\begin{eqnarray}\label{Ansatz_spinor}
     \left(\vbox{\vspace{2.5ex}}\right.\,
            1\,+\,c\,\Gamma_{012345}\,
     \left.\vbox{\vspace{2.5ex}}\right)\,\epsilon
            \,=\,0\quad\Leftrightarrow\quad
     \epsilon~=~ f(x_6,\ldots x_{10})\cdot \left(\vbox{\vspace{2.5ex}}\right.\,
            1\,-\,c\,\Gamma_{012345}\,
     \left.\vbox{\vspace{2.5ex}}\right)\,\epsilon_0
\end{eqnarray}\\

\noindent
\underline{M\,=\,i:} The second bit of the Killing spinor equation, i.e. with $M\,=\,i$ now 
reads
\begin{eqnarray*}
   \hat{D}_i &=& \partial_i\,
                 -\,\frac{1}{4}\,\omega_{i\hat{A}\hat{B}}\,
                  \Gamma^{\hat{A}\hat{B}}
                 -\,\frac{1}{288}
                  \left(\vbox{\vspace{2.5ex}}\right.\,
                          \Gamma^{i_1i_2i_3i_4}{}_{i}
                                  -8\,\Gamma^{i_2i_3i_4}
                                      \delta_i^{i_1}\,
                  \left.\vbox{\vspace{2.5ex}}\right)\,{G}_{i_1i_2i_3i_4}\\
              &=& \partial_i\,
                 +\,\frac{1}{4}\,\frac{2}{3}\,\frac{\partial_{\hat{l}} N}{N}\,
                  \Gamma_i{}^{\hat{l}}
                 -\,\frac{1}{288}
                  \left(\vbox{\vspace{2.5ex}}\right.\,
                          \underbrace{\Gamma^{i_1i_2i_3i_4}{}_{i}}_{(\ast)}
                                  -8\,\underbrace{\Gamma^{i_2i_3i_4}
                                      \delta_i^{i_1}}_{(\ast\ast)}\,
                  \left.\vbox{\vspace{2.5ex}}\right)\,c\,
                  N^{2/3}\,\varepsilon_{i_1i_2i_3i_4i_5}\,\partial^{i_5}N\\
\end{eqnarray*}\\
To simplify ($\ast$) we used\\ 
\begin{eqnarray*}
   \Gamma^{i_1i_2i_3i_4}{}_{i}\,\varepsilon_{i_1i_2i_3i_4i_5}
   &=&
   g_{ik}\,e^{i_1}_{\hat{a}_1}\,e^{i_2}_{\hat{a}_2}\,e^{i_3}_{\hat{a}_3}\,e^{i_4}_{\hat{a}_4}\,e^{k}_{\hat{a}_5}\,\Gamma^{\hat{a}_1\hat{a}_2\hat{a}_3\hat{a}_4\hat{a}_5}\,\varepsilon_{i_1i_2i_3i_4i_5}\\
   &=&g_{ik}\,e^{i_1}_{\hat{a}_1}\,e^{i_2}_{\hat{a}_2}\,e^{i_3}_{\hat{a}_3}\,e^{i_4}_{\hat{a}_4}\,e^{k}_{\hat{a}_5}\,\frac{-1}{(11-5)!}\,\varepsilon^{\hat{a}_1\ldots\hat{a}_5\hat{B}_1\ldots
   \hat{B}_6}\,\Gamma_{\hat{B}_1\ldots
   \hat{B}_6}\,\varepsilon_{i_1i_2i_3i_4i_5}\\
&=&g_{ik}\,\det(e^{-1}_{(5)})\,\frac{-1}{(11-5)!}\,\varepsilon^{i_1i_2i_3i_4k\hat{B}_1\ldots
   \hat{B}_6}\,\Gamma_{\hat{B}_1\ldots
   \hat{B}_6}\,\varepsilon_{i_1i_2i_3i_4i_5}\\
   &=&g_{ik}\,\det(e^{-1}_{(5)})\,\varepsilon^{i_1i_2i_3i_4k}\,\Gamma_{012345}\,\varepsilon_{i_1i_2i_3i_4i_5}\\
   &=& g_{ik}\,\det(e^{-1}_{(5)})\,\Gamma_{012345}\,4!\,\delta^k_{i_5}\\
   &=& 24\,N^{-5/3}\,\Gamma_{012345}\,g_{ii_5}
\end{eqnarray*}\\
and to simplify ($\ast\ast$) we used\\ 
\begin{eqnarray*}
    -8\,\Gamma^{i_2i_3i_4}\,\varepsilon_{ii_2i_3i_4i_5}
    &=&
    -8\,e^{i_2}_{\hat{a}_2}\,e^{i_3}_{\hat{a}_3}\,e^{i_4}_{\hat{a}_4}\,\Gamma^{\hat{a}_2\hat{a}_3\hat{a}_4}\,\varepsilon_{ii_2i_3i_4i_5}\\
    &=&
    -8\,e^{i_2}_{\hat{a}_2}\,e^{i_3}_{\hat{a}_3}\,e^{i_4}_{\hat{a}_4}\,\frac{1}{(11-3)!}\,\varepsilon^{\hat{a}_2\hat{a}_3\hat{a}_4\hat{B}_1\ldots\hat{B}_8}\Gamma_{\hat{B}_1\ldots\hat{B}_8}\,\varepsilon_{ii_2i_3i_4i_5}\\
    &=&
    -8\,e^{i_2}_{\hat{a}_2}\,e^{i_3}_{\hat{a}_3}\,e^{i_4}_{\hat{a}_4}\,\left(\begin{array}{c}8\\2\end{array}\right)\frac{1}{(11-3)!}\,\varepsilon^{\hat{a}_2\hat{a}_3\hat{a}_4\hat{m}\hat{n}\hat{b}_1\ldots\hat{b}_6}\Gamma_{\hat{m}\hat{n}\hat{b}_1\ldots\hat{b}_6}\,\varepsilon_{ii_2i_3i_4i_5}\\
&=&
    -8\,e^{i_2}_{\hat{a}_2}\,e^{i_3}_{\hat{a}_3}\,e^{i_4}_{\hat{a}_4}\,e^{m}_{\hat{m}}\,e^{n}_{\hat{n}}\,\left(\begin{array}{c}8\\2\end{array}\right)\frac{1}{(11-3)!}\,\varepsilon^{\hat{a}_2\hat{a}_3\hat{a}_4\hat{m}\hat{n}\hat{b}_1\ldots\hat{b}_6}\Gamma_{mn\hat{b}_1\ldots\hat{b}_6}\,\varepsilon_{ii_2i_3i_4i_5}\\
&=&
    -8\,\det(e^{-1}_{(5)})\,\left(\begin{array}{c}8\\2\end{array}\right)\frac{1}{(11-3)!}\,\varepsilon^{i_2i_3i_4mn\hat{b}_1\ldots\hat{b}_6}\Gamma_{mn\hat{b}_1\ldots\hat{b}_6}\,\varepsilon_{ii_2i_3i_4i_5}\\
&=&
    -8\,\det(e^{-1}_{(5)})\,\left(\begin{array}{c}8\\2\end{array}\right)\frac{1}{(11-3)!}\,\varepsilon^{i_2i_3i_4mn\hat{b}_1\ldots\hat{b}_6}\Gamma_{mn\hat{b}_1\ldots\hat{b}_6}\,\varepsilon_{ii_2i_3i_4i_5}\\
&=&
    8\,\det(e^{-1}_{(5)})\,\left(\begin{array}{c}8\\2\end{array}\right)
    \frac{2\cdot 3!}{(11-3)!}\,
    \varepsilon^{\hat{b}_1\ldots\hat{b}_6}\Gamma_{ii_5}
    \Gamma_{\hat{b}_1\ldots\hat{b}_6}\\
&=&
    -\,8\,N^{-5/3}\,\left(\begin{array}{c}8\\2\end{array}\right)
    \frac{2\cdot 3!\cdot 6!}{(11-3)!}\,\Gamma_{ii_5}\Gamma_{012345}\\
&=& -\,48\,N^{-5/3}\,\Gamma_{ii_5}\Gamma_{012345}
\end{eqnarray*}\\
Now we put everything together and obtain the following for of the 
Killing spinor equation
\begin{eqnarray*}
   \hat{D}_i &=& \partial_i\,
                 +\,\frac{1}{6}\,\frac{\partial_{\hat{l}} N}{N}\,
                  \Gamma_{i}{}^{\hat{l}}
                 -\,\frac{c}{12}\,
                          \Gamma_{012345}\,
                  \frac{\partial_{i}N}{N}
                 +\frac{c}{6}\,\frac{\partial^{i_5}N}{N}\,
                  \Gamma_{ii_5}\Gamma_{012345}\\
  &=& \partial_i\,+\,\frac{1}{12}\,
                  \frac{\partial_{i}N}{N}\,
                 -\,\frac{1}{12}\,\frac{\partial_{i}N}{N}\,
                  \left(\,1\,+\,c\,\Gamma_{012345}\,\right)\,
                 +\,\frac{1}{6}\,\frac{\partial_{\hat{l}} N}{N}\,
                  \Gamma_{i}{}^{\hat{l}}\,
                  \left(\,1\,+\,c\,\Gamma_{012345}\,\right)\\
\end{eqnarray*} 
Plugging the ansatz for the Killing spinor from (\ref{Ansatz_spinor})
into this equation we obtain 
\begin{eqnarray}
     \left(\,  \partial_i\,+\,\frac{1}{12}\,
                  \frac{\partial_{i}N}{N}\,\right)\,f &=& 0
     \quad\Leftrightarrow\quad f~=~ N^{-\frac{1}{12}}
\end{eqnarray}
In fact the computation we have done is more complicated than
necessary. A significant simplification can be done by choosing 
a representation of gamma matrices, properly adjusted to the 
splitting of the eleven dimensional space into six dimensional 
and five dimensional spaces corresponding to the longitudinal and 
transverse directions of the M5-Brane (cf. exercise \ref{ex_CAM5Brane}).

\subsubsection{BPS-Bound}

Spherical coordinates for the 5-sphere:

\begin{eqnarray*}
     x^5 &=& r\cdot \cos\vartheta^1  
             \hspace{28.5ex} 0\,<\,\vartheta^1\,\leq\, \pi\\
     x^4 &=& r\cdot \sin\vartheta^1\cdot cos\vartheta^2 
             \hspace{21.5ex} 0\,<\,\vartheta^2\,\leq\, \pi\\
     x^3 &=& r\cdot \sin\vartheta^1\cdot sin\vartheta^2\cdot \cos\vartheta^3
             \hspace{14ex} 0\,<\,\vartheta^3\,\leq\, \pi\\
     x^2 &=& r\cdot \sin\vartheta^1\cdot sin\vartheta^2\cdot \sin\vartheta^3\cdot\cos\vartheta^4
             \hspace{6.75ex} 0\,<\,\vartheta^4\,\leq\, 2\pi\\             
     x^1 &=& r\cdot \sin\vartheta^1\cdot sin\vartheta^2\cdot \sin\vartheta^3\cdot\sin\vartheta^4\\
\end{eqnarray*}

\begin{eqnarray*}
     G &=& \frac{1}{4!}\,G_{\alpha_1\ldots\alpha_4}\,dx^{\alpha_1}\wedge
           dx^{\alpha_2}\wedge dx^{\alpha_3}\wedge dx^{\alpha_4}\\
     G &=& \frac{1}{4!}\,G_{\alpha_1\ldots\alpha_4}\,
           \frac{
                  \partial\,(\,x^{\alpha_1},x^{\alpha_2},
                               x^{\alpha_3},x^{\alpha_4}\,)
                }
                {
                  \partial\,(\,\vartheta^1,\vartheta^2,
                            \vartheta^3,\vartheta^4\,)
                }\,d\vartheta^1\wedge
           d\vartheta^2\wedge d\vartheta^3\wedge d\vartheta^4\\
\end{eqnarray*}\\
\begin{eqnarray*}
  Q ~=~  \int\limits_{S^4} G &=& 8\,ac\,\pi^2\hspace{38ex}
\end{eqnarray*}\\
Spacelike part of the metric $g_{ij}~=~\delta_{ij}+h_{ij}$ and computing in 
the flat background $\delta_{ij}$ the asymptotically defined 
energy \cite{ADM}\\
\begin{eqnarray*}
{\cal E} ~=~ \int\limits_{S^4} 
                   \left(~
                           \partial^n\hspace{1pt}h_{mn}-\partial_m h^i{}_i
                   ~\right)
             \,\frac{x^m}{r}\, d\Omega_{S^4}
         ~=~8\,\pi^2\,a\hspace{10.5ex}
\end{eqnarray*}\\
By directly comparing the results for $Q$ and ${\cal E}$ one obtains
the equality\\ 
\begin{eqnarray}
   {\cal E} &=& |\,Q\,|
\end{eqnarray}\\ 
characteristic for solutions preserving supersymmetry.

\subsubsection*{Homework:}

\begin{exercise} Recompute the components of the spin connection
                 (\ref{M5_SC_1}) and (\ref{M5_SC_2}) of the M5 brane 
                 from the equation 
                 \begin{eqnarray*}
                     d\,e^{\hat{A}} \,+\, 
                         \left(\,
                                 \omega_{\raisebox{-.1pt}{\scriptsize
                 $M$}\hat{B}}{}^{\hat{A}}\,dx^M\,
                         \right)\,\wedge\,e^{\hat{B}}~=~0~. 
                 \end{eqnarray*}  
\end{exercise}
\begin{exercise}\label{ex_CAM5Brane}
${\mathbb{V}}$ and ${\mathbb{W}}$ vector spaces, 
${\mathbb{V}}$ even dimensional. 
Let $\Gamma_{D+1} = \Gamma_1\cdot\ldots\cdot \Gamma_{D}$ 
be the chirality operator of the even dimensional Clifford 
algebra ${\rm Cliff}({\mathbb{V}})$. 
Then there is a map $j$, which
establishes the isomorphism below:
\begin{eqnarray*}     
     {\rm Cliff}({\mathbb{V}}\oplus {\mathbb{W}},\eta\oplus g) 
     &\buildrel \cong \over \longrightarrow&
     {\rm Cliff}({\mathbb{V}},\eta)\otimes {\rm Cliff}({\mathbb{W}},g)\\[2ex]
      j(v\oplus w) &=& v\otimes\unity + \Gamma_{D+1}\otimes w  
\end{eqnarray*}
Specialising to the case at hand, one obtains:
\begin{eqnarray*}
     {\rm Cliff}({\mathbb{R}}^{1,5}\oplus 
     {\mathbb{R}}^{5},\eta\oplus g)
     &\buildrel \cong \over \longrightarrow&
     {\rm Cliff}({\mathbb{R}}^{1,5},\eta)\otimes 
     {\rm Cliff}({\mathbb{R}}^5,g)\\[2ex]
      j(v\oplus w) &=& v\otimes\unity + \Gamma_{D+1}\otimes w  
\end{eqnarray*}
Using this representation of the 11d Clifford algebra redo the
computation of the Killing spinor equation of the M5-Brane 
(cf. \cite{Duff:1999rk}).
\end{exercise}
\begin{exercise} Compute the charge $Q$ and the energy ${\mathcal E}$.
\end{exercise}

\vfill\eject
\subsection{M2-Brane}
\label{SubSecM2Brane}

The M2-brane solution \cite{Duff:1991xz} has the form\\ 
\begin{eqnarray}\label{SugraM2}
      ds^2 &=& N^{-2/3}\left(-dt^2+dx_1^2+dx_2^2 \right)
            ~+~N^{1/3}\left(dx_3^2+\ldots +dx_{10}^2\right)\nonumber\\
     G_{012i} &=& c\,\frac{\partial_{i}\,N}{N^2},\hspace{2ex} c~=~\pm1,
     \hspace{5ex} \Delta N~=~0
\end{eqnarray}\\
and it is convenient to consider the fundamental M2-brane solution as 
corresponding  to the choice $N~=~1+\frac{a}{r^6}$ with
$r^2=x_3^2+\ldots +x_{10}^2$.\\

\subsubsection{Ricci Tensor}

To compute the Ricci-Tensor we employ a slightly different technique
compared to the direct one used to obtain the Ricci tensor of the 
M5-Brane in subsection \ref{SubSec_M5_Brane} before. It turns out to 
be much more efficient and can be applied 
to all M- or D- or p-brane solutions occurring in the various 
supergravities (not necessarily eleven dimensional). The algorithm 
is very easy to spell out. First one has to factor out the factor, 
which makes a part of the metric flat. For convenience we have chosen 
the upper part to be the flat one, i.e.\\  
\begin{eqnarray}\label{Metrik_M2}
    g_{\mu\nu} &=& N^{-2/3}\cdot\left(
                         \begin{array}{ccccc}
                             -1 &   &   &   \\
                                & 1 &   &   \\
                                &   & 1 &   \\
                                &   &   & N\cdot\unity_{8} \\
                         \end{array}
                   \right)~.
\end{eqnarray}\\
In Appendix \ref{App_Ricci_conf_trafo} we have computed the properties
of the Ricci tensor with respect to conformal transformations. We can 
apply this transformation rule to generate the Ricci tensor of the 
M2-brane by first computing the Ricci tensor of the metric obtained by
neglecting the overall conformal factor in (\ref{Metrik_M2}) and 
an  additional observation stating that the Ricci tensor
of a product space build out of a flat space $X$ and a curved space
$Y$ is block diagonal as shown below\\
\begin{eqnarray}
              R_{\mu\nu}^{X\times Y}    
              &=& \left(\begin{array}{cc}
                              0 &   \\
                                & R_{ij}^Y \\
                         \end{array}
                   \right)~.
\end{eqnarray}\\  
This in mind we proceed as follows. First we choose $Y$ to be the eight
dimensional subspace $\{x^3,\,\ldots\,,x^{10}\}$ in eq.~(\ref{Metrik_M2})
above, i.e.\\ 
\begin{eqnarray}
     g_{ij}^{(8)} &=& N\cdot \delta_{ij}~.
\end{eqnarray}\\
It can be considered as conformally equivalent to flat space,
whose Ricci tensor vanishes. Applying eq.~(\ref{RescaledRicciT}) once 
with  $D\,=\,d\,(=\,8)$  and the conformal factor taken to be $N$ we obtain:\\
\begin{eqnarray}
  R_{ij}^Y 
  &=& 0~+~ \frac{1}{2}\,
           \left\{\,
                     (1-d)\,\partial_j
                     \left(\,
                             \frac{\partial_i N}{N}\,
                     \right) \,+\, 
                     \partial_i
                     \left(\,
                             \frac{\partial_j N}{N}\,
                     \right) \,-\, 
                     \partial_k
                     \left(\,
                             \frac{\partial^k N}{N}\,
                     \right)\,\delta_{ij}\,
            \right\}\nonumber\\[2ex]
 &&\hspace{2.5ex}+\,\frac{d-2}{2}\,
        \left(\,
                 \frac{\partial_i N\,\partial_j N}{N^2}
                 \,-\,
                 \frac{\partial_k N\,\partial^k N}
                      {N^2}\,\delta_{ij}\,
        \right)\nonumber\\[1ex]
 &=& \frac{2-d}{2}\,\frac{\partial_i\partial_j\,N}{N}\,
     -\,\frac{1}{2}\,\frac{\Delta\,N}{N}\,\delta_{ij}\,
     +\,\frac{3d-6}{4}\,\frac{\partial_i N\,\partial_j\,N}{N^2}\,
     -\,\frac{d-4}{4}\,\frac{\delta^{kl}\,\partial_k\,N\,\partial_l\,N}{N^2}\,\delta_{ij}
\end{eqnarray}\\
where we have to insert $d\,=\,8$ to obtain the result for the eight 
dimensional metric. 
To complete the second step we need the Christoffel symbols of the 
d-dimensional metric produced so far. They are most easily computed 
and simply read\\
\begin{eqnarray}
    \Gamma^i{}_{(jk)} 
    &=& \frac{1}{2}\,g^{il}
        \left(\,
                 g_{lj,k}\,+\,g_{lk,j}\,-\,g_{jk,l}\,
        \right)\nonumber\\[1ex]
    &=& \frac{1}{2}\,
        \left(\,
                   \frac{\partial_j N}{N}\,\delta^i_{k}\,
                +\,\frac{\partial_k N}{N}\,\delta^i_{j}\,
                -\,\frac{\partial^i N}{N}\,g_{jk}\,
        \right)\label{ChrisM2}
\end{eqnarray}\\
and for the contraction of the Christoffel symbol one gets:\\
\begin{eqnarray}\label{ContrChrisM2}
   \Gamma^{i}{}_{ij} &=& \frac{d}{2}\,\frac{\partial_j N}{N}~. 
\end{eqnarray}\\
Now we perform the second conformal rescaling with the conformal
factor taken to be $\Omega=N^{-2/3}$ and $D\,=\,11$. This inserted into
eq.~(\ref{RescaledRicciT}) reads\\
\begin{eqnarray}
  R_{\beta\delta}^{\rm M2} 
  &=& {\tilde R}_{\beta\delta}^{Y} 
      \,+\,\frac{1}{2}\,
           \left\{\,
                     -10\,\tilde\nabla_\delta
                     \left(\,
                             \frac{\partial_\beta\Omega}{\Omega}\,
                     \right) \,+\, 
                     \tilde\nabla_\beta
                     \left(\,
                             \frac{\partial_\delta\Omega}{\Omega}\,
                     \right) \,-\, 
                     \tilde\nabla_\alpha
                     \left(\,
                             \frac{\tilde{\partial}^\alpha\Omega}{\Omega}\,
                     \right)\,\tilde{g}_{\beta\delta}\,
            \right\}\nonumber\\
 &&\,+\,\frac{9}{4}\,
        \left(\,
                 \frac{\partial_\beta\Omega\,\partial_\delta\Omega}{\Omega^2}
                 \,-\,
                 \frac{\partial_\eta\Omega\,\tilde{\partial}^\eta\Omega}
                      {\Omega^2}\,\tilde{g}_{\beta\delta}\,
        \right)
\end{eqnarray}\\
It is useful to split the indices into those from the 3d space along the
M2-brane (in which the unscaled metric is flat), and those 8 directions
orthogonal to the M2-brane. The former we call $a,\,b = 0,\,1,\,2$, the latter
$i,\,j = 3, \,\ldots, 10$. The parts of the Ricci tensor are then calculated
using the rescaling formula. In the flat directions we get, i.e.\\[1ex] 
\underline{$\beta\,=\,a~$ \& $~\delta\,=\,b$:}\\ 
\begin{eqnarray}
   R_{ab} &=& \frac{1}{2}\,\left\{\, 0 \,+\, 0 \,-\, \tilde{\nabla}_i
   \left(\,\frac{\tilde{\partial}^i \Omega}{\Omega}\,\right)\,\eta_{ab}\,\right\} ~-~
   \frac{11-2}{4}\;\frac{\partial_i
   \Omega\,\tilde{\partial}^i\Omega}{\Omega^2}\;\eta_{ab} \,.
\end{eqnarray}\\
The first two entries disappear since the function $N$ does not depend on
these flat coordinates. Straight forward calculation begins by substituting
$\Omega = N^{-2/3}$ and using the correct metric to pull up indices\\
\begin{eqnarray}
   R_{ab} & = & \frac{1}{3}\,\tilde{\nabla}_i
   \left(\,\frac{\tilde{\partial}^iN}{N}\,\right)\,\eta_{ab}  ~-~
   \frac{\partial_i N\,\tilde{\partial}^iN}{N^2}\;\eta_{ab}\nonumber\\ 
          & = & \frac{1}{3}\,\left\{\,\partial_i
   \left(\,\frac{\tilde{\partial}^iN}{N}\,\right) ~+~
   \Gamma^i{}_{ij}\,\frac{\tilde{\partial}^iN}{N}\,\right\}\,\eta_{ab}  -
   \frac{\partial_i N\,\tilde{\partial}^iN}{N^2}\;\eta_{ab}\,.\hspace{16ex} 
\end{eqnarray}\\
Since it is crucial to keep in mind which metric was used to pull up the
indices, we write them down again by inserting the correct factor. This factor
will contribute towards a chain rule, i.e.\\
\begin{eqnarray}
   R_{ab} & = & \frac{1}{3}\,\left\{\,\partial_i
   \left(\,\frac{N^{-1}\partial_iN}{N}\,\right)
   ~+~ 4\;\frac{\partial_j N}{N}\;\frac{N^{-1}\partial_jN}{N}\,
   \right\}\,\eta_{ab}  ~-~
   \frac{\partial_i N}{N}\;\frac{N^{-1}\partial_iN}{N}\;
   \eta_{ab}\nonumber\\
          & = & \frac{1}{3}\,\left\{\,\frac{\Delta N}{N^2}\,-\,2\,
   \frac{\partial_iN\, \partial_iN}{N^3}\, +\,
   4\,\frac{\partial_iN\,\partial_iN}{N^3}\, -\,3\,
   \frac{\partial_iN\,\partial_iN}{N^3}\,\right\}\;\eta_{ab}\nonumber\\
          & = &  \frac{1}{3}\,\left\{\,\frac{\Delta N}{N^2}\,-\, 
   \frac{\partial_iN\,\partial_iN}{N^3}\,\right\}\,\eta_{ab}\,.
   \label{UpperRicciM2} 
\end{eqnarray}\\
From the first to the second line, we have used the result for the contraction
of the Christoffel symbol \eqref{ContrChrisM2}.\\

\noindent 
Next comes the Ricci tensor for the 8d space with indices $i,\, j$ starting
as is usual from the rescaling formula and substituting $\Omega = N^{-2/3}$
(and rearranging terms), i.e.\\[1ex]

\noindent
\underline{$\beta\,=\,i~$ \& $~\delta\,=\,j$:}\\ 
\begin{eqnarray}
    R_{ij} &=& \tilde{R}_{ij} ~+~
    \frac{11-2}{4}\cdot \frac{4}{9}\,
    \left(\,
             \frac{\partial_iN\,\partial_jN}{N^2}\,-\,
             \frac{\partial_kN\,\partial_kN}{N^2}\,\delta_{ij}\,
    \right)\nonumber\\[2ex]
     &&\hspace{4.75ex}-\,\frac{1}{3}\,
        \left\{\,
            -\,10\,\tilde{\nabla}_j
             \left(\,\frac{\partial_iN}{N}\,\right)\,
            +\,\tilde{\nabla}_i
             \left(\,\frac{\partial_jN}{N}\,\right)\, 
            -\,\tilde{\nabla}_k
             \left(\,\frac{\tilde{\partial}^kN}{N}\,\right)\,
    \tilde{g}_{ij}\,
        \right\}\,. 
\end{eqnarray}\\
The term in curly brackets must be treated with care due to the metrics used
to pull up the indices on the differential operators. Hence we explicitly
write the result after pulling down the indices of the differentials in the
curly brackets also using the symmetry in the first two terms\\
\begin{eqnarray*}
   \{\ldots\} & = & -\,\frac{1}{3}\,\left\{\, -\,9\, \partial_j\,
   \left(\,\frac{\partial_iN}{N}\,\right)\,+\, 9\, 
   \tilde{\Gamma}^k{}_{ij}\,
   \frac{\partial_kN}{N}\, -\,  \partial_k\, \left(\,
   \frac{N^{-1}\partial_kN}{N}\,
   \right)\, \tilde{g}_{ij}\, -\, \Gamma^k{}_{kl}\,
   \frac{\tilde{\partial}^lN}{N}\,\tilde{g}_{ij}\,\right\}\,.
\end{eqnarray*}\\
Again we have to make use of \eqref{ContrChrisM2} and we have to pay attention
to the third term. The last expression thus becomes\\
\begin{eqnarray}
   \{\ldots\} & = & -\,\frac{1}{3}\,
   \left\{\vbox{\vspace{3ex}}\right.\, 
   -\,9\, \frac{\partial_j\partial_iN}{N} \,+\,
   9\,\frac{\partial_iN\,\partial_jN}{N^2} \,-\, 
   \frac{\Delta N}{N}\,\delta_{ij} \,+\, 2\,
   \frac{\partial_kN\,\partial_kN}{N^2}\,\delta_{ij} \,-\, 4\,
   \frac{\partial_kN\,\partial_kN}{N^2}\,\delta_{ij} \, \nonumber\\
       && \hspace{7.75ex} +\;9\cdot\frac{1}{2}\left(
   \frac{\partial_iN}{N}\delta^k_j + \frac{\partial_jN}{N}\delta^k_i -
   \frac{\tilde{\partial}^kN}{N}\tilde{g}_{ij}\right)\frac{\partial_kN}{N}\left.\vbox{\vspace{3ex}}\right\}\,.
\end{eqnarray}\\
The full result for the components of the Ricci tensor in 
the transverse directions is:\\
\begin{eqnarray}
   R_{ij} &=& \frac{\partial_i\partial_j\,N}{N}\,
              \left(\,  
                       -3\,+\,3\,
              \right)\,
              +\,\frac{\Delta\,N}{N}\,\delta_{ij}\,
              \left(\,  
                       -\frac{1}{2}\,+\,\frac{1}{3}\,
              \right)\nonumber\\[1ex]
             &&+\,\frac{\partial_i\,N \partial_j\,N}{N^2}\,
              \left(\,  
                       \frac{9}{2}\,+\,1\,
                       -\,\frac{9}{3}\,
                       -\,\frac{9}{3}\,
              \right)\nonumber\\[1ex]
            &&+\,\frac{\partial_k\,N\,\partial_k\,N}{N^2}\,\delta_{ij}\,
              \left(\,  
                      -\,1\,-\,1\,-\,\frac{2}{3}\,+\,
                      \frac{4}{3}\,+\,\frac{3}{2}\,
              \right)\nonumber\\[2ex]
   &=&-\,\frac{1}{6}\,\frac{\Delta\,N}{N}\,\delta_{ij}\,
      -\,\frac{1}{2}\,\frac{\partial_i\,N\, \partial_j\,N}{N^2}\,
      +\,\frac{1}{6}\,\frac{\partial_k\,N\,\partial_k\,N}{N^2}\,\delta_{ij}\,
      \label{LowerRicciM2}
\end{eqnarray}\\
Putting the partial results eq.~(\ref{UpperRicciM2}) and eq.~(\ref{LowerRicciM2}) 
together, the
Ricci-Tensor reads\\
\begin{eqnarray}
    R_{\mu\nu} &=&\left(\begin{array}{cc}
        \left(\,\frac{1}{3}\,\frac{\Delta N}{N^2}\,-\,\frac{1}{3}\,\frac{\partial_i N\partial_i N}{N^{3}}\,\right)\eta_{ab}  & 0 \\
                     0 & -\frac{1}{2}\,\frac{\partial_i N\partial_j N}{N^2}\,
       +\frac{1}{6}\left(\,\frac{\partial_k N\partial_k N}{N^2}\,-\,\frac{\Delta N}{N}\,\right)\, \delta_{ij}
                   \end{array}\right)~.\label{RicciM2}
\end{eqnarray}\\

\subsubsection{Symmetric Field Strength Tensor}

\begin{eqnarray}
   G_{\alpha_1\ldots\alpha_4}G^{\alpha_1\ldots\alpha_4} &=& 4\cdot
   3!\cdot G_{012i}\,G^{012i} 
~=~ 4\cdot 3!\cdot (-N^2)\cdot c^2\cdot \frac{\partial_i
   N}{N^2}\,\frac{\partial^i N}{N^2}\nonumber \\
  &=& -24\cdot c^2\cdot \frac{\partial_i N\,\partial^i N}{N^2}
\end{eqnarray}

\begin{eqnarray}
   G_{\mu\alpha_1\ldots\alpha_3}G_\nu{}^{\alpha_1\ldots\alpha_3} &=& 
   \left\{\begin{array}{cc}
        3!\,c^2\,N^{-8/3}\,\partial_kN\,\partial^k N & {\rm if}~(\mu,\nu)=(0,0)\\
         -\,3!\,c^2\,N^{-8/3}\,\partial_kN\,\partial^k N   & {\rm if}~(\mu,\nu)=(1,1)\\
          ``   & {\rm if}~(\mu,\nu)=(2,2)\\
        -\,3!\,c^2\,N^{-2}\,\partial_i N\,\partial_j N & {\rm if}~(\mu,\nu)=(i,j)
          \end{array}\right.\nonumber\\[2ex]
    &=& -\,3!\,c^2\,N^{-2}\,\left(\begin{array}{cc}
                N^{-2/3}\,\partial_i N \partial^i N\,\eta_{ab} & 0 \\
                 0     & \partial_i N\,\partial_j N
              \end{array}
        \right)
\end{eqnarray}

\begin{eqnarray*}
  \frac{1}{12}\,\left(\,(G^2)_{\mu\nu}\,-\,
   \frac{1}{12}\,(G^2)\,g_{\mu\nu}\,\right)
   &=& -\,\frac{1}{2}\,c^2\,N^{-2}\,\left(\begin{array}{cc}
                \frac{2}{3}\,N^{-2/3}\,\partial_i N \partial^i N\,
                    \eta_{ab} & 0 \\
                 0     & \partial_i N\,\partial_j N\,-\,\frac{1}{3}\,
                       \partial_i N \partial^i N\,g_{ij}
              \end{array}
        \right)
\end{eqnarray*}\\
Comparison of this tensor and the tensor in eq.~(\ref{RicciM2}) while
taking $\Delta\,N\,=\,0$ into account, proves the equality.\\

\subsubsection{Killing Spinor Equation}
The basic steps of the solution of the Killing spinor equation 
can be found in various places, e.g. \cite{Liu}. First of all 
one has to determine the spin connection from\\
\begin{eqnarray}
   de^{\hat{A}}~+~
   \omega_{\hat{B}}{}^{\hat{A}}\,\wedge\,e^{\hat{B}}~=~ 0~.
\end{eqnarray}\\
Evaluation of this identity leads to the spin connection components 
below. For instance setting\\[2ex]
\underline{${\hat{A}}\,=\,{\hat{a}}$~:}\hspace{3ex}$e^{\hat{a}}\,=\,N^{-1/3}\,dx^a\quad\Rightarrow$\\
\begin{eqnarray}
     de^{\hat{a}}  
     ~=~ -\frac{1}{3}\,\frac{\partial_i
     N}{N^{4/3}}\;dx^i\,\wedge\,dx^a
     ~=~ \left(\,-\frac{1}{3}\,\frac{\partial_i
     N}{N}\,\delta^{\hat{a}}_{\hat{b}}\;dx^i\,\right)\,\wedge\,e^{\hat{b}}
     ~=~ -\,\omega_{\hat{B}}{}^{\hat{a}}\,\wedge\,e^{\hat{B}}~,
     \label{Spin1}
\end{eqnarray}\\
i.e.\\
\begin{eqnarray}
    \omega_{i\hat{b}}{}^{\hat{a}} &=&  \frac{1}{3}\,\frac{\partial_i
     N}{N}\,\delta^{\hat{a}}_{\hat{b}}~.\label{M2SC1}
\end{eqnarray}\\
One could has rearranged the terms in (\ref{Spin1}) also this way\\
\begin{eqnarray}
     de^{\hat{a}}  
     ~=~ -\frac{1}{3}\,\frac{\partial_i
     N}{N^{4/3}}\;dx^i\,\wedge\,dx^a
     ~=~ \left(\,+\frac{1}{3}\,\frac{\partial_{\hat{i}}
     N}{N}\,\;e^{\hat{a}}_{c}\,dx^c\,\right)\,\wedge\,e^{\hat{i}}
     ~=~ -\,\omega_{\hat{B}}{}^{\hat{a}}\,\wedge\,e^{\hat{B}}~,
     \label{Spin2}
\end{eqnarray}\\
i.e.\\
\begin{eqnarray}
    \omega_{c\hat{i}}{}^{\hat{a}} &=& -\,\frac{1}{3}\,\frac{\partial_{\hat{i}}
     N}{N}\,e^{\hat{a}}_{c}~.\label{M2SC2}
\end{eqnarray}\\
and similarly\\[2ex]
\underline{${\hat{A}}\,=\,{\hat{i}}$~:}\hspace{3ex}$e^{\hat{i}}\,=\,N^{1/6}\,dx^i\quad\Rightarrow$\\
\begin{eqnarray}
     de^{\hat{i}}~=~ ~=~ \frac{1}{6}\,\frac{\partial_j
     N}{N^{5/6}}\;dx^j\,\wedge\,dx^i
     ~=~ \left(\,\frac{1}{6}\,\frac{\partial_j
     N}{N}\,\delta^{\hat{i}}_{\hat{k}}\;dx^j\,\right)\,\wedge\,e^{\hat{k}}
     ~=~ -\,\omega_{\hat{B}}{}^{\hat{i}}\,\wedge\,e^{\hat{B}}~,
     \label{Spin3}
\end{eqnarray}\\
i.e.
\begin{eqnarray}
    \omega_{j\hat{k}}{}^{\hat{i}} &=& -\,\frac{1}{6}\,\frac{\partial_j
     N}{N}\,\delta^{\hat{i}}_{\hat{k}}~.\label{M2SC3}
\end{eqnarray}\\
Using the method discussed in exercise \ref{ex_CAM5Brane} to write the 
11d Clifford algebra in terms of longitudinal and transverse
directions to the M2-Brane one obtains\\ 
\begin{eqnarray}
     \Gamma^a &=& \gamma^a\otimes\tilde\gamma^{(9)}\qquad
     a\,=\,0,\ldots ,2\\[1ex]
     \Gamma^i &=& \;\unity\;\otimes\tilde\gamma^{i}\qquad
     \hspace{2ex}i\,=\,3,\ldots, 10
\end{eqnarray}\\
and $\tilde\gamma^{(9)}$ is the eight dimensional chirality
operator.\\

\noindent
\underline{M\,=\,a:} Only the spin connection (\ref{M2SC2})
contributes in that case, i.e. \\
\begin{eqnarray}
   \hat{D}_a &=& \partial_a\,
                 -\,\frac{1}{4}\,\omega_{a\hat{A}\hat{B}}\,
                  \Gamma^{\hat{A}\hat{B}}
                 -\,\frac{1}{288}
                  \left(\vbox{\vspace{2.5ex}}\right.\,
                          \Gamma^{\alpha_1\alpha_2\alpha_3\alpha_4}{}_{a}
                                  -8\,\Gamma^{\alpha_2\alpha_3\alpha_4}
                                     \delta_a^{\alpha_1}\,
                  \left.\vbox{\vspace{2.5ex}}\right)\,
                  G_{\alpha_1\alpha_2\alpha_3\alpha_4}\nonumber\\[2ex]
            &=& -\,\frac{1}{4}\,\omega_{a\hat{i}\hat{b}}\,
                  \Gamma^{\hat{i}\hat{b}}
                 -\,\frac{1}{288}
                  \left(\vbox{\vspace{2.5ex}}\right.\,
                          \Gamma^{\alpha_1\alpha_2\alpha_3\alpha_4}{}_{a}
                                  -8\,\Gamma^{\alpha_2\alpha_3\alpha_4}
                                     \delta_a^{\alpha_1}\,
                  \left.\vbox{\vspace{2.5ex}}\right)\,
                  G_{\alpha_1\alpha_2\alpha_3\alpha_4}~.\label{KSPM2a}
\end{eqnarray}\\ 
The other terms simplify as follows:\\
\begin{eqnarray}
   \Gamma^{\alpha_1\alpha_2\alpha_3\alpha_4}{}_{a}\,
   G_{\alpha_1\alpha_2\alpha_3\alpha_4}
   &=& 4!\cdot  \Gamma^{012i}{}_{a}
   G_{012i}
   ~=~ 0.\\
   \Gamma^{\alpha_2\alpha_3\alpha_4}\,\delta^{\alpha_1}_{a}\,
   G_{\alpha_1\alpha_2\alpha_3\alpha_4}
   &=& 3\cdot  \Gamma^{\alpha_2\alpha_3i}
   G_{a\alpha_2\alpha_3i}~=~ 3\cdot\varepsilon_{abc}
   \gamma^{bc}\tilde\gamma^{i}\cdot c\,\frac{\partial_i N}{N^2}
\end{eqnarray}\\
Inserting the partial results back into eq.~(\ref{KSPM2a}) one obtains\\
\begin{eqnarray}
 0 &=& \frac{1}{6}\,\frac{\partial_{\hat{i}}N}{N}e_{a\hat{b}}\,
                  \tilde\gamma^{\hat{i}}\gamma^{\hat{b}}\tilde\gamma^{(9)}
                 +\,\frac{1}{12}\,\varepsilon_{abc}
   \gamma^{bc}\tilde\gamma^{i}\cdot c\,\frac{\partial_i N}{N^2}
\end{eqnarray}\\
and using\\
\begin{eqnarray*}
   \varepsilon_{abc}\gamma^{bc} 
   ~=~\varepsilon_{abc}\,e^b_{\hat{b}}\,e^c_{\hat{c}}\gamma^{\hat{b}\hat{c}}
   &=&\varepsilon_{abc}\,e^b_{\hat{b}}\,e^c_{\hat{c}}\,
      \varepsilon^{\hat{b}\hat{c}\hat{d}}\,\gamma_{\hat{d}}\\
   &=&\varepsilon_{abc}\,e^b_{\hat{b}}\,e^c_{\hat{c}}\,
      \varepsilon^{\hat{b}\hat{c}\hat{d}}\,e^d_{\hat{d}}\,\gamma_{d}\\
   &=& \det(e^{-1}_{(3)})\,\varepsilon_{\hat{b}\hat{c}\hat{d}}\,
       \varepsilon^{\hat{b}\hat{c}\hat{d}}\,\gamma_{d}\\
   &=& -2\,N\,\gamma_{a}
\end{eqnarray*}\\
one finally obtains\\
\begin{eqnarray}
 0 &=& -\,\frac{c}{6}\,\frac{\partial_{\hat{i}}N}{N}\,
                  \tilde\gamma^{\hat{i}}\gamma_{a}
                  \left(\,
                     1\,-\,c\,\tilde\gamma^{(9)}\,
                  \right)~.
\end{eqnarray}\\
To solve this equation we simply have to make the ansatz\\
\begin{eqnarray}
    \epsilon &=& f(x^3,\ldots ,x^{10})\cdot\left(\,
                     1\,+\,c\,\tilde\gamma^{(9)}\,
                  \right)\,\epsilon_0~. 
\end{eqnarray}\\
Note that due to the duality symmetry (\ref{DualityRel}) 
$\gamma^{(9)}$ is identical to $\Gamma_{012}$ and in most 
research papers $\gamma^{(9)}$ is replaced by the latter one.
The evaluation of the second half of the Killing spinor equation
follows the same lines as in the case of the M5-Brane and is 
left as an exercise (exercise \ref{exKSPM2i}). 

\subsubsection*{Homework:}

\begin{exercise} As mentioned before this method is ideal to compute 
the Ricci tensor of all the elementary brane solutions for various 
the supergravities not necessarily eleven dimensional. One simply 
considers the metric\\
\begin{eqnarray}
    g_{\mu\nu} &=& N^{\kappa}\cdot\left(
                         \begin{array}{cc}
                                \eta_{ab}^{(p)} & 0  \\
                                0  & N^\delta\cdot\delta^{D-p}_{ij} \\
                         \end{array}
                   \right)~.
\end{eqnarray}\\ 
Adjusting $p$, $D$, $\kappa$ and $\delta$ according to the needs one 
obtains the Ricci tensor, which must be matched by the suitably chosen
field strengths (not considered here).
\end{exercise}
\begin{exercise}\label{exKSPM2i}
Compute the $M\,=\,i$ part of the Killing spinor equation and
determine the scaling function $f$ in 
$\epsilon\,=\,f(x^3,\ldots x^{10})\,\epsilon_0$. 
\end{exercise}

\vfill\eject
\section{Intersecting Branes}
\label{Sec_Inter_Branes}

All the fundamental brane solutions described so far can be used 
to build more complicated metrics by what is known as intersecting 
branes. The basic example has already been known for some time 
\cite{Ooguri:1996wj,Bershadsky:1996sp}. The solution we are studying 
here was found in
\cite{Papadopoulos:1996uq,Tseytlin:1996bh,Klebanov:1996mh}. 
A review focusing on intersecting branes alone 
is \cite{Gauntlett:1997cv}. 

\subsection{Intersecting M5-M5-Brane Solution}

\begin{eqnarray}
    g_{\mu\nu} &=& \frac{1}{H_1^{1/3}H_2^{1/3}}\left(
                         \begin{array}{cccccccccc}
                             -1 & & & & & & & & &\\
                                      & \ddots &&&&&&&&\\
                                      & & 1 &&&&&\\
                             & & & H_2 & & & & & &\\
                             & & & & H_2 & & & & &\\
                             & & & & & H_1 & & & &\\
                             & & & & & & H_1 & & &\\
                             & & & & & &  & H_1H_2 & &\\
                             & & & & & &  & & \ddots&\\
                             & & & & & &  & & & H_1H_2\\
                         \end{array}
                   \right)
\end{eqnarray}\\
We split the indices $\mu,\nu\,\in\,\{\,x^0,\,\ldots,\,x^{10}\,\}$ by
using the following notation:\\
\begin{eqnarray*}
      x,y &\in& \{\,x^0,\,x^1,\,x^2,\,x^3\,\}\\
      a,b &\in& \{\,x^4,\,x^5\,\}\\
      \tilde{a},\tilde{b} &\in& \{\,x^6,\,x^7\,\}\\
      i,j &\in& \{\,x^8,\,x^9,\,x^{10}\,\} ~=~ \perp
\end{eqnarray*}\\
By direct computation or applying formula (\ref{RescaledRicciT}) as 
illustrated in 
subsection \ref{SubSecM2Brane} one obtains the Ricci tensor.
With $H_1\,=\,H_1(x^8,\,x^9,\,x^{10})$ and $H_2\,=\,H_2(x^8,\,x^9,\,x^{10})$ 
we obtain in non covariant notation\footnote
{
  Summation over equal indices is understood. $\Delta$ denotes the
  Euclidean Laplacian, i.e $\Delta\,=\,\delta^{kl}\partial_k\partial_l$. 
}
%
\begin{eqnarray}
 R_{xy} &=& \frac{1}{6}\,
            \left(\,
                     \frac{\Delta H_1}{H_1^2H_2}\, 
                   +\,\frac{\Delta H_2}{H_1H_2^2}\,
                   -\,\frac{\partial_k H_1\partial_k H_1}{H_1^3H_2}\,
                   -\,\frac{\partial_k H_2\partial_k H_2}{H_1H_2^3}\,
            \right)\,\eta_{xy}\\[1ex]
 R_{ab} &=& \left(\,
                \frac{1}{6}\,
                \left[\,
                      \frac{\Delta H_1}{H_1^2}\,
                   -\,\frac{\partial_k H_1\partial_k H_1}{H_1^3}\,
                \right]\,
             -\,\frac{1}{3}\,
                \left[\,
                      \frac{\Delta H_2}{H_1H_2}\,
                   -\,\frac{\partial_k H_2\partial_k H_2}{H_1H_2^2}\,
                \right]\,
               \right)\,\delta_{ab}\\[1ex]
 R_{\tilde{a}\tilde{b}} 
          &=&  \left(\,
                \frac{1}{6}\,
                \left[\,
                      \frac{\Delta H_2}{H_2^2}\,
                   -\,\frac{\partial_k H_2\partial_k H_2}{H_2^3}\,
                \right]\,
             -\,\frac{1}{3}\,
                \left[\,
                      \frac{\Delta H_1}{H_1H_2}\,
                   -\,\frac{\partial_k H_1\partial_k H_1}{H_1^2H_2}\,
                \right]\,
               \right)\,\delta_{\tilde{a}\tilde{b}}  \\[1ex]
 R_{ij} &=& -\,\frac{1}{2}\,\frac{\partial_i H_1\partial_j H_1}{H_1^2}\,
            -\,\frac{1}{2}\,\frac{\partial_i H_2\partial_j H_2}{H_2^2}\,
            +\,\frac{1}{3}\,
               \left(\vbox{\vspace{2.5ex}}\right.\,
                       \frac{\partial_k H_1\partial_k H_1}{H_1^2}\,
                    +\,\frac{\partial_k H_2\partial_k H_2}{H_2^2}
                    \nonumber\\
                  &&  -\,\frac{\Delta H_1}{H_1}\,
                    -\,\frac{\Delta H_2}{H_2}\,
               \left.\vbox{\vspace{2.5ex}}\right)\,
               \delta_{ij}
\end{eqnarray}\\
The ansatz for the four form field reads\\ 
\begin{eqnarray}
    G_{45ij} &=& \varepsilon_{ijk}\partial_k H_2\\
    G_{67ij} &=& \varepsilon_{ijk}\partial_k H_1
\end{eqnarray}\\
and we obtain\\ 
\begin{eqnarray}
    -\frac{1}{144}\,G^2\,g_{\mu\nu} &=& -\frac{1}{6}\,\left(\,\frac{\partial_kH_1\partial_kH_1}{H_1^3H_2}\,+\,\frac{\partial_kH_2\partial_kH_2}{H_1H_2^3}\,\right)\,(H_1H_2)^{1/3}\,g_{\mu\nu}~.
\end{eqnarray}\\
The other term $(G^2)_{\mu\nu}$ must be computed by splitting the
indices and into longitudinal and transversal ones.\\
  
\noindent 
\underline{$\mu\,=\,i$ \& $\nu\,=\,j$:}~~ Here  $i,j\in \{8,9,10\}$ and one 
obtains\\
\begin{eqnarray*}
   \frac{1}{12}\,G_{i\alpha_1\alpha_2\alpha_3}G_{j}{}^{\alpha_1\alpha_2\alpha_3}&=&
   \frac{1}{2}\left\{\,\frac{1}{H_1^2}\,\left(\,\partial_kH_1\partial_kH_1\,\delta_{ij}
   -\partial_iH_1\partial_jH_1\,\right) \,+\,\frac{1}{H_2^2}\,\left(\,\partial_kH_2\partial_kH_2\,\delta_{ij}
   -\partial_iH_2\partial_jH_2\,\right)\,\right\} 
\end{eqnarray*}\\
and the right hand side of the Einstein equation (\ref{Einstein}) for the indices 
 $(\mu\nu)\,=\,(i,j)$ reads\\
\begin{eqnarray*}
   \frac{1}{12}\,\left(\,(G^2)_{ij}-\frac{1}{12}\,G^2\,g_{ij}\,\right)
   &=& -\,\frac{1}{2}\,\frac{\partial_iH_1\partial_jH_1}{H_1^2}\,-\,\frac{1}{2}\,\frac{\partial_iH_2\partial_jH_2}{H_2^2}\,+\,\left[\,\frac{1}{3}\,\frac{\partial_kH_1\partial_kH_1}{H_1^2}\,+\,\frac{1}{3}\,\frac{\partial_kH_2\partial_kH_2}{H_2^2}\,\right]\,\delta_{ij}
\end{eqnarray*}\\
\underline{$\mu\,=\,a$ \& $\nu\,=\,b$:}~~Here $a,b\,\in\,\{\,4,\,5\,\}$\\
\begin{eqnarray*}
   \frac{1}{12}\,G_{a\alpha_1\alpha_2\alpha_3}G_{b}{}^{\alpha_1\alpha_2\alpha_3}&=&
   \frac{1}{2}\,\frac{\partial_kH_2\,\partial_k H_2}{H_1\,H_2^2}\,\delta_{ab}
\end{eqnarray*}\\
and if one repeats the calculation for $\tilde a,\tilde
b\,\in\,\{\,6,\,7\,\}$
one obtains
\begin{eqnarray*}
   \frac{1}{12}\,G_{\tilde a\alpha_1\alpha_2\alpha_3}G_{\tilde b}{}^{\alpha_1\alpha_2\alpha_3}&=&
   \frac{1}{2}\,\frac{\partial_kH_1\,\partial_k
   H_1}{H_1^2\,H_2}\,\delta_{\tilde a\tilde b}
\end{eqnarray*}\\
\noindent
Proceeding in the same fashion in computing the other tensor
components one obtains finally:\\
{\tiny
\begin{eqnarray*}
    T_{\mu\nu} &=& \left(\begin{array}{cccc}
                      -\frac{1}{6}\,\left(\,\frac{\partial_kH_1\partial_kH_1}{H_1^3H_2}\,+\,\frac{\partial_kH_2\partial_kH_2}{H_1H_2^3}\,\right)\,\eta_{xy} & & & \\[2ex]
            & \hspace{-3.5cm}\left(\,
                -\,\frac{1}{6}\,
                   \frac{\partial_k H_1\partial_k H_1}{H_1^3}\,
             +\,\frac{1}{3}\,
                \frac{\partial_k H_2\partial_k H_2}{H_1H_2^2}\,
               \right)\,\delta_{ab} &  & \\[2ex]
            & &\hspace{-3.5cm}\left(\,
                -\,\frac{1}{6}\,
                   \frac{\partial_k H_2\partial_k H_2}{H_2^3}\,
             +\,\frac{1}{3}\,
                \frac{\partial_k H_1\partial_k H_1}{H_2H_1^2}\,
               \right)\,\delta_{\tilde a\tilde b} & \\[2ex]
            & & & \hspace{-3.5cm}  -\,\frac{1}{2}\,\frac{\partial_iH_1\partial_jH_1}{H_1^2}\,-\,\frac{1}{2}\,\frac{\partial_iH_2\partial_jH_2}{H_2^2}\,+\,\left[\,\frac{1}{3}\,\frac{\partial_kH_1\partial_kH_1}{H_1^2}\,+\,\frac{1}{3}\,\frac{\partial_kH_2\partial_kH_2}{H_2^2}\,\right]\,\delta_{ij}     
                   \end{array}\right)
\end{eqnarray*}
}

\vfill\eject
\section{Exotic solutions}
\label{Sec_Exotic}

Solutions of supergravity theories which preserve 1/2 of the original 
supersymmetry typically occur in the description of brane solutions.
Using the techniques of intersecting brane solutions one can reduce 
this number quite generically to any fraction 1/$2^n$, with $n$ 
an integer. Other fractions are less simple to understand and are called 
exotic.
In \cite{Gauntlett:2002nw,Harmark:2003ud,Miemiec:2003cj} the authors 
discovered a new solution of 11 dimensional supergravity, which
preserved a fraction of $\nu\,=\,5/8$ 
supersymmetries. In a following paper \cite{Gauntlett:2002fz} it was 
shown that this solution fits into a whole class of solutions and the 
questions was raised if further solutions with exotic fractions of 
preserved supersymmetry can be found within this class. In 
\cite{Miemiec:2003cj} it was confirmed in a case by case 
study that this actually is the case.

\noindent
The class of solutions proposed in \cite{Gauntlett:2002fz} are\\ 
\begin{eqnarray}
   ds_{11}^2 &=& -\,(\,dt\,+\,\omega\,)^2 ~+~ ds^2({\mathbb{R}}^{10})
                 \nonumber\\
           G &=& -d\omega\wedge\Omega
\end{eqnarray}\\
with $\Omega$ defined in terms of complex coordinates 
$z^a\,=\,x^{2a-1}+i\,x^{2a}$ of 
$~{\mathbb{C}}^{5}\,=\,{\mathbb{R}}^{10}$,\\
\begin{eqnarray}
  \Omega &=& \frac{i}{2}\,\sum\limits_{a=1}^5 dz^a\wedge d{\bar{z}}^a~,
\end{eqnarray}\\
and $d\omega\,=\,\alpha\,+\,\bar{\alpha}\,\subset\,\Lambda^{2,0}({\mathbb{C}}^{5})\oplus\Lambda^{0,2}({\mathbb{C}}^{5})$.\\

\noindent
The Einstein equation (\ref{Einstein}) and the equation of motion of
the four form field strength (\ref{4Form}) are given in 
section \ref{Sec_MTheory} while the Killing spinor equation 
(\ref{KSP_Eq}) can be written as\\
\begin{eqnarray}\label{KSE}
     \partial_\mu\varepsilon\,-\,
     \frac{1}{4}\omega_{\mu ab}\Gamma^{ab}\,\varepsilon ~+~ \frac{1}{288}
     \left(\, 
              \Gamma_{\mu}\Gamma^{\nu_1\nu_2\nu_3\nu_4} ~-~ 12\,
              \delta_{\mu}^{\nu_1}\Gamma^{\nu_2\nu_3\nu_4}\,
     \right)\,G_{\nu_1\nu_2\nu_3\nu_4}\varepsilon &=& 0~.
\end{eqnarray}\\[-2ex]

\subsection{First Example}

\noindent
The holomorphic two form $\alpha$ we have chosen is\\ 
\begin{eqnarray}
   \alpha &=&\frac{\gamma}{2}\,
                \left(\, 
                         dz^1\wedge dz^2 ~+~ dz^3\wedge dz^4\,
                \right)
\end{eqnarray}\\
which leads to\\ 
\begin{eqnarray}
     \omega&=&\frac{\gamma}{2}\,
                \left(\,
                         -x^3dx^1+x^1dx^3-x^2dx^4+x^4dx^2
                         -y^3dy^1+y^1dy^3-y^2dy^4+y^4dy^2\,
                \right)
\end{eqnarray}\\

\noindent
The equations of motion (\ref{Einstein}) and (\ref{4Form}) are both 
satisfied. The Ricci tensor in a tangent frame is diagonal and given 
by $R_{ab}\,=\,\gamma^2\cdot{\rm diag}(2,1/2,\ldots,1/2,0,0)$ 
while the contractions of the field strength are:\\
\begin{eqnarray}
    G_{c_1c_2c_3c_4}G^{c_1c_2c_3c_4} &=& 288\,\gamma^2\\
    G_{ac_1c_2c_3}G_{b}{}^{c_1c_2c_3}&=& \left\{
                                                \begin{array}{cl}
                                                  0  & (a,b)=(0,0)\\
                                                 30\,\gamma^2 
                                                     & (a,b)=(1,1),\ldots
                                                             ,(8,8)\\
                                                 24\,\gamma^2 
                                                     & (a,b)=(9,9),
                                                       (10,10)\\
                                                  0  & a\neq b
                                                \end{array}
                                         \right.
\end{eqnarray}\\
Obviously eq.~(\ref{Einstein}) is satisfied. 
The second term in the equation of motion for the four form field strength 
(\ref{4Form}) reduces to\\ 
\begin{eqnarray*}
 \frac{1}{2}\,G\wedge G &=& 2\,\gamma^2\,\left(\,2\,dx^{12345678}
                            +dx^{1234569 1\hspace{-1.75pt}0}
                            +dx^{1234789 1\hspace{-1.75pt}0}
                            +dx^{1256789 1\hspace{-1.75pt}0}
                            +dx^{3456789 1\hspace{-1.75pt}0}\,\right)
\end{eqnarray*}\\
and comparing this with $d(\ast G)$ one finds that both add up to zero.\\

\noindent
The Killing spinor equation (\ref{KSE}) can be written more 
symbolically as\\
\begin{eqnarray}\label{SymbKSE}
     \partial_\mu\varepsilon\,-\,{\mathbb{M}}_\mu\varepsilon &=& 0~.
\end{eqnarray}\\
By just solving this equation for the present ansatz one obtains 
12 constant Killing spinors. These 12 solutions exhaust the whole set 
of solutions. 
Now we give the argument from which we draw this conclusion.\\

\noindent
As a preparation we would like to shed some light onto the terms which we
gathered in\\
\begin{eqnarray}
  e_c{}^{\mu}{\mathbb{M}}_\mu &=& \frac{1}{4}\underbrace{\omega_{cab}
       \Gamma^{ab}}_{(3)}
                        ~-~ \frac{1}{288}
                        \left(\vbox{\vspace{2ex}}\right.\, 
                            \Gamma_{c}\underbrace{
                            (\Gamma^{\nu_1\nu_2\nu_3\nu_4}
                          G_{\nu_1\nu_2\nu_3\nu_4})}_{(1)} ~-~ 12\,e_c{}^{\mu}
                            \underbrace{
                              \Gamma^{\nu_2\nu_3\nu_4}G_{\mu\nu_2\nu_3\nu_4}
                            }_{(2)}
                        \left.\vbox{\vspace{2ex}}\right)~.
\end{eqnarray}\\
The components of the field strength $G_{\nu_1\nu_2\nu_3\nu_4}$ are 
constant and those involving the time direction vanish. Also, gamma matrices 
with upper spatial indices are identical to the tangent frame 
gamma matrices.  We conclude that the expressions (1) and (2) are constant. 
The spin connection in (3) turns out to be constant too. So the only source of 
coordinate dependence comes from contraction with the vielbein in the last 
term. 
Since the field strength components which include a time direction vanish, 
the only remaining source for a coordinate dependence is set to zero.\\

\noindent
This example gives a simple necessary condition that each solution 
must satisfy\\
\begin{eqnarray}
      0 ~=~ [\,\partial_\mu,\,\partial_\nu\,]\,\varepsilon
        &=& \partial_\mu(\partial_\nu\varepsilon)
         \,-\,\partial_\nu(\partial_\mu\varepsilon)
        ~=~ \partial_\mu(\,M_\nu\,\varepsilon\,)
         \,-\,\partial_\nu(\,M_\mu\,\varepsilon\,)\nonumber
\end{eqnarray}\\
or transferring the ${\mathbb{M}}$'s to the tangent frame\\
\begin{eqnarray}
     0 ~=~ [\,\partial_\mu,\,\partial_\nu\,]\,\varepsilon    
        &=& \partial_\mu(\,e_\nu{}^a\,M_a\,\varepsilon)
         \,-\,\partial_\nu(\,e_\mu{}^b\,M_b\,\varepsilon)\nonumber\\[2ex]
        &=& e_{\mu}{}^ae_{\nu}{}^b\,
            \left(\,  2\,\omega_{[ab]}{}^c\,{\mathbb{M}}_c ~+~
                      [\,{\mathbb{M}}_b,\,{\mathbb{M}}_a\,]\,
            \right)\,\varepsilon~,\label{necessary}
\end{eqnarray}\\
where  ${\mathbb{M}}_a\,=\,e_a{}^\mu{\mathbb{M}}_\mu$ and
$\omega_{[\mu\nu]}{}^a\,=\,\partial_{[\mu}e_{\nu]}{}^a$. In deriving 
eq.~(\ref{necessary}) we used the fact that in the present case all 
${\mathbb{M}}_a$ are constant. Since the only non vanishing components 
of $\omega_{[\mu\nu]}{}^a$ are\\
\begin{align}
    \omega_{[13]}{}^0 &=\gamma~, & 
    \omega_{[24]}{}^0 &=-\gamma~, &
    \omega_{[57]}{}^0 &=\gamma~, &
    \omega_{[68]}{}^0 &=-\gamma& 
\end{align}\\
equation (\ref{necessary}) evaluated for $(a,b)=(0,1)$ just reduces
to\\ 
\begin{eqnarray}
     0 &=& [\,{\mathbb{M}}_1,\,{\mathbb{M}}_0\,]\,\varepsilon~. 
\end{eqnarray}\\
The kernel is most easily calculated by a computer and in this case 
it turns out to be just the span of the 12 Killing spinors mentioned 
above. So we have already found all solutions.\\

\subsection{Second Example}

We now take the following ansatz for\\ 
\begin{eqnarray}
   \alpha &=& \frac{\gamma}{2}\,
              \left(\,
                  dz^1\wedge dz^2 ~+~ dz^1\wedge dz^3 ~+~ dz^2\wedge dz^3\,
              \right)~.
\end{eqnarray}\\
In this case the Ricci tensor in the tangent frame is no longer diagonal. 
Nevertheless the equations of motion are satisfied both for the metric 
and the four form field strength.\\  
\begin{eqnarray*}
    R_{ab} &=& \left(\,\begin{array}{cccccccccc}
                          3  &  0  &  0  &  0  &  0  &  0  &  0  &
                          0  & \ldots &  0\\
                          0  &  1  &  0  &  1/2  &  0  &  -1/2  &  0  &
                          0  & \ldots &  0\\
                          0  &  0  &  1  &  0  &  1/2  &  0  &  -1/2  &
                          0  & \ldots &  0\\
                          0  & 1/2 &  0  &  1  &  0  &  1/2  &  0  &
                          0  & \ldots &  0\\
                          0  &  0  & 1/2 &  0  &  1  &  0  &  1/2  &
                          0  & \ldots &  0\\
                          0  & -1/2&  0  & 1/2 &  0  &  1  &  0  &
                          0  & \ldots &  0\\
                          0  &  0  & -1/2&  0  & 1/2 &  0  &  1  &
                          0  & \ldots &  0\\
                          0  &  0  &  0  &  0  &  0  &  0  &  0  &
                          0  & \ldots &  0\\
                          \vdots & \vdots &\vdots & \vdots &\vdots & \vdots
                          &\vdots & \vdots & \ddots & \vdots\\
                          0  &  0  &  0  &  0  &  0  &  0  &  0  &
                          0  & \ldots &  0\\
                       \end{array}
               \right)
\end{eqnarray*}
One finds 16 constant Killing spinors $\varepsilon_i$ generating the 
16 dimensional kernels of ${\mathbb{M}}_1$ to ${\mathbb{M}}_6$. 
They are contained in the 20 dimensional kernels of ${\mathbb{M}}_0$ and 
${\mathbb{M}}_7$ to ${\mathbb{M}}_{10}$\footnote
%
{
 Again the dimensions of the kernels are computed using a computer.
}.
%
The form of the 16 constant Killing spinors can be calculated
explicitly and is given for convenience in appendix \ref{AppendixB}. 
The 20 dimensional kernels of ${\mathbb{M}}_0$ and 
${\mathbb{M}}_7$ to ${\mathbb{M}}_{10}$ are all equal and the additional 
four dimensions are spanned by $\vartheta$, which as a consequence of 
eq.~(\ref{necessary}) satisfies\\
\begin{eqnarray*}
     0 ~\neq ~ {\mathbb{M}}_i(\vartheta)  ~\subset~ 
       {\rm span}(\,\varepsilon_1,\ldots \varepsilon_{16}\,)~,
      \quad\quad\quad i~=~1,\ldots , 6~.
\end{eqnarray*}\\
In this situation one can apply the following trick  to
construct a further Killing spinor \cite{Gauntlett:2002nw}.  Setting\\
\begin{eqnarray*}
        \varepsilon &=& \vartheta ~+~ x^i\,{\mathbb{M}}_i\,\vartheta
\end{eqnarray*}\\
the Killing spinor equation (\ref{SymbKSE}) reads:\\
\begin{eqnarray*}
 \mu\,=\,i\,=\,1,\ldots ,6\hspace{3.5ex} &\quad&
 \partial_i\,\varepsilon~-~{\mathbb{M}}_i\,\varepsilon\;~=~\phantom{-\;\;\,}
 {\mathbb{M}}_i\,\vartheta ~-~ {\mathbb{M}}_i\,
 \vartheta\hspace{5.5ex} ~=~ 0,\\[2ex]
 \mu\,=\,a\,=\,0,7,\ldots ,10 &\quad&
 \partial_a\,\varepsilon~-~{\mathbb{M}}_a\,\varepsilon~=~-\,
 {\mathbb{M}}_a\,\left(\,\vartheta\,+\,x^i\,{\mathbb{M}}_i\,\vartheta\,\right) ~=~ 0~.
\end{eqnarray*}\\
Finally we obtain 20 Killing spinors, i.e. $\nu\,=\,5/8$ of the original supersymmetries are preserved.\\

\noindent
In the above we have presented two solutions, which preserve two different 
fractions of supersymmetry. The first example preserves $3/8$, the later 
$5/8$. The first example preserves a different exotic fraction of 
supersymmetry than the example already discovered in \cite{Gauntlett:2002fz}.
We have also tried different ans\"atze for $\alpha$. Their precise form and 
the amount of supersymmetry they preserve can be found in the table below.\\
\begin{center}
\begin{tabular}{|c|l|}
\hline
                       &\\[-1.5ex] 
   $\alpha$            &   $\nu$\\[.5ex]
\hline
                       &\\[-1ex] 
   $dz^1\wedge dz^2$   &    5/8   \\
   $dz^1\wedge dz^2\,+\,dz^1\wedge dz^3$ & 5/8\\
   $dz^1\wedge dz^2\,+\,dz^1\wedge dz^3\,+\,dz^2\wedge dz^3$ & 5/8\\
   $dz^1\wedge dz^2\,+\,dz^3\wedge dz^4$ & 3/8\\
   $dz^1\wedge dz^2\,+\,dz^3\wedge dz^4\,+\,dz^1\wedge dz^5$ & 3/8\\
   $dz^1\wedge dz^2\,+\,dz^2\wedge dz^3\,+\ldots+\,dz^5\wedge dz^1$ & 3/8\\[.5ex]
\hline
\end{tabular}\\
\end{center}
\vskip2ex
In particular, we have only found solutions which preserve either $3/8$ 
or $5/8$ of the original supersymmetry. It would be interesting to see 
whether this ansatz could also produce solutions with other numbers of 
preserved supersymmetries.

\vfill\eject
\section{IIA Action from M-Theory}
\label{Sec_IIA}

The bosonic part of the M-Theory action 
is (cf. section \ref{Sec_MTheory})\\
\begin{eqnarray}\label{BosonicMTheory}
    {\cal L}_{(11)} &=& \phantom{-}\frac{1}{4}\,\sqrt{-g}\,R\, 
                -\,\frac{1}{4\cdot48}\,\sqrt{-g}\,
                  G_{\mu\nu\rho\sigma}G^{\mu\nu\rho\sigma}
                 \nonumber\\
             && +\,\frac{1}{4\cdot144^2}\,\epsilon^{\alpha_1\ldots\alpha_4
                 \beta_1\ldots\beta_4\mu\nu\rho}G_{\alpha_1\ldots\alpha_4}
                 G_{\beta_1\ldots\beta_4}C_{\mu\nu\rho}
\end{eqnarray}\\
with the vielbein $e^a{}_\mu$ defined with respect to the mostly plus 
flat 11d metric.\\

\noindent
We want to perform a Kaluza-Klein reduction on a circle $S^1$ 
\cite{kaluza,klein}, i.e.\ we consider the 11-th direction to  be 
parametrised by the angular coordinate of the circle
$\varphi\,=\,x^{11}/R\,\in\,[0,2\pi)$. Here
$R$ denotes the radius of the $S^1$.\\

\noindent 
This case is of special interest due to the importance of the outcome, 
i.e.\ IIA supergravity, in string theory. Therefore the reduction
shall be performed as explicit as possible.  
To that purpose all fields are expanded in Fourier modes\\ 
\begin{eqnarray}
  T_{\mu\nu\ldots} &=&
  \sum\limits_{n=0}^\infty\,T_{\mu\nu\ldots}^{(i)}\cdot e^{in\varphi}
\end{eqnarray}\\ 
and a truncation to the lowest order in the Fourier expansion is
performed afterwards. From an analytical point of view the Fourier 
expansion is just an orthogonal decomposition of the infinite 
dimensional Hilbert space of square integrable functions  
into a sum of one dimensional subspaces,\\  
\begin{eqnarray}
    {\mathcal{L}}(\mathbb{R}) &=& \bigoplus\limits_{i=0}^\infty 
           {\mathcal{L}}^{(i)}(\mathbb{R})~,
\end{eqnarray}\\
each of which forms an irreducible representation of the group $S^1$.
From a physical point of view this corresponds to a countable set of 
discrete excitations with different mass in the 11-direction. From the 
ten dimensional perspective the 11th direction is invisible as long as 
one can not excite a reasonable amount of these excitations 
apart from the lowest lying one. The ten dimensional theory
corresponds to the truncation to the massless sector.  
That this truncation makes sense is the crucial 
point in the procedure. In the case of an $S^1$-compactification this 
is unambiguous because\\ 
 \begin{eqnarray}
    {\mathcal{L}}^{(i)}(\mathbb{R}) \otimes {\mathcal{L}}^{(j)}(\mathbb{R})
    &=& {\mathcal{L}}^{(i+j)}(\mathbb{R})~.
\end{eqnarray}\\
Thus the subspace ${\mathcal{L}}^{(0)}(\mathbb{R})$ is closed under
the tensor product of representations and one can separate the
massless sector on the level of the action from the massive one. 
For more general
compactification spaces the
truncation must be treated with care due to the complicated way in which 
products of higher excitations may contribute to the massless
excitations (Clebsch Cordon coefficients) \cite{Pope}.\\   
The first step in the reduction is to write the eleven dimensional
metric in terms of a ten dimensional metric, a gauge potential and a 
scalar field, which resemble the degrees of freedom of the
eleven dimensional theory. We make the following choice for the metric, which is still completely
general:\\
\begin{eqnarray}\label{KaluzaKlein}
   g^{\mbox{\tiny (11)}}_{\mu\nu} 
                     &=& e^{\frac{4}{3}\phi}\,
                          \left(\begin{array}{cc}
                            e^{-2\phi}\,g^{\mbox{\sl\tiny st}}_{\mu\nu}
                      ~+~ A_\mu\,A_\nu 
                       & A_\mu\\
                         A_\nu 
                       & 1
                          \end{array}\right).
\end{eqnarray}\\
Note the following: In order to perform the reduction of the eleven
dimensional Ricci scalar it would be advantageous if we could set 
$\phi\,=\,0$. Only in this situation the famous Kaluza-Klein 
formula can be applied, i.e.\\
\begin{eqnarray}\label{KKReduction}
    R_{\mbox{\tiny (11)}} ~=~ R_{\mbox{\tiny (10)}} ~-~ 
                              \frac{1}{4}\;(F^{(2)}_{\mu\nu})^2~.
\end{eqnarray}\\
Here $F^{(2)}$ is the field strength of the potential $A_\mu$, i.e. 
$F^{(2)}_{\mu\nu}\,=\,2\,\partial_{[\mu}A_{\nu ]}$. 
The basic steps of the proof are compiled for the convenience of the 
reader in appendix \ref{AppKaluzaKlein} and exercise \ref{excerKK} 
is devoted to this important formula, too.
In fact one has enough freedom to shift the problem into 
a shape, so that (\ref{KKReduction}) can be applied. We can cope with 
the exponential factors due to the  behaviour of the 
Ricci scalar under a Weyl scaling (cf. \ref{RescaledRicciS}):\\ 
\begin{eqnarray}\label{Weylscaling}
     \tilde{g}_{\mu\nu}~=~e^{2\sigma}\,g_{\mu\nu}~\Rightarrow~
     \tilde{R}_{\mbox{\tiny (d)}} &=& e^{-2\sigma}\,
                   \left[~
                         R_{\mbox{\tiny (d)}}\,-\,2\,(d-1)\,\Delta\sigma
                         \,-\,
                         (d-1)\,(d-2)\,\partial_\mu\sigma\,\partial^\mu\sigma
                   ~\right]~.
\end{eqnarray}\\
A first Weyl scaling is done to get rid of the exponential 
in front of the metric (\ref{KaluzaKlein}). We scale by\\ 
\begin{eqnarray*}
     g^{\mbox{\tiny (11)}}_{\mu\nu} ~=~ e^{\frac{4}{3}\phi}\,
     \bar{g}^{\mbox{\tiny (11)}}_{\mu\nu}
\end{eqnarray*}\\
and the action (\ref{BosonicMTheory}) reads\\
\begin{eqnarray*}
  {\cal L}_{(11)} 
             &=& \phantom{-}\frac{1}{4}\,e^{\frac{22}{3}\phi}\sqrt{-\bar{g}}\,
                  e^{-\frac{4}{3}\phi}\,
                  \left[\vbox{\vspace{3ex}}\right.\, 
                           R \underbrace{
                                          ~-~ 20\cdot\frac{2}{3}\,\Delta\phi
                                        }_{ 
                                          ~+~ \frac{18}{3}\cdot 20\cdot
                                              \frac{2}{3}\,(\partial\phi)^2
                                        }
                             ~-~ \left(\frac{2}{3}\right)^2\cdot10\cdot 9
                                 \cdot (\partial\phi)^2\, 
                  \left.\vbox{\vspace{3ex}}\right]\\ 
             && -\frac{1}{4\cdot48}\,e^{\frac{22}{3}\phi}\sqrt{-\bar{g}}\,
                  e^{-\frac{16}{3}\phi}\,
                  G_{\mu\nu\rho\sigma}G^{\mu\nu\rho\sigma}
                 \nonumber\\
             && +\frac{1}{4\cdot144^2}\,\epsilon^{\alpha_1\ldots\alpha_4
                 \beta_1\ldots\beta_4\mu\nu\rho}G_{\alpha_1\ldots\alpha_4}
                 G_{\beta_1\ldots\beta_4}C_{\mu\nu\rho}~.
\end{eqnarray*}\\
The new metric and its inverse are ($\,\bar{g}^{\mbox{\tiny (11)}}_{\mu\delta}\bar{g}_{\mbox{\tiny (11)}}^{\delta\nu}\,=\,\delta_\mu^\nu\,$):\\
\begin{eqnarray*}
   \bar{g}^{\mbox{\tiny (11)}}_{\mu\delta} 
                     &=& \left(\begin{array}{cc}
                            \tilde{g}_{\mu\delta}
                      ~+~ A_\mu\,A_\delta 
                       & A_\mu\\
                         A_\delta 
                       & 1
                          \end{array}\right)\hspace{3ex}
   \bar{g}_{\mbox{\tiny (11)}}^{\delta\nu}
                     ~=~ \left(\begin{array}{cc}
                            \tilde{g}^{\delta\nu} 
                       & -\,A^\delta\\
                         -\,A^\nu 
                       & 1\,+\,A_\rho A^\rho
                          \end{array}\right)~.\hspace{3ex}
\end{eqnarray*}\\
Now we perform the Kaluza-Klein reduction. The most difficult part is 
to rewrite the terms containing forms due to the split of the metric. 
Luckily there are only two such terms, a contraction of the four form field 
strength and a Chern-Simons term. We start with the contraction of the 
four form field strength now. Since the number of p-forms increases 
significantly, we adopt the usual conventions and denote the degree by a
superscript. For example $A^{(1)}\,=\,A_{\mu}\,dx^{\mu}$. The
Kaluza-Klein reduction 
of $G_{\alpha_1\ldots\alpha_4}G^{\alpha_1\ldots\alpha_4}$ is done 
in four stages, each of which is characterised by the way the 11-direction 
is singled out:\\ 
\begin{eqnarray*}
    G_{\alpha_1\ldots\alpha_4}G^{\alpha_1\ldots\alpha_4} &=&
    g^{\alpha_1\beta_1}\,g^{\alpha_2\beta_2}\,
    g^{\alpha_3\beta_3}\,g^{\alpha_4\beta_4}\,
    G_{\alpha_1\ldots\alpha_4}G_{\beta_1\ldots\beta_4}\\[2ex]
 {\it 1.Term:} &=& 4\cdot g^{a_1b_1}\,g^{a_2b_2}\,
                         g^{a_3b_3}\,
                         \underbrace{g^{11,11}}_{1\,+\,(A^{(1)})^2}\,
                         \underbrace{
                                      G_{a_1\ldots a_3\,11}
                                    }_{
                                      H_{a_1\ldots a_3}^{(3)}
                                    }
                         G_{b_1\ldots b_3\,11}~+~\ldots\\
               &=& 4\cdot (\,1\,+\,(A^{(1)})^2\,)\cdot (H^{(3)})^2 ~+~ \ldots\\
 {\it 2.Term:} &=& {\it 1.Term} ~+~ 2\cdot 
                   \left\{\vbox{\vspace{3ex}}\right.
                        \underbrace{g^{11b_1}}_{-A^{b_1}}\,
                        g^{a_2b_2}\,
                        g^{a_3b_3}\,g^{a_4b_4}\,
                        G_{11\,a_2\ldots a_4}
                        \underbrace{
                                     G_{b_1\ldots b_4}
                                   }_{
                                     F^{(4)}_{b_1\ldots b_4}
                                   }
                   \left.\vbox{\vspace{3ex}}\right\} ~+~ \ldots\\
               &=& {\it 1.Term} ~+~ 2\cdot 4\cdot A^{[b_1}H^{b_2b_3b_4]}\,
                                    F^{(4)}_{b_1\ldots b_4} ~+~ \ldots\\
 {\it 3.Term:} &=& {\it 1.\,-\, 2.Term} ~+~ 
                   \underbrace{
                                 -\,12\;A_{a}\,A^{b}\,H_{bcd}\,H^{acd}
                              }_{
                                   \hbox{\scriptsize 
                                     (see App.\ref{Term3})
                                   }
                              }
                   ~+~\ldots\\[1ex]
 {\it 4.Term:} &=& {\it 1.\,-\,3.Term} ~+~ 
                   F^{(4)}_{a_1\ldots a_4}F^{(4)}{}^{a_1\ldots a_4}\\[1ex]
 {\it finally:}&=& 4\,(A^{(1)})^2\,(H^{(3)})^2
                   ~+~ 8\cdot A^{[b_1}H^{b_2b_3b_4]}\,F^{(4)}_{b_1\ldots b_4}
                   ~+~ F^{(4)}_{a_1\ldots a_4}F^{(4)}{}^{a_1\ldots a_4}\\
               &&  -\,12\;A_{a}\,A^{b}\,H_{bcd}\,H^{acd} ~+~ 4\,(H^{(3)})^2~.  
\end{eqnarray*}\\  
Some short remarks to the numerical factors appearing in the
evaluation of the first three terms:
\begin{enumerate}
\item[{\it 1.}] There are four possibilities to choose a index pair 
      $(\alpha, \beta)$ to be $(11,11)$.
\item[{\it 2.}] The problem is symmetric in the both $G$'s. Therefore 
      we get a first factor of two. Then one observes that 
      the remaining sum can be written compactly as 
      $A^{[b_1}H^{b_2b_3b_4]}$ only, if one includes a factor 
      of $4$ due to the definition of the antisymmetrisation 
      symbol.
\item[{\it 3.}] The evaluation of the 3. Term is included in the 
      appendix (\ref{Term3}). 
\end{enumerate}

\noindent
Finally the first three terms of the last expression can be put into a more compact 
form. This is due to the fact, that the following identity holds:\\
\begin{eqnarray*}
   (\ast) &=&
   \left(\,F^{(4)}_{a_1\ldots a_4}\,+\,c\cdot A_{[a_1}H_{a_2a_3a_4]}\,\right)
   \left(\,F^{(4)}{}^{a_1\ldots a_4}\,+\,c\cdot A^{[a_1}H^{a_2a_3a_4]}\,\right)
   \\[2ex]
   &=& F^{(4)}_{a_1\ldots a_4}F^{(4)}{}^{a_1\ldots a_4} 
       ~+~ 2\,c\,F^{(4)}_{a_1\ldots a_4}\,A^{[a_1}H^{a_2\ldots a_4]}
       ~+~ c^2\,\underbrace{
                             A_{[a_1}H_{a_2\ldots a_4]}\,
                             A^{[a_1}H^{a_2\ldots a_4]}
                           }_{
                             \hbox{\scriptsize 
                                   (see App.\ref{AppTensorIdentity})
                                  }
                           }\\
   &=& \frac{c^2}{4}\,(A^{(1)})^2\,(H^{(3)})^2
       ~+~ 2\,c\,F^{(4)}_{a_1\ldots a_4}\,A^{[a_1}H^{a_2\ldots a_4]}
       ~+~ F^{(4)}_{a_1\ldots a_4}F^{(4)}{}^{a_1\ldots a_4}\\ 
    && -\,12\;\frac{c^2}{4^2}\;A_{a}\,A^{b}\,H_{bcd}\,H^{acd}~.
\end{eqnarray*}\\
Therefore the expression below simplifies to\\ 
\begin{eqnarray*}
    G_{\alpha_1\ldots\alpha_4}G^{\alpha_1\ldots\alpha_4} 
    &=& 4\,(H^{(3)})^2 ~+~ 
        \left(\,
                 F^{(4)}_{a_1\ldots a_4}\,+\,
                 4\cdot A^{(1)}_{[a_1}H^{(3)}_{a_2a_3a_4]}\,
        \right)^2~.
\end{eqnarray*}\\

\noindent
The Chern-Simons term simplifies in the following way\\
\begin{eqnarray*}
  CS &\sim&\varepsilon^{\alpha_1\ldots\alpha_4\beta_1\ldots\beta_4\mu\nu\rho}\,
  G_{\alpha_1\ldots\alpha_4}\,G_{\beta_1\ldots\beta_4}\,C_{\mu\nu\rho}\\
  &=& \varepsilon^{11\,a_2\ldots a_4b_1\ldots b_4ijk}\,
  \underbrace{
                G_{11\,a_2\ldots a_4}
             }_{
                -\,H^{(3)}_{a_2\ldots a_4}
             }\,F^{(4)}_{b_1\ldots b_4}\,A^{(3)}_{ijk}\\
  &&~+~ \varepsilon^{a_1\alpha_2\ldots\alpha_4\beta_1\ldots\beta_4\mu\nu\rho}\,
  G_{a_1\alpha_2\ldots\alpha_4}\,G_{\beta_1\ldots\beta_4}\,C_{\mu\nu\rho}\\
  &\vdots&\\
  &=& -\,8\;\varepsilon^{a_1\ldots a_3b_1\ldots b_4ijk}\,
       \underbrace{
                    H^{(3)}_{a_1\ldots a_3}
                  }_{
                    3\,\partial_{[a_1}B^{(2)}_{a_2a_3]}
                  }\,
        F^{(4)}_{b_1\ldots b_4}\,A^{(3)}_{ijk}\\
  && +\,3\;\varepsilon^{a_1\ldots a_4 b_1\ldots b_4 ij}\,
  F^{(4)}_{a_1\ldots a_4}\,F^{(4)}_{b_1\ldots b_4}\,
       \underbrace{
                    C_{ij\,11}
                  }_{
                    B^{(2)}_{ij}
                  }\\
  &=& +\,6\;\varepsilon^{a_2a_3b_1\ldots b_4a_1ijk}\,
        B^{(2)}_{a_2a_3}\,
        F^{(4)}_{b_1\ldots b_4}\,
        \underbrace{
                      4\,\partial_{[a_1}A^{(3)}_{ijk]}
                   }_{
                       F^{(4)}_{a_1ijk}
                   }\\
  && +\,3\;\varepsilon^{a_1\ldots a_4 b_1\ldots b_4 ij}\,
       F^{(4)}_{a_1\ldots a_4}\,F^{(4)}_{b_1\ldots b_4}\,B^{(2)}_{ij}\\
 &=& 9\;\varepsilon^{a_1\ldots a_4 b_1\ldots b_4 ij}\,
       F^{(4)}_{a_1\ldots a_4}\,F^{(4)}_{b_1\ldots b_4}\,B^{(2)}_{ij}~.
\end{eqnarray*}\\

\noindent
Since we had prepared all terms for the reduction step we can now apply 
the formula (\ref{KKReduction}) for the Ricci tensor and add it to the other 
terms evaluated before to obtain the ten dimensional action:\\
\begin{eqnarray*}
  {\cal L}_{(10)} 
             &=& \frac{1}{4}\,e^{\frac{22}{3}\phi}\sqrt{-\tilde{g}}\,
                  e^{-\frac{4}{3}\phi}\,
                  \left[\, 
                           \tilde{R} ~-~ \frac{1}{4}\,(F^{(2)}_{\mu\nu})^2 
                             ~+~ \left(\frac{2}{3}\right)^2\cdot 90
                                 \cdot (\partial\phi)^2\, 
                  \right]\\ 
             && -\frac{1}{4\cdot48}\,e^{\frac{22}{3}\phi}\sqrt{-\tilde{g}}\,
                 e^{-\frac{16}{3}}\,
                 \left[
                        4\,(H^{(3)})^2 ~+~ 
                        \left(\,
                                 F^{(4)}_{a_1\ldots a_4}\,+\,
                                 4\cdot A^{(1)}_{[a_1}H^{(3)}_{a_2a_3a_4]}\,
                        \right)^2
                 \right]
                 \nonumber\\
             && +\frac{9}{4\cdot144^2}\,\epsilon^{\alpha_1\ldots\alpha_4
                 \beta_1\ldots\beta_4\mu\nu}F^{(4)}_{\alpha_1\ldots\alpha_4}
                 F^{(4)}_{\beta_1\ldots\beta_4}B^{(2)}_{\mu\nu}~.
\end{eqnarray*}\\

\noindent
Note that the metric is still the rescaled one, i.e.\ 
$\tilde{g}_{\mu\nu}\,=\,e^{-2\phi}\,g_{\mu\nu}^{(st)}$.
A second Weyl scaling completes the reduction:\\
\begin{eqnarray*}
  {\cal L}_{(10)} 
             &=&  \frac{1}{4}\,e^{-4\phi}\sqrt{-g_{(st)}}\,
                  \left[\vbox{\vspace{3ex}}\right.\,
                          e^{2\phi}\,
                          \left\{\vbox{\vspace{2.5ex}}\right.\,
                                     R \,+\, \underbrace{
                                                         18\,\Delta\phi
                                                      }_{
                                                         36\,(\partial\phi)^2
                                                      } 
                                       \,-\, 72\,(\partial\phi)^2\,
                          \left.\vbox{\vspace{2.5ex}}\right\} 
                          ~-~ \frac{1}{4}\,e^{4\phi}\,(F^{(2)}_{\mu\nu})^2\\ 
                    &&    \hskip15ex
                          ~+~ \left(\frac{2}{3}\right)^2\cdot 90
                              \cdot e^{2\phi}\,(\partial\phi)^2\, 
                  \left.\vbox{\vspace{3ex}}\right]\\
             && -\frac{1}{4\cdot48}\,e^{2\phi}\cdot e^{-10\phi}\,
                 \sqrt{-g_{(st)}}\,
                 \left[\,
                        4\,e^{6\phi}\,(H^{(3)})^2 ~+~ e^{8\phi}\, 
                        \left(\,
                                 F^{(4)}_{a_1\ldots a_4}\,+\,
                                 4\cdot A^{(1)}_{[a_1}H^{(3)}_{a_2a_3a_4]}\,
                        \right)^2\,
                 \right]
                 \nonumber\\
             && +\frac{9}{4\cdot144^2}\,\epsilon^{\alpha_1\ldots\alpha_4
                 \beta_1\ldots\beta_4\mu\nu}F^{(4)}_{\alpha_1\ldots\alpha_4}
                 F^{(4)}_{\beta_1\ldots\beta_4}B^{(2)}_{\mu\nu}\\[4ex]
             &=& \phantom{-}\frac{1}{4}\,e^{-2\phi}\sqrt{-g_{(st)}}\,
                  \left[\vbox{\vspace{3ex}}\right.\,
                        R \,+\,4\cdot\,(\partial\phi)^2\, 
                          \,-\, \frac{1}{2\cdot 3!}\,(H^{(3)})^2
                  \left.\vbox{\vspace{3ex}}\right]\\
             &&~-~ \frac{1}{4\cdot 4}\,\sqrt{-g_{(st)}}\,(F^{(2)}_{\mu\nu})^2
               ~-~ \frac{1}{4\cdot48}\,\sqrt{-g_{(st)}}\, 
                        \left(\,
                                 F^{(4)}_{a_1\ldots a_4}\,+\,
                                 4\cdot A^{(1)}_{[a_1}H^{(3)}_{a_2a_3a_4]}\,
                        \right)^2
                 \nonumber\\
             &&~+~ \frac{9}{4\cdot144^2}\,
                   \epsilon^{\alpha_1\ldots\alpha_4
                   \beta_1\ldots\beta_4\mu\nu}F^{(4)}_{\alpha_1\ldots\alpha_4}
                   F^{(4)}_{\beta_1\ldots\beta_4}B^{(2)}_{\mu\nu}~.
\end{eqnarray*}\\
This is the action in the  string frame. 
It might be interesting to move from the string to the Einstein frame 
in a last step. This is done by doing the Weyl scaling\\ 
\begin{eqnarray*}
        g_{\mu\nu}^{st} &=& e^{\frac{\phi}{2}}\,g_{\mu\nu}^{E}
\end{eqnarray*}\\
and one obtains\\
\begin{eqnarray*}
  {\cal L}_{(10)} 
             &=& \phantom{-}\frac{1}{4}\,\sqrt{-g_{(E)}}\,
                  \left[\vbox{\vspace{3ex}}\right.\,
                        R ~-~ \frac{1}{2}\,(\partial\phi)^2\, 
                          ~-~ \frac{1}{2\cdot 3!}\,e^{-\phi}\,(H^{(3)})^2
                  \left.\vbox{\vspace{3ex}}\right]\\
             &&~-~ \frac{1}{4\cdot 4}\,e^{\frac{3}{2}\phi}
                   \sqrt{-g_{(E)}}\,(F^{(2)}_{\mu\nu})^2
               ~-~ \frac{1}{4\cdot48}\,e^{\frac{\phi}{2}}\sqrt{-g_{(E)}}\, 
                        \left(\,
                                 F^{(4)}_{a_1\ldots a_4}\,+\,
                                 4\cdot A^{(1)}_{[a_1}H^{(3)}_{a_2a_3a_4]}\,
                        \right)^2
                 \nonumber\\
             &&~+~ \frac{9}{4\cdot144^2}\,
                   \epsilon^{\alpha_1\ldots\alpha_4
                   \beta_1\ldots\beta_4\mu\nu}F^{(4)}_{\alpha_1\ldots\alpha_4}
                   F^{(4)}_{\beta_1\ldots\beta_4}B^{(2)}_{\mu\nu}~.
\end{eqnarray*}\\
Now we rescale all forms by a factor of two to absorb the terms 1/4 in front
of them. This is merely a convention but the one we prefer:\\ 
\begin{eqnarray}
  {\cal L}_{(10)} 
             &=& \phantom{-}\frac{1}{4}\,\sqrt{-g_{(E)}}\,
                  \left[\vbox{\vspace{3ex}}\right.\,
                        R ~-~ \frac{1}{2}\,(\partial\phi)^2\, 
                          ~-~ \frac{1}{3}\,e^{-\phi}\,(H^{(3)})^2
                  \left.\vbox{\vspace{3ex}}\right]\nonumber\\
             &&~-~ \frac{1}{4}\,e^{\frac{3}{2}\phi}
                   \sqrt{-g_{(E)}}\,(F^{(2)}_{\mu\nu})^2
               ~-~ \frac{1}{48}\,e^{\frac{\phi}{2}}\sqrt{-g_{(E)}}\, 
                        \left(\,
                                 F^{(4)}_{a_1\ldots a_4}\,+\,
                                 8\cdot A^{(1)}_{[a_1}H^{(3)}_{a_2a_3a_4]}\,
                        \right)^2
                 \nonumber\\
             &&~+~ \frac{3}{2\cdot(12)^3}\,
                   \epsilon^{\alpha_1\ldots\alpha_4
                   \beta_1\ldots\beta_4\mu\nu}F^{(4)}_{\alpha_1\ldots\alpha_4}
                   F^{(4)}_{\beta_1\ldots\beta_4}B^{(2)}_{\mu\nu}~.\label{IIASUGRA}
\end{eqnarray}

\vspace{1cm}

\subsection*{Homework:}
\begin{exercise}\label{excerKK} 
  Prove formula (\ref{KKReduction}) guided by the steps given in 
  appendix \ref{AppKaluzaKlein}. 
\end{exercise}
\begin{exercise} 
  Derive the equations of motion of the
  Lagrangian (\ref{IIASUGRA}) for all gauge fields. 
\end{exercise}

\vspace{2cm}

\subsection*{Acknowledgement:} We would like to thank Monika
             H{\'e}jjas for her support in finishing 
             the review and M.~J.~Duff for comments. The work 
             of A.M. was financially 
             supported by the 
             Deutsche Forschungsgemeinschaft.

\begin{appendix}
\vfill\eject

\section{Original Action of Cremmer-Julia-Scherk}
\label{App_CJS_Act}

Lagrangian:
\begin{eqnarray}
   {\cal L} &=& -\frac{1}{4\kappa^2}\,eR 
               -\frac{i}{2}\,e\bar{\psi}_\mu\Gamma^{\mu\nu\rho}
                 D_\nu\left(\frac{\omega\,+\,\hat{\omega}}{2}\right)\psi_\rho
                -\frac{1}{48}\,eF_{\mu\nu\rho\sigma}F^{\mu\nu\rho\sigma}
                 \nonumber\\
             && +\frac{\kappa}{192}\,e 
                 \left(
                   \bar{\psi}_\mu\Gamma^{\mu\nu\alpha\beta\gamma\delta}
                   \psi_\nu
                   +12\,\bar{\psi}^\alpha\Gamma^{\gamma\delta}\psi^\beta
                 \right)\, 
                 \left(\,
                          F_{\alpha\beta\gamma\delta}\,+\,
                          \hat{F}_{\alpha\beta\gamma\delta}\,
                 \right)\nonumber\\
             && +\frac{2\kappa}{144^2}\,\epsilon^{\alpha_1\ldots\alpha_4
                 \beta_1\ldots\beta_4\mu\nu\rho}F_{\alpha_1\ldots\alpha_4}
                 F_{\beta_1\ldots\beta_4}A_{\mu\nu\rho}~.
\end{eqnarray}\\
Supersymmetry: 
\begin{eqnarray}
  \delta_{Q}e_\mu{}^a &=& -i\,\kappa\,\bar\epsilon\,\Gamma^a\psi_\mu\nonumber\\
  \delta_{Q}A_{\mu\nu\rho} &=& \frac{3}{2}\,\bar\epsilon\,
                               \Gamma_{[\mu\nu}\psi_{\rho]}\nonumber\\
  \delta_{Q}\psi_\mu{} &=& \frac{1}{\kappa}\,D_\mu(\hat{\omega})\epsilon
                           +\frac{i}{144}
                           \left(
                                 \Gamma^{\alpha\beta\gamma\delta}{}_{\mu}
                                  -8\,\Gamma^{\beta\gamma\delta}
                                     \delta_\mu^\alpha
                           \right)\,\epsilon\,\hat{F}_{\alpha\beta\gamma\delta}
                           ~=~ \frac{1}{\kappa}\,\hat{D}_\mu\epsilon
\end{eqnarray}\\
The signature of the metric is $\eta_{ab}\,=\,(1,-1,\ldots,-1)$ and the 
$\Gamma$-matrices are in a purely imaginary representation of the 
Clifford algebra
\begin{eqnarray*}
         \{\,\Gamma^a,\,\Gamma^b\,\} &=& 2\,\eta^{ab}\,\unity_{32}~.
\end{eqnarray*} 
Abbreviations:
\begin{eqnarray*}
   D_\nu(\omega)\psi_\mu &=& \partial_\nu\psi_\mu\,+\,\frac{1}{4}\,
                             \omega_{\nu ab}\,\Gamma^{ab}\,\psi_\mu\\[1ex]
   F_{\mu\nu\rho\sigma} &=& 4\,\partial_{[\mu}\,A_{\nu\rho\sigma]}\\
   \hat{F}_{\mu\nu\rho\sigma} &=& F_{\mu\nu\rho\sigma} ~-~ 3\,\kappa\,
                                  \bar{\psi}_{[\mu}\Gamma_{\nu\rho}
                                  \psi_{\sigma]}\\[1ex]
   K_{\mu ab} &=& i\,\frac{\kappa^2}{4}\,
                  \left[\,
                           -\bar{\psi}_{\alpha}\Gamma_{\mu ab}{}^{\alpha\beta}
                            \psi_{\beta} ~+~ 2\,
                            \left(\,
                                     \bar{\psi}_{\mu}\Gamma_{b}\psi_{a}\,
                                  -\,\bar{\psi}_{\mu}\Gamma_{a}\psi_{b}\,
                                  +\,\bar{\psi}_{b}\Gamma_{\mu}\psi_{a}\,
                            \right)\,
                  \right]\quad {\rm (\,contorsion\,)}\\
   \omega_{\mu ab} &=& \underbrace{
                                     \omega_{\mu ab}^{(0)}
                                  }_{
                                     \hbox{\scriptsize{Christ}}
                                  } ~+~ K_{\mu ab}\\
   \hat{\omega}_{\mu ab} &=& \omega_{\mu ab} ~+~ i\,\frac{\kappa^2}{4}\,
                             \bar{\psi}_{\alpha}\,
                             \Gamma_{\mu ab}{}^{\alpha\beta}\,
                             \psi_{\beta}
\end{eqnarray*}

%
%

\vfill\eject

\section{Explicit Killing spinors}
\label{AppendixB}

\begin{eqnarray*}
   \varepsilon &=& (\,a_1+a_2+a_3, a_4, a_5, a_6, a_7, a_8, -a_2, -a_4, 
                      a_9, a_6, a_2+a_3+a_{10}, a_4, a_{11}, a_{12}, 
                      a_5+a_{13}+a_{14},\\
                && ~\;a_6, a_5-a_9+a_{13}, -a_6, a_{10}, a_4, a_{15}, -a_{12}, 
                    a_{14}, a_6, a_1, a_4, a_{13}, -a_6, a_{16}, a_8, -a_3, 
                    a_4\,)^T~\\[1ex]
   \vartheta &=& (\,0, a_{17}, 0, a_{18}, 0, a_{19}, 0, -a_{17}, 0, a_{18}, 
                   0, a_{17}, 0, a_{20}, 0, a_{18}, 0, a_{18}, \\
             && ~\;0, -a_{17},0, a_{20}, 0, -a_{18}, 0, -a_{17}, 0, a_{18}, 0, -a_{19}, 0, 
                   -a_{17}\,)^T
\end{eqnarray*}

\vfill\eject

\section{Behaviour  of the Ricci tensor under conformal transformations} 
\label{App_Ricci_conf_trafo}

\begin{eqnarray*}
     G_{\mu\nu} &=& \Omega\,\tilde{g}_{\mu\nu}
\end{eqnarray*}

\begin{eqnarray*}
    \Gamma_{\rho (\nu\lambda)} 
    &=& \frac{1}{2}\,\left(\,G_{\rho\nu,\lambda}+G_{\rho\lambda,\nu} -
        G_{\nu\lambda,\rho}\,\right)
    ~=~ \Omega\,\tilde{\Gamma}_{\rho (\nu\lambda)} \,+\,
        \frac{1}{2}\,\left(\,
                              \tilde{g}_{\rho\nu}\,\Omega_{,\lambda}
                             +\tilde{g}_{\rho\lambda}\Omega_{,\nu} 
                             -\tilde{g}_{\nu\lambda}\Omega_{,\rho}\,
                     \right)\\[2ex]
    \Gamma^{\rho}{}_{(\nu\lambda)} 
    &=& \tilde{\Gamma}^{\rho}{}_{(\nu\lambda)} \,+\,
        \frac{1}{2\Omega}\,\left(\,
                              \tilde{\delta}^{\rho}_{\nu}\,\Omega_{,\lambda}
                             +\tilde{\delta}^{\rho}_{\lambda}\Omega_{,\nu} 
                             -\tilde{g}^{\rho\kappa}\tilde{g}_{\nu\lambda}\Omega_{,\kappa}\,
                     \right)
    ~=~ \tilde{\Gamma}^{\rho}{}_{(\nu\lambda)} \,+\,S^{\rho}{}_{(\nu\lambda)}
\end{eqnarray*}

\begin{eqnarray*}
R^\alpha{}_{\beta\gamma\delta} 
  &=&  \frac{
               \partial (\tilde\Gamma+S)^\alpha{}_{(\beta\delta)}
            }
            {
               \partial x^\gamma
            }
        \,-\,  
        \frac{
               \partial (\tilde\Gamma+S)^\alpha{}_{(\beta\gamma)}
             }
             {
               \partial x^\delta
             }
    \,+\, (\tilde\Gamma+S)^\alpha{}_{(\eta\gamma)}
          (\tilde\Gamma+S)^\eta{}_{(\beta\delta)}
    \,-\, (\tilde\Gamma+S)^\alpha_{(\eta\delta)}
          (\tilde\Gamma+S)^\eta{}_{(\beta\gamma)}\\[2ex]
    &=& \tilde{R}^\alpha{}_{\beta\gamma\delta}
       \,+\,S^\alpha{}_{\beta\gamma\delta}
       \,+\,\tilde{\Gamma}^{\alpha}{}_{\eta\gamma}S^{\eta}{}_{\beta\delta}
       \,+\,S^{\alpha}{}_{\eta\gamma}\tilde{\Gamma}^{\eta}{}_{\beta\delta}
       \,-\,\tilde{\Gamma}^{\alpha}{}_{\eta\delta}S^{\eta}{}_{\beta\gamma}
       \,-\,S^{\alpha}{}_{\eta\delta}\tilde{\Gamma}^{\eta}{}_{\beta\gamma}
\end{eqnarray*}

\begin{eqnarray*}
R_{\beta\delta} 
    &=& \tilde{R}_{\beta\delta}
       \,+\,S^\alpha{}_{\beta\alpha\delta}
       \,+\,\tilde{\Gamma}^{\alpha}{}_{\eta\alpha}S^{\eta}{}_{\beta\delta}
       \,+\,S^{\alpha}{}_{\eta\alpha}\tilde{\Gamma}^{\eta}{}_{\beta\delta}
       \,-\,\tilde{\Gamma}^{\alpha}{}_{\eta\delta}S^{\eta}{}_{\beta\alpha}
       \,-\,S^{\alpha}{}_{\eta\delta}\tilde{\Gamma}^{\eta}{}_{\beta\alpha}
\end{eqnarray*}

\subsubsection*{1. Calculation of $S^\alpha{}_{\beta\alpha\delta}$}

\begin{eqnarray*}
     S^\alpha{}_{\beta\alpha\delta} 
       &=& \frac{\partial S^\alpha{}_{(\beta\delta)}}{\partial x^\alpha} 
        \,-\,  
           \frac{\partial
       S^\alpha{}_{(\beta\alpha)}}{\partial x^\delta}
       \,+\,  S^\alpha{}_{(\eta\alpha)}S^\eta{}_{(\beta\delta)}
        \,-\,
       S^\alpha_{(\eta\delta)}S^\eta{}_{(\beta\alpha)}\\
      &=& \frac{1}{2}\,
          \left\{\vbox{\vspace{3ex}}\right.\,
      \partial_\delta\left(\,\frac{\partial_\beta\Omega}{\Omega}\,\right)\,+\,
      \partial_\beta\left(\,\frac{\partial_\delta\Omega}{\Omega}\,\right)\,-\,
      \partial_\alpha\left(\,\frac{\tilde{\partial}^\alpha\Omega}{\Omega}\,\tilde{g}_{\beta\delta}\,\right)\\
       &&\,-\,D\cdot\partial_\delta\left(\,\frac{\partial_\beta\Omega}{\Omega}\,\right)-\partial_\delta\left(\,\frac{\partial_\beta\Omega}{\Omega}\,\right)+\partial_\delta\left(\,\frac{\partial_\beta\Omega}{\Omega}\,\right)\,
         \left.\vbox{\vspace{3ex}} \right\}
       \,+\,\frac{D}{4}\,\left\{\,2\,\frac{\partial_\beta\Omega\partial_\delta\Omega}{\Omega^2}\,-\,\frac{\partial_\eta\Omega\tilde\partial^\eta\Omega}{\Omega^2}\,\tilde{g}_{\beta\delta}\,\right\}\\
    &&-\,\frac{1}{4}\,\left\{\vbox{\vspace{3ex}}\right.\,
          3\,\frac{\partial_\beta\Omega\partial_\delta\Omega}{\Omega^2}
         \,+\,\frac{\partial_\beta\Omega\partial_\delta\Omega}{\Omega^2}
         \,-\,\frac{\partial_\eta\Omega\tilde\partial^\eta\Omega}{\Omega^2}\,\tilde{g}_{\beta\delta}
         \,+\,D\cdot\,
            \frac{\partial_\beta\Omega\partial_\delta\Omega}{\Omega^2}
         \,-\,\frac{\partial_\beta\Omega\partial_\delta\Omega}{\Omega^2}
         \,-\,\frac{\partial_\beta\Omega\partial_\delta\Omega}{\Omega^2}
         \,-\,\frac{\partial_\eta\Omega\tilde\partial^\eta\Omega}{\Omega^2}\,\tilde{g}_{\beta\delta}\,
         \left.\vbox{\vspace{3ex}}\right\}\\[2ex]
    &=&  \frac{1}{2}\,
          \left\{\vbox{\vspace{3ex}}\right.\,
      (1-D)\,\partial_\delta\left(\,\frac{\partial_\beta\Omega}{\Omega}\,\right)\,+\,
      \partial_\beta\left(\,\frac{\partial_\delta\Omega}{\Omega}\,\right)\,-\,
      \partial_\alpha\left(\,\frac{\tilde{\partial}^\alpha\Omega}{\Omega}\,\tilde{g}_{\beta\delta}\,\right)\,
         \left.\vbox{\vspace{3ex}} \right\}
       \,+\,\frac{D-2}{4}\,\left(\,
           \frac{\partial_\beta\Omega\partial_\delta\Omega}{\Omega^2}
         \,-\,\frac{\partial_\eta\Omega\tilde\partial^\eta\Omega}{\Omega^2}\,\tilde{g}_{\beta\delta}\,\right)
\end{eqnarray*}

\subsubsection*{2. Calculation of cross terms:}

\begin{eqnarray*}
  \tilde{\Gamma}^\alpha{}_{\eta\alpha}\,S^\eta{}_{\beta\delta}
  &=& \tilde{\Gamma}^{\alpha}{}_{\eta\alpha}\,\frac{1}{2\Omega}\,
      \left(\,
             \partial_\beta\Omega\,\delta^\eta_\delta \,+\, 
             \partial_\delta\Omega\,\delta^\eta_\beta \,-\,
             \tilde{\partial}^\eta\Omega\,\tilde{g}_{\beta\delta}\,
      \right)\\[2ex]
  &=& \frac{1}{2\,\Omega}\,
      \left(\,
             \tilde{\Gamma}^\alpha{}_{\delta\alpha}\,\partial_\beta\Omega
             \,+\,\tilde{\Gamma}^\alpha{}_{\beta\alpha}\,\partial_\delta\Omega
             \,-\,\tilde{\Gamma}^\alpha{}_{\eta\alpha}\,
             \tilde{\partial}^\eta\Omega\,\tilde{g}_{\beta\delta}\,
      \right)\\[3ex]
  S^\alpha{}_{\eta\alpha}\,\tilde{\Gamma}^\eta{}_{\beta\delta}
  &=& \frac{1}{2\Omega}\,
      \left(\,
             \partial_\eta\Omega\,\delta^\alpha_\alpha \,+\, 
             \partial_\alpha\Omega\,\delta^\alpha_\eta \,-\,
             \tilde{\partial}^\alpha\Omega\,\tilde{g}_{\alpha\eta}\,
      \right)\,\tilde{\Gamma}^{\eta}{}_{\beta\delta}\,\\[2ex]
  &=& \frac{D}{2\,\Omega}\,\partial_\eta\Omega\,
      \tilde{\Gamma}^\eta{}_{\beta\delta}\\[3ex]
  -\,\tilde{\Gamma}^\alpha{}_{\eta\delta}\,S^\eta{}_{\beta\alpha}
  &=& -\,\tilde{\Gamma}^{\alpha}{}_{\eta\delta}\,\frac{1}{2\Omega}\,
      \left(\,
             \partial_\beta\Omega\,\delta^\eta_\alpha \,+\, 
             \partial_\alpha\Omega\,\delta^\eta_\beta \,-\,
             \tilde{\partial}^\eta\Omega\,\tilde{g}_{\alpha\beta}\,
      \right)\\[2ex]
  &=& \frac{1}{2\Omega}\,
      \left(\,
              -\,\partial_\beta\Omega\,\tilde{\Gamma}^\alpha_{\delta\alpha}
            \,-\,\partial_\alpha\Omega\,\tilde{\Gamma}^\alpha_{\beta\delta}
            \,+\,\tilde{\partial}^\eta\Omega\,\tilde{\Gamma}^\alpha_{\eta\delta}\,
                 \tilde{g}_{\alpha\beta}\,
      \right)\\[3ex]
  -\,S^\alpha{}_{\eta\delta}\,\tilde{\Gamma}^\eta{}_{\beta\alpha}
  &=& -\,\frac{1}{2\Omega}\,
      \left(\,
             \partial_\eta\Omega\,\delta^\alpha_\delta \,+\, 
             \partial_\delta\Omega\,\delta^\alpha_\eta \,-\,
             \tilde{\partial}^\alpha\Omega\,\tilde{g}_{\eta\delta}\,
      \right)\,\tilde{\Gamma}^{\eta}{}_{\beta\alpha}\\[2ex]
  &=& \frac{1}{2\Omega}\,
      \left(\,
              -\,\partial_\eta\Omega\,\tilde{\Gamma}^\eta_{\beta\delta}
            \,-\,\partial_\delta\Omega\,\tilde{\Gamma}^\alpha_{\beta\alpha}
            \,+\,\tilde{\partial}^\alpha\Omega\,\tilde{\Gamma}^\eta_{\beta\alpha}\,
                 \tilde{g}_{\eta\delta}\,
      \right)\\[3ex]
\end{eqnarray*}
And the sum of all the four terms above is 
\begin{eqnarray*}
    \tilde{\Gamma}^\alpha{}_{\eta\alpha}\,S^\eta{}_{\beta\delta}\,+\,\ldots\,
   -\,S^\alpha{}_{\eta\delta}\,\tilde{\Gamma}^\eta{}_{\beta\alpha}
   &=& \frac{1}{2\Omega}\,
       \left(\vbox{\vspace{2.5ex}}\right.\,
               (D-2)\,\partial_\eta\Omega\,\tilde{\Gamma}^\eta_{\beta\delta}
           \,+\,\tilde{\partial}^\eta\Omega\,
             \left[\,
                      {\tilde\Gamma}^\alpha_{(\eta\delta)}\,
                      \tilde{g}_{\alpha\beta}\,+\,
                      {\tilde\Gamma}^\alpha_{(\eta\beta)}\,
                      \tilde{g}_{\alpha\delta}\,-\,
                     {\tilde\Gamma}^\alpha_{(\eta\alpha)}\,
                      \tilde{g}_{\beta\delta}\,
              \right]\,
       \left.\vbox{\vspace{2.5ex}}\right)
\end{eqnarray*}
Putting the previous steps together, one obtains the formula
\begin{eqnarray*}
  R_{\beta\delta} 
  &=& {\tilde R}_{\beta\delta} 
      \,+\,\frac{1}{2}\,
           \left\{\,
                     (1-D)\,\partial_\delta
                     \left(\,
                             \frac{\partial_\beta\Omega}{\Omega}\,
                     \right) \,+\, 
                     \partial_\beta
                     \left(\,
                             \frac{\partial_\delta\Omega}{\Omega}\,
                     \right) \,-\, 
                     \partial_\alpha
                     \left(\,
                             \frac{\tilde{\partial}^\alpha\Omega}{\Omega}\,
                    \tilde{g}_{\beta\delta}\,\right)\,
            \right\}\\
 &&\,+\,\frac{D-2}{4}\,
        \left(\,
                 \frac{\partial_\beta\Omega\,\partial_\delta\Omega}{\Omega^2}
                 \,-\,
                 \frac{\partial_\eta\Omega\,\tilde{\partial}^\eta\Omega}
                      {\Omega^2}\,\tilde{g}_{\beta\delta}\,
        \right)\\
 &&\,+\,\frac{1}{2\Omega}\,
        \left(\,
              (D-2)\,\partial_\eta\Omega\,
              \tilde{\Gamma}^\eta_{(\beta\delta)}
              \,+\,{\tilde \partial}^\eta\Omega\,
              \left[\,
                      {\tilde\Gamma}^\alpha_{(\eta\delta)}\,
                      \tilde{g}_{\alpha\beta}\,+\,
                      {\tilde\Gamma}^\alpha_{(\eta\beta)}\,
                      \tilde{g}_{\alpha\delta}\,-\,
                     {\tilde\Gamma}^\alpha_{(\eta\alpha)}\,
                      \tilde{g}_{\beta\delta}\,
              \right]\,
        \right)
\end{eqnarray*}
or more covariantly
\begin{eqnarray}
  R_{\beta\delta} 
  &=& {\tilde R}_{\beta\delta} 
      \,+\,\frac{1}{2}\,
           \left\{\,
                     (1-D)\,\tilde\nabla_\delta
                     \left(\,
                             \frac{\partial_\beta\Omega}{\Omega}\,
                     \right) \,+\, 
                     \tilde\nabla_\beta
                     \left(\,
                             \frac{\partial_\delta\Omega}{\Omega}\,
                     \right) \,-\, 
                     \tilde\nabla_\alpha
                     \left(\,
                             \frac{\tilde{\partial}^\alpha\Omega}{\Omega}\,
                     \right)\,\tilde{g}_{\beta\delta}\,
            \right\}\nonumber\\
 &&\,+\,\frac{D-2}{4}\,
        \left(\,
                 \frac{\partial_\beta\Omega\,\partial_\delta\Omega}{\Omega^2}
                 \,-\,
                 \frac{\partial_\eta\Omega\,\tilde{\partial}^\eta\Omega}
                      {\Omega^2}\,\tilde{g}_{\beta\delta}\,
        \right)\label{RescaledRicciT}
\end{eqnarray}\\

\subsubsection*{Check:}

\begin{eqnarray}
\label{RescaledRicciS}
  R &=& \frac{1}{\Omega}\,\left\{\,
      {\tilde R} 
      \,-\,(D-1)\,\tilde\Delta(\ln\Omega)\,\,-\, 
        \frac{(D-1)(D-2)}{4}\,
        \partial_\eta(\ln\Omega)\,\tilde{\partial}^\eta(\ln\Omega)\,
        \right\}
\end{eqnarray}

\vfill\eject

\section{Some Details of Kaluza-Klein Reductions}
\label{AppKaluzaKlein}

\subsection*{Basic Kaluza-Klein Formula:}

In order to compute the Ricci tensor one has to compute the 
connection components first. Here the split of the eleven dimensions 
into ten plus one must be made explicit.  
With $M\,=\,(\mu,11)$ and $N\,=\,(\nu,11)$ the metric and its inverse 
read\\
\begin{eqnarray*}
   {g}^{\mbox{\tiny (11)}}_{MN} 
                     &=& \left(\begin{array}{cc}
                            \tilde{g}_{\mu\nu}
                      ~+~ A_\mu\,A_\nu 
                       & A_\mu\\
                         A_\nu 
                       & 1
                          \end{array}\right)\hspace{3ex}
   {g}_{\mbox{\tiny (11)}}^{MN}
                     ~=~ \left(\begin{array}{cc}
                            \tilde{g}^{\mu\nu} 
                       & -\,A^\mu\\
                         -\,A^\nu 
                       & 1\,+\,A_\rho A^\rho
                          \end{array}\right)~.\hspace{3ex}
\end{eqnarray*}\\
and the 11d Christoffel symbols defined as in 
eq.~(\ref{ChristoffelSymbolLower}) split into\\
\begin{eqnarray*}
   \Gamma_{\rho(\mu\nu)} &=& \tilde{\Gamma}_{\rho(\mu\nu)} ~+~
    \frac{1}{2}\,\left(\,
                         A_\mu\,F_{\nu\rho}\,
                      +\,A_\rho\,B_{\nu\mu}\,
                      +\,A_\nu\,F_{\mu\rho}\,
                 \right)\\[1ex]
   \Gamma_{11(\mu\nu)} &=& \frac{1}{2}\,B_{\mu\nu}\\[1ex]
   \Gamma_{\mu (11\nu)} &=& -\,\frac{1}{2}\,F_{\mu\nu}~.
\end{eqnarray*}\\
Here we introduced for shortness $B_{\mu\nu}\,=\,\partial_\mu A_\nu\,+\,\partial_\nu A_\mu$ and $F_{\mu\nu}\,=\,\partial_\mu A_\nu\,-\,\partial_\nu A_\mu$.
All other Christoffel symbols vanish. Only the computation of the
component $\Gamma_{\rho(\mu\nu)}$ contains a tiny difficulty. One must
keep in mind that the metric one has to use in 
eq.~(\ref{ChristoffelSymbolLower}) is
$g_{\mu\nu}\,=\,\tilde{g}_{\mu\nu}\,+\,A_\mu\,A_\nu$. This produces
the additional terms.\\

\noindent
The Christoffel symbols with upper indices defined in eq.~(\ref{ChristoffelSymbolUpper}) read
\begin{eqnarray*}
  \Gamma^{\mu}{}_{(\nu\lambda)}\hspace{2ex} 
  &=&\tilde\Gamma^{\mu}{}_{(\nu\lambda)} \,-\,\frac{1}{2}\,
     \left(\,
               F^\mu{}_\nu\,A_\lambda+F^{\mu}{}_\lambda\,A_\nu\,
     \right) \\
  \Gamma^{\mu}{}_{(11\lambda)}\hspace{1.5ex}  
  &=& -\,\frac{1}{2}\,F^\mu{}_\lambda\\
  \Gamma^{\mu}{}_{(11\,11)} \hspace{.75ex} 
  &=& 0 \\
  \Gamma^{11}{}_{(\nu\lambda)} \hspace{1.5ex} 
  &=& \frac{1}{2}\,B_{\nu\lambda}\,-\,\frac{1}{2}\,A^\mu\,
      \left(\,
               F_\nu{}_\mu\,A_\lambda+F_{\lambda}{}_\mu\,A_\nu\,
      \right)\,-\,A^\mu\,\tilde\Gamma_{\mu(\nu\lambda)} \\
  \Gamma^{11}{}_{(11\mu)}\hspace{1ex}  
  &=& \frac{1}{2}\,A^\nu\,F_\nu{}_\mu \\
  \Gamma^{11}{}_{(11\,11)} &=& 0 
\end{eqnarray*}\\
and for the Ricci tensor defined by eq.~(\ref{RicciTensor}) one obtains\\
\begin{eqnarray}
  R_{\mu\nu}~~ &=& \tilde{R}_{\mu\nu}\,
              +\,\frac{1}{2}\,F_\mu{}^\rho F_{\rho\nu}\,
              +\,\frac{1}{4}\,A_\mu\,A_\nu\,F_{\lambda\rho}
              F^{\lambda\rho}\,
              +\,\frac{1}{2}\,
                 \left(\,
                           A_\nu\,\nabla_\rho\,F_{\mu}{}^\rho\,
                        +\,A_\mu\,\nabla_\rho\,F_{\nu}{}^\rho\,
                 \right)\label{RIJ}\\
  R_{11\mu}~ &=& \frac{1}{2}\,\nabla_\nu\,F{}_{\mu}{}^\nu\,
              +\,\frac{1}{4}\,A_\mu\,F_{\lambda\rho}F^{\lambda\rho}\\
  R_{11\,11} &=& \frac{1}{4}\,F_{\lambda\rho}F^{\lambda\rho}
\end{eqnarray}\\
The last two components are obtained without problems. The
determination of the first one is slightly more involved and  
we fill some of the steps in here. According 
to eq.~(\ref{RicciTensor}) we must compute\\ 
\begin{eqnarray}
   R_{ij} &=& \underbrace{\frac{\partial \Gamma^{k}{}_{(ij)}}{\partial x^k}}_{(1)}\,-\,
              \underbrace{\frac{\partial \Gamma^{M}{}_{(iM)}}{\partial x^j}}_{(2)}\,+
              \underbrace{\Gamma^{M}{}_{(NM)}\Gamma^{N}{}_{(ij)}}_{(3)}\,-
              \underbrace{\Gamma^{M}{}_{(Nj)}\Gamma^{N}{}_{(iM)}}_{(4)}~.
\end{eqnarray}\\
Now we investigate each of the four terms individually.\\
\begin{eqnarray}
    (1) &=& \partial_k\left[\,\Gamma^k{}_{(ij)}\,\right]
        ~=~ \partial_k\left[\,\tilde\Gamma^k{}_{(ij)}\,
             -\,\frac{1}{2}\,\left(\,F^k{}_i\,A_j\,+\,F^k{}_j\,A_i\,\right)\,
            \right]\hspace{12ex}
\end{eqnarray}\\
\begin{eqnarray}
    (2) &=& \partial_j\left[\,\Gamma^N{}_{(iN)}\,\right]
        ~=~ \partial_j\left[\,
              \Gamma^{11}{}_{(i\,11)}\,+\,\Gamma^{k}{}_{(ik)}\,
            \right]\nonumber\\[2ex]
        &=&\partial_j\left[\,
              \frac{1}{2}\,A^l\,F_{li}\,-\,\frac{1}{2}\,A^k\,F^k{}_{i}\,+\,
              \tilde{\Gamma}^k{}_{(ik)}\,
            \right] ~=~ \partial_j\left[\,\tilde{\Gamma}^k{}_{(ik)}\,
            \right]\hspace{14ex}
\end{eqnarray}\\
\begin{eqnarray}
    (3) &=& \Gamma^{M}{}_{(NM)}\Gamma^{N}{}_{(ij)}\nonumber\\ 
        &=&
        \Gamma^{11}{}_{(N\,11)}\Gamma^{N}{}_{(ij)}\,+\,\Gamma^{k}{}_{(N\,k)}\Gamma^{N}{}_{(ij)}\nonumber\\
        &=&\underbrace{\Gamma^{11}{}_{(11\,11)}}_{0}\Gamma^{11}{}_{(ij)}\,+\,\Gamma^{11}{}_{(k\,11)}\Gamma^{k}{}_{(ij)}\,+\,\underbrace{\Gamma^{k}{}_{(11\,k)}}_{0}\Gamma^{11}{}_{(ij)}\,+\,\Gamma^{k}{}_{(l\,k)}\Gamma^{l}{}_{(ij)}\nonumber\\
 &=&
        \frac{1}{2}\,A^l\,F_{lk}\,\left[\,\tilde\Gamma^k{}_{(ij)}\,-\,
        \frac{1}{2}\,\left(\,F^k{}_i\,A_j\,+\,F^k{}_j\,A_i\,\right)\,
        \right]\nonumber\\
 &&~+\,\left(\,-\,\frac{1}{2}\,F^k{}_l\,A_k\,+\,\tilde{\Gamma}^k{}_{(lk)}\,\right)\,\left[\,-\frac{1}{2}\,\left(\,F^l{}_i\,A_j\,+\,F^l{}_j\,A_i\,\right)\,+\,\tilde{\Gamma}^l{}_{(ij)}\,\right]
\end{eqnarray}\\
\begin{eqnarray}
 (4) &=& \Gamma^N{}_{(Mj)}\,\Gamma^M{}_{(iN)}\nonumber\\
     &=&
     \Gamma^{11}{}_{(Mj)}\,\Gamma^M{}_{(i\,11)}\,+\,\Gamma^k{}_{(Mj)}\,\Gamma^M{}_{(ik)}\nonumber\\
     &=&
     \Gamma^{11}{}_{(11\,j)}\,\Gamma^{11}{}_{(i\,11)}\,+\,\Gamma^{11}{}_{(lj)}\,\Gamma^l{}_{(i\,11)}\,+\,\Gamma^k{}_{(11j)}\,\Gamma^{11}{}_{(ik)}\,+\,\Gamma^k{}_{(lj)}\,\Gamma^l{}_{(ik)}\nonumber\\
     &=& \frac{1}{2}\,A^l\,F_{lj}\,\frac{1}{2}\,A^k\,F_{ki}\,+\,
         \left[\,
                  \frac{1}{2}\,B_{kj}\,-\,\frac{1}{2}\,A^l\,
                  \left(\,
                           F_{kl}\,A_j\,+\,F_{jl}\,A_k\,
                  \right)\,-\,A^l\,\tilde{\Gamma}_{l(kj)}\,
         \right]\,\left[\,-\,\frac{1}{2}\,F^k{}_i\,\right]\nonumber\\
    &&~~-\,\frac{1}{2}\,F^k{}_j\,\left[\,
                  \frac{1}{2}\,B_{ik}\,-\,\frac{1}{2}\,A^l\,
                  \left(\,
                           F_{il}\,A_k\,+\,F_{kl}\,A_i\,
                  \right)\,-\,A^l\,\tilde{\Gamma}_{l(ik)}\,
         \right]\nonumber\\
    &&~~+\,\left[\,
                    \tilde{\Gamma}^k{}_{(lj)}\,
                    -\,\frac{1}{2}\,
                    \left(\,
                             F^k{}_l\,A_j\,+\,F^k{}_j\,A_l\,
                    \right)\,
           \right]
           \left[\,
                    \tilde{\Gamma}^l{}_{(ik)}\,
                    -\,\frac{1}{2}\,
                    \left(\,
                             F^l{}_i\,A_k\,+\,F^l{}_k\,A_i\,
                    \right)\,
           \right]
\end{eqnarray}\\
Adding all terms together one arrives at\\ 
\begin{eqnarray}
   R_{ij} &=&
   \tilde{R}_{ij}\,-\,\frac{1}{2}\,\nabla_k\left(\,F^k{}_i\,A_j\,\right)\,-\,\frac{1}{2}\,\nabla_k\left(\,F^k{}_j\,A_i\,\right)\nonumber\\
    &&\hspace{5ex}
   +\,\frac{1}{4}\,F^k{}_i\,B_{kj}\,-\,\frac{1}{2}\,\tilde{\Gamma}^l{}_{(kj)}\,A_l\,F^k{}_i\nonumber\\
    &&\hspace{5ex}
   +\,\frac{1}{4}\,F^k{}_j\,B_{ki}\,-\,\frac{1}{2}\,\tilde{\Gamma}^l{}_{(ik)}\,A_l\,F^k{}_j\nonumber\\
   &&\hspace{5ex} +\,\frac{1}{4}\,F_{lk}\,F^{lk}\,A_i\,A_j~.\label{KKRIJ}
\end{eqnarray}\\
Using now
\begin{eqnarray}
   \nabla_k\left(\,F^k{}_i\,A_j\,\right) 
   &=& \left(\,\nabla_k\,F^k{}_i\,\right)\,A_j\,+\,F^k{}_i
   \nabla_k\,A_j
   ~=~  \left(\,\nabla_k\,F^k{}_i\,\right)\,A_j\,+\,F^k{}_i\,
   \left(\,\partial_k\,A_j\,-\,\tilde{\Gamma}^l{}_{(kj)}\,A_l\,\right)
   \nonumber\\
   &=&  \left(\,\nabla_k\,F^k{}_i\,\right)\,A_j\,+\,F^k{}_i\,
        \left(\,\partial_{[k}\,A_{j]}\,+\,\partial_{(k}\,A_{j)}\,\right)\,-\,\tilde{\Gamma}^l{}_{(kj)}\,A_l\,F^k{}_i\nonumber\\
   &=&  \left(\,\nabla_k\,F^k{}_i\,\right)\,A_j\,+\,\frac{1}{2}\,F^k{}_i\,
        \left(\,F_{kj}\,+\,B_{kj}\,\right)\,-\,\tilde{\Gamma}^l{}_{(kj)}\,A_l\,F^k{}_i
\end{eqnarray}\\
to rewrite eq.~(\ref{KKRIJ}) one obtains exactly the expression for
$R_{ij}$ given in (\ref{RIJ}).\\

\noindent
Performing the last step and computing the Ricci scalar one 
obtains\\ 
\begin{eqnarray*}
  R_{(11)} &=& g^{MN}\,R_{MN} ~=~
  g^{11\,11}\,R_{11\,11}\,+\,2\,g^{11\mu}\,R_{11\mu}\,+\,g^{\mu\nu}\,R_{\mu\nu}\nonumber\\
 &=& \frac{1}{4}\,(1+A_\tau A^\tau)\,F_{\lambda\rho}F^{\lambda\rho}
 ~-~ 2\,A^\mu\,
     \left(\,
     \frac{1}{2}\,\nabla_\nu\,F_\mu{}^\nu\,+\frac{1}{4}\,A_\mu F_{\lambda\rho}F^{\lambda\rho}\,
     \right)\\
 &+& g^{\mu\nu}\,
     \left(\vbox{\vspace{2.5ex}}\right.\,   
             \tilde{R}_{\mu\nu}\,
              +\,\frac{1}{2}\,F_\mu{}^\rho F_{\rho\nu}\,
              +\,\frac{1}{4}\,A_\mu\,A_\nu\,F_{\lambda\rho}
              F^{\lambda\rho}\,
              +\,\frac{1}{2}\,
                 \left(\,
                           A_\nu\,\nabla_\rho\,F_{\mu}{}^\rho\,
                        +\,A_\mu\,\nabla_\rho\,F_{\nu}{}^\rho\,
                 \right)\,
     \left.\vbox{\vspace{2.5ex}}\right)\nonumber
\end{eqnarray*}
\begin{eqnarray}
   R_{(11)} &=&  R_{(10)} ~-~ \frac{1}{4}\,F_{\mu\nu}F^{\mu\nu}~.
\end{eqnarray}

\subsection*{The 3. Term}
\label{Term3}

\begin{eqnarray*}
   {\it 3.Term} &=& 2\cdot
                    \left(
                           \begin{array}{c}
                              4\\ 2
                           \end{array}
                    \right)\cdot
                    g^{11\,b_1}\,
                    g^{a_2\,11}\,g^{a_3b_3}\,g^{a_4b_4}\,
                    G_{11\,a_2a_3a_4}G_{b_1\,11\,b_3b_4}\\
                &=& 2\cdot
                    \left(
                           \begin{array}{c}
                              4\\ 2
                           \end{array}
                    \right)\cdot
                    (-A^{b_1})\,
                    (-A^{a_2})\,g^{a_3b_3}\,g^{a_4b_4}\,
                    (-H_{a_2a_3a_4})\,H_{b_1b_3b_4}\\
                &=& -\,12\,A^{b_1}\,A^{a_2}\,H_{a_2}{}^{b_3b_4}\,H_{b_1b_2b_3}
\end{eqnarray*}
There are 6 possibilities to choose two metric factors out of the four 
$g^{\alpha_i\beta_i}$. And then an additional factor of 2 due to the 
two distributions of the two $11$'s on them.

\subsection*{A tensor identity}
\label{AppTensorIdentity}

\begin{eqnarray*}
       A_{[a_1}H_{a_2a_3a_4]}\,A^{[a_1}H^{a_2a_3a_4]}
  &=& \frac{1}{4}\,
      \left(\,
                   A_{a_1}H_{a_2a_3a_4}\,-\,A_{a_2}H_{a_1a_3a_4}
              \,-\,A_{a_3}H_{a_2a_1a_4}\,-\,A_{a_4}H_{a_2a_3a_1}
      \right)\\
  &&  \frac{1}{4}\,
      \left(\,
                   A^{a_1}H^{a_2a_3a_4}\,-\,A^{a_2}H^{a_1a_3a_4}
              \,-\,A^{a_3}H^{a_2a_1a_4}\,-\,A^{a_4}H^{a_2a_3a_1}
      \right)\\[1ex]
  &=& \frac{1}{4^2}
      \left[\vbox{\vspace{2.5ex}}\right.\;
            (A^{(1)})^2(H^{(3)})^2 
            ~-~ 2\,A_{a_1}\,A^{a_2}\,H_{a_2a_3a_4}\,H^{a_1a_3a_4}\\[-1ex]
  &&\hspace{5.5ex} 
            (A^{(1)})^2(H^{(3)})^2 
            ~-~ 2\,A_{a_1}\,A^{a_3}\,H_{a_2a_3a_4}\,H^{a_2a_1a_4}\\
  &&\hspace{5.5ex} 
            (A^{(1)})^2(H^{(3)})^2 
            ~-~ 2\,A_{a_1}\,A^{a_4}\,H_{a_2a_3a_4}\,H^{a_2a_3a_1}\\
  &&\hspace{5.5ex} 
            (A^{(1)})^2(H^{(3)})^2 
            ~+~ 2\,A_{a_2}\,A^{a_3}\,H_{a_1a_3a_4}\,H^{a_2a_1a_4}\\
  &&\hspace{5.5ex} 
            \phantom{(A^{(1)})^2(H^{(3)})^2} 
            ~+~ 2\,A_{a_2}\,A^{a_4}\,H_{a_1a_3a_4}\,H^{a_2a_3a_1}\\
  &&\hspace{5.5ex} 
            \phantom{(A^{(1)})^2(H^{(3)})^2} 
            ~+~ 2\,A_{a_3}\,A^{a_4}\,H_{a_2a_1a_4}\,H^{a_2a_3a_1}\;
      \left.\vbox{\vspace{2.5ex}}\right]\\
  &=& \frac{1}{4}\;(A^{(1)})^2(H^{(3)})^2
      ~-~\frac{12}{4^2}\;A_{a}\,A^{b}\,H_{bcd}\,H^{acd}
\end{eqnarray*}

\end{appendix}

\vfill\eject

\bibliographystyle{utphys}
\bibliography{mtheory}

\end{document}